\documentstyle[aps,harvard,rmp,twocolumn,epsf]{revtex}

\begin{document}

\title{The Statistical Theory of Quantum Dots}

\author{Y. Alhassid}

\address{ Center for Theoretical Physics, Sloane Physics Laboratory,
     Yale University, New Haven, Connecticut 06520, USA 
\vspace*{4 mm}\\
\parbox{5.5 in}{\rm A quantum dot is a sub-micron-scale conducting device containing up to several
thousand electrons. Transport through a quantum dot at low temperatures is a quantum-coherent process. This review focuses on dots in which the electron's dynamics are chaotic or diffusive, giving rise to statistical properties that reflect the interplay between one-body chaos, quantum interference, and
electron-electron  interactions. The conductance through such dots displays  mesoscopic fluctuations as a function of gate voltage, magnetic field,  and shape deformation.  The techniques used to describe these fluctuations include semiclassical methods, random-matrix theory, and the
supersymmetric  nonlinear $\sigma$ model.
 In open dots, the approximation of noninteracting quasiparticles is justified, and electron-electron interactions contribute indirectly through their effect on the dephasing time at finite temperature. In almost-closed dots, where conductance occurs by tunneling, the charge on the dot is quantized, and electron-electron interactions play an important role.  Transport is dominated by Coulomb blockade, leading to peaks in the conductance that at low temperatures provide information on the dot's ground-state properties.   Several statistical signatures of electron-electron interactions have been identified, most notably in the dot's addition spectrum. The dot's spin, determined partly by exchange interactions, can also influence the fluctuation properties of the conductance. Other mesoscopic phenomena in quantum dots that are affected by the charging energy include the fluctuations of the cotunneling conductance and mesoscopic Coulomb blockade. 
[{\em Published in Rev. Mod. Phys. {\bf 72}, 895 (2000)}]}}

\maketitle

\tableofcontents

\section{Introduction}
\label{intro}

 Recent advances in materials science have made possible the
  fabrication of small conducting devices known as quantum dots, where up to
several  thousand electrons are confined to a region whose linear size is
about  $0.1-1\;\mu$m (Kastner, 1992). Quantum dots are typically made by
forming  a two-dimensional electron gas in the interface region of a semiconductor heterostructure  and applying  an electrostatic potential to  metal gates to
further  confine the electrons to a small region (``dot'') in the interface
plane.  Because the electronic motion is restricted in all three dimensions, a
quantum  dot is sometimes referred to as a zero-dimensional system. The
transport  properties of a quantum dot can be measured by coupling it to leads
and  passing current through the dot. The electron's phase is preserved over
distances  that are large compared with the size of the system, giving rise to
new  phenomena not observed in macroscopic conductors. As the name suggests, conductance through a quantum dot is characterized by quantum coherence.

  Quantum dots belong to a larger class of systems, termed {\em mesoscopic}
by  van Kampen (1981), which are intermediate between  microscopic
systems,  such as nuclei and atoms, and macroscopic bulk matter (Akkermans~{\em et~al.}, 1995).  A system is called  mesoscopic when the electron's phase coherence length $L_\phi$ (the typical distance the electron travels without losing phase coherence) is larger than or comparable to the system's size $L$.  Phase coherence is affected by the coupling of the electron to its environment, and phase-breaking processes involve a change in the state of the environment. In most cases phase coherence is lost in inelastic scatterings, e.g., with other electrons or phonons, but spin-flip scattering from magnetic impurities can also contribute to phase decoherence. Elastic scatterings of the electron, e.g., from impurities, usually preserve phase coherence and are characterized by the elastic mean free path $l$.  $L_\phi$ increases rapidly with decreasing temperature, and for $L \sim 1\;\mu$m, an open system typically becomes mesoscopic below $\sim 100$ mK.  In a mesoscopic sample, the description of transport in terms of local conductivity breaks down, and the whole sample must be treated as a single, coherent entity.

The field of
mesoscopic  physics originated in the study of disordered systems in which the electron's motion is {\em diffusive}, i.e., $l$  is small relative to $L$.  In the late 1980s it became possible to produce high-mobility semiconductor
microstructures that were sufficiently small and free of impurities to ensure that the mean free path $l$ exceeds the system's size $L$. Such devices are
termed  {\em ballistic}. Transport in a ballistic quantum dot is dominated by electronic scattering not from impurities, but from the structure's boundaries. Most experimental research on quantum dots is focused on ballistic dots.

 The coupling between a quantum dot and its leads can be experimentally controlled. In an {\em open dot}, the coupling is strong and the movement of electrons
across  the dot-lead junctions is classically allowed. But when the point
contacts  are pinched off, effective barriers are formed and conductance
occurs  only by tunneling. In these {\em almost-isolated} or {\em closed quantum dots},  the charge on the dot is  quantized, and the dot's low-lying energy levels are discrete, with widths smaller than their spacing. Closed dots have  been called ``artificial atoms'' (Kastner, 1993; Ashoori, 1996) because of
their  discrete excitation spectra.

In the past, experimental studies of quantum phenomena in small systems were limited to natural systems such as atoms and nuclei. Quantum dots are man-made structures small enough to be governed by the laws of quantum mechanics. The advantage of these artificial systems is that their transport properties are readily measured, with
the  strength of the dot-lead couplings, the number of electrons in the dot,
and  the dot's size and shape all under experimental control. Furthermore, effects of time-reversal symmetry breaking are easily measured by applying a magnetic field.

 Quantum dots are not the only miniature structures whose transport
properties  have been measured. Similar experiments have been performed
recently  on even smaller structures such as very clean metallic nanoparticles
(Ralph, Black, and Tinkham, 1997; Davidovi\'{c} and Tinkham, 1999),  $C_{60}$ molecules deposited on gold
substrate (Porath and Millo, 1996),  and carbon nanotubes (Bockrath~{\em et~al.}, 1997; Tans~{\em et~al.}, 1997; Cobden~{\em et~al.}, 1998).
 Some of the phenomena  observed in these systems are striking similar to those seen  in quantum dots, suggesting that quantum dots are
   generic systems for exploring the physics of small, coherent quantum
   structures.

\subsection{Fabrication and physical parameters of quantum dots}

\begin{figure}
\epsfxsize=8 cm
\centerline{\epsffile{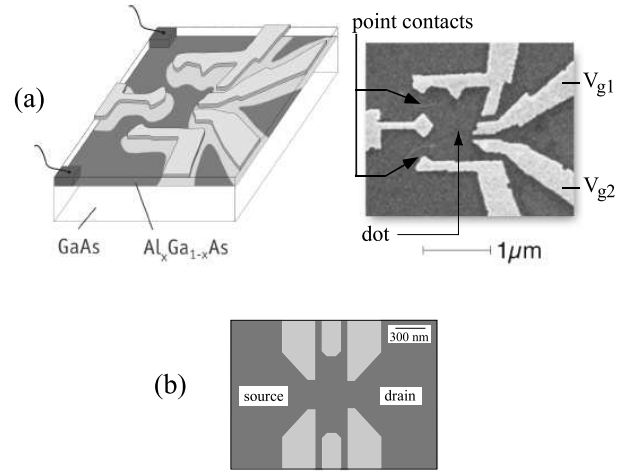}}
\vspace*{2 mm}
\caption
{Quantum dots: (a) a quantum dot used by Folk~{\em et~al.} (1996).   On the right is a scanning electron micrograph
of  the dot (top view), and on the left is a schematic drawing of the device. Electrons are trapped vertically in the interface of a
GaAs/AlGaAs  heterostructure, and form a 2D electron gas (darker area). Their lateral confinement to the dot region is achieved by
applying  a negative voltage to the top metal gate (lighter shade), depleting the electrons
underneath.   The dot is coupled to two leads (source and drain) through point contacts. Two gate
voltages  $V_{g1}$ and $V_{g2}$ can be varied to change the shape and area of
the  dot.
(b) A diagram of a micrograph of another dot by
Oosterkamp~{\em et~al.} (1997).  The darker area includes the dot region
(center)  and the two large 2D electron-gas areas on the left and right (source and drain
regions).  The lighter shade represents the metal gates. The dot's size
is  controlled by the middle pair of gates, and its tunnel barriers  can be
varied  by the pairs of gates on the left and on the right.
}
\label{fig:dots}
\end{figure}

 There are two  types of quantum dots, corresponding
to  lateral and vertical geometries. In the more common lateral dot,  the current flows within the plane to which the electrons are
confined; in a vertical dot (Reed~{\em et~al.}, 1988), the current flows
perpendicular  to the plane. Quantum dots are produced by several  techniques. A typical example of a lateral dot and its fabrication method are illustrated in Fig. \ref{fig:dots}(a): on the right is an electron micrograph of the dot and on the left is a schematic  drawing.  A layer of AlGaAs is grown on top of a layer of
 GaAs by molecular-beam epitaxy. Electrons
accumulate at the GaAs/AlGaAs interface to form a two-dimensional electron gas (their motion in the vertical direction is confined to the lowest state of a quantum well).   Metal gates (lighter regions in the micrograph) are created at the top of the structure by
electron-beam  lithography. A negative bias applied to the top metal gate
depletes  the electrons under the gate and restricts them to a small region
(the  dark central region in the micrograph). The dot is coupled to the bulk 2D electron-gas regions by two individually adjustable point contacts.  A voltage $V_{sd}$ applied between the source and the drain drives a current $I$ through the device. The linear conductance
is  determined from $G=I/V_{sd}$ in the limit of small $V_{sd}$.  The shape and size of the dot can be controlled by voltages $V_{g1}$ and $V_{g2}$ applied to two shape-distorting gates.  The ability to control the dot-lead couplings as well as the dot's shape and area allows us to study a continuous range of physically interesting situations.

A schematic view of another lateral dot (Oosterkamp~{\em et~al.}, 1997) is shown in Fig. \ref{fig:dots}(b). The lighter areas represent the metal
gates. The darker area contains the electrons: the central region is the dot
itself,  connected by point contacts to the large 2D electron-gas regions on the left and
right.  The left and right pairs of gates control the dot's barriers (i.e., its
degree  of openness), while the central pair of gates is used to vary its
shape and size.

 The confined electrons are typically $\sim 50 - 100$ nm below the surface. The
effective  mass of an electron in GaAs is rather low: $m^\ast=0.067\;m_e$. A
typical  sheet density of $n_s \sim 4 \times 10^{11}\;$cm$^{-2}$ corresponds
to  a Fermi wavelength of $ \lambda_F =(2\pi/n_s)^{1/2}\sim 40\;$nm (about two
orders  of magnitude larger than in a metal) and Fermi energy of $E_F \sim 14
\;$meV.
The mobility of GaAs/AlGaAs heterostructures is in the range $\mu_e \sim
10^{4}-10^{6}\;$cm$^2$/V$\cdot$s,  leading to a typical mean free path of
$l=v_F  m^\ast\mu_e/e \sim 0.1 -10 \;\mu$m ($v_F$ is the Fermi velocity). Electron transport in submicron dots with the higher mobility values is thus ballistic. To observe quantum coherence effects it is usually necessary to have a mean single-particle level spacing $\Delta$ in the dot that is comparable to or larger than the temperature.   For  a dot with an effective area of ${\cal A}\sim
0.3\;\mu$m$^2$,  the spacing $\Delta =\pi \hbar^2/m^\ast{\cal A} \sim 11 \;\mu$eV can be resolved at temperatures of $\sim 100\;$mK (corresponding to $kT =
8.6\;\mu$eV).   The lowest effective electron temperatures attained using
dilution  refrigerators are $\sim 50 \;$mK.

To observe charge quantization in the dot, two conditions must be satisfied.   First,
the  barriers must be large enough that the transmission is small. This gives
the  condition $G \ll e^2/h$ (i.e., the dot is almost isolated). Second, the temperature must be low enough that
the  effects of charge quantization are not washed out. The dot's ability
to  hold charge is described classically by its average capacitance $C$.
Since  the energy required to add a single electron is $\approx e^2/C$ per electron in the dot, we have the
condition   $kT \ll e^2/C$. A typical charging energy of a GaAs disk of radius
$0.2\;\mu$m      is $E_C = e^2/C \sim 1000\; \mu$eV, and the condition $kT \ll
e^2/C$  is always satisfied at the low temperatures used in experiments.
The  tunneling of an electron into the dot is usually blocked by
the  classical Coulomb repulsion of the electrons already in the dot, and the
conductance  is small. This phenomenon is known as Coulomb blockade. But by changing the gate voltage $V_g$ we can
compensate  for this repulsion, and at the appropriate value of $V_g$ the
charge  on the dot will fluctuate between ${\cal N}$ and ${\cal N}+1$
electrons,  leading to a maximum in the conductance.  This leads to so-called Coulomb-blockade oscillations of the conductance as a function of the gate voltage. At sufficiently low temperatures these oscillations turn into sharp peaks [see, for example, Fig.  \ref{fig:closed-dot}(c)] that are spaced almost uniformly in $V_g$ by an amount essentially proportional to the charging energy $E_C$.

\begin{figure}
\epsfxsize= 8 cm
\centerline{\epsffile{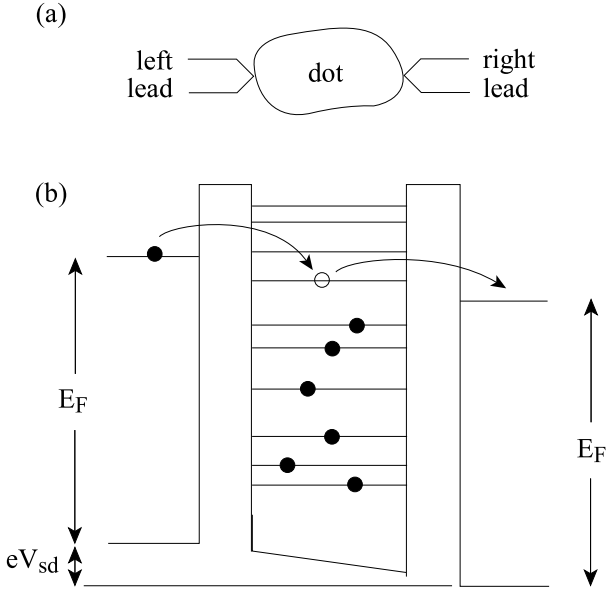}}
\vspace*{2 mm}
\caption
{Conductance by resonant tunneling of a single electron through a quantum
dot:
(a) Schematic view of an isolated quantum dot that is weakly coupled to
left  and right leads.  The point contacts
effectively  create tunnel barriers between the dot and the leads.
(b) Side view. When the Fermi energy of the electron in the source and
drain  reservoirs matches the first unoccupied level in the dot, the electron can tunnel across the barrier into the dot. A current will flow in response to a small source-drain voltage $V_{sd}$. The potential in the dot is controlled by the gate voltage $V_g$. The effect of Coulomb charging energy is not shown here and is illustrated in Fig. \protect\ref{fig:CB}.
}
\label{fig:tunnel-dot}
\end{figure}

Coulomb blockade was first observed in tunnel junctions containing a small
metallic  particle (see, for example, Giaever and Zeller, 1968) in the  classical
regime\footnote{A  metallic particle in 3D has a much smaller mean level
spacing  than a 2D dot of the same size because of the differences in
dimensionality  and effective mass.} $\Delta \ll kT \ll e^2/C$, where tunneling
occurs  through a large number ($\sim kT/\Delta$) of levels. 
Kulik and Shekhter (1975) introduced a transport theory for this classical regime. 
Early theoretical work on Coulomb blockade effects in a single junction was done by Ben Jacob and Gefen (1985), Likharev and Zorin (1985), and Averin and Likharev (1986).  The first
controlled  experiment on a single-electron tunneling
device was by Fulton and  Dolan (1987).  The first observation of Coulomb
blockade  in a semiconductor device was by {Scott-Thomas~{\em et~al.} (1989).
Low-temperature  experiments in semiconductor quantum dots can probe the
 quantum  Coulomb-blockade regime $kT \ll \Delta \ll e^2/C$, where tunneling occurs  through a single resonance in the dot.  Resonant tunneling  is illustrated in Fig. \ref{fig:tunnel-dot}.

\subsection{From ``regular'' to chaotic dots}

\begin{figure}
\epsfxsize= 7.8 cm
\centerline{\epsffile{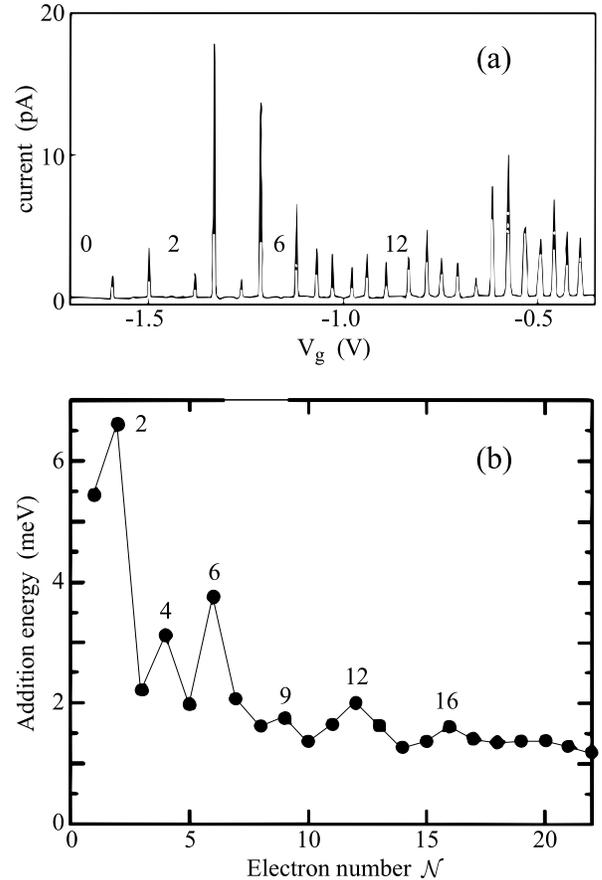}}
\vspace*{2 mm}
\caption
{Shell structure observed in the addition energy of a small vertical quantum
dot.  The dot has the shape of a 2D disk with a harmonic-like confining
potential.  (a) Coulomb-blockade peaks in the current vs gate voltage
$V_g$; (b) Addition
energy [extracted from the peak spacings shown in panel (a)] as a
function  of electron number ${\cal N}$ in the dot. The maxima correspond to
filled  (${\cal N}=2,6,12$) or half-filled (${\cal N}=4,9,16$) spin-degenerate
harmonic-oscillator shells. The half-filling follows Hund's rule favoring the
filling  of the valence shell with parallel spins.  From Tarucha~{\em et~al.} (1996).
}
\label{fig:harmonic-dot}
\end{figure}

 Vertical dots are suitable for spectroscopic studies of a dot with few
electrons  (${\cal N} \alt 20$). Such dots can be prepared in regular shapes,
such  as a disk, where the confining potential is harmonic and the
single-particle  levels are arranged in shells. This shell structure is
observed  by measuring the Coulomb-blockade peaks as a function of the number
of  electrons in the dot (Tarucha~{\em et~al.}, 1996; Kouwenhoven, Oosterkamp~{\em et~al.}, 1997); see, for example, Fig. \ref{fig:harmonic-dot}(a). The spacings between adjacent peaks
are  not uniform and can be converted into an addition spectrum, shown in Fig. \ref{fig:harmonic-dot}(b).  The addition spectrum exhibits
clear maxima at ${\cal N}=2,6$, and $12$, corresponding to completely filled
shells  of a 2D spin-degenerate harmonic oscillator. Additional maxima can be
seen   at ${\cal N}=4, 9$, and $16$, describing half-filled shells with
parallel  spins, in agreement with Hund's rules from atomic physics. The
single-particle  spectrum is sensitive to a magnetic field (as in atoms), and configuration rearrangements are seen at avoided
crossings  of single-particle levels. Overall, a simple
single-particle model plus constant charging energy, supplemented by a
perturbative  treatment of the exchange interaction, can explain qualitatively
the  observed pattern of addition energies versus magnetic field. A good
quantitative  agreement is obtained when compared with Hartree-Fock
calculations  of a few-electron system.

Hartree-Fock calculations -- feasible in small dots -- become impractical for  dots with several hundred electrons. Moreover, many of the lateral dots with ${\cal N} \agt 50$ electrons often have no particular symmetry.  Scattering of an electron from the  irregular boundaries of such dots leads to single-particle dynamics that are mostly chaotic.   Measured quantities such as the dot's conductance and addition spectrum display ``random'' fluctuations when  various parameters (e.g., shape and magnetic field) are varied.  We are entering
the  statistical regime, in which new kinds of questions are of
interest.  For example, rather then trying to calculate the precise, observed
sequence  of conductance peaks in a specific dot, we can study
the  statistical properties of the dot's conductance sampled from
different shapes and applied magnetic fields.

 Classical chaos, i.e., the exponential sensitivity of the time evolution of a dynamical system to initial conditions, is
well  understood not only in closed systems but also in open scattering
systems  (e.g., quantum dots) assuming that the particle spends sufficient
time  in the finite scattering regime (e.g., the dot) before exiting into the
asymptotic  regime (e.g., the leads).  In describing transport through
coherent  systems, we are interested in the quantum manifestations of classical
chaos.  The link between classical and quantum chaos was first established in
1984  with the Bohigas-Giannoni-Schmit (BGS) conjecture (Bohigas, Giannoni, and  Schmit, 1984) that the statistical
quantal  fluctuations of a classically chaotic system are described by random-matrix  theory (RMT). These authors found that the statistical
properties  of $\sim 700$ eigenvalues of the Sinai billiard -- a 2D
classically  chaotic system --  follow the predictions of RMT.  Figure \ref{fig:NNS}(b) compares the nearest-neighbor
spacing  distribution of the Sinai billiard's eigenvalues (histogram) with the
same  distribution calculated from RMT (solid line).

\begin{figure}
\epsfxsize= 7.5 cm
\centerline{\epsffile{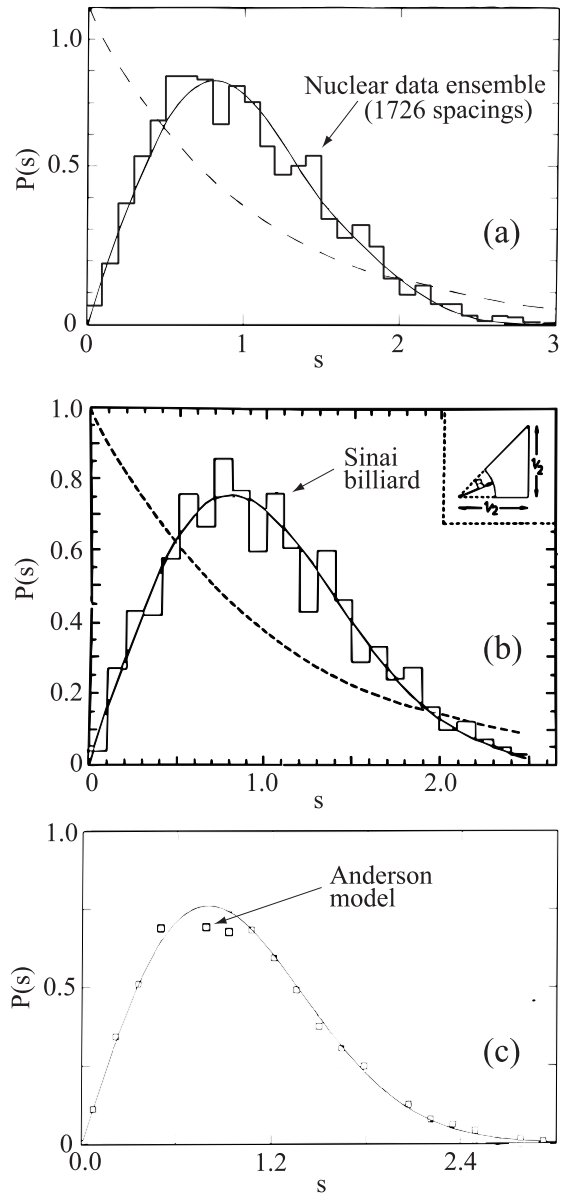}}
\vspace*{2 mm}
\caption
{The universality of RMT.  The nearest-neighbor
level-spacing  distribution $P(s)$ (where $s$ is the spacing in units of the mean level spacing) in (a) a  compound
nucleus, (b) a 2D chaotic system,  and (c) a disordered system, compared with the Wigner-Dyson distribution (solid lines)
predicted  by RMT. Dashed lines show the Poisson
distribution  describing $P(s)$ for a random sequence of levels.
Panel (a) shows $P(s)$ for the nuclear data ensemble -- 1726 neutron
and  proton resonances measured in several heavy nuclei.  From
Bohigas, Haq and Pandey (1983).  Panel (b) is $P(s)$ 
for  700 eigenvalues of the Sinai billiard, a classically chaotic system. The
eigenfunctions vanish at the boundaries indicated by the
inset. From Bohigas, Giannoni, and Schmit (1984). Panel (c) illustrates  $P(s)$ for a 3D Anderson model (open squares) in its diffusive regime with on-site disordered potential $w/t=2$ (see Sec. \ref{disordered}). From
Dupuis and Montambaux (1991).
}
\label{fig:NNS}
\end{figure}

Random-matrix theory  differs in a fundamental way  from the conventional
statistical  approach. Rather than 
 declaring ignorance with respect to the
microscopic dynamical state of the system, we declare ignorance with respect
to  the Hamiltonian itself (Balian, 1968). The only relevant information is
the  system's fundamental space-time symmetries, and otherwise the Hamiltonian
can  be chosen ``at random.''   This revolutionary idea was
introduced  by Wigner in the 1950s to explain the spectral properties of a
complex  many-body system, the compound nucleus, and was developed by
Dyson,  Mehta and others in the early 1960s. Since RMT has no scale (its only physical 
parameter is determined by the mean level spacing $\Delta$ which scales out if all energies are measured in units of $\Delta$), it leads to {\em universal} predictions.   For
example,  neutron and proton resonances measured in heavy
nuclei and collected in the so-called nuclear data ensemble (Bohigas, Haq and Pandey, 1983)
were  found to obey the predictions of RMT. In particular, the resonances'
spacing  distribution, represented by the histogram in Fig.
\ref{fig:NNS}(a),  is well described by the Wigner-Dyson distribution of RMT
(solid  line), just as the eigenvalues of the Sinai billiard are [Fig.~\ref{fig:NNS}(b)].

\begin{figure}
\epsfxsize= 8 cm
\centerline{\epsffile{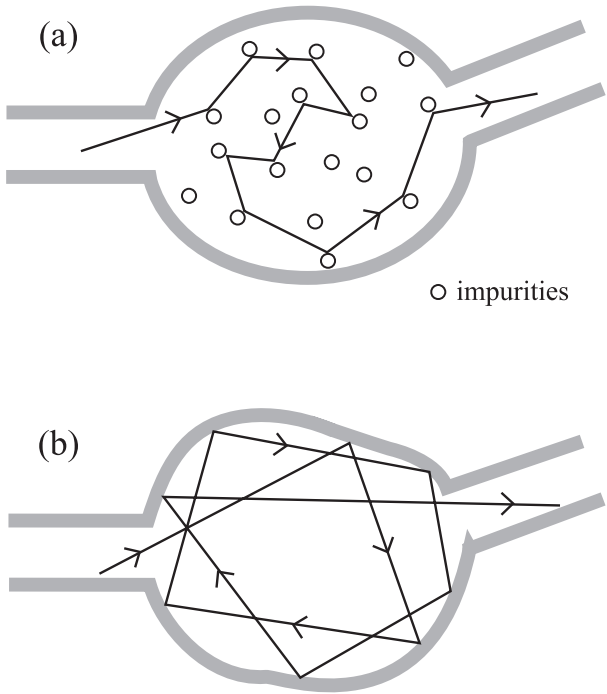}}
\vspace*{2 mm}
\caption
{Schematic drawing of a diffusive dot and a ballistic dot attached to two
leads: (a) an impurity-rich diffusive dot, showing an electron's trajectory with
elastic  scatterings from the impurities;  (b) a
ballistic dot. There are few impurities and the electron moves
ballistically.   Typically, an electron is scattered several times from the
dot's  boundaries before exiting the dot.
}
\label{fig:schematic-dots}
\end{figure}

The usefulness of RMT for neutron resonances was justified by the complexity
of  the compound nucleus above the neutron threshold. In contrast, the Sinai
billiard  is a relatively simple but chaotic 2D system. According to the BGS
conjecture,  classical chaos is a sufficient condition for the
applicability  of RMT.  We shall see that RMT is indispensable for
understanding  the universal statistical properties of chaotic quantum dots.
Random-matrix theory  also links chaotic ballistic dots to mesoscopic disordered metals, as we
discuss  next.

\subsection{From disordered to chaotic ballistic dots}\label{disordered-to-ballistic}

A disordered, impurity-rich quantum dot -- similar to the diffusive systems
that  were the original focus of mesoscopic physics -- is shown schematically
in  Fig. \ref{fig:schematic-dots}(a). An electron enters the dot
through a lead and scatters elastically from the impurities. Here $l \ll L$, and the transport is
diffusive ($l$ is the mean free path and $L$ is the linear size of the dot).  The characteristic time scale is $\tau_D$, the time for the
electron  to diffuse across the dot. The associated energy scale is
 known as the {\em Thouless energy}
$E_c  =\hbar/\tau_D$.  Figure \ref{fig:schematic-dots}(b)
illustrates  the more common ballistic dot. There is relatively little disorder, and
transport  is dominated by scattering from the dot's boundaries. When the
boundaries  are irregular, the electron's dynamics is mostly chaotic.  The
relevant  time scale in ballistic dots is the {\em ergodic time} $\tau_c$,
which  is roughly the time of flight across the dot.  The related energy scale
$E_T=\hbar/\tau_c$  is termed the {\em ballistic Thouless energy}.
 The chaotic nature of the classical motion inside the dot can be revealed in
the conductance only if the electron scatters off the boundaries at
least  several times before escaping through a lead. We therefore limit our
discussion  to dots in which $\tau_{\rm escape} \gg \tau_c$, where $\tau_{\rm escape}$
is  the mean escape time of the electron into the leads. Equivalently, the
average  width $\Gamma$ of a level in the dot (to decay into the leads) must
be  small compared with $E_T$ ($\Gamma \ll E_T$). For a diffusive dot, a
similar  condition $\Gamma \ll E_c$ is required.

 One of the
important  consequences of quantum coherence is the interference of waves
describing  an electron propagating along different paths between the incoming
and  outgoing leads. We can observe these interference effects in the conductance  by changing a phase-sensitive parameter in the system, for example, the electron's Fermi
momentum or the external magnetic field.  The conductance through diffusive or ballistic open
dots  thus exhibits aperiodic but reproducible fluctuations as a function of a
 parameter [see, for example,  Fig.
\ref{fig:open-dot}(c)].

Early theoretical studies of disordered conductors were based mostly on
weak-disorder perturbation theory (i.e., the diagrammatic approach). Two important phenomena were discovered:
(i) The average conductance is smaller in the absence of a magnetic field
than  in its presence, an effect known as weak localization (see
Bergmann, 1984, and references therein). This quantum interference effect requires $L_\phi >> l$ and occurs already in macroscopic conductors ($L > L_\phi$). (ii) The rms fluctuations of the conductance in a mesoscopic conductor
are  of the order $e^2/h$, independent of the size of the average conductance,
a  phenomenon known as {\em universal conductance fluctuations}  
(Altshuler, 1985; Lee and Stone, 1985).  A natural question that arises is whether similar
mesoscopic phenomena can be observed in  ballistic dots. Given the BGS
conjecture,  the application of RMT to describe the universal features of the
conductance  through chaotic ballistic dots was a logical step.  As long as
the electron dynamics in the dot are chaotic, the conductance fluctuations are independent of the dot's
geometry,  although they do depend on the properties of the leads (such as the
number  of modes and the average transparency of each mode). Only in the limit
of  a large number of fully transmitting modes does the size of the
fluctuations  become truly universal in the sense of universal conductance fluctuations.

  The link between universal conductance fluctuations in disordered systems and RMT was first postulated by
Altshuler and Shklovskii (1986) and by Imry (1986a). Earlier work by Thouless
had  pointed to the close connection between the conductance and spectral
properties:  the conductance expresses the sensitivity of the energy levels to
a  change in the boundary conditions (Edwards and Thouless, 1972; Thouless,
1974, 1977). This relation was used by Altshuler and Shklovskii (1986) to argue that
RMT  spectral correlations are at the origin of universal conductance fluctuations.  Their diagrammatic
calculations  also showed that spectral correlations in disordered systems are
{\em nonuniversal}  for energy scales that exceed the Thouless energy. At such
energy  scales the electron diffuses for times that are short compared with
$\tau_D$  and therefore does not have enough time to reach the system's
boundaries.   An analogous  breaking of RMT universality in chaotic systems
was  demonstrated by Berry (1985) for time scales that are shorter
than  the ergodic time.
These observations suggested that the universal regime of RMT is applicable
to  energy scales that are smaller than the Thouless energy in disordered
systems,  and smaller than the ballistic Thouless energy in ballistic chaotic
systems.

A proof of RMT universality in the disordered case was achieved using the
{\em supersymmetry}  method, a field-theoretical approach to disordered systems
(Efetov, 1983).  The supersymmetry method is a technique to carry out the
ensemble  average of a product of Green's functions, where the original
disordered  problem is mapped onto a supersymmetric nonlinear $\sigma$ model. Below the Thouless energy this field-theoretical model reduces to 0D, where it is equivalent to RMT.
  The RMT universality in weakly disordered systems is demonstrated in Fig. \ref{fig:NNS}(c), where the nearest-neighbor level-spacing
distribution,  calculated numerically for a disordered metal, is seen to
follow  the Wigner-Dyson distribution of RMT (solid line).

\subsection{From open to closed dots}

 The strength of the
dot-lead  coupling affects
the width of a typical resonance in the dot to decay into the leads. A simple
expression  for the width can be obtained from an argument due to
Weisskopf (1937).  Imagine a wave packet near the entrance to a channel
$c$  in one of the leads.  The wave packet evolves in time and returns after
the  recurrence time or Heisenberg time $\tau_H =h/\Delta$, where $\Delta$ is
the  mean level spacing.\footnote{In the original Weisskopf argument,  $\Delta$ is the many-body mean level spacing. However, we are mostly interested in the immediate vicinity of the ground state of the dot where (for not-too-strong interactions)  the many-particle mean level spacing is of the order of the single-particle level spacing.} The probability to decay if close to a channel $c$ is
given  by the transmission coefficient $T_c$. The decay rate 
$\Gamma_c/\hbar$  of a level into channel $c$ is then given by the frequency
of  attempted decays times the transmission coefficient (i.e., $\tau_H^{-1}
T_c$),  and the total width $\Gamma =\sum_c \Gamma_c$ of a level to
decay  into any of the channels is
\begin{equation}\label{Weisskopf}
\Gamma = {\Delta \over 2\pi} \sum\limits_c T_c \;.
\end{equation}

 In an open dot, the width ${\cal W}$  of the dot-lead interface is much
larger  than the Fermi wavelength, so that the lead supports a large number of
open  modes ($\Lambda={\rm Int}[k_F {\cal W}/\pi] \gg 1$) with transmission
coefficients  of order $\sim 1$ (no tunnel barriers). From Eq. (\ref{Weisskopf})
we  immediately see that $\Gamma \gg \Delta$. Thus in open dots the resonances
are  strongly overlapping. This is  analogous to the compound-nucleus regime
of  many overlapping resonances, which occurs at several MeV above the neutron
threshold.   Ericson (1960, 1963) predicted that in this regime the coherent
superposition  of a large number ($\sim \Gamma/\Delta$) of resonance amplitudes
would  cause ``random'' but reproducible fluctuations of the nuclear-reaction
cross section  as a function of the reaction energy.  Ericson fluctuations
were  observed a few years after his prediction in light-ion reactions
(Ericson and Mayer--Kuckuk, 1966).  An example is shown in Fig.
\ref{fig:open-dot}(a),  where the measured differential cross section for the
reaction $p$ + $^{35}$Cl $\to \alpha$ + $^{32}$S is plotted as a function of the
proton  energy. Ericson argued that the energy autocorrelation function of the
cross section  should be a Lorentzian with a width $\Gamma$ that is just the width of a typical
resonance  in the nucleus. This is shown in Fig.
\ref{fig:open-dot}(b),  where the autocorrelation function of the cross section
(solid  line) is fitted to a Lorentzian (dashed line).  Similar fluctuations
were observed in the conductance $G$ of a ballistic open dot as a function of
the  Fermi momentum $\hbar k_F$, as shown in Fig.
\ref{fig:open-dot}(c).  A convenient way of analyzing the fluctuations is to
consider  the  Fourier transform of the autocorrelation function of the conductance,
i.e.,  the power spectrum $S(f_k)$ of  $G=G(k_F)$. If the autocorrelation of the conductance vs energy is a Lorentzian 
of  width $\Gamma$, then $S(f_k) \propto e^{-2 \pi \Gamma |f_k|/\hbar v_F}$ (where $v_F$ is the Fermi velocity). This exponential behavior is demonstrated in  Fig. \ref{fig:open-dot}(d) for $S(f_V)$ , the power spectrum of the conductance vs gate voltage (changing the gate voltage is equivalent to changing the Fermi energy). For ballistic dots we can give a semiclassical
interpretation  of $\Gamma$: in an open chaotic system the classical escape
time  is distributed exponentially, with a characteristic mean escape time of
$\tau_{\rm escape}=\hbar/\Gamma$.

\begin{figure}
\epsfxsize= 8 cm
\centerline{\epsffile{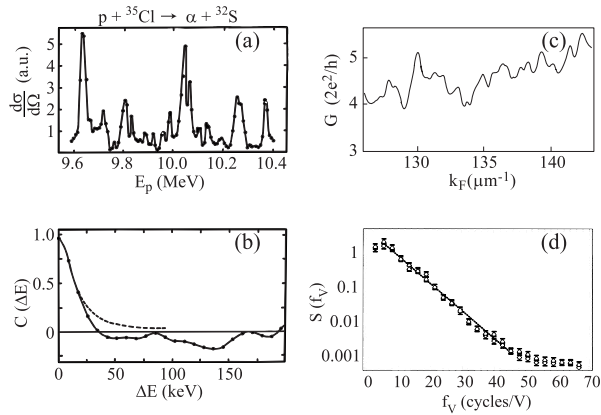}}
\vspace*{2 mm}
\caption{
Ericson fluctuations in the compound nucleus [panels (a) and (b)] and in an open quantum
dot  [panels (c) and (d)]: (a) the measured differential
cross section  $d\sigma/d\Omega$ at $\theta=170^\circ$ for the reaction $p+
^{35}$Cl  $\to
\alpha + ^{32}$S vs the incoming proton energy $E_p$. This reaction 
proceeds through the compound nucleus $^{36}$Ar at an excitation energy of
$\sim  18.5\;$MeV;   (b) the energy autocorrelation function
$C(\Delta E)$ (solid  line) of the cross section in panel (a)
(von Brentano~{\em et~al.}, 1964), where the dashed line is a Lorentzian fit whose width $\Gamma$ corresponds to the lifetime $\hbar/\Gamma$ of the
excited  compound nucleus $^{36}$Ar (from Ericson and Mayer--Kuckuk, 1966); 
(c) the conductance $G$ measured in an open chaotic dot
as  a function of the electron's Fermi momentum $k_F$ (Keller~{\em et~al.}, 1996);  the average conductance increases as a function of $k_F$, but the magnitude
of  the fluctuations is constant of the order $e^2/h$; 
   (d) the power spectrum
$S(f_V)$ (Chang~{\em et~al.}, 1995),  i.e., the Fourier transform of the conductance autocorrelation
function  vs gate voltage (changing the gate voltage is equivalent to
changing  the Fermi energy).
 The dashed
line is a fit to an exponential (which corresponds to a Lorentzian form of
the conductance autocorrelation function). A log-linear scale is used.}
 \label{fig:open-dot}
\end{figure}

 The cross section of a nuclear reaction is proportional to the squared $S$-matrix  element between the entrance and exit channels. An analogous situation
exists  in coherent transport through open quantum dots.
The formulation of conductance in coherent systems was pioneered by
Landauer (1957, 1970)  and refined by Imry (1986b) and
B\"{u}ttiker (1986a).  They described the conductance as a scattering
process  and  expressed it directly in terms of the total transmission through
the  sample.\footnote{It is interesting to note that Landauer's formula can be
derived  from Weisskopf's formula (\ref{Weisskopf}) by applying the latter to
the  leads instead of the dot (where the leads are considered as decaying
quantum  systems emitting electrons); see Bertsch (1991).} The total
transmission  is the sum  over  squared $S$-matrix elements between all
entrance  and exit channels. The average conductance is then expected to
increase  linearly with the number of open channels, as can be  seen in  Fig. \ref{fig:open-dot}(c).
However, the magnitude of the fluctuations is $\sim e^2/h$, independent of
the  average conductance. The nuclear cross-section fluctuations are also
universal,  although since the measured cross section corresponds to a
specific  selection of exit and entrance channels, the size of the
fluctuations  is comparable to the average.

In recent years the experimental focus has shifted from open
to  closed dots, where the statistical behavior of
individual wave functions can be probed.  In closed dots the transmission coefficients are small, $T_c\ll
1$,  and according to Eq. (\ref{Weisskopf}), $\Gamma \ll \Delta$ (assuming a small number of channels). This is the
regime  of isolated resonances, analogous to the compound-nucleus regime of
isolated  neutron resonances just above the neutron threshold. Such narrow
resonances  were observed in the total cross section to scatter
thermal neutrons from heavy nuclei.  Figure
\ref{fig:closed-dot}(a)  shows such resonances for the reaction  $n
+^{232}$Th.  The distribution of the widths of these resonances is shown on a log-linear scale in  Fig. \ref{fig:closed-dot}(b)  and
 is well  described by the so-called Porter-Thomas
  distribution (solid line) predicted  by  RMT. It is given by $P(\hat \Gamma) \propto \hat\Gamma^{-1/2}
e^{-\hat\Gamma/2}$,  where $\hat \Gamma$ is the width measured in units of the
average  width.

In closed dots the conductance is not a smooth function of the gate voltage
as  in open dots, but instead exhibits Coulomb-blockade peaks [see, for example,  Fig. \ref{fig:closed-dot}(c)].  The spacings between these peaks
are  almost uniform because they are dominated by the large charging energy,
in  contrast to the nuclear case, where the spacings between the observed
resonances  fluctuate widely.
Moreover,  the observed conductance peak widths in closed dots are all $\sim
kT$  because of thermal broadening. However, the peak heights exhibit
order-of-magnitude  fluctuations, as can be seen in Fig.
\ref{fig:closed-dot}(c).
 These peak fluctuations
are determined by the spatial fluctuations of the individual resonance
wave functions  in the vicinity of the leads. The
statistical  approach to  Coulomb-blockade peak heights was developed by
Jalabert, Stone, and Alhassid (1992).  They used $R$-matrix theory -- originally introduced
by  Wigner and Eisenbud (1947) for nuclear reactions -- to relate the Hamiltonian of the closed system to the
scattering  resonances of the weakly open system, and then applied an RMT
approach  to quantify the wave-function fluctuations. The conductance peak-height distributions  were found to be universal and sensitive only to the space-time
symmetries  of the dot.   These distributions were measured a few years later
(Chang~{\em et~al.}, 1996; Folk~{\em et~al.}, 1996)  and were in agreement with the theoretical predictions.
The  distribution of the conductance peak heights in the absence of magnetic
field  is shown in Fig. \ref{fig:closed-dot}(d). This is
the  case of conserved time-reversal symmetry, analogous to the
neutron-resonance  statistics in Fig. \ref{fig:closed-dot}(b).

\begin{figure}
\epsfxsize= 8 cm
\centerline{\epsffile{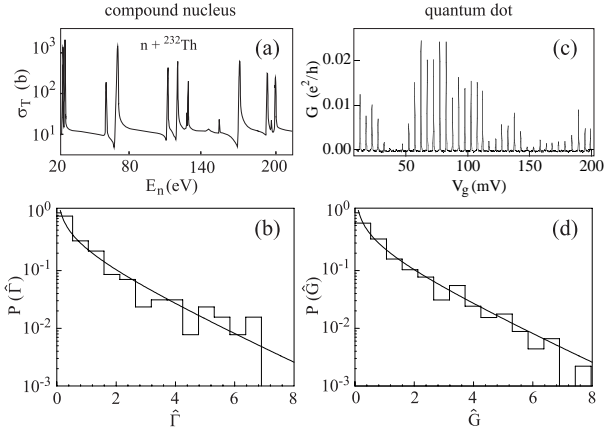}}
\vspace*{2 mm}
\caption
{Neutron-resonance-width statistics in the compound nucleus [panels (a) and (b)] and
Coulomb-blockade peak statistics in closed quantum dots [panels (c) and (d)]: (a) a series of neutron resonances in the total cross section
$\sigma_T$  (in barns) of n$+^{232}$Th as a function of the incoming neutron
energy  $E_n$ (in eV) (from Brookhaven
National  Laboratory, 1964); (b) distribution of the normalized neutron resonance
widths  $\hat\Gamma=\Gamma/\bar\Gamma$ using a log-linear scale  [$223$ neutron
resonances  in $^{233}$Th are included from the measurements of
Garg~{\em et~al.} (1964)]; the solid line is the Porter-Thomas distribution
$P(\hat\Gamma)  \propto \hat\Gamma^{-1/2} e^{-\bar\Gamma/2}$
  predicted by RMT;
(c) a series of Coulomb-blockade peaks observed in the
conductance  $G$ of closed GaAs/AlGaAs dots (with ${\cal N} \sim 1000$
electrons)  as a function of gate voltage $V_g$  at zero magnetic field
(Folk~{\em et~al.}, 1996); (d) distribution of the
normalized conductance  peak heights $\hat G=G/\bar G$ (histogram) on a
log-linear  scale; 600 peaks are included, of which only $\sim 90$ are
statistically  independent. The solid line is the
theoretical  prediction based on RMT (Jalabert, Stone and Alhassid, 1992) and contains no
free  parameters. Notice the agreement with the experiment over almost three
orders  of magnitude.
}
\label{fig:closed-dot}
\end{figure}

\subsection{From noninteracting to interacting electrons}

Studies of statistical fluctuations in open dots have generally ignored
electron-electron  interactions. Electron-electron inelastic scattering
reduces  the coherence time of the electrons at finite temperature, but
otherwise  the excitations around the Fermi energy are assumed to be
noninteracting  quasiparticles.  Landau's Fermi-liquid theory treats the
electrons in good metals as weakly interacting quasiparticles whose lifetime
near  the Fermi surface is large compared to $\hbar/kT$.  In open dots with many open channels
the  average conductance is large, and a noninteracting picture, similar to
that  of metals, is justified. But as the strength of the dot-lead couplings
is  reduced, interaction effects become important in transport. The simplest way to take
interactions  into account is to include only the long-range component of the
Coulomb  interaction, namely the average interaction among ${\cal N}$
electrons  in the dot. This is known as the constant-interaction  model. This
model  has the advantage of simplicity, and it does explain some of the
observed  phenomena. However, discrepancies with other recent experiments --
notably  the measured peak-spacing distribution -- indicate that
electron-electron  interactions can have significant effects beyond the
constant-interaction  model.  An important
parameter  is the dimensionless gas parameter $r_s$ measuring the ratio
between  a typical Coulomb interaction and the average kinetic energy. When
$r_s$  is small, it is possible to use a combination of mean-field approximations and the known statistics of the noninteracting limit.  However, in semiconductor
quantum  dots, $r_s \sim 1-2$, and deviations are expected.
 The interplay between one-body chaos and
many-body  interactions in quantum dots is a fascinating open problem and is
currently the main focus of the field.

\subsection{Methods}

The statistical theory of quantum dots is based on four main approaches:
diagrammatic methods,  semiclassical methods, random-matrix theory,
and  supersymmetry. The first approach (i.e., the diagrammatic method) was
developed in early work on disordered conductors. More recent progress -- based primarily on the last three methods and driven in part by a deeper
understanding of their interrelations -- has brought the statistical approach
to  a new level of maturity.
While several reviews and books have been written on the semiclassical, RMT,
and  supersymmetry approaches, they are usually discussed separately and often
not  in relation to quantum dots. A section of this review describes
these methods in the context of the statistical theory of
quantum  dots, emphasizing the relationships among the methods. Many of the
results of that section will be used in subsequent sections.

\subsection{Outline of the review}

We begin by reviewing transport theory in quantum dots (Sec.
\ref{transport}).  An important result is
the  Landauer formula expressing the conductance in terms of
the  $S$ matrix of the device.  We discuss resonant tunneling and Coulomb blockade in closed dots, and  processes that
dominate  the off-resonance conductance, such as cotunneling.  The results of this section will be used in developing the statistical theory of quantum dots.

Section \ref{statistical-theory} covers the main principles and tools  of the
statistical  fluctuation theory: the
semiclassical  approach
(demonstrated  for disordered metals), random-matrix theory, and the supersymmetry method.

Section \ref{open-dots} examines mesoscopic conductance fluctuations in open
dots using random-matrix and semiclassical approaches.
We discuss the main mesoscopic phenomena in fully coherent open quantum
dots  -- weak localization and conductance fluctuations, and how they are
affected by a finite coherence time.

Section \ref{closed-dots} reviews the statistical theory of closed dots.
 The statistics and parametric correlations of the peak heights are
derived both for temperatures that are small and that are comparable to the mean level spacing.   The constant-interaction 
model can explain some, but not all, of the observed statistics.  Adding electrons to the dot changes or ``scrambles'' the
single-particle  spectrum and wave functions because of
charge  rearrangement.  This statistical scrambling is discussed in the framework of RMT.

 Section \ref{interactions} summarizes recent progress in understanding the
effects  of electron-electron interactions on the statistical fluctuations.
 The focus is on peak-spacing
statistics,  where discrepancies with the results of the CI
model  have been seen experimentally.  The Hartree-Fock
approximation  offers an intermediate framework that describes interaction effects within a single-particle theory.
  Spin effects are also discussed.

Finally, in Sec. \ref{charging-energy-effects} we consider various charging energy effects on the mesoscopic fluctuations,  such as the fluctuations of the off-resonance conductance where the dominating conductance mechanism is cotunneling, and the phenomenon of mesoscopic
Coulomb  blockade due to the backscattering of
electrons  into an open lead.

Throughout the review we have made an effort to adhere to a uniform notation,
which  sometimes required changing the notation of the original papers. We
have  also tried to keep to standard notation except in cases where the same
symbol  is used for different quantities. The topics covered in this review
are too broad to permit discussion of all related work, and we apologize to
those whose work could not be included.

\section{Electron Transport Through Quantum Dots}
\label{transport}

  Conductance through quantum
dots is a phase-coherent process. In this section we review the
  formalism of coherent transport.

 In Sec. \ref{classical-transport} we remind the reader of the
quasiclassical  theory of transport in a metal where phase
coherence is ignored.  At low temperatures and in small conductors, however, the
coherence  length of the electron can be larger than the system's size, and
 conductivity is no longer a local quantity. Section \ref{LIB} describes
 conductance as a coherent scattering process; its main result
is  the Landauer formula expressing the conductance in terms of the $S$ matrix.
It is often useful to relate the $S$ matrix of
the  open system to the eigenfunctions and eigenvalues of the closed system.
This  relation is derived in $R$-matrix theory,
 discussed in Sec. \ref{R-matrix-formalism}. Section
\ref{resonant-tunneling}  describes an important limit of the theory, the weak-coupling  limit, where conductance occurs by resonant tunneling.

 Section \ref{Coulomb-blockade} reviews Coulomb blockade, a central
phenomenon  in closed quantum dots with large tunnel barriers.
Coulomb  blockade is essentially a classical phenomenon, observed at
temperatures  that are small compared with the charging energy. The
discreteness of the dot's levels becomes important in the quantum Coulomb-blockade regime when the temperature drops below the mean level spacing $\Delta$.

Section \ref{cotunneling} considers the off-resonance conductance in
Coulomb-blockade dots, which is dominated by cotunneling, i.e.,  the
virtual  tunneling of an electron (hole) through a large number of
intermediate  levels in the dot. Finally,
Sec.  \ref{non-linear-transport} explains how conductance measurements in
the  nonlinear regime provide information on excited states in the dot.

\subsection{Quasiclassical description of conductivity}
\label{classical-transport}

 In a perfect periodic potential in a crystal, the motion of the electrons is
described  by extended Bloch states. In real samples, however, there is some
disordered  potential due to impurities, defects, dislocations, etc. The
electrons  scatter from the impurities elastically (i.e., the electron's
energy  is conserved and only its momentum is reoriented), and repeated
scatterings  lead to  diffusive motion. In the quasiclassical description of diffusion,  the
electron  is assumed to lose phase coherence after each collision with
an  impurity, i.e., $L_\phi \sim l$, and the conductivity can be introduced as a {\em local} intensive quantity (see below).

The diffusion current of electrons is given by
$\bbox{J}_d = - D \nabla n_s$,
where $n_s=n_s(\bbox{r},t)$ is the electron density and $D$ is the diffusion
constant.
Using the continuity equation $\nabla \cdot \bbox{J}_d = -\partial
n_s /\partial  t$, we obtain the diffusion equation
\begin{equation}\label{diffusion-equation}
{\partial n_s \over \partial t} = D \nabla^2 n_s \;.
\end{equation}
   Describing the diffusion (in $d$ dimensions) as a random walk, we can relate the diffusion
constant  $D$ to the elastic mean free path $l$ through
$D = v_F l/d$.

Special solutions to the diffusion equation, $n_{\bbox{q}} \propto e^{i
\bbox{q}\cdot  \bbox{r}} e^{-Dq^2 t}$, are known as the classical
diffusion  modes.
The classical diffusion propagator, describing the time evolution of the
single-electron  density, can be expanded in these modes:
$D(\bbox{r}^\prime,\bbox{r};t)= [\sum_{\bbox q}e^{i \bbox{q}\cdot
(\bbox{r}^\prime  - \bbox{r})} e^{-Dq^2 t}] \theta(t)$.
 Its Fourier transform is
$D(\bbox{q},\omega) = 1 /( -i \omega + D q^2)$.

  The electrical conductivity $\sigma$ is defined by Ohm's law, $\bbox{J}_e =
\sigma  \bbox{E}$, describing the linear relation between the electric current
density   $\bbox{J}_e$ and the applied electric field $\bbox{E}$.
To relate the conductivity to the mean free time $\tau$ ($\tau=l/v_F$, where $v_F$ is the Fermi velocity), we assume that after
each  collision the electron's velocity is ``randomized.'' Between collisions
the  electrons are accelerated in response to the electric field, and their
average  velocity is $\overline{\bbox{v}} = (e/m) \bbox{E} \tau$. Since the
current  is given by $\bbox{J}_e = n_s e \overline{\bbox{v}}$, we recover Ohm's
relation  with
\begin{equation}\label{Drude}
\sigma = {e^2 \over m} n_s \tau \;,
\end{equation}
 also known as Drude's formula. When combined with $D=v_F^2\tau/d$ and $d n_s/2E_F=\nu$ ($\nu$ is the density of single-particle states per unit volume at the Fermi energy), Drude's formula (\ref{Drude}) leads to Einstein's relation
\begin{eqnarray}\label{Einstein}
\sigma = e^2  \nu  D \;.
\end{eqnarray}
In a 2D electron gas, $\nu=m/\pi \hbar^2$ is constant, and $\sigma =  {e^2 \over h} k_F l $.  In a good
metal,  $k_F l \gg 1$, and the conductivity is much larger than the
quantum  unit $e^2/h$.

 The quantity that is measured directly is the conductance,
defined  as the ratio between the current and the applied voltage.  In a
macroscopic  conductor the conductivity is an intensive quantity,  and
the  conductance $G$ is related to the conductivity by
$ G \sim  \sigma ({\cal S}/ L)$,
where $L$ and ${\cal S}$ are the length and transverse cross section of the
conductor,  respectively.

\subsection{Conductance in mesoscopic systems and the Landauer formula}
\label{LIB}

   In a mesoscopic system
$L_\phi \agt L$, and the local description of conductivity breaks down.  It is then meaningful  to discuss only the measurable quantity -- the conductance -- or alternatively,  to introduce conductivity as a nonlocal tensor.  For
recent  books discussing transport in mesoscopic systems, see, for example,
Datta (1995),  Imry (1996), and Ferry and Goodnick (1997).

 The transport properties of a mesoscopic structure are characterized by
 conductance coefficients that are sample specific.
Several  probes (leads) attached to electron reservoirs are connected to the
system.  A current is passed through the structure and the voltage is measured
at  the different probes. Denoting by $V_n$ the voltage of probe $n$ and by $I_n$ the current
through  probe $n$, we expect in the limit of small voltages  a set of linear relations
\begin{equation}\label{conductances}
I_{n'} = \sum\limits_n G_{n' n} V_n \;.
\end{equation}
The coefficient $G_{n' n}$ is the conductance between leads $n$ and $n'$. It
follows  from Kirchhoff's law $\sum_n I_n =0$ that $\sum_n G_{n' n} =
\sum_{n'}  G_{n' n} =0$.
For the special case of a two-lead dot,  $G_{12}=G_{21}=-G_{11}=-G_{22}\equiv
G$,  and $G=I_1/(V_2 -V_1)$.

 A formulation that takes into account phase coherence as well as the
geometry  of the structure was developed by Landauer (1957, 1970) and
Imry (1986b).  The basic idea was to relate the conductance to an
underlying  scattering matrix (i.e., transmission coefficients). The first
derivations  of such transmission formulae from linear-response theory were by Economou and Soukoulis (1981)  and Fisher and Lee (1981).   The work of Landauer and Imry was generalized by B\"{u}ttiker (1986a, 1988a) to a general
configuration  of probes. We first discuss the
macroscopic  approach of B\"{u}ttiker and then describe a simplified
 derivation of the  Landauer formula from statistical reaction theory by
Bertsch (1991).
   For illuminating  discussions of the Landauer approach and its
applications, see Stone and Szafer (1988), Baranger and Stone (1989),
  Datta (1995), Stone (1995), and
Imry (1996). For a recent review, see Imry and Landauer (1999).

  We assume a mesoscopic structure connected to several leads $n$,  each
supporting  $\Lambda_n$ propagation modes of the electrons. We denote by
$S^{n'  n}_{c' c}$
the scattering amplitude of the structure to scatter an electron from channel
$c$  in lead $n$ to channel $c'$ in lead $n'$. A scattering amplitude between
different  leads ($n' \neq n$) is called a transmission amplitude $t$, while a
scattering  amplitude between channels that belong to the same lead ($n'=n$)
is  called a reflection amplitude $r$. The total transmission from lead $n$ to
lead  $n'$ is $T^{n' n}= \sum_{c' c}|t^{n' n}_{c' c}|^2$ (where the sum is over all channels $c$ in lead $n$ and $c'$ in lead $n'$),
and  the total reflection in lead $n$ is  $R^{n n}= \sum_{c' c}|r^{n n}_{c' c}|^2$ ($c$ and $c'$ are channels in lead $n$). A voltage $V_n$ in probe $n$ causes a current
$I^{\rm inject}_{c  n}= e v_c (d n_c^+/ d\epsilon) e V_n$ to be injected in each mode $c$, where
$v_c$  and $d n_c^+/ d\epsilon$  are the longitudinal velocity and density of ingoing states (per unit length of the lead) in
mode  $c$.  Assuming noninteracting electrons in the lead, $v_c (d n_c^+/ d\epsilon) =1/h$
and  the injected current is $I^{\rm inject}_{c n} = (e^2/h) V_n$. The electrons
injected  in lead $n$ are scattered into lead $n' \neq n$ and produce there a
current  $I_{n' n} =\sum_{c' c} |t^{n' n}_{c' c}|^2 I^{inject}_{c
n}=(e^2/h)  T^{n' n} V_n$. Electrons are also backscattered into the same lead
$n$  giving a net current of $I_{nn} = (e^2/h)(R^{n n}- \Lambda_n) V_n$, where
the  injected current has been subtracted.  The total current in lead $n'$ is
$I_{n'}  = \sum_{n} I_{n' n}$. We obtain Eq. (\ref{conductances}) with
\begin{eqnarray}\label{Landauer}
G_{n' n} = \left\{ \begin{array}{ll} 2{e^2 \over h} T^{n' n} & \; ({\rm for} \;\;  n' \neq n)  \\
 2{e^2 \over h} (R^{n n} -\Lambda_n) & \; ({\rm for}\;\; n'=n )
\end{array} \right. \;,
\end{eqnarray}
where a factor of $2$ was included to take into account the spin degeneracy
of  the electrons.

A simple derivation of Landauer's formula at finite
temperature is provided  by Bertsch (1991). The
starting point is Weisskopf's formula (\ref{Weisskopf}),  first
applied in nuclear physics to neutron evaporation
(Weisskopf, 1937),  and often used in the transition state
theory of chemical reactions.  We assume for simplicity that the
mesoscopic structure is connected  to two leads only.   Rather
than applying Eq. (\ref{Weisskopf}) to the structure  itself, we
apply it to the leads, considering each lead as an equilibrated,
decaying system emitting electrons. At zero temperature, the
decay  rate $\Gamma/\hbar$ from each lead is given by
\begin{equation}\label{decay-rate}
{\Gamma \over \hbar} = {1 \over 2\pi \hbar \rho(E)} \sum T \;,
\end{equation}
where $\rho(E)$ is the density of states in that lead and $\sum T$ represents
a  sum over transmission coefficients in all open channel states. At finite
temperature  the current due to electrons emitted by the lead is $I = e
\sum_\lambda  (\Gamma(E_\lambda)/\hbar)f(E_\lambda - \mu)$,  where
$f(E_\lambda  - \mu)$ are Fermi-Dirac occupation probabilities ($\mu$ is the
chemical  potential in the lead).
Converting the sum into  an integral $\sum_\lambda \to \int dE
\rho(E)$  and using Eq. (\ref{decay-rate}), we find that the level density cancels out
and  we obtain
\begin{equation}\label{lead-current}
I = {e \over 2\pi \hbar} \int dE \left(\sum T \right) f(E-\mu) \;.
\end{equation}
  To find the conductance we apply a small source-drain voltage $V_{sd}$  and
calculate  the current. Each lead emits a current given by
Eq. (\ref{lead-current}).  However, the two leads have different chemical
potentials  $\mu_1$ and $\mu_2$, with $\mu_1-\mu_2=eV_{sd}$. The net current
$I=I_1-I_2$  is thus nonzero:
\begin{equation}
I = {e \over 2\pi \hbar} \int dE \left(\sum T \right) [f(E-\mu_1) - f(E
-\mu_2)]  \;,
\end{equation}
where we have assumed that the total transmission $\sum T$ is independent of
the  direction of flow of the current. Indeed, for $V_{sd}=0$ the whole system is
equilibrated  ($\mu_1=\mu_2$) and the net current must be zero, leading to the
equality  of $\sum T$ for both leads.  In the linear regime, $f(E-\mu_1) - f(E
-\mu_2)  \approx - e V_{sd} f'(E-E_F)$, where $E_F$ is the finite-temperature  Fermi
energy  (i.e., chemical potential) of the equilibrated system, and we find
\begin{equation}\label{finite-T-Landauer}
G = 2{e^2 \over h} \int dE [-f'(E-E_F)] \left( \sum T\right) \;.
\end{equation}
The factor of 2 accounts for spin degeneracy. Equation (\ref{finite-T-Landauer}) is the finite-temperature Landauer formula. According to
transition state theory, $\sum T$ represents a sum over all
available  channels. Since the electron can decay from any mode $c$ in the
first  lead to any mode $c'$ in the second lead, $\sum T  = T^{2,1}$, where $T^{2,1}$ is the total transmission between the two leads.  In the limit of zero temperature, $-f'(E-E_F) \to \delta
(E-E_F)$,  and Eq. (\ref{finite-T-Landauer}) reduces to the first case of Eq.
(\ref{Landauer}).

 A more microscopic derivation of Landauer's formula using linear-response theory is based on a Kubo formula (Kubo, 1957; Greenwood, 1958); see, for example, 
  Baranger and Stone (1989). The generalization of Landauer's formula  (\ref{finite-T-Landauer})
to the multileads case is
\begin{equation}\label{finite-T-LIB}
G_{n' n}= 2 {e^2 \over h} \int_0^\infty dE [-f'(E - E_F)] \sum\limits_{c' c}
|S^{n' n}_{c' c}(E)|^2 \;,
\end{equation}
 where the sum is over all entrance channels in lead $n$ and all exit
channels  in lead $n'$.

  The Landauer formula works well in open structures where the picture of noninteracting quasiparticles is valid and interactions contribute only through finite dephasing times. In closed dots, where charging energy is important, it usually does not hold except in special cases (see Sec. \ref{Coulomb-blockade}). There were attempts to generalize Landauer's formula to the
interacting  case. Meir and Wingreen (1992) used the nonequilibrium Keldish
formalism  to derive a formula for the current in an interacting-electron
region.  At the limit of zero temperature, where only elastic processes are
allowed,  their formula reduces for the linear-response  conductance reduces to a Landauer-type formula. However, at finite temperature or in the nonlinear
regime, such Landauer-type formula does not hold because of inelastic processes.

The symmetry properties of the conductance coefficients were discussed by
B\"{u}ttiker (1988a).
The time-reversal properties of the $S$ matrix in an external magnetic
field  $B$, $S^T(-B) = S(B)$, lead to
$G_{n' n}(B) = G_{n n'}(-B)$.
These relations do not imply that $G$ is a symmetric matrix except when
$B=0$.  Note, however, that for a two-lead dot $G$ is always symmetric [see the discussion following Eq. (\ref{conductances})].

\subsection{$R$-matrix formalism}
\label{R-matrix-formalism}

  $R$-matrix theory relates the $S$ matrix of the dot to the  discrete eigenvalues and eigenstates of
 the closed system, and was introduced by Wigner and Eisenbud (1947)
 in nuclear-reaction theory. The methods were generalized and reviewed
by Lane and Thomas (1958), and a pedagogical treatment is available in Blatt and Weisskopf (1952).  An equivalent formulation that expresses the
$S$ matrix  in terms of an effective non-Hermitian Hamiltonian of the open
system  was presented by Mahaux and Weidenm\"{u}ller (1969) and is also derived in a recent review
on  scattering in chaotic systems  by Fyodorov and Sommers (1997).  The $R$-matrix
formalism  was adapted to quantum dots by Jalabert, Stone, and Alhassid (1992).

We consider a 2D cavity
in the region ${\cal A}$ of the $x$-$y$ plane and assume left and right
leads  (along the $x$ axis) that are
 attached to the dot at the lines of contact $x=x_l,x_r$
(denoted  by ${\cal C}_l$ and ${\cal C}_r$).   We denote by $H$ the dot's Hamiltonian and consider the eigenvalue
problem
\begin{eqnarray}\label{Hdot}
H \Psi_\lambda = E_\lambda \Psi_\lambda \;,
\end{eqnarray}
 where $\Psi_\lambda(\bbox{r})$  vanishes at the
walls  and  satisfies a general homogeneous boundary condition at  the dot-lead  interfaces:
   $\partial\Psi_\lambda/\partial n-h_{l,r}\Psi_\lambda=0$ ($\hat{n}=\pm\hat{x}$ is the normal to each interface
and $h_{l,r}$ are
constants).    A scattering
 solution $\Phi(\bbox{r})$ at energy $E$ can be expanded
$\Phi(\bbox{r})=\sum_\lambda  a_\lambda\Psi_\lambda(\bbox{r})$.  Since
$\Phi$  and $\Psi_\lambda$ are solutions of the
 Schr\"odinger equation inside the dot at energies $E$ and $E_\lambda$,
respectively, we find
\begin{eqnarray}\label{ExpCoeff}
   a_\lambda & = &\int\limits_{\cal A}dA\;\Psi_\lambda^\ast\Phi \nonumber \\ & = &
   {\hbar^2\over 2m} {1\over E_\lambda-E} \int\limits_{\cal C}dl 
  \left(\Psi_\lambda^\ast{\partial\Phi\over\partial n}
   -\Phi{\partial\Psi_\lambda^\ast\over\partial n}\right).
\end{eqnarray}
We denote by $\phi^i_{c}$
a complete set of  transverse wave functions, where $c$ is a channel
index and $i=l,r$. For an open channel,
$\phi^i_{c} (y)= \sqrt{2/{\cal W}_i} \sin (\kappa^i_c y)$,
where ${\cal W}_i$ is the width of the lead $i$ and $\kappa_c^i=c\pi/{\cal
W}_i\;\;\;  (c=1,2,\ldots,\Lambda_i$) are the quantized transverse momenta.
 Inside each lead we can expand
   $\Phi(\bbox{r})=\sum_{c=1}^{\Lambda_i}u_c^i(x)\phi_c^i(y)$.
Using Eq. (\ref{ExpCoeff}) and the boundary conditions satisfied by the dot's eigenfunctions, the longitudinal
components $u_c^i(x)$ of the scattering solution at the contact points
$x=x_{l,r}$ are given by
  $ u_c^i(x_i)=\sum_{i^\prime c^\prime}R_{i c; i^\prime c^\prime}
   \left[\partial u_{c^\prime}^{i^\prime}/\partial n
   -h_{i^\prime}u_{c^\prime}^{i_\prime}\right]_{x=x_{i^\prime}}$, where
\begin{eqnarray}\label{R-matrix}
   R_{i^\prime c^\prime; i c}(E)=
     \sum\limits_\lambda
     {y_{c^\prime\lambda}^{i^\prime}  y_{c \lambda}^{i\;\ast}\over
   E_\lambda-E} \;
\end{eqnarray}
is the $R$ matrix defined in terms of $y_{c\lambda}^i=\sqrt{\hbar^2/2m}
   \int_{\cal C} dl\;
   \phi_c^{i\;\ast}\Psi_\lambda$, the {\it  reduced} partial-level width amplitude
 for the decay from level
$\lambda$  into channel $c$ in lead $i$.  In the following, we shall omit the
lead  label $i$ and assume it is included in the channel label $c$.

The $S$ matrix can be expressed in terms of the $R$ matrix. To this end,
 it is convenient to define the $K$ matrix
by $K  =  (k P)^{1/2} R ( k P)^{1/2}$. Here  $k P$ is a diagonal matrix
 where $k_c$ is the longitudinal channel momentum
($\hbar^2k_c^2/2m + \hbar^2\kappa_c^2/2m = E$),
and $P_c$  is the penetration factor to tunnel through
the barrier in channel $c$ [$P_c=k_c^{-1}{\rm Im} (\partial\ln u_c^+/\partial n)|_{x=x_i}$, where $u_c^+$ is the outgoing wave component of $u_c$; $P_c=1$ in the absence of a barrier and $P_c \ll 1$ in the presence of a barrier].   In terms of the $K$ matrix (see, e.g., Blatt and Weisskopf, 1952) 
\begin{eqnarray}\label{K-to-S}
 S  =  { 1 + i K \over 1 - iK}  \;.
\end{eqnarray}
 Equation (\ref{R-matrix}) can be rewritten  for the $K$ matrix
\begin{eqnarray}\label{K-matrix}
   K_{c^\prime; c}(E)= \frac{1}{2}
     \sum\limits_\lambda
     {\gamma_{c^\prime \lambda}  \gamma_{c \lambda} ^\ast \over
   E_\lambda-E} \;,
\end{eqnarray}
where
\begin{eqnarray}\label{partial-amplitude}
\gamma_{c\lambda}= \sqrt{\hbar^2 k_c P_c \over m}   \int\limits_{\cal C} dl\;
   \phi_c^{\ast}\Psi_\lambda
\end{eqnarray}
  are known as the partial-width amplitudes.

The relation (\ref{K-matrix}) for the $K$ matrix can also be written in an
arbitrary fixed basis of wave functions (not eigenfunctions) $\rho_j$ in the
dot:
\begin{eqnarray}
 K = \pi W^\dagger {1 \over H - E } W \;.
\end{eqnarray}
Here $W$ is a
$N  \times \Lambda$ matrix of coupling constants
between the leads and the dot:
\begin{eqnarray}\label{W}
 W_{j c} \equiv \left( \hbar^2 k_c P_c \over 2\pi m \right)^{1/2}
\int\limits_{\cal  C}dl\;
   \rho^\ast_j \phi_c \;.
\end{eqnarray}

Equations (\ref{K-to-S}) and  (\ref{K-matrix}) for the $S$ matrix
can be rewritten in terms of the Green's function of an effective Hamiltonian:
\begin{eqnarray}\label{S-matrix}
S = 1 - 2\pi i W^\dagger { 1 \over E - {\cal H}_{\rm eff}} W \;,
\end{eqnarray}
where ${\cal H}_{\rm eff} = H - i \pi W W^\dagger$.
${\cal H}_{\rm eff}$ is non-Hermitean; its real part
is  the dot's Hamiltonian $H$ and its imaginary part is the  ``width'' matrix
$\Gamma  \equiv \pi W W^\dagger$. It generally describes the Hamiltonian of an
open  system that is coupled to an external region. The eigenvalues of ${\cal
H}_{\rm eff}$  are complex (${\cal E}_\lambda - i \Gamma_\lambda/2$) and describe
resonances  of energies ${\cal E}_\lambda$ and widths $\Gamma_\lambda$.

\subsection{Resonant tunneling}
\label{resonant-tunneling}

 In the weak-coupling limit a typical resonance
width  in the dot is much smaller than the average spacing $\Delta$ between
resonances.  In this limit only the
 resonance $\lambda$  that is closest to the scattering  energy $E$
 contributes to the $K$ matrix  (\ref{K-matrix}). This leads
  (for $c \neq c'$) to a Breit-Wigner resonance  formula
\begin{equation}\label{Breit-Wigner}
|S_{c' c}|^2 = { \Gamma_{c' \lambda} \Gamma_{c \lambda} \over (E -
E_\lambda)^2  + \left({\Gamma_\lambda \over 2} \right)^2} \;,
\end{equation}
where
$\Gamma_{c \lambda} \equiv |\gamma_{c\lambda}|^2$ is the partial  width of the resonance level
$\lambda$  to decay into channel $c$, and
$\Gamma_{\lambda} \equiv \sum_{c} \Gamma_{c \lambda}$
 is the total width  of the level.
Equation (\ref{Breit-Wigner}) can also be obtained directly from
Eq. (\ref{S-matrix}).   In the weak-coupling limit the width matrix $W W^\dagger$
is  small (compared with $\Delta$), and we can diagonalize ${\cal H}_{\rm eff}$ in
perturbation  theory. To leading order, the eigenfunctions are the same as
those  of $H$, and their widths (defined in terms of the
eigenvalues  of ${\cal H}_{\rm eff}$) are
$\Gamma_\lambda = 2 \pi \langle \lambda | W W^\dagger |\lambda\rangle=
\sum_c |\gamma_{c \lambda}|^2$ (where we have used $\gamma_{c \lambda}
=  \sqrt{2\pi} \sum_j W^\ast_{j c} \langle j | \lambda\rangle$).

 The zero-temperature conductance in the tunneling regime is obtained by substituting the Breit-Wigner form
[Eq. (\ref{Breit-Wigner})] in Landauer's formula:
\begin{equation}\label{G-Breit-Wigner}
G (E) = { e^2 \over h}  { \Gamma^l_\lambda \Gamma^r_\lambda \over (E -
E_\lambda)^2  + \left({\Gamma_\lambda \over 2} \right)^2} \;,
\end{equation}
where $\Gamma^{l(r)}_\lambda=\sum_{c \in \;l(r)}\Gamma_{c \lambda}$ is
the  width of the level $\lambda$ to decay into the left (right) lead and we have ignored spin degeneracy.

 At finite temperature the conductance is calculated by convoluting Eq. 
(\ref{G-Breit-Wigner}) with the derivative $f^\prime$ of the
Fermi-Dirac  distribution according to Eq. (\ref{finite-T-LIB}). In typical experiments with closed dots, $\bar \Gamma \ll T \ll \Delta$, and the thermal smearing factor $f'$  does not change much over
the  resonance width. $G$ is then proportional to the integral of the  Breit-Wigner form [Eq. (\ref{G-Breit-Wigner})],
\begin{equation}\label{conductance-peak}
G (E_F,T) \approx
G_\lambda^{\rm peak} {1 \over \cosh^2\left({E_\lambda -E_F \over 2kT}\right)}
\;.
\end{equation}
Here, $E_F$ is the Fermi energy in the leads and
\begin{mathletters}\label{peak-height}
\begin{eqnarray}
G^{\rm peak}_\lambda = {e^2 \over h} {\pi \bar \Gamma \over 4 kT} g_\lambda \label{G-to-g}\\\noindent  {\rm where} \;\;\;\; g_\lambda = {2 \over \bar\Gamma} {\Gamma^l_\lambda
\Gamma^r_\lambda  \over \Gamma^l_\lambda + \Gamma^r_\lambda } \;. \label{g-peak}
\end{eqnarray}
\end{mathletters}
Equation (\ref{conductance-peak}) describes a conductance peak of width $\sim kT$ that is centered at $E_F=E_\lambda$ and has a peak-height amplitude of $G^{\rm peak}_\lambda$.

\subsection{Coulomb blockade}
\label{Coulomb-blockade}

Coulomb blockade occurs when an ``island'' of  electrons is weakly coupled to
two  leads via tunnel junctions. When the coupling is weak, the conductance
drops  below $e^2/h$ and the charge on the dot becomes quantized.  The linear
conductance  of the dot oscillates as a function of the gate voltage
with a period that corresponds to the addition of a single electron to the
dot. For reviews on Coulomb blockade, see Averin and  Likharev (1991) and
van Houten, Beenakker, and  Staring (1992).  For recent experimentally oriented reviews on
transport  in Coulomb-blockade quantum dots, see Meirav and Foxman (1995),
Kouwenhoven, Marcus~{\em et~al.} (1997),  and Kouwenhoven and McEuen (1998).

 Coulomb-blockade oscillations were first observed in a metallic grain
(Giaever and Zeller, 1968),  where $\Delta \ll kT$
and  the spectrum could be treated as a quasicontinuum. A transport theory for
this  classical regime was developed by
Shekhter (1972) and Kulik and Shekhter (1975).
The total classical electrostatic energy of a Coulomb island with ${\cal
N}$  electrons is
\begin{equation}\label{electrostatic-energy}
U({\cal N}) = -Q V_{\rm ext} + Q^2/2C = - {\cal N} e V_{\rm ext} + {\cal N}^2 e^2 /2C
\;,
\end{equation}
where $V_{\rm ext}$ is the potential difference between the electron gas and the
reservoir  (due to a gate voltage), and $C$ is the total
capacitance  between the island and its surroundings.
By defining an externally induced charge variable $Q_{\rm ext}\equiv C V_{\rm ext}$,
we  can write Eq. (\ref{electrostatic-energy}) as $U({\cal N}) = (Q -
Q_{\rm ext})^2/2C$  up to an additive constant. $Q_{\rm ext}$ can be varied
continuously  by changing $V_{\rm ext}$. The number of electrons in the dot for a
given  $V_{\rm ext}$ is determined by minimizing $U({\cal N})$.  When
$Q_{\rm ext}={\cal  N}e$, the minimum is obtained for the state with charge
$Q={\cal  N} e$, and the energy of the states with $Q=({\cal N} \pm 1)e$ is
higher  by $e^2/2C$. As a result, the tunneling density of states has a gap of $E_C=e^2/C$ around the Fermi energy, blocking the flow of electrons into the island. This situation is demonstrated in panels (a) and (c) of  Fig. \ref{fig:CB}. However,
when  $Q_{\rm ext}=({\cal N}+1/2)e$, both states $Q={\cal N}e$ and $Q=({\cal
N}+1)e$  are degenerate (Glazman and Shekhter, 1989), allowing the tunneling of one more charge into the metallic particle [see panels (b) and (d) of Fig. \ref{fig:CB}]. The
conductance is maximal at this degeneracy point.

  In semiconductor quantum dots the mean level spacing is much larger than in
metal  grains of a similar size, and  experiments can easily probe the quantum
Coulomb-blockade  regime $T < \Delta \ll e^2/C$.  A simple Hamiltonian for the dot can be written by assuming electrons in a one-body confining potential plus an electrostatic energy  [Eq. (\ref{electrostatic-energy})]:
\begin{equation}\label{CI}
H_{\rm dot} = \sum_\lambda (E_\lambda-e \alpha V_g) a_\lambda^\dagger
a_\lambda^{}  + e^2 \hat{\cal N}^2/2C \;,
\end{equation}
where $a_\lambda^\dagger | 0 \rangle$ is a complete set of single-particle
eigenstates  in the dot with energies $E_\lambda$, and $\hat{\cal N}=\sum_\lambda
a^\dagger_\lambda  a_\lambda$ is the electron number operator in the dot.
The external potential of Eq. (\ref{electrostatic-energy}) $V_{\rm ext}= \alpha V_g$ is written in terms of a gate voltage $V_g$ and $\alpha
=C_g/(C_g  + C_{\rm dot})$, where $C_g$ is the gate-dot capacitance and $C_{\rm dot}$
is  the dot-leads capacitance.
The Hamiltonian (\ref{CI}) is known as the constant-interaction  model, since only the average constant part of the electron-electron interaction ($\hat{\cal N}^2 e^2/2C$)  is taken into account.

 When $T \ll \Delta$,
conductance  is possible only by resonant tunneling through a corresponding
quantized  level in the dot.  Resonant tunneling of the ${\cal N}$th electron
occurs  when the total energy before and after the tunneling event is
conserved:
$E_F + U({\cal N}-1) = E_{\cal N} +  U({\cal N})$.
Using Eq. (\ref{electrostatic-energy}) we find that the effective Fermi energy
$\tilde E_F \equiv E_F + e \alpha V_g$ satisfies $\tilde E_F  = E_{\cal N}
 +\left( {\cal N}-\frac{1}{2}\right)(e^2 / C)$.  The spacing between
 Coulomb-blockade  peaks is now given by
\begin{equation}\label{peak-spacing}
\Delta_2 \equiv \Delta \tilde E_F =  (E_{{\cal N} + 1} - E_{\cal N}) + e^2/C \;.
\end{equation}
 The charging energy is usually  much larger than the mean level
spacing  in the dot so that the Coulomb-blockade peaks are almost equidistant.
   Figure \ref{fig:CB} illustrates the phenomenon
of Coulomb blockade in the quantum  regime.

\begin{figure}
\epsfxsize= 8 cm
\centerline{\epsffile{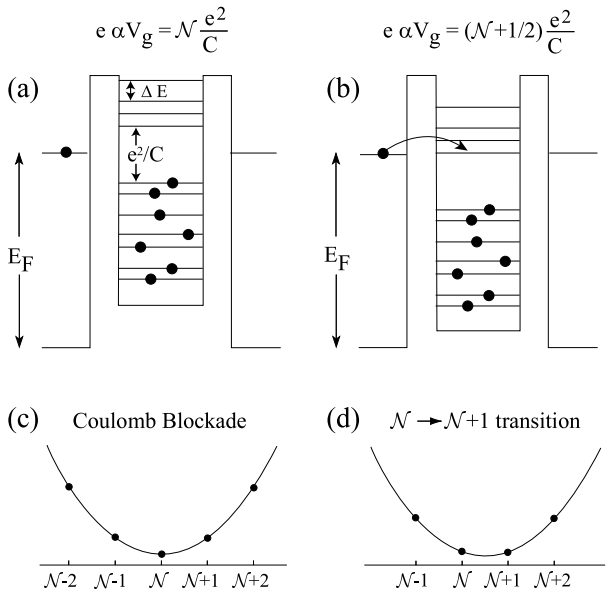}}
\vspace*{2 mm}
\caption{Schematic illustration of Coulomb blockade in an almost-closed dot. When the gate
voltage is tuned to a value $e\alpha V_g={\cal  N}e^2/C$ [panel (a)]
there is a charging energy gap in the single-particle spectrum on both sides of the Fermi energy,  blocking the tunneling of electrons into the dot.
 On the other hand, when the gate voltage increases to $e\alpha  V_g=({\cal N}+1/2)e^2/C$  [panel (b)], it compensates for the Coulomb repulsion, and the charging energy
gap for adding an electron to the dot vanishes.  When the Fermi energy in the leads  matches the
first unoccupied single-particle state in the dot,
resonant  tunneling of an electron into the dot occurs.
 Panels (c) and (d) show the total electrostatic  energy $U({\cal N})$
of the dot vs the number of electrons  [see Eq.
(\protect\ref{electrostatic-energy})]:
(c) for $e\alpha V_g={\cal
N}e^2/C$, the energy of the dot is minimal for ${\cal N}$
electrons  in the dot, leading to charge quantization;  (d) when
$e\alpha V_g=({\cal  N}+1/2)e^2/C$, the energies of a dot 
 with ${\cal N}$ and ${\cal  N}+1$ electrons are equal and
the dot's charge can fluctuate between ${\cal  N}e$ and $({\cal
N}+1) e$. This is known as the degeneracy point. }
\label{fig:CB}
\end{figure}

 At $T \sim \Delta$, several resonances contribute to the
conductance  peak.  A finite-temperature theory for the on-resonance conductance  was
derived  by Beenakker (1991) and Meir, Wingreen, and Lee (1991) using linear
response  theory, and by Averin, Korotkov, and Likharev (1991) in the nonlinear $I$-$V$ regime.

For $kT \gg \Gamma$ (where $\Gamma$  is a typical level width), the coherence between the electrons in the leads and in the dot can be ignored, and a master-equations approach is valid. Beenakker (1991) used the constant-interaction model (\ref{CI}), where an
eigenstate  of the dot is described  by a set of occupation numbers $\bbox
n\equiv  \{n_\lambda\}$ of the single-particle levels, and its
energy  is ${\cal E}(\bbox n) = \sum_\lambda E_\lambda n_\lambda + U({\cal
N})$.  The dot is
connected  to left and right reservoirs that are in thermal equilibrium at
temperature  $T$ and Fermi energy  $E_F$. At equilibrium
$P_{eq}(\bbox n) = {\cal Z}^{-1} \exp\left\{ -[ {\cal E}(\bbox n) - {\cal N} E_F]/kT
\right\}$,  where
 ${\cal Z}$ is the grand-canonical partition function. A small source-drain
voltage  $V_{sd}$ is applied between the left and right reservoirs, causing a
current  $I$ through the dot (see Fig. \ref{fig:tunnel-dot}), and the
linear  conductance is calculated from $G=I/V_{sd}$. The voltage drop is $\eta V_{sd}$
across  the left barrier, and $(1-\eta) V_{sd}$ across the right barrier. The
current  through the left lead is described by the net flux of electrons that
tunnel  in or out of the dot across the left barrier
\begin{eqnarray}\label{current}
I =  - e \sum\limits_\lambda \sum\limits_{\bbox n} & & P(\bbox n)
{\Gamma^l_\lambda \over \hbar} \left\{\delta_{n_\lambda, 0} f(\tilde
E_\lambda({\cal  N}+1) + \eta e V_{sd}) \right. \nonumber \\ && \left. - \delta_{n_\lambda, 1}[1- f(\tilde
E_\lambda({\cal  N})+ \eta e V_{sd})]\right\} \;,
\end{eqnarray}
where $f(x)=[1 + \exp(x/kT)]^{-1}$ is the Fermi-Dirac distribution in the
reservoirs, $\tilde E_\lambda({\cal N}) \equiv E_\lambda -(\tilde E_F
- ({\cal N}-1/2)e^2/C)$, and $\Gamma_\lambda^l/\hbar$ is the tunnel rate from level $\lambda$ to the left lead. 
 The first term on the right-hand side of Eq. (\ref{current}) describes
the  tunneling of electrons from a level in the left lead with filling $f$
into  an empty level $\lambda$ in the dot, while the second term represents
the  tunneling of an electron from an occupied state $\lambda$ in the dot to
a  level in the left lead with emptiness of $1-f$. Only energy-conserving transitions are taken into account, and
the summation over $\lambda$ accounts for the possibility of tunneling through different levels $\lambda$.

  The {\em  nonequilibrium  distribution}  $P(\bbox n)$ of electrons in the dot
satisfies  a set of master equations
\begin{equation}\label{master}
{\partial P(\bbox n) \over\partial t} = \sum\limits_{\bbox m} [ P(\bbox m)
{\cal  W}_{{\bbox m} \to \bbox n} - P(\bbox n){\cal W}_{\bbox n \to \bbox m}]
\;,
\end{equation}
where ${\cal W}_{\bbox m \to \bbox n}$ is the transition rate from state
$\bbox  m$ to state $\bbox n$. Only states $\bbox m$ that differ by one
electron  from the state $\bbox n$ appear in the sum in Eq. (\ref{master}), and
the  respective transition rates are
\begin{eqnarray}\label{transition-rates}
{\cal W}_{1_\lambda \to 0_\lambda} = & {\Gamma^{l}_\lambda \over \hbar} &
\left[1-f(\tilde  E_\lambda({\cal N}+1)+\eta e V_{sd})\right]  \nonumber \\ & + &
{\Gamma^{r}_\lambda  \over \hbar} \left[1-f(\tilde E_\lambda({\cal
N}+1)-(1-\eta)  e V_{sd})\right] \;, \nonumber \\
{\cal W}_{0_\lambda \to 1_\lambda}  = & {\Gamma^{l}_\lambda \over \hbar} &
f(\tilde  E_\lambda({\cal N})+\eta e V_{sd}) \nonumber \\ & + & {\Gamma^{r}_\lambda \over
\hbar}  f(\tilde E_\lambda({\cal N})-(1-\eta) e V_{sd})\;,
\end{eqnarray}
where ${\cal N}=\sum_\mu n_\mu$ is the number of electrons in the final state
$\bbox  n$. In the notation used in Eq. (\ref{transition-rates}), $1_\lambda \to 0_\lambda$, for example, 
denotes  the transition $\bbox m \to \bbox n$ with $m_\lambda=1, n_\lambda=0$, 
and  $m_\mu=n_\mu$ for all $\mu \neq \lambda$. The term ${\cal W}_{1_\lambda \to
0_\lambda}$  describes the tunneling of an electron from a filled $\lambda$
level  in a dot with ${\cal N}+1$ electrons into a level in the left and right
leads  with emptiness of $1-f$, while ${\cal W}_{0_\lambda \to 1_\lambda}$
corresponds  to the tunneling of an electron from a level in the left or right
lead  with filling $f$ into a dot with ${\cal N}-1$ electrons.

 We are interested only in the stationary state $\partial P(\bbox n)/\partial t
=0$,  leading to detailed balance equations
$P(1_\lambda){\cal W}_{1_\lambda \to 0_\lambda}= P(0_\lambda) {\cal
W}_{0_\lambda  \to 1_\lambda}$. A solution to the detailed
balance  equations in the linear response regime  (Beenakker, 1991) is  of
the  form $P(\bbox n) \approx P_{eq}(\bbox n)[1 + (eV_{sd}/kT)L(\bbox n)]$,
where
$L(\bbox n) = \sum_\lambda \left[\Gamma^r_\lambda /
(\Gamma^l_\lambda  + \Gamma^r_\lambda) -\eta \right]n_\lambda + {\rm const.}$
Calculating the current (\ref{current}) to first order in $V_{sd}$ using the solution for $P(\bbox n)$, one obtains the
conductance  $G(T,\tilde E_F)$ as a function of the temperature and the
effective  Fermi energy
\begin{eqnarray}\label{finite-T-G}
\begin{array}{lll}
 G (T,\tilde{E}_F) = \frac{e^2}{h}\, \frac{\pi \bar{\Gamma}}{4 kT} g\;,\;\;
{\rm where}
\;\; g = \sum_\lambda w_\lambda(T,\tilde{E}_F) g_\lambda \;
\end{array}
\end{eqnarray}
is the dimensionless conductance expressed as a thermal average over the
level  conductances
  $g_\lambda =  2 \bar{\Gamma}^{-1}  \Gamma_\lambda^l \Gamma_\lambda^r
   /( \Gamma_\lambda^l + \Gamma_\lambda^r)$.
The thermal weights are given by
\begin{eqnarray}\label{thermal-weight'}
w_\lambda =  & \sum_{\cal N} & 4 P_{\cal N} \langle n_\lambda \rangle_{_{\cal N}} \nonumber \\ 
& & \times \left[
1 - f\left(E_\lambda + ({\cal N}- 1/2){e^2 \over C}  - \tilde{E}_F \right)
\right]  \;,
\end{eqnarray}
where $P_{\cal N}$ is the probability that the dot has ${\cal N}$ electrons,
and  $\langle n_\lambda \rangle_{_{\cal N}} $ is the {\em canonical occupation}  of a level $\lambda$. The
contribution  to $w_\lambda$ in Eq. (\ref{thermal-weight'}) from a fixed number of
electrons  ${\cal N}$ corresponds to the product of the probability that level
$\lambda$  is occupied in a dot with ${\cal N}$ electrons and the probability
that  a state in the lead is empty at the same total energy.
 The probability  $P_{\cal N}$ is given by $P_{\cal N}= \exp[-\Omega({\cal
N})/kT]/\sum_{\cal  N'} \exp[-\Omega({\cal N'})/kT]$, where $\Omega({\cal
N})  \equiv F({\cal N}) +U({\cal N})- {\cal N} E_F$ and $F({\cal  N})$ is the canonical noninteracting free energy.

In typical experiments, $T,\Delta \ll e^2/C$, and only one term in the sum of  Eq. 
(\ref{thermal-weight'})  contributes to a given conductance peak. For gate
voltages  that are tuned to a conductance peak between ${\cal N}-1$ and ${\cal
N}$  electrons in the dot, we have
\begin{eqnarray} \label{thermal-weight}
w_\lambda = 4   f(\Delta F_{\cal N}- \tilde{E}_F)
\langle n_\lambda \rangle_{_{\cal N}} \left[
1 - f\left(E_\lambda  - \tilde{E}_F \right) \right]
\;,
\end{eqnarray}
where  $\Delta F_{\cal N} \equiv F({\cal N}) - F({\cal N}-1)$, and we have
used  $P_{\cal N} \approx f(\Omega({\cal N}) -\Omega({\cal N}-1))$.  In Eq. (\ref{thermal-weight}) and in the following, $\tilde E_F$ (or equivalently $e \alpha V_g$) is measured relative to $({\cal N}-1/2)e^2/C$.

In the limit $k T \ll \Delta$  and spin-nondegenerate levels,
 only one level $\lambda={\cal N}$ contributes to
Eq. (\ref{finite-T-G}).  Furthermore $\Delta F_{\cal N} \approx E_{\cal N}$, and
using  $f(x)[1-f(x)]=-kT f'(x)$, Eq. (\ref{thermal-weight}) becomes
$w_{\cal N}(kT \ll \Delta) = \cosh^{-2}
\left[(E_{\cal  N} - \tilde E_F) / 2kT\right]$,
so that Eq. (\ref{finite-T-G}) reduces to Eqs. (\ref{conductance-peak}) and
  (\ref{peak-height}). However, if the electron tunnels into an empty level $\lambda$ that is spin degenerate (with spin $1/2$), we find a conductance peak $G(\tilde E_F,T) = (e^2 / h) (2\pi \bar \Gamma / kT) (2 /\bar \Gamma) \left[\Gamma_\lambda^l \Gamma_\lambda^r / (\Gamma_\lambda^l + \Gamma_\lambda^r)\right] \left[1 -f(E_\lambda - \tilde E_F)\right]\\ \times [2 +  e^{(E_\lambda - \tilde E_F)/k T}]^{-1}$ (Glazman and Matveev, 1988). The conductance maximum is shifted to $\tilde E_F = E_\lambda - (kT \ln 2)/2$, and the peak height in Eq. (\ref{peak-height}) is rescaled by a factor of  $8 (\sqrt{2} -1)^2 \approx 1.37$. In this case, the scattering approach of Sec. \ref{resonant-tunneling} leads to a conductance peak as in (\ref{conductance-peak}), with a peak height (\ref{peak-height}) scaled by a factor of $2$.  This result is wrong since it ignores charging energy.

Similarly, if the method of Sec.  \ref{resonant-tunneling} (valid in the absence of charging energy) is applied at
temperatures $T \sim \Delta$,  we obtain Eq. (\ref{finite-T-G}),  but with weights
\begin{equation}\label{LIB-weights}
w^{(0)}_\lambda = -4kT f'(E_\lambda-\tilde E_F) \;.
\end{equation}
Indeed in the limit $e^2/C \to 0$, all terms in Eq. (\ref{thermal-weight'})
contribute.  The factor $1-f$ becomes independent of ${\cal N}$,
$\sum_{\cal  N} P_{\cal N} \langle n_\lambda \rangle_{\cal N} \equiv
f(E_\lambda  - \tilde E_F)$, and the weights $w_\lambda$ reduce to Eq. 
(\ref{LIB-weights}). However, for $e^2/C \gg \Delta$,
 only  $P_{\cal N}$ and $P_{{\cal N}-1}$ are non-negligible,  as states with number
  of electrons different from ${\cal N}$ and ${\cal N}-1$ are pushed
away  by the charging energy.  We conclude that, in the presence of charging energy,  the approach based on the Breit-Wigner and Landauer formulas breaks down at finite
temperature and Eq.  (\ref{LIB-weights}) must be replaced by Eq. (\ref{thermal-weight}).

  The discreteness of the spectrum is unimportant at high temperatures
$k T  \gg \Delta$ where the canonical occupations can be
approximated  by the Fermi-Dirac distribution.  Assuming that the tunneling
rates  $\Gamma^{l(r)}/\hbar$ depend only weakly on energy, one finds (for
$\Delta  \ll kT \ll e^2/C$)
\begin{eqnarray}\label{conductance-peak-T}
& & G (\tilde E_F,T) \approx  G_\lambda^{\rm peak} {1 \over \cosh^2\left({\mu - \tilde
E_F  \over 2.5 kT}\right)}\;; \nonumber \\&  & G_\lambda^{\rm peak} =  {e^2 \over h}  {\pi
\over  \Delta} {\Gamma^l \Gamma^r \over \Gamma^l + \Gamma^r} \;,
\end{eqnarray}
 where $\Gamma^{l(r)}$ are the energy-averaged partial widths.
  Comparing with the corresponding formulas in
the  quantum limit $k T \ll \Delta$ [Eqs. (\ref{conductance-peak}) and
(\ref{peak-height})],  we see that for the same temperature, the line shape is
similar  but with a somewhat larger width in the classical limit. The peak
  height in the classical regime is temperature independent since $\sim
kT/\Delta$  levels contribute, canceling
the $1/T$ dependence of the peak height in the quantum regime.  Furthermore,
 the widths $\Gamma^{l(r)}$ in the classical
formula (\ref{conductance-peak-T}) are energy averaged, and thus lead to smooth
 variation of the peak heights  as a function of $V_g$. In contrast, the peak
  heights exhibit strong fluctuations in the quantum regime [see, for example, the
   peak series in Fig. \ref{fig:closed-dot}(c)].

If $kT \gg e^2/C$, the charging energy is  not important and
Landauer's formula reproduces the correct
conductance
$G^{(0)} = 2 (e^2 / h) (\pi / \Delta) \Gamma^l \Gamma^r / (\Gamma^l + \Gamma^r)$.
Coulomb-blockade oscillations disappear and the conductance is  large as its peak height in the classical Coulomb-blockade regime.

\subsection{Cotunneling}
\label{cotunneling}

 The off-resonance conductance can be calculated using perturbation theory (in the dot-leads coupling).
First-order  processes are forbidden by energy conservation because of the
charging  energy gap, so the leading-order contribution
is  second order.   The main second-order tunneling mechanism is known as
cotunneling  (Averin and Nazarov, 1990; Glazman and Matveev, 1990a).  In {\em inelastic cotunneling}, an electron that tunnels from the left lead into a state in the dot is followed by an electron that tunnels from a different state of the dot into the right lead. In {\em elastic cotunneling}, an electron tunnels into the dot (from the left lead) and out of the dot (into the right lead) through the same intermediate state of the dot.   Also contributing to cotunneling
are  holes that move from the right to the left lead, describing first an
electron  that tunnels out of the dot into the right lead, followed by an
electron  that tunnels from the left lead into the dot.  The transitions
involved  in cotunneling are non-energy-conserving and therefore {\em
virtual}.  The intermediate single-particle states in the dot are separated
from  the Fermi energy in the leads by a gap $E_e$  for the virtual tunneling
of  an electron to a state above the Fermi energy, and by a gap $E_h$ for the
virtual  tunneling of a hole to a state below the Fermi energy ($E_e+
E_h=E_C=e^2/C$).

We restrict the discussion to low temperatures $kT < \sqrt{E_C \Delta}$, where elastic cotunneling dominates (Averin and Nazarov, 1990).
The dot-leads Hamiltonian is described by
\begin{eqnarray}\label{Hamiltonian}
H = && + \sum_{k, c \in l,r} E_{k c} c_{k  c}^{\dagger}
c_{k  c}^{} \nonumber \\& & +  \sum_{k, c \in l,r \atop \lambda} (V_{k c,
\lambda}^{} c_{k c}^{\dagger} a_\lambda^{} + H. c.) \;,
\end{eqnarray}
where $H_{\rm dot}$ is the dot's Hamiltonian (\ref{CI}), $c_{k c}^\dagger$
 creates an electron with
wave number $k$ in channel $c$ in either the left ($l$) or right ($r$)
lead with energy $E_{k c}$, and
$V$  is a tunneling matrix element between the left or right lead and the dot.
 The elastic cotunneling conductance is
 \begin{eqnarray}\label{cotunneling-G}
G   =  {e^2 \over h} \sum_{c\in l, c^\prime \in r} |{\cal T}_{c^\prime c}|^2
\;,
\end{eqnarray}
where
\begin{eqnarray}\label{cotunneling-amplitude}
  {\cal T}_{c^\prime c}  =   -\sum_{E_\lambda > E_F} && {\gamma^{r \ast}_{c^\prime
  \lambda}{\gamma^{l}_{c \lambda}}  \over
 |E_\lambda-E_F| + E_e} \nonumber \\ && + \sum_{E_\lambda \le E_F} {\gamma_{c^\prime
\lambda}^{r  \ast}\gamma_{c \lambda}^l
 \over  | E_F- E_\lambda |+ E_h}
\end{eqnarray}
is the elastic cotunneling amplitude from channel $c$ in the left lead to channel
$c^\prime$  in the right lead.
We have defined  $\gamma^{l(r)}_{c \lambda}= \sqrt{2 \pi \rho_c}
V^{l(r)}_{c  \lambda}$ [$\rho_c(E) = \sum_k  \delta(E - \epsilon_{kc})$
is  the lead density of states in channel  $c$] to be the partial-width
amplitude  of an electron in a state $\lambda$ to decay into channel $c$ in
the  left (right) lead. The cotunneling amplitude contains contributions from both particle ($E_\lambda >E_F$) and hole
($E_\lambda  \leq E_F$) states. Each term in the sum over
particle  (hole) states is the amplitude for an electron (hole) to tunnel from
the  left (right) lead to the right (left) lead through an intermediate
 state $\lambda$. Expression (\ref{cotunneling-amplitude}) assumes that both $E_e$ and $E_h$ are of the order of $E_C$, and thus $\gg \Delta$.
In contrast to the conductance at the peak, a large number of excited states in the dot contribute to the
off-resonance  conductance.

\subsection{Nonlinear transport}
\label{non-linear-transport}

 Thus far we have considered only the linear conductance. This corresponds to
a source-drain voltage $eV_{sd}$ that is smaller than a typical level spacing
$\Delta$  in the dot. The electron can then tunnel only through the lowest
unoccupied  level, and the observed Coulomb-blockade oscillations provide
information  on the ground states of the dot with increasing number of
electrons.  More generally, the current depends on the number of available
states in the dot between the chemical potentials in the left and right
leads.  Since the difference between the chemical potentials is $eV_{sd}$, we
expect  that as $V_{sd}$ increases additional states in the dot become
available  for tunneling. A nonlinear transport theory in quantum dots was
developed  by Averin and Korotkov (1990) and Averin, Korotkov, and Likharev (1991). In the
classical  regime, the current increases in steps as a function of $V_{sd}$
(Coulomb  staircase), corresponding to the increase in the number of available
charge  states in the dot.  In the quantum regime, the current depends on the
number  of available excited levels in the dot through which a fixed number of
electrons  can tunnel (Johnson~{\em et~al.}, 1992). Thus nonlinear-transport measurements
in  the quantum regime provide information on the excitation spectrum of a dot
with  a fixed number of electrons.

  In practice, the current through the dot can be measured as a function of
both $V_{sd}$ and a gate voltage $V_g$. For a
fixed  $e V_{sd}$ below the first excited state in the dot, the
current versus $V_g$ displays the usual Coulomb-blockade peaks of the
linear  regime. However, when $eV_{sd}$ is above the first excited state in
the  dot, the electron can tunnel through two states in the dot, and each
single  peak develops into a double peak. Similarly, as $e V_{sd}$ increases
above  the second excited state, each Coulomb-blockade oscillation is composed
of  three peaks. The differential conductance $dI/dV_{sd}$ displays a peak
when  a level in the dot matches the chemical potential in one of the leads.
From  the spacings among the peaks in each oscillation, it is possible to
infer  the excitation spectrum of the dot. The differential conductance forms a diamond-shaped diagram
in the $V_{sd}$-$V_g$ plane (McEuen~{\em et~al.}, 1993).   Each diamond corresponds to a Coulomb-blockaded region with
a  certain number of electrons on the dot.

\section{Statistical Theory: from Disordered Metals to Ballistic Dots}
\label{statistical-theory}

 The phase coherence of transport in
mesoscopic  structures leads to quantum interference effects. Consequently,  the conductance in these structures exhibits fluctuations as a function of experimentally
controllable  parameters, such as magnetic field, gate voltage, or sample.
  Such fluctuations were first observed in disordered metals. To describe the statistical properties of these fluctuations, we
assume that the impurity configurations in the sample constitute an ensemble. Similar samples fabricated by similar methods differ from
each  other in the details of their impurity configurations, and each can be
thought  of as a different member of the ensemble.
For early reviews on quantum interference effects in mesoscopic structures,
 see Altshuler, Lee, and Webb (1991),
Beenakker and van Houten (1991),  and Washburn and Webb (1993). See also the books by
Datta (1995)  and  Imry (1996).

In the last decade it has become possible to produce relatively clean
high-mobility GaAs quantum dots. These ballistic devices also exhibit conductance
fluctuations.  In irregularly shaped dots, where the classical dynamics of the
electron  is chaotic, the fluctuations are universal and depend only on the
symmetry  class and the transmission properties of the leads.  The
physical  origin of the fluctuations in ballistic dots is similar to that in
diffusive  structures, namely quantum interference effects. The statistical properties of these fluctuations are also assumed to be described by an appropriate ensemble. In practice, the concept of an ensemble (for both diffusive and ballistic systems) is justified by the ergodic hypothesis. These systems' statistical properties result from averaging over energy, shape, magnetic field, etc.

 The diagrammatic and semiclassical methods played a key role in
our  theoretical understanding of disordered structures and ballistic dots,
respectively.  Two additional powerful approaches contributed to recent
progress  in the field: random-matrix theory (RMT) and the supersymmetry
method.   Random-matrix theory originated in nuclear physics (Wigner, 1958; see also Wigner, 1951, 1955, 1957), and was later
conjectured  to describe the universal quantal fluctuations in systems whose
associated  classical dynamics is chaotic (Bohigas, Giannoni, and Schmit, 1984).  The supersymmetry
method  was originally conceived as a method for carrying out exact ensemble
averages  in disordered systems (Efetov, 1983). In a certain regime it is
equivalent  to RMT, but more generally it can  be used to derive
nonuniversal corrections.

 Section \ref{disordered-ballistic} introduces the relevant
scales  and commonly used models for disordered and ballistic quantum dots.
The  following sections review the principal methods of the
statistical  approach.  Section \ref{semiclassical} outlines
the  semiclassical approach.  The semiclassical treatment of
disordered  systems has mostly reproduced results originally derived in
impurity  perturbation theory, but it offers a more
intuitive  approach.  The diagrammatic approach is not
discussed  here; its main lines can be found in Abrikosov, Gor'kov, and Dzhyaloshinskii (1963) and
Altshuler and Simons (1995).  Section \ref{RMT} reviews the RMT approach, and Sec. \ref{supersymmetry} gives a
brief introduction to the supersymmetry method.

\subsection{Disordered metals and ballistic dots}
\label{disordered-ballistic}

\subsubsection{Scales in the diffusive regime}
\label{scales}

In a mesoscopic structure, the coherence length exceeds the
system's  size: $L_\phi > L$ (see Sec. \ref{intro}). Other relevant length
scales in disordered systems are the mean free path $l$ and the Fermi wavelength $\lambda_F$.
The {\em diffusive regime} corresponds to  $l \ll L$.
In the {\em metallic} or {\em weakly
disordered  regime}, the Fermi wavelength is much smaller
than  the elastic mean free path:
$\lambda_F \ll l$ (i.e., $k_F l \gg 1$).

Another important length scale in disordered systems is the localization length $\xi$ over which the electron's wave function is localized (for a review of localization
theory see Lee and Ramakrishnan, 1985). In the metallic regime, the localization
length is large, and we shall restrict our discussions to the nonlocalized regime where $ \xi \gg L$.

 The time the electron takes to diffuse across the length $L$ of a disordered sample is $\tau_D=L^2/D$, where $D$ is the diffusion coefficient. The associated energy scale
\begin{equation}\label{Thouless-energy}
E_c \equiv {\hbar \over \tau_D} =  {\hbar D \over L^2}
\end{equation}
is known as the Thouless energy (see Sec. \ref{disordered-to-ballistic}). The Thouless energy can be directly related to the conductance.
 The conductance $G$ of a homogeneous conductor in $d$ dimensions is $G \sim \sigma L^{d-2}$, where $\sigma$ is the conductivity.
Using  Einstein's relation (\ref{Einstein}) for $\sigma$, we can write
\begin{equation}\label{G-estimate}
G \sim {e^2 \over \hbar} (\nu L^d) \left( {\hbar D \over L^2}\right) =  {e^2
\over  \hbar} {E_c \over \Delta}\;.
\end{equation}
  The
dimensionless Thouless conductance\footnote{The dimensionless Thouless conductance is usually denoted by $g$. In this review we denote it by $g_T$ since $g$ is used to denote the dimensionless conductance peak height in a closed dot [see, e.g., Eq. (\ref{finite-T-G})] and the dimensionless conductance in an open dot (see, e.g., Sec. \ref{conductance-distributions}).} $g_T$ is defined by  $G \equiv
(e^2/\hbar)g_T$,  and according to Eq. (\ref{G-estimate})
 \begin{equation}
g_T = {E_c \over \Delta}  \equiv N(E_c)\;,
\end{equation}
i.e., $g_T$ measures the number of levels in an energy interval $E_c$.
  In 2D systems, $g_T = l/ \lambda_F$, and
the dimensionless conductance is therefore large in the metallic regime.

 The relations among relevant length scales can be translated into relations among the corresponding energy and time scales. The following relations are useful:
\begin{eqnarray}
\frac{\hbar/\tau}{E_c} = d \left({L \over l} \right)^2 \;;\;\;\;\;
\frac{E_F}{\hbar/\tau} = \frac{1}{2} k_F l \;,
\end{eqnarray}
where $\tau$ is the mean free time.
It follows that the diffusive regime is characterized by $E_c \ll {\hbar/ \tau}$, i.e., the time to diffuse across the sample is large compared with the
mean free time: $\tau_D \gg \tau$.  In a good metal ${\hbar/\tau} \ll E_F$ and $E_c \gg \Delta$.

In summary, the following inequalities hold in a disordered metal in its diffusive regime:
\begin{mathletters}
\begin{eqnarray}
\Delta \ll E_c \ll \hbar/\tau \ll E_F \;, \\
\tau_H \gg \tau_D \gg \tau \gg \hbar/E_F \;,
\end{eqnarray}
\end{mathletters}
where $\tau_H=h/\Delta$ is the Heisenberg time.
  The energy region above $\hbar/\tau$ corresponds to ballistic motion of the
electron  since the corresponding time scale is shorter than the average time between scatterings from the impurities.  In the energy range between $E_c$ and
$\hbar/\tau$,  the dynamics are diffusive, but there is not sufficient time
for  the electron to reach the boundaries. Energy scales below
$E_c$  correspond to time scales in which the electron has reached the system's
boundaries  and the diffusive motion has explored the full length of the structure. We shall see (e.g., in Sec. \ref{semiclassical:spectral}) that this is the regime where the fluctuations are universal (known as the ergodic regime).

\subsubsection{Scales in the ballistic regime}
\label{ballistic}

   For weaker disorder and/or a smaller sample,
the  mean free time $\tau$ increases and/or the time $\tau_D$ to diffuse across the dot
 decreases, and eventually the Thouless energy $E_c$  exceeds $\hbar/\tau$. In this limit the system can be
considered  clean, and the dynamics across its length are {\em
ballistic}.   In terms of length scales, the ballistic regime is defined by
$L \ll l$.

In ballistic structures $\tau_D$ is meaningless and another time scale becomes relevant: the {\em ergodic time}
$\tau_c$,  which is of the order of the time of flight across the sample. The
ergodic  time plays the same role in ballistic systems that the diffusion time
$\tau_D$  plays in disordered systems.  The quantity analogous to the Thouless
energy  is $E_T \equiv\hbar/\tau_c$, sometimes called the ballistic
Thouless  energy.  The ballistic  dimensionless conductance is $g_T=E_T/\Delta =
\tau_H/2\pi  \tau_c$.
In 2D  we can estimate $g_T$ from $\tau_c \sim L/v_F$ and $\tau_H=h
\nu  {\cal A}$ to be $g_T \sim  {\cal N}^{1/2}/2\pi$ (where ${\cal N}$ is the number of electrons in the dot).

  Within the ballistic regime (where $\hbar/\tau \ll \hbar/\tau_c$), it is possible to distinguish two cases depending on the relation between $\hbar/\tau$ and $\Delta$ (Altland and Gefen, 1993).  When $\Delta \ll \hbar/\tau \ll \hbar/\tau_c$, a typical impurity matrix element can mix many levels, while for $\hbar/\tau \ll \Delta, \hbar/\tau_c$ the disorder is very weak and can be treated in low-order perturbation theory.

\subsubsection{Models of disordered structures and ballistic dots}
\label{disordered}

  The dynamics of a single electron in d dimensions
 is described by
Schr\"odinger's equation
\begin{equation}
{1\over{2m^\ast}}\left(\bbox{p}+{e\over c}\bbox{A}\right)^2\Psi+V\Psi=E\Psi\;,
\end{equation}
where $V(\bbox{r})$ is a one-body potential and
$\bbox{A}(\bbox{r})$ is a vector potential describing a magnetic field.  The disorder is modeled by an ensemble of
random potentials $\{V(\bbox{r})\}$. Often this ensemble is taken to be Gaussian with
\begin{eqnarray}
\overline{V(\bbox{r})} =  0 \;; \;\;\;\;
\overline{V(\bbox{r}) V(\bbox{r}^\prime)}= {\hbar \over 2\pi \nu \tau}
\delta(\bbox{r}-\bbox{r}^\prime)\;.
\end{eqnarray}
 The parametrization of the strength of the disorder in terms of the
impurity  scattering rate $1/\tau$ is obtained in the
Born  limit. A discretized version of this
model is known as the tight-binding Anderson model (Anderson, 1958).
In second-quantized form
\begin{eqnarray}\label{AndModel}
   H = -\sum\limits_{\langle m,n\rangle} \left( t_{mn}e^{i \theta_{m n}}
 a^\dagger_m a_n + {\rm h.c.} \right) + \sum\limits_{m} V_m a^\dagger_m a_m
\;,
\end{eqnarray}
 where $a^\dagger_m$ creates an electron at site $m$,
 $t_{mn} = \hbar^2/2m^\ast a^2$ ($a$ is the lattice spacing) is a hopping
matrix  element between nearest neighbors $\langle m,n\rangle$, and
$\theta_{mn}  = {e \over\hbar c} \int_m^n\bbox{A}\cdot d \bbox{\ell}$ is an
Aharonov-Bohm  phase.
 $V_m$ is the disorder potential at site $m$, often assumed to be uniformly
distributed  over the interval
$[-w/2,w/2]$. The disorder parameter $w$ determines the elastic mean free path, and in the Born approximation $k_Fl \propto (w/t)^{-2}$.
  The Anderson model has become
the  standard model for describing single-particle dynamics of disordered
mesoscopic systems. In 3D, $w/t$ has a critical value above which all states
become  localized and the conductivity falls to zero, corresponding to a
metal-insulator  transition (Lee and Ramakrishnan, 1985). One-parameter renormalization-group analysis has shown that in 1D and 2D all states are localized.
 However, in 2D the localization length $\xi$ is
exponentially  large for weak disorder, and most states are extended over the dimension of the system.

Ballistic dots are often modeled as cavities.  Billiard models are popular
because  their mathematical properties are best known and there are efficient
methods to solve them numerically. By changing shape
parameters in billiard models, it is possible to describe systems with
classical  motion ranging from integrable to fully chaotic. A good example  is the conformal billiard
(Robnik, 1983), whose shape is determined by the image of the
unit  circle in the complex $z$ plane under the conformal mapping $w(z) =
(z+bz^2+c\mbox{e}^{i\delta}z^3)/\sqrt{1+2b^2+3c^2}$.

\subsection{The semiclassical approach}
\label{semiclassical}

 As a coherent phenomenon, transport in mesoscopic systems should be
described  by quantum mechanics. A classical treatment, on the
other  hand, has the advantage of physical intuition. The semiclassical
approach  is a bridge that seeks to  describe quantum-mechanical phenomena in the language of classical
physics.  The semiclassical approach played an
important  role in the development of the mesoscopic theory of
 open ballistic dots where the approximation of non-interacting quasiparticles holds (Sec. \ref{open-dots}).
 Applications to disordered systems have, for the most part, confirmed results derived
earlier  in the diagrammatic approach.

 Most of the applications are based on an expansion of the Green function in
terms  of classical trajectories. The retarded Green
function  $G^R(\bbox r, \bbox r';t) \equiv \langle \bbox r'| e^{-iH t/\hbar}|
\bbox  r\rangle$ propagates the particle from $\bbox r$ at $t=0$ to $\bbox r'$
at  a later time $t$. In Feynmann's path-integral representation, $G^R(\bbox
r,  \bbox r';t)= \int_{\bbox x(0)=\bbox r}^{\bbox x(t)=\bbox r'}
D[\bbox  x] e^{i S[\bbox x]/\hbar}$ is described as a functional integral over all trajectories
$\bbox   x(t)$ that connect $\bbox x(0)=\bbox r$ to $\bbox x(t)=\bbox r'$,
where $S[\bbox x]= \int_0^t d\tau[ m (d\bbox x/d \tau)^2/2 -
V(\bbox  x)]$ is the action. In the limit $\hbar
\to  0$ one can use the stationary phase approximation, leading to $\delta S =0$, Hamilton's variational
principle  for the {\em classical} trajectories $\bbox x_{\alpha}$ between
$(\bbox  r,0)$ and $(\bbox r',t)$. Small quantal fluctuations around each of
the  classical solutions $\bbox x_{\alpha}$ are included by expanding the
action  to second order $S[\bbox x]\approx S[\bbox x_{\alpha}] + \delta^2
S[\bbox  x_{\alpha}]/2$ and doing the Gaussian integral. The result is Van
Vleck's  formula,
\begin{equation}\label{sc-Green-time}
G^R(\bbox r, \bbox r';t) \approx \sum\limits_{\alpha \in \; \{\bbox r, \bbox
r';  t\}} A_\alpha e^{i S_\alpha /\hbar}\;,
\end{equation}
where the sum is over all classical paths $\alpha$ between $(\bbox r, 0)$ and
$(\bbox  r',t)$ with action $S_\alpha = S[\bbox x_\alpha]$ and amplitude
$A_\alpha$  given by
\begin{eqnarray}
 A_\alpha & = & \left( {1 \over 2 \pi i \hbar} \right)^{d\over 2}
\left| \det \left( -
{\partial^2 S_\alpha \over \partial \bbox r' \partial \bbox r} \right)
\right|^{1/2}
 e^{-i {\pi \over 2} \nu_\alpha} \;; \nonumber \\
S_\alpha & = & \int_{\bbox x(0)=\bbox r}^{\bbox x(t)=\bbox r'} (\bbox p \cdot d
\bbox  x - H d t)
\;.
\end{eqnarray}
The classical action $S_\alpha(\bbox r,0;\bbox r',t)$ is a function
of  the initial $(\bbox r , 0)$ and final $(\bbox r', t)$. In $d$ degrees of
freedom,  $-\partial^2 S_\alpha/\partial \bbox r' \partial \bbox r$ is a $d\times
d$  matrix, and the integer phase index $\nu_\alpha$  is the number of its
negative  eigenvalues (equal to the number of conjugate points along the
path). Since $-\partial S_\alpha/\partial\bbox r = \bbox
p_\alpha$  is the initial momentum, this matrix can also be written as
$\partial  \bbox p_\alpha/\partial \bbox r'$.

  The energy representation of the retarded Green's function is the Fourier
transform  of Eq. (\ref{sc-Green-time}). Doing the time integral by stationary phase, we find
\begin{eqnarray}\label{sc-Green-energy}
 G^R(\bbox r, \bbox r' ;E)& = & \int_0^\infty dt e^{iEt/\hbar}
G^R(\bbox  r, \bbox r', t) \nonumber \\ &\approx &\sum\limits_{\alpha \in \; \{\bbox r, \bbox
r';  E\}} \tilde A_\alpha e^{i \tilde S_\alpha /\hbar}\;,
\end{eqnarray}
where now the sum is over all classical paths $\alpha$ with energy $E$ that
begin  at $\bbox r$ and end at $\bbox r'$.  The modified action $\tilde
S_\alpha  = \tilde S_\alpha(\bbox r,\bbox r';E) = S_\alpha + E t$ and amplitude $\tilde A_\alpha$ are
\begin{eqnarray}\label{sc-amplitude-action}
\tilde A_\alpha & =  & {1 \over i\hbar}\left( {1 \over 2 \pi i \hbar} \right)^{d-1
\over  2}
\left| \det \left(
{\partial \bbox{p}_\alpha \over \partial \bbox{r}'} \right) \right|^{1/2}
\left|{dT_\alpha \over dE} \right|^{1/2} e^{-i {\pi \over 2}
\tilde\nu_\alpha}
 \;; \nonumber \\
\tilde{S}_\alpha& =& \int_r^{r'} \bbox{p}\cdot d \bbox{r}
 \;,
\end{eqnarray}
where $T_\alpha = \partial \tilde S_\alpha/\partial E$ is the duration of orbit
$\alpha$, the derivative matrix $-\partial \bbox p_\alpha/\partial\bbox r'$ is evaluated at $t=T_\alpha$, and $\tilde\nu_\alpha$ is a modified phase
index (see, e.g., Reichel, 1992).

Equation (\ref{sc-Green-energy}) is the starting point of the semiclassical
approximation for quantities that can be expressed in terms of
Green's  functions.  For example, the density of states  $\rho(E) = \sum_i
\delta(E-E_i)$  can be written as $\rho(E) = -\pi^{-1} \int\; d \bbox r
 {\rm Im}\; G^R(\bbox r, \bbox r;E)$.
Using Eq. (\ref{sc-Green-energy}) and integrating
over  $\bbox r$ in the saddle-point approximation leads to a sum over periodic  orbits (Gutzwiller 1967, 1969, 1970, 1971; Balian and Bloch, 1972). The level density is
decomposed  into a smooth average part (Weyl's term) and a fluctuating part:
$\rho  = \bar{\rho} + \rho_{\rm fluct}$. In a fully chaotic system, the periodic orbits are isolated, and $\rho_{\rm fluct}$ can be written in terms of Gutzwiller's trace formula (Gutzwiller, 1971)
\begin{equation}\label{trace-formula}
\rho_{\rm fluct}(E) = {1 \over \pi \hbar} \sum\limits_{{\rm p.o.}\;\alpha}
{T_\alpha  \over |\det(\tilde M_\alpha - I)|^{1/2}} \cos \left({\tilde
S_\alpha  \over \hbar} - \sigma_\alpha {\pi \over 2}\right)\;,
\end{equation}
where the sum is over periodic orbits $\alpha$. The phase $\sigma_\alpha$ is the Maslov index (containing the phase
index  $\tilde \nu_\alpha$) and $\tilde M_\alpha$ is a $(2d-2)$-dimensional
stability  matrix of the orbit. $\tilde M_\alpha$ is a submatrix of the
$2d$-dimensional  monodromy matrix $M_\alpha$ that describes the linear
relation  between a small change in the initial and final (i.e., after one
period)  $\delta \bbox r$ and $\delta \bbox p$.

\subsubsection{Spectral correlations in chaotic and disordered systems}
\label{semiclassical:spectral}

Spectral properties of metallic grains have long been of interest.
Gor'kov and  Eliashberg (1965)  studied the electrical polarizability of small metallic
grains  by assuming RMT spectral fluctuations (see Sec.
\ref{RMT}).   More recently, it became possible to do spectroscopy of
low-lying  states in quantum dots (Sivan~{\em et~al.}, 1994).  Here we discuss briefly the semiclassical calculation of spectral correlation in both chaotic and disordered systems. For reviews on chaos see, for example, Gutzwiller (1990) and Giannoni, Voros,  and Zinn-Justin (1991). For a recent
review  of spectral correlations in disordered systems, see
Dittrich (1996). In general, semiclassical methods are valid at energy scales that are large compared with the mean level spacing $\Delta$, namely, at time scales below the Heisenberg time $\tau_H$.

An important statistical measure of spectral correlations is
the two-point correlation function of the density of states, measuring the
correlations  of $\rho$ at two different energies, $E$ and $E+\Omega$.
Measuring  energies in units of the mean level spacing (i.e., $\epsilon \equiv
E/\Delta$  and $\omega \equiv\Omega/\Delta$), the correlator becomes
dimensionless:
 \begin{equation}\label{K-omega}
K(\omega) \equiv \overline{\rho(\epsilon) \rho(\epsilon +
\omega)}  - \overline{\rho}^2 \;.
\end{equation}
The Fourier transform  $K(t) = \int d\omega K(\omega) e^{i \omega t}$ is
known  as the spectral form factor.

We first derive a semiclassical  expression for  $K(t)$ (Berry, 1985) in a chaotic system using the periodic orbit expansion (\ref{trace-formula}).
Measuring  time in units of  $\hbar/\Delta$ (so that the Heisenberg time
is $\tau_H = 2\pi$)  and using
$\tilde  S_\alpha(E + \Omega)\simeq  \tilde S_\alpha(E) + \hbar T_\alpha
\omega$,  one finds
$K(t) \sim (2\pi)^{-1} \overline{\sum_{\alpha,\beta} {\cal A}_\alpha
{\cal  A}_\beta e^{{i \over \hbar}(\tilde S_\alpha - \tilde S_\beta)}
\delta\left[t  - (T_\alpha + T_\beta)/2 \right]}$,
where ${\cal A}_\alpha \equiv T_\alpha/|\det(\tilde M_\alpha - I)|^{1/2}$, and the average is over the energy $E$. In the {\em diagonal approximation},  only pairs of orbits with $\alpha =\beta$ are taken into
account,  and
\begin{equation}\label{K-t-sc}
K_{\rm sc}(t) \sim  (2\pi)^{-1}\overline{\sum\limits_{\alpha} {\cal
A}_\alpha^2  \delta(t-T_\alpha)} \approx  |t|/2\pi  \;.
\end{equation}
In the last step in deriving Eq. (\ref{K-t-sc}), we have used the classical sum rule of  Hannay and Ozorio de Almeida (1984) for ergodic systems. This sum rule is valid for times that are long compared with a typical period of the short periodic orbits ($\sim \tau_c$), 
but  much shorter than the Heisenberg time.
 We shall see that this result is also the universal RMT result below the
Heisenberg time. In deriving Eq. (\ref{K-t-sc}) we have assumed a
system  in which time-reversal symmetry is broken. For conserved time-reversal
symmetry  we must also consider pairs of orbits that are time reversals of each
other, and this will increase $K(t)$ by a factor of $2$.

  The semiclassical approach to disordered metals can be found in Argaman, Imry and Smilansky (1993) and Montambaux (1997).
For $t \ll \tau_H$, the form factor $K(t)$ can be related semiclassically
to  the return probability $P(t) = |\langle \bbox{r} | e^{-i H t/\hbar} |
 \bbox{r}\rangle |^2$ (which measures the
probability  that the electron will return to its original starting point
$\bbox{r}$  after time $t$): $K(t) = (2 \pi)^{-1} t P(t)$.
For diffusive motion the classical return probability is calculated from
\begin{equation}\label{return-prob}
P_{\rm cl}(t) = D(\bbox{r},\bbox{r};t) = \sum\limits_{\bbox{q}} e^{-D q^2 t} \;.
\end{equation}
In the case of conserved time-reversal symmetry there is an
additional contribution from the constructive interference of orbits that are
time-reversed pairs. We thus have $P(t) = (2/\beta) P_{cl}(t)$, where
$\beta=1$  for conserved time-reversal symmetry and $\beta=2$ for broken
time-reversal  symmetry.
The Fourier transform of Eq. (\ref{return-prob}) is $P(\omega) = \sum_{\bbox{q}}
(-i  \omega + D q^2)^{-1}$, and the semiclassical two-point correlation
function  is given by
\begin{equation}\label{K-classical}
K_{sc}(\omega)= {1 \over \beta \pi^2} {\rm Im} {\partial P \over \partial
\omega}  = -{1 \over \beta \pi^2} {\rm Re} \sum\limits_{\bbox{q}} {1 \over (-i
\omega  + D q^2 + \gamma)^2}\;,
\end{equation}
where an additional broadening
$\gamma \sim \Gamma_\phi=\hbar/\tau_\phi$ is introduced to take into account the electron's finite coherence
time (if $\tau_\phi \gg \tau_H$ one chooses $\gamma \sim \Delta$,
since  the semiclassical approximation breaks down at the Heisenberg time).
Equation (\ref{K-classical}) was first derived by Altshuler and Shklovskii (1986) using diagrammatic techniques. The diagonal ``classical'' contribution is
known  as the {\em diffuson}, while the interference contribution (for conserved time-reversal symmetry) is described by the
{\em cooperon}, obtained by summation of maximally crossed diagrams (Altshuler and Simons, 1995).

Inspecting Eq. (\ref{K-classical}), we can distinguish two regimes. For
$\Omega=\omega  \Delta < E_c$, the $\bbox{q}=0$ diffusion mode dominates, and
we obtain the ergodic limit where the electron samples the whole  dot. On the
other  hand, for $\Omega \gg E_c$, the electron samples only a small fraction
of  the dot, and we can consider the limit of diffusion in an infinite system,
where  the summation in Eq. (\ref{return-prob}) over all modes $\bbox{q}$ can be
performed  exactly to give $P_{cl}(t)= L^d/(4\pi D t)^{d/2}$. In these two
limits  it is found that (Braun and Montambaux, 1995; Montambaux, 1997)
\begin{eqnarray}\label{K-sc-disordered}
K(\omega) \approx \left\{ \begin{array}{cc}
- {\rm Re}\;{1 \over \beta \pi^2(\omega + i \gamma)^2}\;\;\;({\rm for}\; 1 \ll \omega <
g_T)  \\
-{1 \over \beta \omega^2} \left(\frac{\omega}{g_T}\right)^{d/2} \cos\left({\pi
d  \over 4} \right)\;\;\;({\rm for}\; \omega \gg g_T)
\end{array} \right. \;.
\end{eqnarray}
For $1 \ll \omega < g_T$, $K(\omega)$ in Eq. (\ref{K-sc-disordered}) is
universal. It coincides (for $\gamma=0$) with the
semiclassical result of Berry (1985) in chaotic systems for $ \omega$ below the ballistic $g_T$.
 In Sec. \ref{RMT} we shall see that this universal semiclassical result
describes  the smooth part of the RMT prediction for $ 1 \ll
\omega  < g_T$. Random-matrix theory is nonperturbative and provides the exact universal
results  in a  broader regime $\omega < g_T$ (i.e., not requiring $\omega \gg 1$). On
the  other hand, according to Eq. (\ref{K-sc-disordered}),
the regime  $\omega \gg g_T$ is nonuniversal and the correlator depends on
dimensionality  and size (through $g_T$). Although this Altshuler-Shklovskii
nonuniversal  power law was first predicted in 1986, it was observed
numerically  in disordered metals only in 1995 (Braun and Montambaux, 1995).

\subsubsection{Conductance fluctuations in disordered metals}

 A semiclassical approach to conductance fluctuations in disordered systems was discussed by Argaman (1996). The conductivity tensor of noninteracting electrons can be expressed in terms of Green's functions, and  a semiclassical expression can be obtained using Eq. (\ref{sc-Green-energy}).
 In the diagonal approximation one recovers the energy-averaged classical conductivity $\bar \sigma$, i.e., Drude's formula.
 Quantum corrections result from interference of
classical  trajectories. Only pairs of trajectories that are related
by  symmetry can survive the averaging process.  In particular, in systems with
 time-reversal  symmetry we can consider the interference between a path
that forms a closed loop and a path that follows a similar trace except in the loop segment, which it crosses in the opposite direction.  These paths are time-reversed partners within the loop segment and their constructive
 interference leads to a decrease $\Delta \bar \sigma$ in the  average conductivity, an effect known
  as  weak localization or coherent backscattering.  Quantum corrections to the conductance due to self-intersecting time-reversed paths were first noticed by Langer and Neal (1966) using diagrammatic methods.
A semiclassical expression for the weak-localization correction can be found
 semiquantitatively by estimating the total number of closed loops. The probability for one closed
orbit  of period $t$ is the return probability $P(t)$. Since the initial point
$\bbox{r}$  can be anywhere along the path and the position of the electron is uncertain
within  $\lambda_F$ from its deterministic classical orbit, we  multiply this probability by the volume of a tube of length $v_F t$ and
thickness  $\lambda_F$. We obtain the following estimate (for a large conductor of volume $V$):
\begin{equation}\label{sigma-WL}
{\Delta \bar\sigma / \bar\sigma} \simeq -v_F \lambda_F^{d-1} V^{-1}
\int_{\tau}^{\tau_\phi}  dt P_{cl}(t) \;,
\end{equation}
 where $V^{-1} P_{\rm cl}(t)= (4\pi D t)^{-d/2}$ (see Sec. \ref{semiclassical:spectral}). The above picture is valid only for times when the motion is diffusive ($t
>  \tau$) and coherent ($t < \tau_\phi$), hence the range of integration in Eq. 
(\ref{sigma-WL}). (Here we define the effective dimension $d$ of a conductor to be the number of dimensions for which the sample's extension is larger than $L_\phi$.) Integrating Eq. (\ref{sigma-WL}), we find that the weak-localization correction to the average conductance per unit length is finite in 3D, $\Delta \bar G \simeq -(e^2/h)[(3/\pi)^{1/2} l^{-1} - (D\tau_\phi)^{-1/2}]$, but increases logarithmically with $\tau_\phi$ in 2D, $\Delta \bar G \simeq -(e^2/h) \ln(\tau_\phi/\tau)$, and linearly with $L_\phi$ in 1D, $\Delta \bar G = -(e^2/h) 2\pi (D\tau_\phi)^{1/2}$.   The weak-localization correction to the conductivity was originally  derived in the framework of disorder perturbation theory (Gorkov, Larkin and Khmelnitskii, 1979; see also  Bergmann, 1984, and references therein; Khmelnitskii, 1984).  For a quasiclassical approach to weak
localization  in disordered systems see Chakravarty and Schmid (1986). A magnetic field breaks time-reversal symmetry and can destroy the weak-localization correction (Altshuler~{\em et~al.}, 1980).

  The conductance in  a mesoscopic conductor fluctuates  as a function of Fermi energy and applied magnetic
field.  The magnitude of these fluctuations was found to be universal and of
order  $e^2/h$ (Altshuler, 1985; Lee and Stone 1985; Stone, 1985;  Lee, Stone and Fukuyuma, 1987). This is the phenomenon
of  universal conductance fluctuations.  Semiclassically, these fluctuations  are  related to
the  return probability through
${\sigma^2(G) / \bar{G}^2} \propto \int_0^\infty dt \;t \; P(t)$
(Argaman, 1996; Montambaux, 1997).

\subsection{The universal regime: random-matrix theory}
\label{RMT}

Random-matrix theory describes the statistical
fluctuations in the universal regime (i.e., at energy scales below the Thouless energy).   It was introduced by Wigner (1951, 1955, 1957, 1958) to explain the
statistical  fluctuations of neutron resonances in the compound nucleus.
Rather than trying to explain individual eigenfunctions, RMT addresses
questions  about their statistical behavior. Its original
justification was our lack of knowledge of the exact
Hamiltonian;  RMT assumes maximal ignorance regarding the system's Hamiltonian
except  that it must be consistent with the underlying symmetries. The theory
proceeds  to construct ensembles of Hamiltonians classified by their
symmetry.  Wigner's ideas were followed by those of Porter and Rosenzweig (1960) and 
Mehta and Gaudin (1960, 1961).  Using group-theoretical methods developed by
Wigner (1959),  Dyson (1962a) showed that
there  are three classes of random-matrix ensembles. In a seminal paper entitled ``The
Threefold  Way,'' Dyson (1962d) proved that the most general kind of
matrix  ensemble is a direct
product  of irreducible ensembles that belong to one of the three classes.
Most  of the early developments in the late 1950s and early 1960s are
collected in Porter (1965).  An early extensive review of RMT and its
applications  in nuclear physics was written by  Brody~{\em et~al.} (1981). A detailed account of RMT can be found in the book by Mehta (1991).

 Two major, independent developments in the early 1980s considerably
broadened  the range of validity of RMT. One was  the BGS conjecture
(Bohigas, Giannoni, and  Schmit, 1984)  linking the quantal fluctuations in chaotic systems to RMT.
Berry (1985) understood that the universality of
RMT in chaotic systems holds for time scales that are  longer than
the  shortest periodic orbits (i.e., the ergodic time). For time
scales  that are also much shorter than the Heisenberg time, the RMT results
coincide  with the semiclassical approach. However, RMT also provides the universal
results at longer times where the diagonal semiclassical approximation
fails.  The BGS conjecture was confirmed in a large number of numerical
studies.  Applications of RMT to
chaotic  systems were reviewed by Bohigas (1991).

The second major development was Efetov's supersymmetry method
(Efetov, 1983),  which made possible a nonperturbative treatment of the
single-particle  disorder problem by mapping it onto the supersymmetric
nonlinear $\sigma$ model. For weakly disordered systems and
below  the Thouless energy, this supersymmetric theory is in 0D
and  can be shown to be equivalent to RMT.

  Random-matrix theory also seems to describe the statistical properties of interacting systems at high enough excitation (e.g., the compound nucleus), but its applicability to the ground-state properties
   of closed dots (where interactions are important) is
    not yet fully understood
(see  Sec. \ref{interactions}). Random-matrix theory has many applications in quantum
physics;  a comprehensive review emphasizing common concepts was written
recently by Guhr, M\"{u}ller-Groeling,  and Weidenm\"{u}ller (1998).

  Section \ref{Gaussian-ensembles}
covers  the most common ensembles of random matrices, the Gaussian ensembles.
 Section \ref{crossover-ensembles} discusses
 the crossover ensembles, which are useful for describing the transition between
different  symmetry classes, e.g., the effects of a time-reversal
symmetry-breaking  magnetic field.  In Sec. \ref{Gaussian-processes}, we
generalize  the Wigner-Dyson ensembles to Gaussian processes, an  appropriate framework  for describing the universal statistical properties of a system
 that depends on an external parameter.  We end with another
type  of ensemble -- Dyson's circular ensemble, suitable for describing
statistical $S$ matrices and useful in the
statistical  theory of open quantum dots (Sec. \ref{RMT:open}).

\subsubsection{Gaussian ensembles}
\label{Gaussian-ensembles}

 The basic premise of RMT is that the statistical fluctuations of
certain  quantum systems can be described by an {\em ensemble} of $N\times N$
``random''  matrices $H$.  Since the matrix elements of  the
Hamiltonian  of a physical system vanish between states with different good
quantum  numbers, it is only the Hamiltonian matrix in a subspace with
fixed  values of good quantum numbers that is assumed to be
``random.''   Dyson (1962a, 1962d) found that there are only three
types  of ensembles, depending on the underlying space-time symmetries of the
system.  If the system is invariant under time reversal and under rotations,
there is a basis where the Hamiltonian operator is represented
by  a {\em real symmetric matrix} (systems with
time-reversal  symmetry and broken rotational invariance but with integer total 
angular  momentum also belong to this ensemble).
If  time-reversal symmetry is broken, irrespective of rotational invariance,
then  the Hamiltonian matrix is {\em complex Hermitean} . A third ensemble corresponds
to  systems that conserve time-reversal symmetry but are not rotationally
invariant  and have half-integer total angular momentum. The matrix elements of such
systems  are {\em real quaternion}. A quaternion $q$ is a $2\times2$ matrix expressed in terms of a linear combination of the unit matrix  $I$ and the three Pauli matrices $\sigma_j$, i.e., $q = a_0 I + i \sum_j a_j \sigma_j$. The quaternion is real when the coefficients $a_0$ and $a_j$ are real. Each of the three classes
has  a different number of independent real components $\beta$ that
characterizes  a matrix element. We have $\beta=1,2$ and $4$ for the ensembles of real
symmetric,  complex Hermitean, and  real quaternion matrices, respectively.

  The matrices that represent the same physical Hamiltonian in two different
bases  are related by a similarity transformation
$H^\prime = W^{-1} H W$,
where $W$ is the matrix connecting the two bases. We consider only
transformations  $W$ that preserve the ``type'' of the matrix $H$ that belongs
to  the given ensemble. For example, for $\beta=1$, the different bases can be chosen as ``real''
and  the matrix $W$ must be  orthogonal. Similarly, $W$ must be 
 unitary  for $\beta=2$ and  {\em symplectic} (i.e., unitary matrices with real
quaternion  elements) for $\beta=4$. The three ensembles are thus
called  orthogonal ($\beta=1$), unitary ($\beta=2)$ and symplectic
($\beta=4$).

 As a physical example, consider a single electron moving in a disorder or confining potential that is not rotationally invariant. If time-reversal symmetry and the electron spin are conserved, $\beta=1$ (since the motion is restricted to orbital space where the angular momentum assumes integer values). If time-reversal symmetry is broken, e.g., by a magnetic field, $\beta=2$. Finally, if time-reversal symmetry is conserved but spin-rotation symmetry is broken, e.g., by strong spin-orbit scattering, $\beta=4$ (in this case the orbital and spin spaces are coupled and the total angular momentum has half-integer values).

 We denote  by $P(H) dH$ the probability of finding a matrix $H$ whose elements
 are  in an interval $d H_{ij}$ around $H_{ij}$
(for $\beta=1$,  $dH \equiv \prod_{i \leq j} dH_{ij}$). There
are  various methods of deriving the distribution $P(H)$. Porter and Rosenzweig (1960) require that
the probability measure satisfy the following two properties (for the
orthogonal  case):
(i) Invariance; $P(H^\prime) = P(H)$ under any similarity  transformation
 with an orthogonal matrix $W$. Indeed, in complex or chaotic
systems  all bases chosen to represent the Hamiltonian should be statistically
equivalent  to each other.
 (ii) Statistical independence; all independent matrix elements  are statistically independent, $P(H) = \prod_{i\leq j}
P_{ij}(H_{ij})$.
The most general ensemble to satisfy both conditions is
\begin{equation}\label{Gaussian-ens}
P(H) \propto e^{-{\beta \over 2 a^2} {\rm Tr} H^2} \;.
\end{equation}
The distribution (\ref{Gaussian-ens}) is manifestly invariant under
any  orthogonal transformation of $H$.

The measures for the other ensembles are similarly derived.  We find the same
expression  (\ref{Gaussian-ens}) for all three ensembles. The quantity 
$\beta$ is introduced in the measure (\ref{Gaussian-ens}) for convenience
only (the average level density becomes
independent  of $\beta$). The corresponding three Gaussian ensembles are called the Gaussian
orthogonal  ensemble (GOE), Gaussian unitary ensemble (GUE), and the Gaussian
symplectic  ensemble (GSE) for $\beta=1,2$, and $4$, respectively.

 The Gaussian ensembles can also be defined by their first two moments.  For
the  orthogonal ($\beta=1$) and unitary  ($\beta=2$) cases we have
\begin{mathletters}\label{GE-m}
\begin{eqnarray}
   \overline{H_{ij}} & = & 0 \;;\;\;\;\;
   \overline{H_{ij}H_{kl}}= {{a^2}\over{2\beta}}  g^{(\beta)}_{ij,kl}\;; \label{GEmoments} \\
   g^{(\beta=1)}_{ij,kl} & = & \delta_{ik}\delta_{jl}+\delta_{il}\delta_{jk}
   \;;\;\;\;\;
   g^{(\beta=2)}_{ij,kl} = 2\delta_{il}\delta_{jk}\;. \label{gbeta}
\end{eqnarray}
\end{mathletters}
In the orthogonal case, where all matrix elements are real, the variance of
each  diagonal element is $a^2$, while that of each off-diagonal element is
$a^2/2$.  In the unitary case, the diagonal elements are real with variance
$a^2/2$,  while the off-diagonal elements are complex and their real and
imaginary  parts each have a variance of $a^2/4$.
 
 Another derivation of the  random-matrix ensembles was proposed by
Balian (1968).  His approach is based on information theory
(Shannon, 1948),  where the missing information
(or entropy) associated with a distribution $P(H)$ is defined by
$S\left[P(H) \right] = - \int d H P(H) \ln P(H)$ and measures
 the amount of missing information required to determine uniquely the
system's  Hamiltonian. Two constraints must be imposed on $P(H)$: $\int dH
P(H)=   1$ (normalization) and $\int dH {\rm Tr} H^2 P(H)  =  {\rm const}$
(to ensure that the Hamiltonian's eigenvalues are bounded).  We then choose the distribution $P(H)$ that is consistent with the
constraints   but is otherwise least biased, i.e., that  maximizes the missing
information.
The solution is given by Eq. (\ref{Gaussian-ens}). This construction  exemplifies
the  essence of RMT: it is the most ``random'' ensemble that is consistent
with  the underlying symmetries. The Gaussian ensembles lead to local correlations that are universally valid. In particular,  their correlators coincide, after proper scaling, with the correlators of non-Gaussian ensembles (Br\'{e}zin and Zee, 1993; Hackenbroich and Weidenm\"uller, 1995).

The  distribution of eigenvalues and eigenvectors can be calculated from Eq. (\ref{Gaussian-ens}). We diagonalize
 the matrix $H$,
\begin{equation}\label{diagonalize}
W^{-1} H W = E\;,
\end{equation}
and transform the variables $H_{ij}$ to new variables that consist of the $N$
eigenvalues  $E_\lambda$ and a set of $\beta N(N-1)/2$ variables
parametrizing  the diagonalizing matrix $W$. This requires calculation of
the  Jacobian $J$ of the transformation. We present here a simple derivation of $J$ in
the  GOE case (Bohr and Mottelson, 1969).
When $J \neq 0$, the transformation (\ref{diagonalize}) is one to one and  the eigenvectors must be uniquely determined from $H$. However, when two
eigenvalues  are degenerate, $E_\lambda=E_\mu$, this is not the case since the
degenerate  eigenvectors are determined only up to a linear combination, and
therefore  $J=0$. Furthermore, since the transformation (\ref{diagonalize}) is
linear  in $E_\lambda$, $J$ is a polynomial of degree $N(N-1)/2$ in the
$E_\lambda$'s.  These properties of $J$ determine its
dependence on the eigenvalues: $J \propto \prod_{\lambda < \mu}
|E_\lambda  - E_\mu|$.
The general result for any of the three ensembles is $J \propto
\prod_{\lambda  < \mu} |E_\lambda - E_\mu|^\beta$,
where the proportionality constant depends on the eigenvector parameters
alone.   We conclude that the eigenvalues are uncorrelated from the eigenvectors and are distributed according to
\begin{equation}\label{eigenvalue-dist}
P_N(E_1, E_2, \ldots, E_N) \propto \left(\prod\limits_{\lambda < \mu}
|E_\lambda  - E_\mu|^\beta\right) e^{-{\beta \over 2 a^2} \sum\limits_\nu
E_\nu^2}\;.
\end{equation}

\noindent {\em a. Spectral statistics}\\

The average level density in RMT is given by Wigner's semicircle
$\bar\rho=\sqrt{2a^2  N-E^2}/\pi a^2$. Random-matrix theory is not expected to reproduce global
level  densities of realistic systems but only to describe the local
fluctuations  of the spectrum. To compare the statistical properties of a
given spectrum with RMT, one first unfolds it, i.e., transforms it into
one  with constant average level density
(Bohigas and Giannoni, 1984).  An assumption implicit in most
applications of RMT is that of ergodicity: the ensemble average is
equivalent to the running average over a given spectrum. Given a physical
system,  one can collect statistics from different parts of the spectrum and
then  compare them with the RMT ensemble average.

 There are several useful statistical measures of spectral fluctuations:\\

\noindent (i)  Nearest-neighbor level-spacing distributions $P(s)$.
Their asymptotic forms for large $N$ cannot be written in a simple form, but
they  are surprisingly well approximated by the simple expressions obtained
for  $N=2$ (Wigner's surmise)
\begin{eqnarray}\label{Wigner}
P_{\rm WD}(s) = \left\{\begin{array}{lll}{\pi \over 2} s e^{-{\pi \over 4}s^2} \;\;\;
&({\rm  GOE}) \\
 {32 \over \pi^2} s^2 e^{-{4 \over \pi}s^2} \;\;\; &({\rm GUE}) \\
{2^{18} \over 3^6\pi^3} s^4 e^{-{64 \over 9\pi}s^2} \;\;\; &({\rm
GSE})\end{array}  \right.\;,
\end{eqnarray}
where the spacing $s$ is measured in units of $\Delta$.  Equations (\ref{Wigner}) are often called the Wigner-Dyson distributions. The GOE and GUE distributions are shown in Fig. \ref{fig:RMT}(a).
 Level repulsion is stronger in the GUE than in the GOE, as is
seen  from the small spacing behavior $P_{\rm WD}(s) \propto s^\beta$.

\begin{figure}
\epsfxsize= 8 cm
\centerline{\epsffile{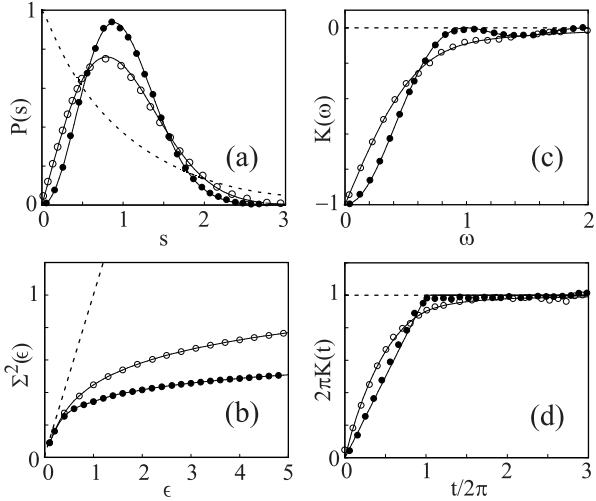}}
\vspace*{2 mm}
\caption{Spectral measures of a metal in the diffusive regime, compared with the
predictions of RMT. The circles are from numerical
simulations  of the Anderson model in its weakly disordered regime: open  circles, without magnetic flux; solid circles, with  magnetic flux.  The solid lines are
the  predictions of RMT using GOE for conserved time reversal symmetry (no flux) and GUE for broken time-reversal symmetry (with flux).(a) Nearest-neighbor  level-spacing distribution P(s); (b) 
number-variance  statistics $\Sigma^2(\epsilon)$ defined by
(\protect\ref{number-var}); (c)  two-point level-density  correlation function $K(\omega)$ defined by Eq.
(\protect\ref{K-omega})  (excluding a self-correlation $\delta$-function term);
(d) the  form factor $K(t)$, i.e., the Fourier transform of $K(\omega)$.  The agreement with the RMT predictions is
nearly  perfect. The dotted lines correspond to the Poisson statistics where
the  energy levels are assumed to be random. Adapted from
Braun and Montambaux (1995)  and Montambaux (1997).
}
\label{fig:RMT}
\end{figure}

\noindent (ii) The $n$-point cluster functions
$Y_n(\epsilon_1,\ldots,\epsilon_n)$ of $n$
levels  $\epsilon_\lambda = E_\lambda/\Delta$. Dyson (1962b) defines  the  $n$-point correlation function $R_n(E_1,\ldots,E_n)= \left[ N!/(N
-n)!\right]\int  \left(\prod_{i>n}dE_i\right) P_N(E_1,\ldots,E_N)$ as the
probability  density of finding the $n$ levels  $E_1,\ldots,E_n$ irrespective
of  the location of all other levels.  $Y_n$ is the
$n$th-order cumulant constructed from the $R_m$ ($m=1,\ldots,n$). For
example,  the two-level cluster function is defined by
$Y_2(\epsilon_1,\epsilon_2) = \Delta^2\left[ R_1(E_1)R_1(E_2)
-R_2(E_1,E_2)\right]$,
and depends only on the energy difference
$\omega\equiv\epsilon_2-\epsilon_1$.   For the Gaussian ensembles
 (Dyson, 1970; Mehta, 1971)
\begin{eqnarray}\label{cluster}
Y_2(\omega) =\left\{\begin{array}{ll}
\left( {\sin \pi \omega \over \pi \omega }\right)^2 -  \left[ {\rm Si} (\pi
\omega)-  \frac{1}{2}\pi {\rm sgn} (\omega) \right] & \\ \;\;\;\;\; \;\;\;\;\;\;\times  \left[ {\cos \pi \omega
\over  \pi \omega } - { \sin \pi \omega \over (\pi \omega)^2} \right] &
({\rm  GOE}) \\
 \left( {\sin \pi \omega \over \pi \omega }\right)^2  & ({\rm GUE}) \\
\left( {\sin 2\pi \omega \over 2\pi \omega }\right)^2 -  {\rm Si} (2\pi
\omega) & \\ \;\;\;\;\; \;\;\;\;\;\times \left[{\cos 2\pi \omega \over 2\pi \omega } - { \sin 2\pi \omega
\over  (2\pi \omega)^2} \right]  &({\rm GSE})  \end{array}
 \right.
\end{eqnarray}
where ${\rm Si}(x) = \int_0^x dt \sin t/t$. 
It is also useful to transform to the time domain, where  $b_2(t)
 \equiv \int_{-\infty}^\infty d\omega e^{2\pi i \omega t}
Y_2(\omega)$  is known as the two-level form factor (Brody~{\em et~al.}, 1981).

The two-point level-density correlation function $K(\omega)$ defined by Eq.
(\ref{K-omega})  is directly related to $Y_2$
by $K(\omega)= \delta(\omega) - Y_2(\omega)$,
while the associated spectral form factor is $K(t) =1 -
b_2(t/2\pi)$.  The RMT result for times $t \ll \tau_H=2\pi$ is
 $K(t) \approx -(2/\beta)(|t|/2\pi)$, in agreement
with  the diagonal semiclassical approximation (\ref{K-t-sc}).  Similarly, for
$\omega  \gg 1$, Eqs. (\ref{cluster}) give $K(\omega) \approx
-1/\beta\pi^2\omega^2$,  in agreement with the semiclassical results [see, for example, Eq.
(\ref{K-sc-disordered}) in a diffusive system below the Thouless energy].
 The GOE and GUE results for  $K(\omega)$ and
$K(t)$ are shown in Figs. \ref{fig:RMT}(c) and \ref{fig:RMT}(d), respectively.

\noindent (iii) The number variance $\Sigma^2(\epsilon)=
\overline{n^2(\epsilon)}  - \overline{n(\epsilon)}^2$. This  measures the variance of
the  number of levels $n(\epsilon)$ in an energy interval of length
$\epsilon$.  Since $n(\epsilon)=\int_0^\epsilon d\epsilon' \rho(\epsilon')$ we
have  the relation
\begin{equation}\label{number-var}
\Sigma^2(\epsilon) = 2\int_0^\epsilon d\omega ( \epsilon - \omega) K(\omega)
=  \epsilon - 2\int_0^\epsilon d\omega (\epsilon-\omega)Y_2(\omega)\;.
\end{equation}
 The GOE and GUE number variances are shown in Fig. \ref{fig:RMT}(b).
  For small
$\epsilon$,  $\Sigma^2(\epsilon) \approx \epsilon$, but of particular interest
is  the large-$\epsilon$ logarithmic behavior
$\Sigma^2(\epsilon) \approx (2 / \beta \pi^2) \ln \epsilon
 + {\rm const} + {\em O}(\epsilon^{-1})$,
where the constant is $\beta$ dependent.  For $ 1 \ll \epsilon <
g_T$ the diagrammatic (or semiclassical) result for the number variance in disordered metals coincides with the RMT results.
  However, for $\epsilon \gg g_T$, the diagrammatic result $\Sigma^2(\epsilon) \approx -\beta^{-1} \left(\epsilon/g_T\right)^{d/2}$ is
nonuniversal.  The electron diffuses for a time $t$ that
is  much shorter than $\tau_D$ and covers only an area of linear dimension
$\sqrt{D  t}$ that is much smaller than  $L$. The number
variance  is then proportional to the number of such ``areas''  $(L/\sqrt{D
t})^d  \sim (\epsilon/g_T)^{d/2}$ contained within the total area of the system.

\noindent (iv) The Dyson-Mehta $\Delta_3$ statistics (Dyson and Mehta, 1963). A
straight  line is fitted by least squares to the staircase function (defined as the number of levels below a given energy) in an interval
of  length $\epsilon$. Here $\Delta_3(\epsilon)$ is the  least-squared
deviation  from this best linear fit. The ensemble average of $\Delta_3$ is
related  to the number variance by (Pandey, 1979; Mehta, 1991)
\begin{equation}\label{Delta_3}
\overline{\Delta}_3 (\epsilon) = {2 \over \epsilon^4} \int_0^\epsilon d\omega
(\epsilon^3  - 2\epsilon^2 \omega + \omega^3) \Sigma^2(\omega) \;.
\end{equation}
In RMT,  $\Delta_3$ starts as
$\epsilon/15$  and behaves asymptotically as
$\overline{\Delta}_3(\epsilon) \approx (1 /\beta \pi^2) \ln \epsilon + {\rm
const}$.

  Gaussian orthogonal ensemble spectral correlations were found in the statistical analysis of the
 nuclear data  ensemble, which consists of 1726 measured resonances
 in various compound  nuclei (Haq, Pandey, and Bohigas, 1982; Bohigas, Haq, and Pandey, 1985).
Random-matrix theory also describes the universal regime of disordered
metals.  This is confirmed in Fig. \ref{fig:RMT}, which compares Anderson
model  calculations with and without magnetic flux (Dupuis and Montambaux, 1991; Braun and Montambaux, 1995) to
RMT  predictions.

\noindent {\em b. Eigenfunction statistics}

In RMT,  the probability distribution of an eigenvector's components $\psi_i$
($i=1,\ldots,N$)  is  determined from the orthogonal (unitary)
invariance  of the ensemble (Brody~{\em et~al.}, 1981),
\begin{equation}\label{vec-dist}
P(\psi_1,\psi_2,\ldots,\psi_N) \propto \delta(\sum_i |\psi_i|^2 -1)\;,
\end{equation}
where the metric is given by ${\cal D}[\bbox{\psi}] =
\prod_{i=1}^{N} d\psi_i$ for $\beta=1$ and
$\prod_{i=1}^{N} (d\psi^\ast_i d\psi_i / 2\pi i)$ for $\beta=2$.

To find the distribution of a finite number of components
$\psi_1,\ldots,\psi_\Lambda$,  we integrate Eq. (\ref{vec-dist}) over all other
$N-\Lambda$  components to find $P(\psi_1,\ldots,\psi_\Lambda)
\propto  \left( 1 - \sum_{i=1}^{\Lambda}|\psi_i|^2 \right)^{ \beta
(N-\Lambda)/2  -1}$. In the asymptotic limit $N \to \infty$, this distribution is a Gaussian $P(\bbox{\psi})
 \propto \exp \left[-(\beta N/2) \sum_{i=1}^\Lambda
|\psi_i|^2 \right]$.
Of particular interest is the distribution of the intensity of a single
component  $y\equiv |\psi_i|^2$,
\begin{equation}
P(y) = \left({\beta \over 2 \overline{y}}\right)^{\beta/2}{1 \over (\beta/2 -1)!} y^{\beta/2-1} e^{-\beta y \over 2\bar{y}} \;,
\end{equation}
which is just the $\chi^2$ distribution in $\beta$ degrees of freedom. For the GOE ($\beta=1$) this is the Porter-Thomas distribution (Thomas and Porter, 1956) describing  the neutron resonance widths in the compound nucleus -- see,
for example, Fig. \ref{fig:closed-dot}(b).

 More generally, for $n$ eigenvectors ($n \ll N$), 
$P(\bbox{\psi}_\lambda)  \propto \prod_{\lambda =1}^{n}
\exp \left[-(\beta N/2) \sum_{i=1}^{\Lambda_i} |\psi_{i\lambda}|^2 \right]$,
and components that belong to different eigenvectors are to leading order
uncorrelated.

\subsubsection{Crossover ensembles}
\label{crossover-ensembles}

 In some applications we are interested in the fluctuation properties of
systems  in the crossover regime between two different symmetries.  The statistics in the
crossover  regime  between GOE and GUE can be described by the Mehta-Pandey ensemble (Pandey and Mehta, 1983; Mehta and Pandey, 1983; Mehta, 1991; Bohigas, 1991)
\begin{eqnarray}\label{transition-ens}
  H=S+i\alpha A  \;,
\end{eqnarray}
where $S$ and $A$ are, respectively, symmetric and antisymmetric real
matrices
and $\alpha$ is a real parameter. The matrices $S$ and $A$ are uncorrelated
 and chosen from Gaussian ensembles of the same variance.
  Similar ensembles can be
constructed to describe the crossover between the GUE and GSE or between the GOE and GSE (Mehta, 1991).

We are interested in the asymptotic limit $N \to\infty$, where the proper
transition  parameter is given by a typical symmetry-breaking matrix element
measured  in units of $\Delta$ (French and Kota, 1982; French~{\em et~al.}, 1985, 1988):
$\zeta = (\overline{H^2}_{\rm break})^{1/2}/\Delta = {\alpha
\sqrt{N}  / \pi}$.
For a  fixed $\zeta$, the statistics of the ensemble
(\ref{transition-ens})  become independent of $N$ in the limit $N \to
\infty$.  The crossover parameter can also be expressed as
$2  \pi \zeta = \sqrt{\tau_H/\tau_{\rm mix}}$,
 where $\tau_{\rm mix}$ is the mixing time defined
 in terms of the spreading width  $\hbar/\tau_{\rm mix} = 2 \pi
\overline{H^2}_{\rm break}/\Delta$
 of the time-reversal symmetry-breaking interaction (Pluha\u{r}~{\em et~al.}, 1994).
 The spectral statistics of the transition ensemble (around the middle of the spectrum) make the complete
crossover  for $\zeta \sim 1$.

 In the transition ensembles, the eigenvalues and eigenvectors are no longer
uncorrelated.   The spectral statistics  were
derived  by Pandey and Mehta (1983; see also Mehta and Pandey, 1983 and
Mehta, 1991), but
until recently less was known about the statistics of the eigenvectors in
the  transition ensembles. Fal'ko and Efetov (1994) used supersymmetry to derive the distribution of
the  wave function intensity, as well as the joint
distribution of the wave function intensity at two distant spatial points  (Fal'ko and Efetov, 1996).  The recent work of
van Langen, Brouwer, and Beenakker (1997) and Alhassid, Hormuzdiar, and Whelan (1998), based on earlier
work  of French~{\em et~al.} (1988), leads to a closed expression for the joint
distribution of any finite number of the eigenvector's components.

The components of an eigenvector $\psi$ are complex: $\psi_i= \psi_{iR} + i\psi_{iI}$.  The eigenvector is determined only up to a phase $e^{i
\theta}$, which can be fixed by
rotating  to a principal frame where (French~{\em et~al.}, 1985, 1988)
\begin{eqnarray}\label{invariants}
\sum\limits_{i=1}^N {\psi}_{i R}{\psi}_{i I} = 0 \;;\;\;\;\;\;
\sum\limits_i {\psi}_{i I}^2/\sum\limits_i {\psi}_{i R}^2 \equiv t^2 \;.
\end{eqnarray}
 The eigenvector's components are distributed in the complex plane to form an
ellipsoid  whose semiaxes define the principal frame. The parameter $t$ in Eq. (\ref{invariants}) ($0\leq t \leq 1$) determines the
shape  of this ellipsoid and is found to fluctuate in the
crossover  regime. Earlier theories
(Zyczkowski and  Lenz, 1991; Kogan and Kaveh, 1995; Kanzieper and Freilikher, 1996)  ignored these fluctuations.

 We now consider eigenvectors with a fixed shape parameter $t$.  Under
an orthogonal transformation $O$,
the  real and imaginary parts of $\psi$ transform like $\psi_R \to O \psi_R$
and  $\psi_I \to O \psi_I$ and do not mix with each other. Consequently,
Eq.  (\ref{invariants}) and the probability distribution of the transition  ensemble (\ref{transition-ens}) are invariant under an orthogonal transformation
$O$. Thus the 
conditional  probability distribution of the components of an eigenvector with
a  fixed ``shape''  $t$ is given by
  $P(\psi_1, \ldots, \psi_N | t) \propto
    \delta \! \left( \sum_{i=1}^N  \psi_{i R}^2 - 1/(1+t^2) \right)
   \delta \! \left( \sum_{i=1}^N  \psi_{i I} ^2 -  t^2/(1+t^2) \right) \\
     \times  \delta \! \left( \sum_{i=1}^N \psi_{i R} \psi_{i I} \right)$. In the limit $N \to \infty$, the
 joint conditional distribution of a finite number of components $\Lambda \ll N$ becomes a Gaussian:
\begin{eqnarray}\label{transition-comp-dist}
 P( \psi_1,&& \ldots,\psi_\Lambda | t)  \nonumber \\  = &&\left( {N \over 2\pi} { 1 + t^2 \over
t}  \right)^\Lambda \nonumber \\
\times\; && \exp \left( - N { 1 + t^2 \over 2} \sum_{i=1}^\Lambda
\psi_{i R}^2 - N { 1 + t^2 \over 2 t^2}  \sum_{i=1}^\Lambda\psi_{i I}^2
\right)   \;.
\end{eqnarray}
The full distribution  is computed by
averaging  the conditional distribution (\ref{transition-comp-dist}) over the
distribution  $P_\zeta(t)$ of the shape parameter:
$P_\zeta(\psi_1,\ldots,\psi_\Lambda) = \int_0^1 dt P_\zeta(t)
 P(\psi_1,\ldots,\psi_\Lambda |t)$.

 It is still necessary to determine $P_\zeta(t)$. This can be done (Alhassid, Hormuzdiar, and Whelan, 1998) by calculating the
distribution  of the square of a single component using Eq. 
(\ref{transition-comp-dist}) and comparing
it  with the supersymmetry calculation by Fal'ko and Efetov (1994) of the
wave-function-intensity  distribution. It is found that
\begin{eqnarray}\label{shape-dist}
&& P_\zeta(t) = \pi^2 { 1-t^4 \over t^3} \zeta^2
 e^{-{\pi^2 \over 2} \zeta^2 \left(t- 1/t \right)^2} \nonumber \\
 & & \times  \left\{ \phi_1(\zeta) + \left[ \frac{1}{4} \left( t + \frac{1}{t}
\right)^2  -
{1 \over 2\pi^2 \zeta^2 } \right]  \left[ 1 - \phi_1(\zeta) \right]
\right\}  \;,
\end{eqnarray}
where $\phi_1(\zeta)= \int\limits_0^1
 e^{ - 2 \pi^2 \zeta^2 (1 - y^2)} dy$.

van Langen, Brouwer, and Beenakker (1997) calculated $P_\zeta(t)$ directly in the framework
of  RMT  using a result of Sommers and Iida (1994) for the joint distribution
of  an eigenvalue and its associated eigenvector for a Hamiltonian in the
ensemble  (\ref{transition-ens}). Rather than $t$, an equivalent ``phase
rigidity''  parameter $\rho\equiv |\sum_i \psi_i^2|^2=
[(1-t^2)/(1+t^2)]^2$  is used.

 The crossover distribution of a single component of the eigenfunction $y
\equiv  |\psi_i|^2=\psi_{iR}^2 + \psi_{iI}^2$ is found from
Eq. (\ref{transition-comp-dist})  to be (see also Fal'ko and Efetov, 1994)
\begin{equation}\label{intensity-dist}
P_\zeta(y) =\left\langle {1\over 2 \bar y}\left(t + \frac{1}{t}\right)
e^{-(t+1/t)^2  y/4\bar y} I_0\left( { 1-t^4\over 4t^2}{y \over \bar y}
\right) \right\rangle \;,
\end{equation}
 where $I_0$ is the modified Bessel function of order zero, and $\langle \ldots\rangle$ denotes an average over the distribution in Eq. (\ref{shape-dist}).

\subsubsection{Gaussian processes}
\label{Gaussian-processes}

 Consider a chaotic system that depends on an external parameter  and whose
symmetry  class is the same for all values of the parameter. An interesting question is whether any universality can be found in the fluctuations of the system properties versus this parameter.

 A semiclassical  theory for a statistic that measures the correlation of energy
levels  at different values of an external parameter was proposed by
Goldberg~{\em et~al.}~(1991).  Szafer and Altshuler (1993) and
Simons and Altshuler (1993a, 1993b) discovered that certain parametric  spectral correlators of
disordered  systems are universal after an appropriate scaling of the parameter.
Beenakker (1993) and Narayan and Shastry (1993)  suggested that the parametric correlators
can  be studied in the framework of Dyson's Brownian-motion model (Dyson, 1962c).

The universality of eigenfunction correlators was shown by
Alhassid and Attias (1995) through the generalization of Wigner-Dyson
random-matrix  ensembles to random-matrix processes describing the statistical
 properties  of systems that depend on an external parameter. The supersymmetry
 approach  was used to obtain the universal form of the oscillator
strength-function correlator (Taniguchi, Andreev, and Altshuler, 1995). A Brownian-motion model for
the  parametric evolution of matrix elements of an operator between
eigenstates  was introduced by Wilkinson and Walker (1995).

  We first discuss the generalization of the Gaussian ensembles  to
Gaussian  processes. We assume a system whose Hamiltonian $H(x)$  depends
on  an external parameter $x$. The system is chaotic or weakly disordered for
all  values of $x$, and its underlying symmetry class is independent of $x$.
 To generalize the statistical description of
RMT  to include such a parametric dependence, we assume a Gaussian process $H(x)$ characterized  by
\begin{eqnarray} \label{GP-correlations}
   \overline {H_{ij} (x) }  =  0 \;;\;\;\;
   \overline{ H_{ij} (x) H_{kl} (x^\prime) }   =  {a^2 \over {2\beta}}
   f(x,x^\prime) g^{(\beta)}_{ij,kl} \;,
\end{eqnarray}
where the coefficients $g^{(\beta)}_{ij,kl}$ are defined in Eq. (\ref{gbeta}).
The  process correlation function is assumed to be stationary and symmetric: 
$f(x,x^\prime)=f(|x-x^\prime|)$.  It is normalized to $f(x,x)=1$ so that the
matrix  elements of $H(x)$ satisfy  the Gaussian ensemble relations
(\ref{GE-m})  for each  fixed value of $x$.

  The Gaussian process can also be defined in terms of its probability distribution
\begin{eqnarray}\label{GPmeasure}
   P\left[H(x'')\right]\propto &&
   \exp \{-\left(\beta/{2a^2}\right)
   \int dxdx^\prime  \nonumber \\
&&  \times {\rm Tr}\left[H(x)K(x,x^\prime)H(x^\prime)\right] \} \;,
\end{eqnarray}
where the metric is ${\cal D} H \equiv \prod_x dH(x)$ with $dH$ being the
usual  Gaussian ensemble metric. Equation (\ref{GPmeasure}) is a direct generalization of the Gaussian ensemble 
measure  (\ref{Gaussian-ens}) to include a parametric dependence of the
random-matrix  Hamiltonian. The kernel $K$ is the functional inverse of $f$.

Eq. (\ref{GP-correlations}) with  $f(x-x') = \delta(x-x')$ 
describes  a white-noise random-matrix process $\Phi(x)$.
Wilkinson (1989)  introduced the process
$H(x) = \int d x' w(x-x') \Phi(x')$
in the study of the statistics of avoided crossing in chaotic systems. It is
a  Gaussian process with a correlation function $f(x) = \int dx' w(x-x')w(-x')$.

 Depending on the symmetry class of the matrices $H(x)$, there are three
types  of processes: the Gaussian orthogonal process ($\beta=1$), the Gaussian
unitary  process ($\beta=2$), and the Gaussian symplectic process ($\beta=4$).  A Gaussian process 
is characterized by a correlation function $f$,  but
 in the  asymptotic limit $N \to \infty$, only the short-distance behavior
  of $f$ (in  parameter space) is important. We therefore expand $f$ to leading
order  in $x-x'$: $f(x-x') \approx 1 - \kappa |x-x'|^\eta$,
where $\kappa$ and $\eta>0$ are constants, and classify the Gaussian processes according to
the value of $\eta$.

 Of particular interest are two-point parametric correlators of observables $O(x)$:
$c_{\cal O}(x-x^\prime) = \overline {\delta O(x)\delta O(x^\prime)}/
[\sigma(O(x)) \sigma(O(x'))]$,
where $\delta O(x) = O(x) - \bar O(x)$ and  $\sigma^2(O(x))=
\overline{(\delta  O(x))^2}$.
To calculate  $c_{\cal O}(x-x')$ we only need to know
the  joint distribution of $H(x)$ and $H(x')$.  The Gaussian process has the useful property
that  the joint distribution of any finite number of  matrices
$H(x),H(x'),H(x''),\ldots$  is Gaussian. In particular (Attias and Alhassid, 1995)
\begin{eqnarray}\label{2matrix-dist}
   P\left[H(x),H(x^\prime)\right] & & \propto \exp \{-(\beta/{2a^2})
   {\rm Tr}\left[H(x)^2+H(x^\prime)^2 \right. \nonumber \\ & & \left. -2fH(x)H(x^\prime)\right]
   /(1-f^2) \} \;,
\end{eqnarray}
where $f \equiv f(x-x')$.

The Gaussian process parametric correlators depend on $a$, $f$, and $N$, but become universal upon an appropriate scaling of $x$.  The dependence on $a$ is eliminated by unfolding the energies $E_i
\to  \epsilon_i=E_i/\Delta$.  To eliminate the dependence on $f$, we calculate the mean-squared
parametric  change $\overline{(\Delta \epsilon_i)^2}$ of a given level using first-order  perturbation theory,
\begin{eqnarray}\label{level-diffusion}
   \overline{\Delta\epsilon_i^2}= D_\epsilon \mid\Delta x\mid^\eta
   +{\cal O}\left(\mid\Delta x\mid^{2\eta}\right)\;,
\end{eqnarray}
where $D_\epsilon=\lim_{\Delta x\rightarrow 0}
   {\overline{\Delta\epsilon_i^2}/{\Delta x^\eta}} =
{{4N\kappa}/{\pi^2\beta}}$.
Equation (\ref{level-diffusion}) suggests that the energy levels undergo
short-range  ``diffusion'' as a function of the parameter $x$ (characterized
by  an exponent $\eta$), with $D_\epsilon$ playing the role of the diffusion constant.
 For a Gaussian process with $\eta < 2$, the
  levels show irregular behavior as a function of $x$ which becomes smooth in the limit
   $\eta=2$. This is demonstrated in Fig. \ref{fig:univ-corr}(a).
   Gaussian processes with $\eta=2$ are the only
{\em  differentiable} Gaussian processes (Attias and Alhassid, 1995); namely, they have the
property  that almost every one of their members has a continuous derivative
$dH/dx$.  Since in most physical applications the Hamiltonian is an analytic
function  of its parameter, we are interested only in differentiable Gaussian processes,  i.e., $\eta=2$, and in the following we assume $\eta=2$.
Equation (\ref{level-diffusion}) suggests the  scaling (Simons and Altshuler, 1993a, 1993b)
\begin{eqnarray}\label{scaling}
   x\rightarrow\bar{x}\equiv \sqrt{D_\epsilon} x  =
\left[\overline{\left({{\partial\epsilon_i}/{\partial
x}}\right)^2}\right]^{1/2}  x \;,
\end{eqnarray}
under which the Gaussian process correlation function becomes independent
of  the non-universal constant $\kappa$:
   $f\approx 1- (\beta\pi^2/4N)\mid\bar{x}-\bar{x}^\prime\mid^2$.
   Analytic calculations of certain correlators as well as
numerical  simulations support the conjecture that, for large $N$, all
two-point  correlators depend on $N$ and $f$ only through
the  combination $N(1-f)=
   ({\beta\pi^2}/ 4)\mid\bar{x}-\bar{x}^\prime|^2$. We conclude that all
parametric  correlators are universal as a function of
$\mid\bar{x}-\bar{x}^\prime\mid$. The universality can also be demonstrated by relating the Gaussian process to Dyson's Brownian motion model (Mitchell, Alhassid, and Kusnezov, 1996).  The scaled $\bar x$ measures $x$ in units of the average parametric distance between avoided level crossings.

\begin{figure}
\epsfxsize= 8 cm
\centerline{\epsffile{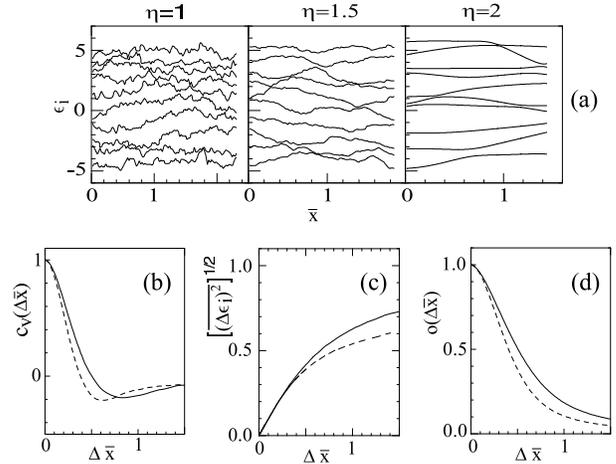}}
\vspace*{2 mm}
\caption{Gaussian processes and parametric correlators: (a) typical parametric variation of the energy
levels of Gaussian  processes (\protect\ref{GP-correlations})
with correlation functions $f(x) = \exp(-|x|^\eta)$ as a function of the scaled parameter $\bar x$. Shown are
processes with $\eta=1, 1.5$ and $2$.  $\eta=2$ is the
limit where the parametric variation of the energy levels
becomes smooth. Panels (b) -- (d) show three parametric correlators
in the GOE (solid lines) and GUE (dashed lines) symmetries:   (b)
the level-velocity correlator $c_v(\Delta  \bar x)$; (c)  the level-diffusion correlator
$\left[\overline{(\Delta\epsilon_i)^2}\right]^{1/2}$; and (d) the
wave-function-overlap parametric correlator $o(\Delta \bar x)$
of Eq. (\protect\ref{overlap-corr}).  The correlators are
calculated using the Gaussian process (\protect\ref{simple-GP}). The level-velocity correlator coincides  with the one calculated by
Simons and Altshuler (1993a, 1993b) in the Anderson  model. Adapted
from Attias and Alhassid (1995).}
\label{fig:univ-corr}
\end{figure}

Since the parametric correlators are universal, any Gaussian process can be used to
calculate  them. A particularly simple Gaussian process is given by
(Austin and Wilkinson, 1992; Alhassid and Attias, 1995)
\begin{equation}\label{simple-GP}
H(x) = H_1 \cos x + H_2 \sin x \;,
\end{equation}
where $H_1$ and $H_2$ are two independent Gaussian matrices that belong to
the  appropriate symmetry.  The process (\ref{simple-GP}) is characterized by
 $f(x-x') = \cos (x-x')$, and is therefore an $\eta=2$
process.   It is stationary and particularly useful for numerical
simulations.

  Several spectral parametric correlators were calculated and their
universality  demonstrated in models of disordered dots  as well as chaotic
dots.
The correlator $c_v(x-x')$ of the level velocity
 $v_\lambda(x) \equiv \partial
E_\lambda/\partial  x$ was calculated by Simons and Altshuler (1993a, 1993b) using the
Anderson  model.  Its universal form, calculated using the Gaussian process
(\ref{simple-GP}),   is shown in Fig.
\ref{fig:univ-corr}(b)  for both the orthogonal (solid line) and unitary (dashed
line)  symmetries. The Gaussian unitary process correlator  decorrelates faster than the Gaussian orthogonal process correlator. Also shown in Fig.
\ref{fig:univ-corr}(c) is the level-diffusion
correlator  $\left[\overline{(\Delta \epsilon_i)^2}\right]^{1/2} = \left[\overline{(\epsilon_i(x) -
\epsilon_i(x'))^2}\right]^{1/2}$,  describing the root mean square (rms) of the change of a given energy
level  over a finite parametric distance $\Delta \bar x$ (Attias and Alhassid, 1995).
The level-density correlator
$k(\omega,\Delta x) \equiv \overline{\delta \rho(\epsilon,x)\delta
\rho(\epsilon  + \omega, x + \Delta x)}$,
where $\rho(\epsilon,x) = \sum_\lambda
\delta(\epsilon-\epsilon_\lambda(x))$  [$\epsilon_\lambda$ are the unfolded
energy  levels measured in units of $\Delta$], can be calculated in
closed  form using the supersymmetry method (Simons and Altshuler, 1993a, 1993b).

 Parametric correlators that depend on the eigenfunctions, such as the
averaged  parametric overlap correlator (Alhassid and Attias, 1995; Attias and Alhassid, 1995) and the
strength-function correlator (Taniguchi, Andreev, and Altshuler, 1995), were also found to be
universal.  The overlap correlator measures the squared overlap of an
eigenfunction  at different values  of the external parameter:
\begin{eqnarray}\label{overlap-corr}
   o(x-x^\prime)=
\overline{\mid\langle\psi_\lambda(x)\mid\psi_\lambda(x^\prime)\rangle\mid^2}
   \;.
\end{eqnarray}
This correlator was calculated numerically using the Gaussian process (\ref{simple-GP}) and is shown in
 Fig. \ref{fig:univ-corr}(d). The wave functions are observed
to  decorrelate faster in the unitary case. This parametric overlap correlator
is  well fit by a Lorentzian in the orthogonal case ($\beta=1$) and by a
squared  Lorentzian in the unitary case ($\beta=2$):
\begin{eqnarray}\label{approx-overlap-corr}
   o(\bar{x}-\bar{x}^\prime)\approx
\left[ {1\over{1+(\bar{x}-\bar{x}^\prime)^2/\alpha_\beta^2}}\right]^\beta
 \;,
\end{eqnarray}
with $\alpha_1=0.48\pm 0.03$ in the orthogonal case and
$\alpha_2=0.64\pm  0.04$ in the unitary case.

\subsubsection{Circular ensembles}
\label{circular-ensembles}

 The Gaussian ensembles are well suited to the Hamiltonian approach, in which
the  Hamiltonian is assumed to be ``random'' and the statistics of various
physical  quantities of interest are then calculated from their relation to
the  Hamiltonian. However, transport properties of electrons through quantum
dots  can be related directly to the $S$ matrix (see Sec.
\ref{LIB}),  and it is sometimes possible to derive the
fluctuations  of the conductance by assuming the $S$ matrix to be the
fundamental  ``random'' object (see Sec. \ref{RMT:open}).  The  $S$ matrix
associated  with a Hermitian Hamiltonian is unitary.  Random-matrix ensembles
of  unitary matrices are the {\em circular ensembles} introduced by Dyson (1962a).

There are three types of circular
ensembles:  orthogonal, unitary, and symplectic. For systems
with  time-reversal invariance, the $S$ matrix is symmetric unitary, while for
systems  with broken time-reversal symmetry, the $S$ matrix is an arbitrary
unitary  matrix. Finally, the
symplectic  ensemble is composed of self-dual unitary quaternion $S$ matrices
(Mehta, 1991).   The measure of the circular ensembles is  uniform over
the  unitary group of $S$ matrices, subject to the consistency condition that
the  symmetry properties of the $S$ matrix are preserved. More explicitly, the
measure  is required to satisfy
\begin{equation}\label{circular-ens}
d\mu(S) = d \mu(U S V)
\end{equation}
for any unitary matrices $U$ and $V$ that preserve the symmetry properties as
$S$  under the transformation $S \to U S V$. The measure satisfying Eq. 
(\ref{circular-ens})  is known as the Haar measure.

  A unitary matrix $S$ that belongs to any of the three circular ensembles
can  be diagonalized
$U^{-1} S U =E$,
where the eigenvector matrix $U$ is orthogonal, unitary, or symplectic for
$\beta=1,2$,  or $4$, respectively. The eigenvalues  $e^{i\theta_i}$  define
the  eigenphases of the $S$ matrix. The
eigenvector  and eigenvalue distributions are uncorrelated, and
\begin{equation}\label{circular}
P(\theta_1,\ldots,\theta_N) \propto
\prod\limits_{i\leq j}|e^{i \theta_i} - e^{i\theta_j}|^\beta\;,
\end{equation}
 where the measure is $\prod_j d\theta_j$. In the symplectic case the
eigenvalues  are doubly degenerate and only the nondegenerate values are
taken  in Eq. (\ref{circular}).
The density of the eigenphases is uniform on the circle, and
there  is no need for unfolding. In the limit $N \to \infty$, various spectral correlations and cluster functions coincide with those of
the  Gaussian ensembles.

\subsection{The supersymmetry method}
\label{supersymmetry}

   The supersymmetry method provides a technique for calculating   ensemble
averages  over products of Green's functions, where the Hamiltonian is either a random
matrix or describes a single particle in a random potential.
This  is a nontrivial problem since the random quantities appear in the denominator.  The method
maps  the original problem -- after performing the ensemble average -- onto a
field-theoretical model, the supersymmetric nonlinear $\sigma$ model. For
random  matrices, the $\sigma$ model is in 0D, while for a weakly
disordered system in $d$ dimensions, the $\sigma$ model is in $d$ dimensions.  However, for energy scales below the
Thouless  energy, it reduces to the 0D $\sigma$-model, explaining the RMT
universality  in weakly disordered systems. The supersymmetry method was
introduced  for disordered systems by Efetov (1983), and its use for random matrices was discussed by
Verbaarschot, Weidenm\"{u}ller, and Zirnbauer (1985).   A variety of physical applications are
included in a review by Altshuler and Simons (1995).  A book by
Efetov (1997)  discusses the method and its numerous applications.  The  method was used extensively in the calculation of spectral and eigenfunction statistics in disordered systems; see a recent review by Mirlin (2000).

We first describe briefly the supersymmetry method for
averaging over random-matrix ensembles.
  The starting point is to write the Green's 
function  as a multidimensional Gaussian integral where the ensemble average
is  straightforward.
 In general, the inverse of a matrix $K$ can be written as
$\left(K^{-1}\right)_{\lambda \mu} = (\det K) \int d\bbox{s}\;  s_\lambda
s^\ast_\mu  e^{- \bbox{s}^\dagger K \bbox{s}}$,
 where $s_\nu$ are commuting complex variables and $d\bbox{s}
\equiv  \prod_\nu (d s^\ast_\nu d s_\nu/ 2 \pi i)$. The determinant of $K$ can also be written as a Gaussian integral:
$\det K = \int d\bbox{\chi}\; e^{-\bbox{\chi}^\dagger K \bbox{\chi}}$, but
over  Grassmanian (anticommuting) variables $\chi_\nu$ with $d \bbox{\chi} \equiv
\prod_\nu  (d\chi_\nu^\ast d \chi_\nu)$. Combining both
representations,  we can  write the inverse of a matrix as a pure Gaussian
integral over both commuting and anti-commuting variables:
\begin{equation}\label{inverse-Gauss}
\left(K^{-1}\right)_{\lambda \mu} =\int d \bbox{s} d \bbox{\chi}\; s_\lambda
s^\ast_\mu  e^{- \bbox{s}^\dagger K \bbox{s} -\bbox{\chi}^\dagger K
\bbox{\chi}} \;.
\end{equation}
Equation (\ref{inverse-Gauss}) can be used to express the advanced and retarded
Green's  functions of an $N\times N$ Hamiltonian matrix $H$ as a
Gaussian  integral in the set $(\bbox{s},\bbox{\chi})$  by choosing $K=\mp
i(E^\pm  -H)$.  In the GOE case the quadratic form in the exponent of Eq. (\ref{inverse-Gauss}) is
rewritten  by doubling the number of components (the doubling is not needed for the GUE case):
\begin{eqnarray}\label{Green-Gauss}
   G(E^\pm)=&& \mp i\int d\bbox\Psi \;(s s^\dagger) \nonumber \\
   &&\times \;\exp\left\{\pm{i\over 2}\bbox\Psi^\dagger[(E^\pm-H)\times
I_4]\bbox\Psi\right\}
   \;.
\end{eqnarray}
Here $\bbox{\Psi} \equiv (\bbox{s}, \bbox{s^\ast},  \bbox\chi,
\bbox\chi^\ast)^T$  is a $4N$-dimensional vector and the measure is
$d\bbox\Psi\equiv  d\bbox{s} \; d \bbox \chi$, while $M\otimes I_n$ denotes a
block  diagonal matrix with n blocks $M$.

Equation (\ref{Green-Gauss}) can be generalized to products of Green's functions by
introducing  a separate vector $\bbox\Psi_j$ for each Green's  function.
 We describe below the calculation of quantities that involve the
product  of two Green's functions at two energies $E$ and $E'$ and for
Hamiltonians  $H(x)$ and $H(x')$ taken at two different values of an external
parameter.  For example, the evaluation of the parametric level-density correlator $k[(E-E')/\Delta,x-x']$ requires the calculation of terms like $C_k \equiv \overline{{\rm Tr}G(E^-,x){\rm
Tr}G(E^{\prime +},x^\prime)}$.  We have
\begin{eqnarray}\label{2-Green-Gauss}
  && C_{k} =\int d\bbox\Psi \; p_{k}(\bbox\Psi) \nonumber \\
   && \times\; \exp\left[{i\over 2}\bbox\Psi^\dagger(-E\Lambda+{\Omega/
2}+i\delta)\bbox\Psi\right]
   \overline{\exp\left({i\over 2}\bbox\Psi^\dagger \Lambda {\cal H}
\bbox\Psi\right)}
   \;,
\end{eqnarray}
  where $\bbox\Psi \equiv \left(\bbox\Psi_1 \atop \bbox\Psi_2 \right)$ is an
$8N$-dimensional  vector, and $\cal H$ and $\Lambda$ are $8N
\times  8N$ matrices. Here ${\cal H}$ has a $2\times 2$ block-diagonal form, where the
two diagonal blocks are $H(x)\otimes I_4$ and  $H(x')\otimes I_4$. Moreover, $\Lambda$
has  a similar structure with diagonal blocks of $I_N\otimes I_4$ and
$-I_N\otimes  I_4$. We have also defined $\bar E = (E+E')/2$,
$\Omega = E-E'$, and $p_k(\bbox\Psi)
\equiv  (\bbox s_1^\dagger \bbox  s_1)(\bbox s_2^\dagger \bbox s_2)$.
The calculation of Eq. (\ref{2-Green-Gauss}) can be mapped onto the 0D
supersymmetric  nonlinear $\sigma$ model as follows:

\noindent (i)  The average in Eq. 
(\ref{2-Green-Gauss})  is taken using the second-order cumulant in the
exponent.   We obtain an integral representation of $C_k$ with an
action  {\em quartic} in  $\bbox\Psi$.

\noindent (ii) The action is converted to one that is {\em quadratic}
in $\bbox \Psi$  by introducing an auxiliary integration variable
$\sigma$ that is an $8\times  8$ graded matrix (the Hubbard-Stratonovich transformation).

\noindent (iii) The Gaussian integral over $\bbox \Psi$ is done exactly,
leading  to an effective action in $\sigma$.

\noindent (iv) The $\sigma$ integral is evaluated by the saddle-point method
in  the limit $N \to \infty$ and $\Delta \propto 1/N \to 0$ (the quantities
$\omega  \propto \Omega/\Delta$, $\tilde\delta \propto \delta/\Delta$, and
$|\bar  x - \bar x'| \propto N^{1/2} |x-x'|$ are all kept constant in this limit).
This leads to a nonlinear equation for $\sigma$ that is a supersymmetric
generalization  of the nonlinear $\sigma$ model.

\noindent (iv) The solutions $Q$ of the saddle-point equation form a manifold
(satisfying  $Q^2 =I$). Here
 $C_k$ can be expressed as an integral over this saddle-point
manifold, 
\begin{equation}\label{2-Green-Q}
   C_{k}=\int dQ \; p_{k}(Q) \exp[-{\cal F}(Q)] \;,
\end{equation}
where the ``free energy'' is given by
\begin{eqnarray}\label{free-RMT}
  {\cal F}(Q) = && i (\pi \omega/ 4 + i\tilde\delta){\rm Trg}(Q\Lambda) \nonumber \\
  && -(\pi^2/64) \mid \bar x- \bar x^\prime\mid^2 {\rm Trg}([Q,\Lambda]^2)
\end{eqnarray}
and ${\rm Trg}$ is a supertrace. The integration over $Q$ is performed using a  parametrization
of  the saddle-point manifold by Efetov (1983).  The integration can be done
over  all but three commuting variables that appear in the final analytic
 result for $k(\omega,x)$; see Simons and Altshuler, 1993a, 1993b.

The supersymmetry method for disordered systems follows along similar
 lines except that the discrete label $n$ ($n=1,\ldots,N$) of
the  supervector $\bbox{\Psi}$ becomes the continuous spatial label
$\bbox{r}$.   The auxiliary  supermatrices in the Hubbard-Stratonovich
transformation  acquire a
spatial  dependence $\sigma(\bbox{r})$.  The saddle-point approximation
requires  a large parameter -- $k_F l$, which  plays a role analogous to $N$ in RMT. By minimizing the action
one  finds a spatially uniform saddle point $Q_0 =\Lambda$ that has the
same  form as in RMT. However, the saddle-point manifold now  consists of matrix fields $Q(\bbox{r})$ that describe {\em
long}-wavelength fluctuations.
By expanding the action around the saddle point to leading order in $1/k_F
l$, we find (for the case without an external parametric dependence of the
Hamiltonian)  a free energy of the form
\begin{equation}\label{free-diffusive}
{\cal F}[Q] = {\pi \nu \over 8} \int d \bbox{r} [ \hbar D {\rm Trg}(\nabla
Q(\bbox  r))^2 + 2 i \Omega {\rm Trg}(Q\Lambda)]\;,
\end{equation}
where $D$ is the classical diffusion coefficient and $\nu$ is the density of
states  (per unit volume).  The physical quantities  are evaluated 
according to expressions of the form of Eq. 
(\ref{2-Green-Q})  but where now the integration is over $d Q(\bbox{r})$ at
each  $\bbox{r}$.

The gradient term
in Eq. (\ref{free-diffusive}) is of the order of the Thouless energy
$E_c=\hbar  D/L^2$.
In the limit $\Omega \ll E_c$, one can ignore the $\bbox{r}$ dependence of $Q$ and the gradient term vanishes.  In this case, the volume integration of the second term in Eq. (\ref{free-diffusive}) gives exactly the first term in the RMT free energy (\ref{free-RMT}).   Deviations
 from the universal behavior are important when $\Omega$ exceeds the Thouless
energy: the electron does not have enough time to
diffuse  to the boundaries and ``senses'' the dimensionality of the system.
 Such nonuniversal corrections were derived by Andreev and Altshuler (1995) for the
  asymptotic spectral correlator $K(\omega)$ in terms of the
  nonzero eigenvalues of the diffusion operator.

 A supersymmetric $\sigma$ model for ballistic
  chaotic systems was suggested by Muzykantskii and Khmelnitskii (1995), using disorder averaging when $l > L$, and by
     Andreev~{\em et~al.} (1996)  using energy averaging in a pure Hamiltonian system. The eigenvalues of the diffusion operator in the disordered case are replaced by those of the Perron-Frobenius operator in the ballistic case (Agam, Altshuler,  and Andreev, 1995). The first nonzero eigenvalue sets the scale for the ergodic time above which we expect the RMT universality. However, the mathematical difficulties involved in implementing this approach seem so far to be insurmountable.
 Similar methods were used to investigate spectral and
wave function  fluctuations in billiards with diffusive surface scattering
(Blanter, Mirlin, and Muzykantskii, 1998).

\section{Mesoscopic Fluctuations in Open Dots}
\label{open-dots}

  In this section we apply the methods of Sec.
\ref{statistical-theory}  to open quantum dots, where there are
usually  several conducting channels in each lead, and the conductance is
typically  much larger than $e^2/h$.  Open  dots with a large number of channels are  characterized by
many  overlapping resonances, i.e., $\bar\Gamma \gg \Delta$, where $\bar\Gamma$ is an average width of a resonance level in the dot.
 When the single-electron dynamics in the dot are chaotic and the electron spends enough time in the dot before it escapes (i.e., $\tau_{\rm escape}
\gg  \tau_c$), the conductance exhibits universal mesoscopic fluctuations as a function of gate voltage or magnetic  field, independent of the dot's size and
shape.  However, the fluctuations do depend on the number of modes in the leads
and  their transmission coefficients.  In the limit of a large number of open
channels,  the fluctuations become similar to the universal conductance
fluctuations known from disordered conductors.
In the universal regime, RMT can be used to characterize the
conductance fluctuations.  For a comprehensive review of the random-matrix theory of quantum transport, including applications to open dots,  see
Beenakker (1997).

The semiclassical approximation becomes useful in the limit of a large number
of  modes in the leads.  It can predict the magnitude of certain dynamical
quantities  that cannot be calculated in RMT, e.g., the correlation length of
the  conductance fluctuations versus magnetic field. But the semiclassical
approach is not suitable
for  dots with  fewer than $\sim 3$ modes per lead, or for calculating
quantities  such as the conductance distribution. For reviews of the
semiclassical  approach to transport in open ballistic microstructures, see
Baranger, Jalabert, and Stone (1993b), Stone (1995), and Baranger (1998).
Our  presentation in this Section integrates the random-matrix and
semiclassical approaches to open dots.

  The experimental results, while confirming the expected phenomena, disagree
quantitatively  with theory. The main reason is that the coherence length of the electron is
finite at any nonzero temperature.  When dephasing as well as thermal smearing are included in the
statistical  approach, good agreement is found between theory and experiment.
This  point was nicely demonstrated in a recent experiment (Huibers, Switkes, Marcus, Brouwer, {\em et al.}, 1998) in dots with single-mode leads.

Although the quantum dots used in experiments are high-mobility ballistic
structures,  many of the derivations in this section are also applicable to
diffusive dots. Indeed, as long as $\tau_{\rm escape} \gg \tau_D$, the same RMT universality  predicted in ballistic dots is expected for a diffusive dot.

Section \ref{RMT:open} presents two random-matrix approaches to
conductance  fluctuations in open dots, while Sec. \ref{semiclassical:open}
describes  the semiclassical approach. In Sec. \ref{mesoscopic:open} we
quantify  the mesoscopic fluctuations of the conductance in open dots,
including  the conductance distributions, weak-localization effect, and
fluctuations  versus energy (Ericson fluctuations) or versus an experimentally
controllable  parameter.  Finally, Sec. \ref{phase-breaking} discusses dephasing and its effects on the mesoscopic fluctuations.
Throughout  the Section we include comparison to experimental results, with an
emphasis  on more recent results. Additional experimental results can be found
in  the review of Westervelt (1998).

\subsection{The random-matrix approach}
\label{RMT:open}

  Historically, the random-matrix approach to scattering proceeded along two
main  directions: the Hamiltonian approach, in which the system's Hamiltonian is assumed to belong to a Gaussian ensemble (Sec.
\ref{Gaussian-ensembles}), and the $S$-matrix approach, in which the $S$ matrix
itself is assumed to belong to a certain ensemble (e.g., the circular ensemble of Sec. \ref{circular-ensembles}).

 The Hamiltonian approach in the regime of many overlapping resonances can be
traced  back to the statistical theory of nuclear reactions (see, for example,
Hauser and Feshbach, 1952, and Feshbach, 1992).  Ericson fluctuations
(Ericson, 1960, 1963; Ericson and Mayer--Kuckuk, 1966)  of the cross section of statistical nuclear
reactions  as a function of energy are a manifestation of the interference of
a  large number of overlapping resonances.  The Hamiltonian
approach  was followed by Verbaarschot, Weidenm\"{u}ller, and Zirnbauer (1985) in calculating
the  autocorrelation function of $S$-matrix elements, and by
Fyodorov and Sommers (1996a, 1996b, 1997)  in calculating various statistical properties of the
$S$ matrix  in chaotic systems.

 The random-$S$-matrix approach was introduced in the theory of statistical
nuclear  reactions by Mello, Pereyra, and  Seligman (1985) and by Friedman and Mello (1985), through  the maximal entropy approach. A similar approach was applied to quantum transport in
disordered  metals; see, for example, the review by Stone~{\em et~al.} (1991). The
connection  between the statistics of the eigenphases in chaotic scattering
and  the circular ensembles was suggested by Bl\"{u}mel and Smilansky (1988, 1989, 1990); for a
review  see Smilansky (1990). Direct applications of the random-$S$-matrix  approach to transport in open dots were initiated by
Baranger and Mello (1994) and Jalabert, Pichard, and Beenakker (1994); for reviews see Mello (1995), Beenakker (1997), and Mello and Baranger (1999). Both approaches turn out
to  be equivalent when the external parameters of the system are fixed, as proven by Brouwer (1995). Parametric correlations  can be successfully derived in the Hamiltonian approach but not in the $S$-matrix approach.

 In the Hamiltonian approach,   we
assume a Hamiltonian $H$ that is taken
from the appropriate Gaussian ensemble of random matrices. The statistical
properties  of the $S$ matrix and of the conductance can then be inferred
from Eq. (\ref{S-matrix}).
Various moments and correlation functions of the $S$ matrix depend only on
combinations  of the dot-leads coupling coefficients $W_{\mu c}$ that are invariant under the transformations
 that leave the ensemble distribution invariant. Such invariant combinations are
\begin{eqnarray}\label{M-matrix'}
 M_{c c^\prime} \equiv {2 \pi \over N} \sum_\mu W^\ast_{\mu c} W_{\mu
c^\prime}  \;,
\end{eqnarray}
measuring the degree of correlation among the open channels.
For example, the average $S$ matrix is given by
\begin{equation}\label{average-S}
\overline{S} = {I - \pi M/2\Delta \over I + \pi M/2\Delta}\;,
\end{equation}
where $M$ is the matrix defined in Eq. (\ref{M-matrix'}).
It follows from Eq. (\ref{average-S}) that the matrix $M=2 \pi W^\dagger W/N$ is
completely  determined by $\bar{S}$, and therefore all moments and
correlations  of the statistical $S$ matrix are functions of $\bar S$ only.

 The average $S$ matrix is not unitary. The eigenvalues $T_c$ of $1- \bar S \bar S^\dagger$ measure the unitary deficiency of the $S$-matrix ($0\leq T_c \leq 1$) and are called the
{\em transmission  coefficients}.
In a set of  ``eigenchannels'' for which the
 $M$ is diagonal
($M_{c c^\prime} = w^2_c \delta_{c c^\prime}$),
 $\bar S$ is diagonal too and $T_c \equiv 1 - |\bar{S}_{c
c}|^2 = { ({2 \pi w_c^2 / \Delta} )/ \left( 1 + { \pi w_c^2 / 2 \Delta}
\right)^2}$.

  As long as we work at a fixed energy and given values of the external
parameters,  it is possible to convert the original probability density
(\ref{Gaussian-ens})  for the Hamiltonian $H$ to a probability density $P(S)$
in  the space of unitary $S$ matrices of the respective symmetry.
It was shown by Brouwer (1995) that for $\Lambda_{\rm tot}=\Lambda_1
+  \Lambda_2$ open channels
\begin{equation}\label{Poisson}
P(S) \propto \left|\det(1-\bar{S}^\dagger S) \right|^{-\beta(\Lambda_{\rm
tot}-1)  -2} \;,
\end{equation}
where $\bar{S}$ is given by Eq. (\ref{average-S}) and the measure $d \mu (S)$ is
that  of the corresponding circular ensemble.  For $\bar S \neq 0$, Eq. (\ref{Poisson}) is known as the Poisson kernel
(Hua, 1963).  It was derived by
Pereyra and Mello (1983) in nuclear physics through maximizing the entropy ${\cal S}[P(S)] = -\int d \mu (S) P(S) \ln P(S)$
of  an arbitrary distribution $P(S)$ subjected to the constraints
$\overline{S^n}  = \bar S^n$ (for any positive integer $n$).
For $\bar{S} =0$, we recover
Dyson's  circular ensemble for the scattering matrices (see Sec.
\ref{circular-ensembles}).  Indeed, owing to the invariance
(\ref{circular-ens})  of the circular ensemble's measure, the average
$S$ matrix  must satisfy $\bar S = \overline{U S V} = U \bar S V$ for
arbitrary  unitary matrices $U$ and $V$, a condition that can only be met by
$\bar  S=0$.  This corresponds to the case of ideal leads where all
$T_c  =1$.

Thus far we have not distinguished between channels belonging to the left and
right  leads. This distinction becomes important in calculating the
conductance  through the dot.  In general, the $S$ matrix
is defined by the linear relation between the incoming and outgoing amplitudes. It can be written in the form $S = \left( {r \atop t} {t' \atop r'} \right)$,
where $r$ and $r'$ are the reflection matrices on the left and on the right,
respectively,  while $t$ and $t'$ are transmission matrices from left to right
and  from right to left, respectively. For $\Lambda_1$ ($\Lambda_2$) channels
 in  the left (right) lead, $r$ and $r'$ are square matrices of dimension
  $\Lambda_1\times  \Lambda_1$ and $\Lambda_2 \times \Lambda_2$, while $t $
   and  $t'$  are rectangular matrices of order $\Lambda_1\times\Lambda_2$
    and $\Lambda_2\times\Lambda_1$.

 Here $tt^\dagger$ and $t't'^\dagger$ share the same number ${\rm min}
(\Lambda_1,\Lambda_2)$  of nonzero transmission eigenvalues $\tau_a$.
 The zero-temperature conductance
 is proportional to the total transmission $T$:
\begin{equation}
T = {\rm Tr}\; (t t^\dagger) = {\rm Tr}\; (t' t'^\dagger) = \sum\limits_a
\tau_a  \;.
\end{equation}
For ideal leads (i.e., $\bar S=0$), the distribution of the transmission eigenvalues is given by (Baranger and Mello, 1994; Jalabert, Pichard, and Beenakker, 1994)
\begin{equation}\label{transmission-dist}
P_\beta(\tau_1,\tau_2,\ldots) \propto \left( \prod\limits_{a < b} |\tau_a -
\tau_b  |^\beta \right) \prod\limits_c \tau_c^{\beta -2 \over 2} \;.
\end{equation}
  In the more
 general case of nonideal leads, the Poisson kernel
(\ref{Poisson})  should be taken into account beyond Eq. 
(\ref{transmission-dist}); see Brouwer, 1995.

\subsection{The semiclassical approach}
\label{semiclassical:open}

  The semiclassical approach to open dots is useful in the limit of a large
number  of modes $\Lambda \to \infty$ (equivalent to $\hbar \to 0$).
The  starting point is an expression for the transmission amplitude $t_{c'
c}$  from mode $c$ in the left lead to mode $c'$ in the right lead  in terms
of  the retarded Green's function $G^R$. Assuming the dot-lead interfaces are at
$x=x^l$  and $x=x^r$, we have
\begin{eqnarray}
t_{c' c} = && -i\hbar (v_{c'} v_c)^{1/2} \int dy' dy {\phi_{c'}^r}^\ast (y') \nonumber \\
&& \times\;G^R(x^l,y;x^r,y';E)  \phi^l_c(y) \;,
\end{eqnarray}
where $\phi^{l(r)}_c(y)=\sqrt{2/{\cal W}}\sin(c \pi y/{\cal W})$  $(c=1,\ldots,\Lambda)$ are the transverse channel wave functions in the left
(right)  lead (${\cal W}$ is the width of each lead). The retarded Green's function is then approximated by a sum over
classical  paths [Eq. (\ref{sc-Green-energy})].  The integrals over the
lead-dot  cross sections are done by stationary phase, leading to
(Jalabert, Baranger, and Stone, 1990)
\begin{eqnarray}\label{semiclassical-t}
t_{c' c} =&&  -\left( \pi i  \over 2 k {\cal W} \right)^{1/2}
\sum\limits_{\alpha(\bar c' \bar c)}
  {\rm sign}(\bar c'){\rm sign}(\bar c) \nonumber \\  && \times\;\tilde A_\alpha
e^{{ i \over \hbar}\tilde S_\alpha(\bar c', \bar c, E) - i {\pi \over 2}
\tilde  \mu_\alpha} \;,
\end{eqnarray}
where $\bar c=\pm c$ and $\tilde \mu_\alpha$ is a modified Maslov index (given by Baranger, Jalabert and Stone, 1993b). The sum is taken over paths
$\alpha$  that start on the left at angle $\theta$ and end on the right at
angle  $\theta'$,  determined by equating the initial and final transverse
momenta  of the electron to the quantized values of the momenta in modes $c$
and  $c'$, respectively (i.e., $\sin \theta = {\pi \bar c / k {\cal W}}$ and $\sin \theta' = {\pi \bar c' / k {\cal W}}$).
The modified amplitude and action are given by
\begin{mathletters}
\begin{eqnarray}
\tilde A_\alpha (\theta,\theta') & = & \left[ {1 \over {\cal W} |\cos \theta'|} \left| \left(
\partial  y \over \partial \theta'\right)_\theta \right| \right]^{1/2} \;,
\label{amplitude}\\
\tilde S_\alpha (\bar c', \bar c, E)&= & S_\alpha(y'_0,y_0,E)  +  {\pi \hbar
\bar  c y_0 / {\cal W}} -
{\pi \hbar \bar c' y'_0 / {\cal W}} \;, \label{action}
\end{eqnarray}
\end{mathletters}
where $y_0$ and $y'_0$ are determined from the stationary phase conditions for the angles $\theta$ and $\theta'$. Note that the energy dependence in the sum of Eq. 
(\ref{semiclassical-t}) appears only in the action $\tilde S_\alpha = k
\tilde  L_\alpha$ (where $\tilde L_\alpha$ is the effective length of path
$\alpha$).

Expression (\ref{semiclassical-t}) holds only for ``chaotic'' isolated
trajectories  that scatter from the boundary of the dot before exiting on the
right.  Direct trajectories give rise to nonuniversal
effects,  and the geometry of the dot is usually chosen so as to minimize
these  effects.

\subsection{Mesoscopic fluctuations of the conductance}
\label{mesoscopic:open}

\subsubsection{Conductance distributions}
\label{conductance-distributions}

 According to Landauer's formula, the zero-temperature dimensionless
conductance  $g$ [$g=G/(2e^2/h)$] is the total transmission
$T=\sum_a  \tau_a$, and its distribution for ideal leads can be
determined  from the joint distribution (\ref{transmission-dist}) of the
transmission  eigenvalues.
For single-mode ideal leads ($\Lambda_1=\Lambda_2=1$; Baranger and Mello, 1994; Jalabert, Pichard, and Beenakker, 1994),
\begin{equation}\label{g-dist:open}
P(g) = {1\over 2} \beta g^{\beta -2 \over 2}\;\;\;(0<g<1) \;.
\end{equation}
 The GOE and GUE distributions are shown by solid lines in the upper panels
of  Fig. \ref{fig:g-dist}. In the absence of a magnetic field ($\beta=1$), it
is  more probable to find smaller conductances; in the presence of a magnetic
field  ($\beta=2$), all allowed values  are equally probable; and in the
presence  of spin-orbit scattering ($\beta=4$), larger conductances are more
probable.

\begin{figure}
\epsfxsize= 7 cm
\centerline{\epsffile{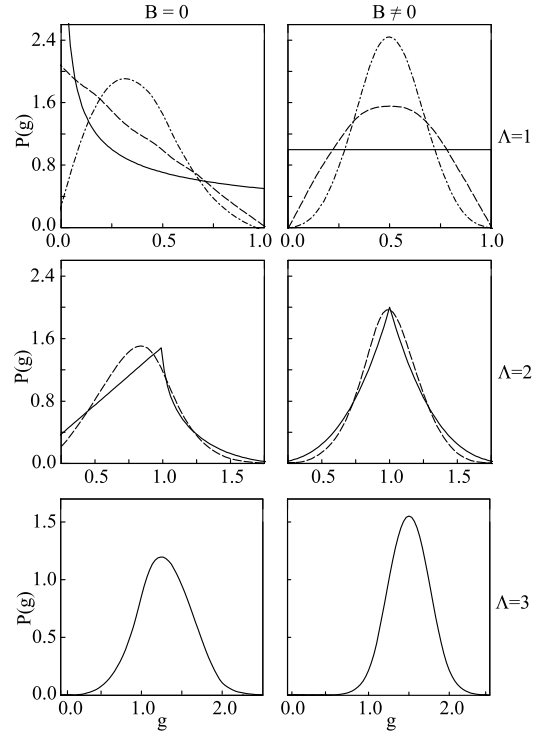}}
\vspace*{2 mm}
\caption{Conductance distributions in open dots with $\Lambda=1,2$ and $3$ channels
in  each lead for conserved (left) and fully broken (right) time-reversal
symmetries.  The distributions are calculated using Dyson's circular random
matrix  ensembles: solid lines, the distributions in the absence of
phase breaking; dashed lines (shown for $\Lambda=1,2$ only),  the results of
the  voltage-probe model with a single-mode phase-breaking lead
($\Lambda_\phi=1$); dash-dotted lines (shown for $\Lambda=1$ only), $\Lambda_\phi=2$. Adapted from Baranger and Mello (1994, 1995).
 }
\label{fig:g-dist}
\end{figure}

 The conductance distributions are also sensitive to the number of channels
in  each lead and were calculated  for $\Lambda_1=\Lambda_2= \Lambda\leq 3$
 (Baranger and Mello, 1994) using
the  circular ensembles. The results are shown by the solid
lines  in Fig. \ref{fig:g-dist}.  Already for $\Lambda=3$ they are quite close
to  a Gaussian, which is their exact asymptotic limit ($\Lambda \to \infty$).
  The average and variance  are given by
\begin{mathletters}\label{g-moments}
\begin{eqnarray}
 \bar g = &&  {\Lambda \over 2} -{\Lambda \over 2(2\Lambda -1
+2/\beta)}\left({2\over  \beta} -1 \right) \nonumber \\ && \to {\Lambda \over 2}
-\frac{1}{4}  \left({2\over \beta} -1 \right) \;, \label{average-g} \\
 \sigma^2(g) = &&  {2 \over \beta} { \Lambda^2 (\Lambda-1+2/\beta)^2 \over
(2\Lambda+  2 +2/\beta)(2\Lambda-1+4/\beta)(2\Lambda-1+2/\beta)^2} \nonumber \\ & & \to
{1\over  8\beta}\;,
\label{variance-g}
\end{eqnarray}
\end{mathletters}
where the limit is $\Lambda\to \infty$. The asymptotic values were derived
earlier  by Iida, Weidenm\"{u}ller, and Zuk (1990a, 1990b) in the Hamiltonian approach.

 The dominating term $\Lambda/2$ in the average conductance (\ref{average-g})  is just its classical value, while the second term is the quantum
weak-localization correction (see below). Its asymptotic value is $-1/4$ and $0$ for
conserved and broken time-reversal symmetry, respectively.

 One conclusion from Eq. (\ref{variance-g}) is that the rms of the
conductance  fluctuations is of order ${e^2/ h}$, irrespective of the size of
the  average conductance $\Lambda {e^2 / h}$ or the size of the system (the
number  of modes  $\Lambda = {\rm int}[k {\cal W}/\pi]$ is size dependent).
This  is a manifestation of the phenomenon of universal conductance
fluctuations, known from disordered metals. In the
limit $\Lambda \to \infty$,  the conductance variance  is twice as large
 in the absence of magnetic field  than in its presence.

The case of nonideal leads was treated in the Hamiltonian approach  by
Iida, Weidenm\"{u}ller, and Zuk (1990a, 1990b).   The
$S$-matrix  approach (where the appropriate ensemble is the Poisson kernel) was followed by
Brouwer and Beenakker (1994). For equivalent
channels  ($T_c=T$), and in the limit $\Lambda T \gg 1$, one can integrate over
the  ensemble by diagrammatic methods (Brouwer and Beenakker, 1996).

Direct
paths  connecting the leads can also lead to $\bar S \neq 0$. This situation
is  analogous to nuclear reactions in which both direct and equilibrated
components  contribute to the reaction cross section (Feshbach, 1992).
Baranger and Mello (1996)  extracted an average ``optical'' $S$ matrix through
an  energy average of the numerical data (in cavities). The distributions
predicted  by the corresponding Poisson kernel (\ref{Poisson}) were then found
to  be in good agreement with the direct numerical calculations.

 We can gain additional physical insight into the behavior of the average
conductance  and its weak-localization correction in the semiclassical
approximation (Baranger, Jalabert and Stone, 1993a). Using  Eq. (\ref{semiclassical-t}) in the
diagonal  approximation (see Sec. \ref{semiclassical:spectral}) and
replacing  $\sum_c \to \int_{-1}^1 d \sin \theta$, we obtain
 for the average transmission
$\bar T = \Lambda {\cal T}$,
where $\cal T$ is the classical transmission probability per incoming mode (Beenakker and van Houten, 1989).   This result
agrees  with the leading-order term of the RMT result (\ref{average-g}) if
${\cal  T} =1/2$. Indeed, for a fully chaotic system and ideal leads, the
electron  injected into the dot forgets its origin and has equal probability
of exiting on the left or on the right.

  The quantum correction to the average transmission is of order unity and
cannot  be calculated directly in the semiclassical approach. Instead we can
evaluate  such a correction to the average of the total reflection $R =
\sum_{c'  c} |r_{c' c}|^2$. To leading order $\bar R \approx \Lambda
{\cal  R}$, where ${\cal R}$ is the classical reflection probability.  However, if time reversal is
a  good symmetry, a correction to the diagonal part of the reflection $\bar
R_D  =\overline{ \sum_{c} |r_{c c}|^2}$ comes from pairs of
 time-reversed orbits -- they have the same action and Maslov index,
and their contribution to the diagonal reflection does not average
to  zero.    We find that the
correction  to $\bar R_D$ is
$\delta \bar R_D = \frac{1}{2} \int_{-1}^1 d \sin \theta
\sum_{\alpha(\theta,  \pm \theta)} |\tilde A_\alpha(\theta,\theta)|^2$, where $A_\alpha(\theta,\theta)$ is given by Eq. (\ref{amplitude}) for $\theta=\theta'$, and the sum is over all classical paths $\alpha$ that enter and exit at angle $\theta$. 
 For chaotic trajectories the function $P(\theta,\theta')\equiv
\sum_{\alpha(\theta,\theta')} |\tilde A_\alpha(\theta,\theta')|^2$
satisfies  approximately a uniformity condition $P(\theta,\theta)\approx
\int_{-1}^1  d \sin \theta' P(\theta,\theta')$ (Baranger, Jalabert, and Stone, 1993b), 
leading to the weak-localization correction (for conserved time-reversal symmetry)
\begin{equation}\label{delta-R}
\delta \bar R_D = {\cal R}\;.
\end{equation}
In the absence of time-reversal  symmetry, the return probability to the same channel is
$\overline{  |r_{c c}|^2}=
 \overline{\sum_{c'} |r_{c' c}|^2}/\Lambda \approx {\cal R} /\Lambda$.
When  time-reversal symmetry is conserved, this probability increases by
$\delta  \bar R_D/\Lambda \approx {\cal R} /\Lambda$, so its value doubles to
$2{\cal  R} /\Lambda$.

 How does the weak-localization correction to the reflection affect the
conductance?   We use the unitarity condition of the $S$ matrix
$\overline{|S_{c  c}|^2} + \sum_{c'\neq c}\overline{|S_{c'c}|^2} =1$,
and  assume that all off-diagonal
probabilities
$\overline{|S_{c'c}|^2}$ ($c'\neq c$) are equal. An increase in  the return
probability  $\overline{|r_{c c}|^2}=\overline{|S_{c c}|^2}$ by ${\cal R}/\Lambda$ must then be
compensated for by a decrease $\approx  {\cal R}/2\Lambda^2$ in  each individual off-diagonal probability. The change in the total average transmission
$ \bar T = \sum_{c'\in r, c \in l}\overline{|S_{c'c}|^2}$ is then
$\delta \bar T \approx - {\cal R}/2$.
For ideal leads, ${\cal R} =1/2$, and $\delta \bar T = -1/4$, in agreement
with Eq. (\ref{average-g}).

  The enhancement of the mean-square diagonal $S$-matrix element
relative  to the mean-square off-diagonal element due to time-reversal
symmetry  is well known in the statistical theory of nuclear reactions, where
it  is called the elastic enhancement factor. It was also observed in quantum
chaotic  scattering (Doron, Smilansky, and Frenkel, 1990; Bl\"{u}mel and Smilansky, 1992).

\subsubsection{Weak localization}
\label{weak-localization:open}

 In Sec. \ref{conductance-distributions} we saw that the average
conductance  increases when time-reversal symmetry is fully broken. In
this section we  discuss the dependence of $\bar{g}$ on
a magnetic field $B$ as time-reversal symmetry is broken gradually.

 The semiclassical expression for $\bar g$ versus $B$ was derived by
Baranger, Jalabert, and Stone (1993a).   For a weak magnetic field  the classical
trajectories  are approximately the same as for $B=0$, but the wave function
acquires  an Aharonov-Bohm phase. The contribution to the action in the
presence  of a field $B$ derived from a vector potential $\bbox A$  is
$(e/c)\int  \bbox{A}\cdot \bbox{d \ell}$. Under time reversal,  this action
changes  sign, and the action difference between paths that are time-reversed
pairs  is nonzero:
$(\tilde S_\alpha - \tilde S_{\alpha^T})/\hbar =
(2 e / \hbar c) \int \bbox{A}\cdot \bbox{d \ell} =
 2{ \Theta_\alpha B / \Phi_0}$,
where $\Theta_\alpha$ is the area (times $2\pi$) swept by the classical
trajectory  $\alpha$.  Following the method that led to Eq. (\ref{delta-R}), we now find
\begin{equation}\label{WL}
\delta \bar R_D(B) = {\cal R} \int d \Theta P(\Theta) e^{2 i {\Theta B\over
\Phi_0}}\;,
\end{equation}
   where $P(\Theta)\equiv \frac{1}{2} \int d \sin \theta \;d \sin
\theta' \sum_{\alpha(\theta,\theta')} |\tilde A_\alpha|^2 \delta(\Theta -
\Theta_\alpha)$ is the distribution of the effective area $\Theta$ enclosed by the electron trajectory. This area distribution is exponential
 in chaotic systems
(Doron, Smilansky, and Frenkel, 1990; Jalabert, Baranger, and Stone, 1990)  but satisfies only a power law in nonchaotic
systems. Using an exponential form $P(\Theta)
 \propto e^{-\alpha_{cl} |\Theta|}$ in  Eq. (\ref{WL}), and the unitarity argument discussed at the end of Sec. \ref{conductance-distributions}, we
find
\begin{mathletters}\label{semiclassical-WL}
\begin{eqnarray}
\delta g(B) \equiv \bar g^{\rm GUE} - \bar g (B) & = & {{\cal R}\over 2} {1 \over 1 + (B/B_{cr})^2} \;, \label{WL-chaotic}\\
B_{cr}& = & \alpha_{cl} \Phi_0/2 \;. \label{crossover-B}
\end{eqnarray}
\end{mathletters}
Here $B_{cr}$ is the crossover field measuring a typical field required to
suppress the weak-localization correction.
 $\delta g(B)$ is largest at $B=0$ (note that the average conductance itself has a dip at $B=0$).

 In a chaotic system $\alpha_{cl}
\propto  \tau_{\rm escape}^{-1/2}$ (Jensen, 1991). The basic argument is that area accumulation  in a chaotic dot is diffusive  and $\overline{\Theta^2}$ is
 therefore linear in  time.  Since
the  average time spent by the electron in the dot is $\tau_{\rm escape}$, we
have  $(\overline{\Theta^2})^{1/2}/{\cal A}\propto
(\tau_{\rm escape}/\tau_c)^{1/2}$  (${\cal A}$ is the dot's area and $\tau_c$ is
the  ergodic time). From $\alpha_{cl} \propto (\overline{\Theta^2})^{-1/2}$
and  using Eq. (\ref{crossover-B}), we obtain for the crossover flux $\Phi_{cr}= B_{cr} {\cal A}$
\begin{equation}\label{crossover-Phi}
{\Phi_{cr} \over \Phi_0} =
\frac{1}{2}\kappa  \left(2\pi{\tau_c \over \tau_{\rm escape}}\right)^{1/2}=
\frac{1}{2}\kappa  \left(2\pi{\bar \Gamma \over E_T}\right)^{1/2}
\;,
\end{equation}
where $E_T = \hbar/\tau_c$ is the ballistic Thouless energy and $\bar
\Gamma=\hbar/\tau_{\rm  escape}$ is the mean escape width of a level in the
dot. The quantity 
$\kappa$ is a geometrical factor that depends on the device. In a dot with a
total  of $2\Lambda$ ideal open channels, $\bar \Gamma = (\Delta/2\pi) 2
\Lambda$,  and
${\Phi_{cr} / \Phi_0} = (\kappa/2) g_T^{-1/2} \sqrt{2\Lambda}$,
where $g_T\equiv E_T/\Delta$ is the ballistic Thouless conductance.
In particular, $B_{cr} \propto \sqrt{\Lambda}$.

 The exact weak-localization line shape $\delta g(B)$ for a small number of channels was
calculated  in the Hamiltonian approach using the crossover ensemble
(\ref{transition-ens})  for $H$ in Eq. (\ref{S-matrix}) for the
$S$ matrix  (Pluha\u{r}~{\em et~al.}, 1994). The exact expression (evaluated by supersymmetry)
is  a complicated triple integral [see, for example, Eq. (6) of
Pluha\u{r}~{\em et~al.}, 1994],  but it is well approximated by a Lorentzian
\begin{equation}\label{RMT-WL}
 \delta g(B) \approx \delta g(0) {1 \over 1 + (\zeta/\zeta_{cr})^2} \;,
\end{equation}
where $\delta g(0) = \Lambda / (4\Lambda +2)$, in agreement with the circular
ensemble  result (\ref{average-g}), and
$\zeta_{cr} = (\sqrt{\Lambda} / 2\pi) \left(1- 1/{2\Lambda}
\right)^{-1/2}$.
 This crossover scale of the conductance is different from the crossover scale of the
spectral  statistics ($\sim 1$), and  can be understood in
terms  of a competition between the mixing time
$\hbar/\tau_{\rm mix}=2\pi \Delta \zeta^2$ and the decay time of a typical
resonance  in the dot $\hbar/\tau_{\rm escape}= (\Delta/2 \pi)\Lambda$. The
conductance  statistics make the crossover when $\tau_{\rm escape}/\tau_{\rm
mix}  = 4 \pi^2 \zeta^2/\Lambda \sim 1$ or $\zeta_{cr} \sim
\sqrt{\Lambda}/2\pi$.

 Comparing the RMT result [Eq. (\ref{RMT-WL})] with the semiclassical result
[Eq. (\ref{semiclassical-WL})],  we conclude that if the time-reversal symmetry is
broken  by a magnetic field, then $\zeta \propto B$. This result is
confirmed  in billiard-model calculations (Bohigas~{\em et~al.}, 1995; Alhassid, Hormuzdiar, and Whelan, 1998).
 A semiclassical expression can be derived for the
proportionality  constant (Bohigas~{\em et~al.}, 1995; Pluha\u{r}~{\em et~al.}, 1995).  Obviously $\zeta_{cr}$
and  $B_{cr}$ satisfy the same proportionality relation, confirming our
earlier  conclusion that $B_{cr} \propto \sqrt{\Lambda}$.

We have seen in Sec. \ref{conductance-distributions} that the conductance
 variance in open dots is reduced by a factor of 2 when time-reversal
symmetry  is broken. Using the transition random-matrix ensemble
(\ref{transition-ens}),  Frahm (1995) derived the complete dependence
of  the conductance variance on the transition parameter $\zeta$. In the
limit $\Lambda \gg 1$,
\begin{equation}\label{variance-B}
 \sigma_\zeta^2(g) - \sigma_{\rm GUE}^2(g)
 \approx  \left[ {\delta \sigma^2(g) \over 1 +
(\zeta/\zeta_{cr})^2}\right]^2   \;,
\end{equation}
where $\sigma_{\rm GUE}^2=1/16$ and $\delta \sigma^2(g) \equiv \sigma^2_{\rm
GOE}(g)  - \sigma_{\rm GUE}^2(g)=1/16$.

 The weak-localization effect has been observed in open quantum dots by several
experimental  groups (Marcus~{\em et~al.}, 1992;  Berry~{\em et~al.}, 1994a, 1994b; Chang~{\em et~al.}, 1994; Keller~{\em et~al.}, 1994, 1996; Chan~{et~al.}, 1995).
Methods  employed to find the average conductance include averaging  over energy
(Berry~{\em et~al.}, 1994a, 1994b; Keller~{\em et~al.}, 1996),  over shapes (Chan~{et~al.}, 1995), and over an ensemble of
dots  (Chang~{\em et~al.}, 1994).

Figure \ref{fig:WL-Chang} shows $-[\bar G(B) -\bar G(0)]= \delta G(B) - \delta
G(0)$  as a function of magnetic field $B$ for two different ballistic
cavities  -- a stadium [panels (a) and (c)] and a circle [panels (b) and (d)].  Panels (a) and (b) display
the  experimental result of Chang~{\em et~al.} (1994) using
an  array of $6 \times 8$ ``identical'' quantum dots.  The ensemble average
shows  a clear weak-localization effect.  While the line shape for
the  stadium is a Lorentzian [Eq. (\ref{semiclassical-WL})], expected for a chaotic
cavity,  the line shape for the circle is {\em triangular}, consistent with a power-law area distribution that is expected
for  an integrable system.
   Figures \ref{fig:WL-Chang}(c) and \ref{fig:WL-Chang}(d) show the results of numerical
calculations for the same 2D geometries used in the experiment.

\begin{figure}
\epsfxsize= 8 cm
\centerline{\epsffile{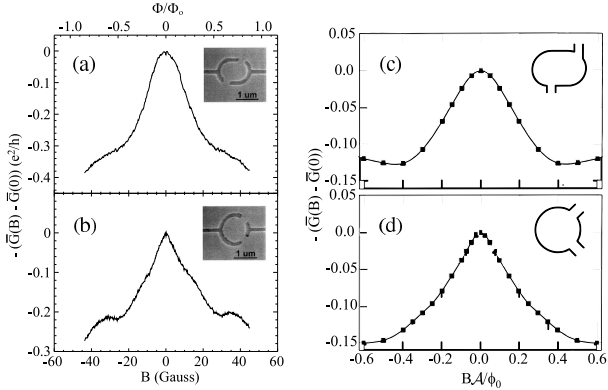}}
\vspace*{2 mm}
\caption{
Weak-localization effect in open quantum dots:  
experiment [panels (a) and (b)] vs theory [panels (c) and (d)]. The experimental results  for  $-[\bar G(B) - \bar G(0)]$ (averaged over 48 dots) are shown for (a) a stadium
 and (b) a circle. The observed line shape is  a Lorentzian for the stadium (a chaotic structure)
 and triangular for the circle (a
regular  system). The insets are electron micrographs of the
fabricated  microcavities.
Numerical results of the weak-localization effect are shown for (c) a stadium and (d) a circle. The circle's cavity includes a weak
disorder  potential to account for possible impurities in the experimental
structures.   Note the similarity between experiment and theory for both
geometries.  Adapted from Chang~{\em et~al.} (1994).
}
\label{fig:WL-Chang}
\end{figure}

  Chan~{\em et~al.} (1995) fabricated a dot whose shape could be distorted and collected statistics at each value of the magnetic field $B$ from
various  shapes of the dot.   The measured  weak-localization effect and the
variances  of the conductance fluctuations (for $\Lambda=2$ channels in each lead) are well fitted to Eqs. (\ref{RMT-WL})
and  (\ref{variance-B}), respectively ($\zeta/\zeta_{cr}=
B/B_{cr}=2B/\alpha  \Phi_0$ was used in these formulas).  However, the
parameters  found in the fit to the data  $\delta g= 0.15$ (where $g$ is the
conductance  in units of $2e^2/h$) and $\delta \sigma^2(g) = 0.00475$ are
significantly  smaller than the RMT values of  $\delta g=0.2$ and $\delta
\sigma^2(g)  \approx 0.0229$ for $\Lambda=2$ [see Eqs. (\ref{g-moments})].
This  can be explained by dephasing effects that will be
 discussed in Sec. \ref{phase-breaking}.
The characteristic inverse area parameter $\alpha$ is found to be
$\alpha = 0.14 \pm 0.01\;\mu$m$^{-2}$ and
$\alpha = 0.11 \pm 0.01\;\mu$m$^{-2}$ from  fits to the weak-localization
line shape (\ref{RMT-WL}) and variance curve (\ref{variance-B}), respectively. An independent measurement of
this  parameter is from the power spectrum of the conductance fluctuations
versus  magnetic field (see Sec. \ref{parametric-correlations:open}).

\subsubsection{Ericson fluctuations}
\label{Ericson-fluct}

 One of the prominent features of stochastic nuclear reactions is the rapid
fluctuation  of the cross section $\sigma$ versus the total reaction energy
$E$ [see, for example, Fig. \ref{fig:open-dot}(a)].
Ericson (1960, 1963)  quantified these fluctuations in terms of the energy
autocorrelation  function
\begin{equation}\label{Ericson}
c_\sigma(\Delta E)=\overline{\sigma(E + \Delta E)\sigma(E)} -
\overline{\sigma(E)}^2
\propto { \Gamma^2 \over (\Delta E)^2 +\Gamma^2} \;,
\end{equation}
where the average is taken over $E$. According to
Ericson,  this  autocorrelation function is a Lorentzian whose width  measures
the  average resonance width $\Gamma$ in the compound nucleus.

 Do Ericson fluctuations also occur in quantum dots?  The experimental
results  in Fig. \ref{fig:open-dot}(c), showing the conductance
as  a function of the Fermi momentum, exhibit fluctuations that are indeed
similar  to those observed in nuclear reactions.   The quantity analogous to Eq. 
(\ref{Ericson})  is the conductance autocorrelation function versus the Fermi
momentum  $k$:
$c_g(\Delta k) \equiv \overline{g (k+ \Delta k) g(k)} -\overline{g}^2$.
 This correlator can be estimated semiclassically (Jalabert, Baranger, and Stone, 1990). Similar
work  for the  autocorrelations of the $S$-matrix elements in chaotic systems
was  done by Bl\"{u}mel and Smilansky (1988, 1989, 1990).  Using the Landauer
formula and the semiclassical
expression  (\ref{semiclassical-t}) for the transmission amplitudes, we find
\begin{eqnarray}\label{k-correlations}
c_g(\Delta k) & = & \left| \int_0^\infty dL e^{i \Delta k L} P(L) \right|^2
\;,
\end{eqnarray}
 where $P(L) \equiv \frac{1}{2} \int d \sin \theta d \sin
\theta'
\sum_{\alpha(\theta,\theta')} |\tilde A_\alpha|^2 \delta(L - \tilde
L_\alpha)$
is the classical distribution of path lengths (the sum is over all trajectories that originate and end at the same angles $\theta$ and
$\theta'$). In a chaotic system the length
distribution  is exponential: $P(L) \propto e^{-\gamma_{cl} L}$ and
$c_g(\Delta  k)$ is a Lorentzian, 
\begin{equation}\label{k-correlation-sc}
c_g(\Delta k) \propto {1 \over 1 + (\Delta k /\gamma_{cl})^2}\;.
\end{equation}
Thus the scale for the conductance variation as a function of the Fermi
momentum  is set by $\gamma_{cl}$, the inverse average length of a chaotic
trajectory  traversing the dot.  To relate Eq. (\ref{k-correlation-sc}) to Ericson's formula (\ref{Ericson}), it is necessary to convert $\gamma_{cl}$  to a
correlation  length in energy  $E_{cor}= \hbar v_F \gamma_{cl}$. Using $\gamma_{cl} = 1/v_F \tau_{\rm escape}$,
 we find  $E_{\rm cor} \approx \hbar/\tau_{\rm escape} = \bar \Gamma$, in agreement with Ericson's formula.

 Fluctuations of the conductance as a function of the Fermi momentum were studied experimentally  by Keller~{\em et~al.} (1996). A typical measurement of $G$ versus
$k_F$  is shown in Fig. \ref{fig:open-dot}(c).  While the average conductance increases linearly
with  the number of modes $k{\cal W}/\pi$, the fluctuations are of universal
size  $\sigma^2(g) \sim 1$. The power spectrum $S_g(f_k)$ of $g(k)$ [$S_g(f_k) = |\int dk e^{2\pi i f_k} g(k) |^2$] is the
Fourier  transform of the conductance autocorrelator $c_g(\Delta k)$, and for
a  Lorentzian line shape [Eq. (\ref{k-correlation-sc})] we expect an exponential
power  spectrum $S_g(f_k) = S_g(0) e^{-2\pi \gamma_{cl} |f_k|}$ [see, for example, Fig. \ref{fig:open-dot}(d)].
 Numerical calculations give good agreement between the classical value of
$\gamma_{cl}$  and its value extracted from fitting an exponential to the
quantum  calculations of the power spectrum (Jalabert, Baranger,  and Stone, 1990).

 The autocorrelation function $c_g^\zeta(\Delta E)$ for a fixed value of
the  transition parameter $\zeta$ in the crossover regime between GOE and
GUE  was calculated in RMT for ideal leads with $\Lambda \gg 1$
(Frahm, 1995): 
\begin{equation}\label{E-correlation-cr}
c_g^\zeta(\Delta E) = \frac{1}{16}\left\{ {1 \over \left[ 1 +\left({\zeta
\over  \zeta_{cr}}\right)^2 \right]^2 + \left({\Delta E \over
\bar\Gamma}\right)^2  } +   { 1 \over 1 + \left({\Delta E \over
\bar\Gamma}\right)^2}  \right\} \;.
\end{equation}
 Equation (\ref{E-correlation-cr}) reduces to a Lorentzian in both
the  GOE ($\zeta=0$) and GUE ($\zeta \gg 1$) limits, in agreement with the semiclassical result (\ref{k-correlation-sc}).

\subsubsection{Parametric correlations}
\label{parametric-correlations:open}

In this section we discuss the mesoscopic fluctuations of the conductance as
a  function of an external parameter.  Note that the energy should not be
considered  an external parameter since it enters in a special way in Eq.  (\ref{S-matrix}) for the $S$ matrix, while a dependence on a
generic  external parameter $x$ is introduced through $H=H(x)$.

 Of particular experimental interest is the parametric correlator when the
parameter  varied is a magnetic field:
$c_g(\Delta B) \equiv \overline{g(B + \Delta B) g(B)} - \bar{g}^2$. Following
the  semiclassical approach of Sec. \ref{Ericson-fluct}, we now obtain
\begin{eqnarray}
c_g(\Delta B)  =  \left| \int_{-\infty}^\infty d \Theta e^{i \Theta \Delta
B/\Phi_0  } P(\Theta) \right|^2 \;, \label{B-correlation-2}
\end{eqnarray}
where $P(\Theta)$ is the classical area distribution swept by the electron in the dot. In a chaotic  dot,
$P(\Theta) \propto e^{-\alpha_{cl} |\Theta|}$ (see Sec.
\ref{weak-localization:open}),  and  $c_g(\Delta B)$  is a  squared
Lorentzian:
\begin{equation}\label{correlation-B-sc}
c_g(\Delta B) \propto \left[ 1 + (\Delta B /2B_{cr})^2 \right]^{-2} \;.
\end{equation}
The correlation field $B_c$ is thus twice the crossover field
$B_{cr}$ [Eq. (\ref{crossover-B})], because the phase involved in the correlator is proportional to the difference of
areas as opposed to their sum in the weak localization case.

 Typical conductance fluctuations versus magnetic
field and the conductance correlator $c_g(\Delta B)$ in an open stadium dot are shown in Fig. \ref{fig:B-corr}(a) and (b), respectively (Jalabert, Baranger, and Stone, 1990). The
calculated correlator [solid line in panel (b)] compares well with
the  semiclassical prediction (\ref{correlation-B-sc}) [dashed line in panel (b)].  The power spectrum of $g(B)$ [$S_g(f_B) \equiv \overline{|\int
d  B e^{i 2\pi f_B B} g(B)|^2}$] is just the Fourier transform of $c_g(\Delta B)$ and is given by (Marcus~{\em et~al.}, 1993b)
\begin{equation}\label{power-sl}
S_g(f_B) = S_g(0) (1 + 2\pi \alpha \Phi_0 |f_B|) e^{-2\pi\alpha \Phi_0 |f_B|} \;,
\end{equation}
where we have used $B_{cr}=\alpha \Phi_0/2$ [see Eq. (\ref{crossover-B})]. 
The inset of Fig. \ref{fig:B-corr}(b) shows the calculated power spectrum  $S_g(f_B)$ (solid line) in comparison with a best fit to Eq. 
(\ref{power-sl})  (dashed line). Good agreement between the classical value of
$\alpha$  (determined from the area distribution in the stadium) and
the  quantum value of $\alpha$ [determined from a fit of (\ref{power-sl}) to
the  power spectrum] is found over a variety of the stadium dot's parameters.

\begin{figure}
\epsfxsize= 8 cm
\centerline{\epsffile{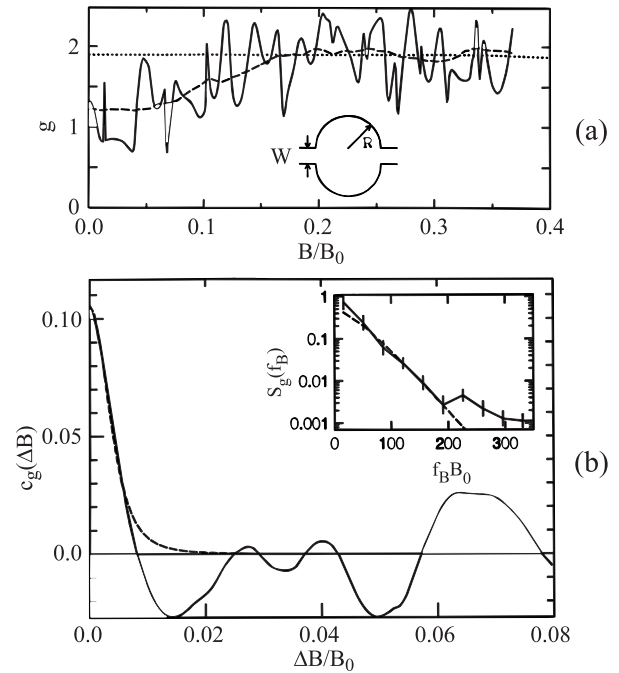}}
\vspace*{2 mm}
\caption
{Conductance fluctuations vs magnetic field in an open chaotic dot:
(a) dimensionless conductance  $g$ vs $B/B_0$ (where $B_0= mcv_F/e{\cal W}$) for an open stadium with $R/{\cal W}=2$ (see inset)  and
$k_F  {\cal W}/\pi=4.5$; solid line, the quantum calculation;  dashed
line, the smoothed quantum result; dotted line, the classical
result. (b) The calculated conductance correlator $c_g(\Delta B)$ vs $\Delta  B /B_0$ for the open stadium (solid
line)  and its semiclassical prediction (\protect\ref{correlation-B-sc})
(dashed  line). The inset shows the smoothed power spectrum $S_g(f_B)$ of
$g(B)$  [i.e., the Fourier transform of $c_g(\Delta B)]$. The
dashed  line is the best fit to Eq. (\protect\ref{power-sl}). From
Jalabert, Baranger and Stone (1990).
}
\label{fig:B-corr}
\end{figure}

  The power spectrum  $S_g(f_B)$ was determined in
the  experiment of Chan~{et~al.} (1995) by averaging over individual spectra
from  the measurement of $g(B)$ at different shapes. A two-parameter [$S_g(0)$
and  $\alpha$]  fit to (\ref{power-sl}) reproduces the data well with $\alpha
=  0.14\pm \;\mu$m$^{-2}$, in agreement with the value derived from the
weak-localization effect (see Sec. \ref{weak-localization:open}).
In the same experiment the conductance $g$ was also measured versus a
shape-distortion  voltage $V$.   Figure
\ref{fig:open-dot}(d)  shows the power spectrum $S_g(f_V)$, averaged at ten
different  measurements of $g(V)$ collected at various fixed values of the
magnetic  field.  The dotted line is a fit to
$S_g(V) = S_g(0) e^{-2\pi \xi f_V}$,
suggesting a Lorentzian line shape for the conductance correlator versus
shape:  $c_g(\Delta V) \propto 1/[1 + (\Delta V/\xi)^2]$.

Using RMT, it is possible to calculate the correlation function in both
energy and a time-reversal-symmetry-breaking parameter $\zeta$.  For
ideal  leads with $\Lambda \gg 1$ and for completely broken time-reversal symmetry (Frahm, 1995),
\begin{equation}\label{ideal-leads}
c_g(\Delta \zeta,\Delta E) = \frac{1}{16}\; {1 \over \left[ 1
+\left({\Delta  \zeta / 2\zeta_{cr}}\right)^2 \right]^2 + \left({\Delta E
/  \Gamma}\right)^2 } \;.
\end{equation}
For $\Delta E =0$, Eq. (\ref{ideal-leads}) reduces to the squared Lorentzian
of  Eq. (\ref{correlation-B-sc}).
Efetov (1995) calculated this correlator for nonideal leads.
Asymptotic expressions for the conductance correlator $c_g(\Delta \zeta)$ in RMT were derived for any number of channels $\Lambda$ by Gossiaux, Pluha\u{r}, and Weidenm\"{u}ller (1998) and by Pluha\u{r} and Weidenm\"{u}ller (1999). The limits considered were small and large $\Delta \zeta$ at fixed $\Lambda$, as well as large $\Lambda$ at fixed $\Delta \zeta/\Lambda$.

\subsection{Conductance fluctuations at finite temperature}
\label{finite-T:open}

 Temperature affects the conductance fluctuations in two ways, through
thermal  smearing and through the increase of the dephasing rate with
temperature.  Both effects reduce the fluctuations.
In this section we discuss the first effect, i.e., the role of temperature in
mesoscopic  fluctuations in the limit of complete phase coherence.

Efetov (1995) derived the conductance correlator versus magnetic
field  $c_g(\Delta B;T)$ at finite temperature $T$.  Using Eq. 
(\ref{finite-T-LIB})  for the finite-temperature conductance, this correlator
can  be related to the zero-temperature correlator $c_g(\Delta B, \Delta E)$
in  both magnetic field and energy (see Sec.
\ref{parametric-correlations:open}):
 $c_g(\Delta B;T) = T^2 \int_{-\infty}^\infty {d \over dT} \left[2T
\sinh\left({\Delta  E / 2 T}\right)\right]^{-2} c_g(\Delta B, \Delta E) d
\Delta  E$.
 At low temperatures $T \ll \Gamma/2\pi$, this correlator reduces to
the  squared Lorentzian
of Eq. (\ref{correlation-B-sc}), while at high temperatures  $T \gg
\Gamma/2\pi$,  this correlator is a Lorentzian, 
\begin{equation}\label{high-T-corr}
c_g(\Delta B;T) = {\pi \Gamma \over 96 T} { 1 \over 1 + (\Delta B/2B_{cr})^2}
\;.
\end{equation}
The high-temperature limit
of  the conductance variance $\sigma^2(g)= \pi \Gamma / 96 T$ can be seen
to  be much smaller than the zero-temperature variance of $1/16$. Thus the
conductance  fluctuations are reduced significantly when the temperature
exceeds  a typical level width $\Gamma$.

\subsection{Dephasing}
\label{phase-breaking}

The magnitudes of the observed weak-localization correction and conductance fluctuations in open dots are often reduced compared to the theoretical expectations.
 This discrepancy is explained by dephasing times
that  are comparable with typical escape times in the dot.  Dephasing can be
caused  by a voltage probe (since an electron absorbed by the probe is
reinjected  into the dot with an uncorrelated phase) or by inelastic processes
 in the dot such as electron-electron collisions.

There are two important issues concerning dephasing. The first is to describe
quantitatively how mesoscopic fluctuations are affected by finite
dephasing  rates (Sec. \ref{dephasing-models}). The second is the dependence
of  the dephasing rate on temperature (Sec. \ref{T-dephasing}).
Recent experimental results in dots with single-mode leads are described in Sec. \ref{dist-dephasing}.

 \subsubsection{Models for dephasing}
\label{dephasing-models}

  The dephasing voltage-probe model was introduced by
B\"{u}ttiker (1986a, 1986b) and applied to the conductance fluctuations in open
dots  by Baranger and Mello (1995) and by Brouwer and Beenakker (1995). A third dephasing lead is added and its voltage $V_\phi$ is adjusted
to  keep $I_\phi=0$, thus conserving the average number of electrons in the
dot.  The effective two-lead conductance $2(e^2/h) g =
I_1/(V_2  - V_1)$ can be expressed in terms of the
 three probe conductance coefficients:
\begin{equation}\label{dephasing-lead}
g = g_{21} + { g_{2 \phi} g_{\phi 1} \over g_{\phi 2} + g_{ \phi 1}} \;.
\end{equation}
Each of the first two leads has $\Lambda$ equivalent modes, and the third lead
is  assumed to have $\Lambda_\phi$ modes. The dephasing rate
$\Gamma_\phi$  is related to the total number $\Lambda_\phi$ of dephasing
modes  by
\begin{equation}\label{dephasing-rate}
\Lambda^{\rm eff}_\phi \equiv \Lambda_\phi T_\phi = 2 \pi {\Gamma_\phi / \Delta}
\;,
\end{equation}
where $T_\phi$ is the transmission probability per mode in the third lead.
The  parameter $\Lambda_\phi^{\rm eff} \equiv \Lambda_\phi T_\phi$  measures the
effective  number of ideal dephasing modes.

 The entire $(2\Lambda +\Lambda_\phi)\times
(2\Lambda  +\Lambda_\phi)$ $S$ matrix is assumed to be described by the respective
circular ensemble.  The fluctuations of the measured conductance $g$ are then
calculated  from Eq. (\ref{dephasing-lead}).
Using the relations  $g_{21} + g_{22} + g_{2 \phi}=0$,  $g_{1 1} + g_{2 1} +
g_{\phi  1}=0$ and $g_{12}+g_{22}+ g_{\phi 2}=0$ (see Sec.
\ref{LIB}),  it is possible to eliminate from Eq. (\ref{dephasing-lead})  all conductance coefficients that are related to the
third  lead. The distribution of $g$ can then be inferred from the known
distribution of the sub-$S$ matrix  (Pereyra and Mello, 1983; Friedman and Mello, 1985)
that corresponds to the two physical leads (Brouwer and Beenakker, 1995).

 Figure \ref{fig:g-dist} shows conductance distributions calculated by
Baranger and Mello (1995)  for $\Lambda_\phi=1$ dephasing mode in open dots with
single-  or double-mode leads (dashed lines) and for $\Lambda_\phi=2$ in dots
with  single-mode leads (dashed-dotted lines).
They also calculated analytically (to leading order in $1/\Lambda_\phi$) the
average  and variance of the conductance for an ideal, multimode voltage probe,
 \begin{mathletters}\label{dephasing-g-moments}
\begin{eqnarray}
\bar g &= & {\Lambda \over 2} + \left({\Lambda \over
\Lambda_\phi}\right){\beta  -2 \over 2 \beta} + {\cal O}(\Lambda_\phi^{-2})
\label{dephasing-average-g}  \\
\sigma^2(g) &= & \left({\Lambda \over \Lambda_\phi}\right)^2{2\Lambda + 2
-\beta  \over 4 \beta \Lambda} + {\cal O}(\Lambda_\phi^{-3})
\label{dephasing-variance-g}  \;,
\end{eqnarray}
\end{mathletters}
and found an interpolation formula for the
 weak-localization correction (at $B=0$): 
\begin{equation}\label{interpolation-WL}
\delta g \equiv - \left(\bar g(B=0) - \frac{\Lambda}{2}\right) \approx -\frac{1}{4}\;{1\over 1  + \Lambda_\phi/2\Lambda} \;.
\end{equation}
We  see from Eq. (\ref{interpolation-WL}) that the universal asymptotic result for the
weak-localization correction ($-1/4$) is valid only if $\Lambda_\phi \ll 2\Lambda$.

 The voltage-probe model describes {\em localized} dephasing (at the point
contact  between the dot and the third lead) and is less suitable to describe
dephasing  due to inelastic processes that occur through the
whole  dot. A way to introduce dephasing uniformly over the dot
is  to add an imaginary part $-i\Gamma_\phi/2$ to the dot's Hamiltonian.  This
model  does not conserve the number of electrons and was modified by
Brouwer and Beenakker (1997) to conserve electron number by mapping it on the
voltage-probe  model.  Compared with the standard voltage-probe model, they found conductance
distributions that were narrower and with strongly
 suppressed tails.  This is because the number-conserving imaginary-potential model
is  more effective in dephasing than the localized ideal voltage-probe model.
Nevertheless,  the asymptotic results for the average and variance coincide
exactly  with Eqs. (\ref{dephasing-g-moments}).  Fig.
\ref{fig:dephasing}(b)  shows the dephasing rate $\Gamma_\phi$ [calculated from
$\Lambda^{\rm eff}_\phi$  via Eq. (\ref{dephasing-rate})] versus the weak-localization correction $\delta g$ for
both  dephasing models.

The complete form of the weak-localization line shape was calculated in the
(non-number-conserving)  imaginary-potential model (Efetov, 1995): $\delta g(B)\equiv \bar g^{\rm GUE} - \bar g(B) 
= (1/4) \; [1 + ( B/ B_{cr})^2 + \Lambda^{\rm eff}_\phi/ 2 \Lambda]^{-1}$,
and  found to have no explicit dependence on temperature. This temperature independence
is  used to determine the dephasing rate from the measured weak-localization
effect at finite temperature, as is discussed in the next section.

\subsubsection{Temperature dependence of dephasing}
\label{T-dephasing}

  Dephasing rates were determined experimentally in disordered 2D (Choi, Tsui, and Alavi, 1987) and 1D (Kurdak~{\em et~al.}, 1992) semiconductors, in disordered 1D metals (Lin and Giordano, 1986; Echternach, Gersheson and Bozler, 1993), and more
recently  in open quantum dots (Marcus~{\em et~al.}, 1993a, 1994; Clarke~{\em et~al.}, 1995; Huibers, Switkes, Marcus, Campman, and Gossard, 1998).  To
determine  $\tau_\phi$ experimentally, it is best to measure a quantity that
is  sensitive to dephasing, yet is not affected explicitly by thermal
smearing,  e.g., the weak-localization effect in the magnetoconductance. This was recently implemented by Huibers, Switkes, Marcus, Campman, and Gossard (1998)
in  open quantum dots using three independent methods:

\noindent (i) Measuring the weak-localization correction $\delta g$ (at $B=0$). From
$\delta  g$ one can determine the number of effective dephasing channels
$\Lambda^{\rm eff}_\phi$  using one of the phenomenological dephasing models [e.g.,  Eq. (\ref{interpolation-WL})] and therefore determine $\Gamma_\phi$
from  Eq. (\ref{dephasing-rate}). The  inset in Fig. \ref{fig:dephasing}(b)
shows  $\Gamma_\phi$ versus $\delta g$ in the two dephasing models discussed
in  Sec. \ref{dephasing-models}.

 Figure \ref{fig:dephasing} shows the recent measurements of
Huibers, Switkes, Marcus, Campman, and Gossard (1998).  Panel (a) is the measured $\delta g$ for four
different  dots as a function of temperature.  The smaller values of $\delta
g$  for larger dots are consistent with Eq. (\ref{dephasing-rate}) since
$\Delta  \propto {\cal A}^{-1}$ and $\Lambda^{\rm eff}_\phi \propto {\cal A}$ for a  fixed value of the dephasing rate (i.e., at fixed temperature). The
extracted  values of $\tau_\phi$, shown in  Fig.
\ref{fig:dephasing}(b) (symbols),  are approximately independent of the dot's area.

\noindent (ii) Measuring the width of the weak-localization line shape.  According
to  Eqs. (\ref{semiclassical-WL}) or (\ref{RMT-WL}), $\delta g(B)$ is a Lorentzian characterized by a crossover
field  $B_{cr}$ that is proportional to the square root of the total number of
open  channels. In the presence of
phase  breaking, this total number of channels should include the number of
dephasing  channels
\begin{equation}\label{field:dephasing}
{B_{cr} {\cal A}/ \Phi_0} = (\kappa/2) g_T^{-1/2} \sqrt{2 \Lambda +
\Lambda^{\rm eff}_\phi}
 \;,
\end{equation}
where $\kappa$ is a geometrical constant of the dot and $g_T$ is the Thouless conductance.  Equation 
(\ref{field:dephasing})  can be used to determine $\Lambda^{\rm eff}_\phi$ from
the  measured $B_{cr}$. This equation contains an additional unknown parameter
$\kappa$,  but it is temperature independent and can be determined, for
example,  by a best fit to the $\delta g$ data.

\noindent (iii) Measuring the power spectrum $S_g(f_B)$ of the conductance
fluctuations  versus magnetic field (Marcus~{\em et~al.}, 1993a, 1994; Clarke~{\em et~al.}, 1995). At higher
temperatures,  the conductance correlator in open dots is given by a Lorentzian
(\ref{high-T-corr})  and its Fourier transform is exponential:
$S_g(f_B) = S_g(0) e^{-2\pi i B_c |f_B|}$ (where $B_c=2 B_{cr}$).  Using a two-parameter fit to the
measured  power spectrum it is possible to extract the correlation field
$B_c$  and then use Eq. (\ref{field:dephasing}) to determine the dephasing
time,  as in the second method above.

\begin{figure}
\epsfxsize= 8.5 cm
\centerline{\epsffile{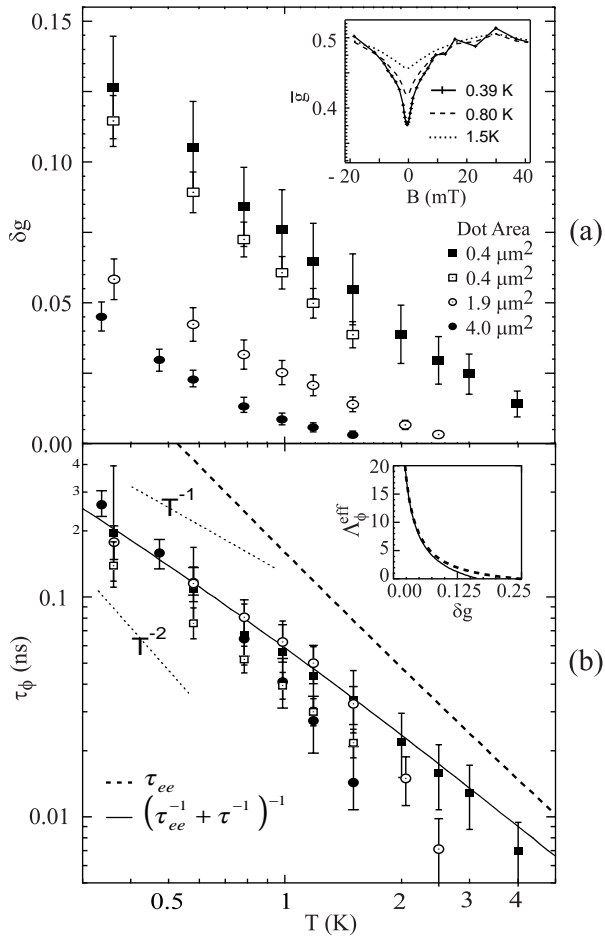}}
\vspace*{2 mm}
\caption
{The measured temperature dependence of the dephasing time in open quantum
dots:  (a) measured weak-localization correction $\delta g$ in four
dots  as a function of temperature $T$; note the decrease in $\delta g$ for
the  larger dots [in agreement with Eq. (\protect\ref{dephasing-rate}), see text]; inset,
 $\bar g$ vs magnetic field at several temperatures for the first
device;
(b) symbols, the extracted dephasing time vs temperature; dashed
line, $\tau_{ee}$ from Eq. (\protect\ref{interpolation-WL});   solid
line,  a fit to the theoretical $\tau_\phi=(\tau_{ee}^{-1} +
\tau^{-1})^{-1}$   of a 2D disordered system (see text) with $l
=0.25\;\mu$m;  inset,  the effective number of dephasing channels
$\Lambda_\phi^{\rm eff}$  vs $\delta g$ in the voltage-probe model (dashed line)
and  in the number-conserving imaginary-potential model (solid line). Adapted from
Huibers, Switkes, Marcus, Campman, and Gossard (1998).}
\label{fig:dephasing}
\end{figure}

  The dependence of the dephasing time on temperature in disordered conductors is theoretically
understood (Altshuler and Aronov, 1985; for a recent review see Aleiner, Altshuler, and Gershenson, 1999). At low temperatures, the
electron-electron scattering rate dominates the electron-phonon rate. There are two contributions to the electron-electron
dephasing  rate in 2D: a large-energy-transfer contribution quadratic in $T$ that is
characteristic of clean metals, 
$\tau_{ee}^{-1} = [(\pi k T)^2 / 2 h E_F] \ln \left( {E_F / kT}\right)$  (see, for example, Pines and Nozi\`{e}res, 1966), and
a  small-energy-transfer ($\alt kT$) contribution (Nyquist rate) linear in $T$,
$\tau^{-1} = (k T \lambda_F / h l) \ln\left( {\pi l /
\lambda_F}\right)$  (Altshuler and Aronov, 1985; Imry, 1996).  The total rate is approximately $\tau_\phi^{-1} \approx
\tau^{-1}  + \tau^{-1}_{ee}$. At low temperatures the Nyquist rate dominates, and it vanishes as $T \to 0$. Some experiments (e.g., Mohanty, Jariwala, and  Webb, 1997)  find an apparent saturation of the dephasing rate as $T \to 0$ for a reason that is not yet understood.

In ballistic quantum dots (0D systems) the situation is less clear.
Microscopic  estimates in closed ballistic dots by Sivan~{\em et~al.} (1994) and
Altshuler~{\em et~al.} (1997)  give $\tau_\phi^{-1} \propto T^2$ for $kT \gg
\Delta$.  However, no theoretical estimates are available for open quantum
dots.  The results of  Huibers, Switkes, Marcus, Campman, and Gossard (1998) offer a puzzle: they
suggest a temperature dependence of the rate that is characteristic of 2D
disordered  systems and not just a  $T^2$ dependence [see  Fig.
\ref{fig:dephasing}(b)].  More  experiments and theoretical work  may be necessary to resolve this issue.

\subsubsection{Conductance distributions}
\label{dist-dephasing}

    Using the experimentally determined dephasing rates, one can apply the phase-breaking
models  described in Sec. \ref{dephasing-models}  to
predict  the conductance distributions and compare them with the
experimental  distributions. Realistic calculations should include effects
due both to dephasing and to thermal smearing.

 A detailed comparison between experiment and theory for single-mode leads
was  reported by Huibers, Switkes, Marcus, Brouwer, {\em et~al.} (1998).  The measured
weak-localization  correction was used to determine $\tau_\phi(T)$ as in
Sec.  \ref{T-dephasing}, and found to be consistent with the results shown
in  Fig. \ref{fig:dephasing}(b). The number-conserving imaginary-potential model
(Sec.  \ref{dephasing-models}) was then used together with a thermal smearing  procedure to calculate the conductance
distributions.   Figure \ref{fig:dist-open} compares some of these theoretical
distributions  with the measured ones for both conserved (left) and fully
broken  (right) time-reversal symmetry. The distributions measured at the
lower  temperature [panels (a) and (c)] are clearly asymmetric.
The  dotted lines in panels (a) and (c) of Fig. \ref{fig:dist-open} are the
predicted  RMT distributions (\ref{g-dist:open}) without any phase breaking
and  at $T=0$. The dashed lines include the correct dephasing rate at the
corresponding  temperature but ignore thermal smearing effects. Finally, the
solid  lines include both dephasing and explicit thermal effects  and are in
good  agreement with the data.  The results show that both direct thermal
effects  and finite dephasing rates play an important role.

\begin{figure}
\epsfxsize= 8  cm
\centerline{\epsffile{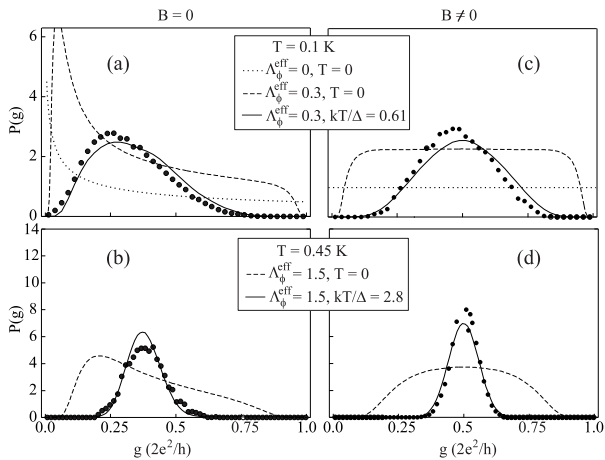}}
\vspace*{2 mm}
\caption{Finite-temperature conductance distributions in
single-mode open dots without a magnetic field [panels (a) and (b)] and  with a
magnetic field [panels (c) and (d)].Panels (a) and (c) are for $kT=0.61\; \Delta$  and panels (b) and (d) are for $kT=2.8\;\Delta$.  The solid circles  are the experimental results.
 The dotted  lines in
panels (a) and (c) are the fully phase-coherent $T=0$  distributions
(\protect\ref{g-dist:open}). The dashed lines in all panels are the
theoretical calculations without thermal smearing but with the
experimentally  determined effective number of dephasing modes
at the relevant temperatures  -- $\Lambda^{\rm eff}_\phi=0.3$ in panels (a) and (c), and $\Lambda^{\rm eff}_\phi=1.5$  in panels (b) and (d). The solid lines in all panels are the theoretical  predictions when both the
finite dephasing rate and thermal smearing  effects are
accounted for.  Adapted from Huibers, Switkes, Marcus, Brouwer, {\em et~al.} (1998). }
\label{fig:dist-open}
\end{figure}

\section{Mesoscopic Fluctuations in Closed Dots}
\label{closed-dots}

 Recent studies of statistical fluctuations in quantum dots have focused on
almost-isolated  or ``closed'' dots that are separated by barriers from the
leads.   Our main interest is in the quantum Coulomb-blockade regime, where
the  temperature is comparable to or smaller than the mean level spacing in
the  dot.   Experimentally, a series of peaks is observed in the conductance
versus  gate voltage;  see, for example,  Fig.
\ref{fig:closed-dot}(c).  The width of these peaks is thermally broadened, but
their  heights fluctuate strongly.  Whereas in open dots the conductance
fluctuations  originate from the interference of many overlapping resonances,
in  closed dots the peak-height fluctuations result from the spatial
fluctuations  of individual resonance wave functions at the dot-lead
interfaces.

   A statistical theory of the conductance peaks in Coulomb-blockade dots was
introduced  by Jalabert, Stone and Alhassid (1992).  The peak-height distributions
were  derived from RMT, and found to be universal and
sensitive  only to the symmetry class. The sensitivity of a
Coulomb-blockade  peak to an experimentally controlled parameter can be studied by changing a magnetic field or the shape of the dot.  Alhassid and Attias (1996a) derived the
 peak-height parametric correlation function and showed that it is universal
  once the parameter is scaled.
The predicted statistics of the peak heights for both conserved and broken
time-reversal  symmetry were observed by two experimental groups
(Chang~{\em et~al.}, 1996; Folk~{\em et~al.}, 1996)  using dots of different design and size. The latter
group  also confirmed the predicted functional form of the peak-height
autocorrelation  function in a magnetic field.

 In a closed dot, the charge on the dot is quantized and the Coulomb interactions cannot be ignored. The simplest model is the
constant-interaction model (\ref{CI}), which takes into account only the average Coulomb energy of
the dot's electrons. Interaction effects beyond the constant-interaction model are probably less relevant for the fluctuations of sufficiently highly excited states. However, interactions may cause deviations from RMT for the low-lying part of the spectrum, which is the region of interest in this section.  We shall see that the constant-interaction model can explain some, but not all,  of the observed statistical fluctuations. The main evidence for the breaking of the constant-interaction model comes from the
peak-spacing  fluctuations.  In the constant-interaction model,  these
fluctuations reflect the spacings between the single-particle levels in the
dot and are expected to follow a Wigner-Dyson distribution.  However, the observed distribution is more symmetric and closer to a Gaussian
(Sivan~{\em et~al.}, 1996; Simmel, Heinzel and Wharam, 1997; Patel, Cronenwett, {\em et~al.}, 1998).  A discussion of interaction effects beyond the constant-interaction model is deferred to Sec. \ref{interactions}.

Charging energy affects the statistics of the Coulomb-blockade peaks at
finite  temperature.  An outstanding issue has been the observed peak-to-peak correlations: temperature enhances these correlations but the experimental correlations at low temperatures are stronger than expected.
An  interesting phenomenon is the saturation of peak-to-peak
 correlations as a function of temperature (Patel, Stewart, {\em et~al.}, 1998). This effect can be explained
by  the statistical scrambling of the dot's spectrum  as electrons are added to the dot.

 Section \ref{statistical-theory:closed}  describes the main elements of the
statistical  theory of closed dots. Sections \ref{conductance-peaks} and
\ref{conductance-correlator}  review the statistics and parametric
correlations respectively, of the conductance peaks. The crossover statistics
in  the presence of a time-reversal symmetry-breaking field are discussed in Sec.
\ref{crossover:closed}.  The peak-spacing statistics in the framework of a
single-particle model plus constant charging energy are reviewed in Sec.
\ref{peak-spacing:closed}.  Finally, a statistical theory at finite
temperature,  and the use of temperature to probe the statistical scrambling
of  the dot's spectrum versus electron number, are presented in Secs. 
\ref{finite-T:closed}  and \ref{scrambling}, respectively.

\subsection{Statistical theory at low temperatures}
\label{statistical-theory:closed}

We first discuss Coulomb-blockade dots at $T \ll \Delta$.  Typically these low temperatures are
still  much larger than a typical resonance width and the observed conductance
peaks  are thermally broadened (see Sec. \ref{resonant-tunneling}).
The interesting information is carried by the conductance
peak height $G^{\rm peak}_\lambda \propto \Gamma^l \Gamma^r/(\Gamma^l + \Gamma^r)$ [see Eq. (\ref{peak-height})], where $\Gamma^{l(r)}$ is the width of a resonance level to decay in to the left (right) lead. Here $\Gamma^{l(r)} = \sum_c
|\gamma_{c}^{l(r)}|^2$, where $\gamma_{c}^{l(r)}$ is the partial
amplitude to decay into channel $c$ in the left (right) lead. The quantity $\gamma_{c \lambda}$ is given by Eq. 
(\ref{partial-amplitude}), and  can be expressed  as a 
scalar product  of the resonance wave function
$\bbox{\psi}_\lambda= (\psi_{\lambda 1},\psi_{\lambda
2},\ldots)$ and the channel wave function $\bbox{\phi}_c =
(\phi_{c  1}, \phi_{c 2}, \dots)$: 
\begin{equation}\label{scalar-product}
\gamma_{c \lambda} = \langle \bbox{\phi}_c |
 \bbox{\psi}_\lambda \rangle \equiv \sum\limits_j \phi^\ast_{c j}
\psi_{\lambda  j} \;.
\end{equation}
Here we expanded the wave function $\Psi_\lambda = \sum_j
\psi_{\lambda  j} \rho_j$ in a fixed basis $\rho_j$ in the dot and
defined  the channel vector
$\phi_{c j} \equiv \left( {\hbar^2 k_c P_c / m } \right)^{1/2}
\int_{\cal  C}dl\;
   \rho^\ast_j(\bbox{r}) \phi_c(\bbox{r}) = \sqrt{2\pi} W_{j c}$ [$W$ is the
dot-lead  coupling matrix introduced in Eq. (\ref{W})]. The scalar product in Eq. (\ref{scalar-product})
is defined over the dot-lead interface and differs from the
usual scalar product in the Hilbert space of the dot.

 Another modeling of a quantum dot assumes pointlike contacts, and
each  lead is composed of several such point contacts
$\bbox{r}_c$ (Prigodin, Efetov, and Iida, 1993; Mucciolo, Prigodin, and Altshuler, 1995).   Each point contact
constitutes one channel,  and the  partial width of a level
$\lambda$ to decay into it is $\gamma_{c\lambda} = (\alpha_c
{\cal A} \Delta /\pi)^{1/2}\,
     \Psi_\lambda(\bbox{r}_c)$,
where $\alpha_c$ is a  dot-lead coupling parameter and ${\cal A}$ is the area
of  the dot.   The partial width can still be expressed as a scalar product
[Eq. (\ref{scalar-product}] but now the channel vector is defined by
  $\phi_{c j} \equiv  (\alpha_c {\cal A} \Delta /\pi)^{1/2}
 \rho^*_j (\bbox{r}_c)$.

Fluctuations in $G^{\rm peak}_\lambda$ arise from fluctuations of the widths $\Gamma^l$ and
$\Gamma^r$. These widths are determined by the partial width
amplitudes $\gamma_{c \lambda}$, which in turn are  expressed by Eq. (\ref{partial-amplitude}). The penetration
factor $P_c$ in Eq. (\ref{partial-amplitude}) is a smooth function
of the Fermi energy (or gate voltage), and fluctuations can only arise from the
overlap integral $\int_{\cal  C} d l \phi_c^\ast
\Psi_\lambda$, i.e., from the spatial fluctuations  of
$\Psi_\lambda(\bbox{r})$ across the dot-lead interface ${\cal
C}$.  For a chaotic ballistic dot,  these fluctuations are described by RMT.   A key relation for connecting the physical quantity (i.e.,
width)  to RMT is Eq.
(\ref{scalar-product}), expressing  the partial-width
amplitude as a {\em projection} of the random-matrix
eigenfunction $\bbox{\psi}_\lambda$ on a fixed channel vector
$\bbox{\phi}_c$.

\subsection{Conductance peak statistics}
\label{conductance-peaks}

 The main goal of this section is to derive the statistical distributions of
the  conductance peak heights.  These distributions were derived by Jalabert, Stone, and Alhassid (1992) using RMT
and  by Prigodin, Efetov, and Iida (1993) using supersymmetry. The case of
correlated  channels in two-channel leads was treated by
Mucciolo, Prigodin and Altshuler (1995)  for broken time-reversal symmetry, and the general case of multimode leads with possibly inequivalent and
correlated channels was discussed by Alhassid and Lewenkopf (1995).

\subsubsection{Partial-width amplitude distribution}
\label{joint-distribution}

The  joint distribution of the partial-width amplitudes $\bbox{\gamma} =(\gamma_1, \gamma_2, \ldots, \gamma_\Lambda)$ of a resonance $\lambda$ can be computed from the RMT wave-function statistics.
Using  Eqs. (\ref{vec-dist}) and
(\ref{scalar-product}),
 \begin{eqnarray}\label{joint-dist-def}
  P(\bbox{\gamma}) = && \frac{\Gamma(\beta N/2)}{\pi^{\beta N/2}}
   \int \! D[\bbox{\psi}] \left[\prod_{c=1}^{\Lambda}
   \delta ( \gamma_c - \langle\bbox{\phi}_c|\bbox{\psi}\rangle) \right] \nonumber \\
  && \times\, \delta \!\left( \sum_{\mu=1}^N |\psi_\mu |^2 -1 \right) \;,
\end{eqnarray}
where $D[\bbox{\psi}] \equiv \prod_{\mu=1}^{N}{d\psi_\mu}$ for
the GOE and $D[\bbox{\psi}] \equiv \prod_{\mu=1}^{N}{d\psi^*_\mu
d\psi_\mu/2\pi i}$ for the GUE. The integral
(\ref{joint-dist-def}) can be evaluated following the methods of
Ullah (1967) and transforming to a new set of orthonormal
channels. For $\Lambda \ll N$ and $N\rightarrow \infty$, we recover a Gaussian distribution
(Alhassid and Lewenkopf, 1995),
\begin{eqnarray}
\label{joint-dist}
 P(\bbox{\gamma}) = (\det M)^{-\beta/2}
         \, \mbox{e}^{- \frac{\beta}{2}
         \bbox{\gamma}^\dagger M^{-1} \bbox{\gamma} }\;.
\end{eqnarray}
The distributions (\ref{joint-dist}) are normalized with the measure
$D[\bbox{\gamma}] \equiv \prod_{c=1}^{\Lambda}
 \left({d\gamma_c/\sqrt{2\pi}}\right)$
for the GOE and $D[\bbox{\gamma}]\equiv\prod_{c=1}^{\Lambda}
\left({d\gamma^*_c d\gamma_c/2\pi i}\right)$ for the GUE.  They
  can also be derived (Krieger and Porter, 1963; Ullah, 1963) from the
requirement  that their form be invariant under orthogonal (unitary)
transformations  for the GOE (GUE). The quantity $M$ in Eq. (\ref{joint-dist})
is the channel correlation  matrix
\begin{eqnarray}
\label{M-matrix}
  M_{cc^\prime} =  \overline{ \gamma_c^* \gamma_{c^\prime}} =
\frac{1}{N} \langle \bbox{\phi}_c
 |\bbox{\phi}_{c^\prime}\rangle \;
\end{eqnarray}
and is identical with the matrix $M=2\pi
W^\dagger  W/N$  defined in Sec. \ref{RMT:open} [see Eq.
(\ref{M-matrix'})].  The eigenvalues $w_c^2$ of $M$ are just the average
partial  widths $\bar \Gamma_c$ [see Eq. (\ref{M-matrix})], and for the present case of isolated resonances  $\bar \Gamma_c \approx (\Delta / 2\pi ) T_c$.
For a general set of channels, the matrix $M$ can be nondiagonal (describing
correlated  channels) and have different diagonal elements (corresponding to inequivalent channels).

 We note that in chaotic systems the joint distribution of  an
eigenfunction's  amplitudes at $\Lambda$ spatial points $\bbox
r_1,\ldots,\bbox  r_\Lambda$ is a special case of Eq. (\ref{joint-dist}). Indeed, 
the  partial-width amplitudes of the point-contact model are
proportional  to the wave-function amplitudes at a set of fixed spatial points
(see also Srednicki, 1996).

 In the point-contact model, $M_{c' c}$ is a measure of
the  spatial wave-function correlations at two different spatial points
$\overline{\Psi^\ast(\bbox{r})  \Psi(\bbox{r'})}$. Expanding the eigenfunction
in  the fixed basis $\Psi(\bbox{r}) = \sum_{j}
\psi_{j}\rho_j(\bbox{r})$,  and using the RMT relation
$\overline{\psi^\ast_{j}  \psi_{j'}} = \delta_{j j'}/N$, we find
$\overline{\Psi^\ast(\bbox{r}) \Psi(\bbox{r'})}= \sum_j
\rho_j^\ast(\bbox{r})  \rho_j(\bbox{r'})/N$.
 The fixed basis $\rho_j(\bbox{r})$ is chosen such that the
eigenfunction's components are distributed randomly on the unit sphere in
$N$ dimensions.   Random-matrix theory is expected to describe fluctuations in a chaotic system
on  a local energy scale. Therefore, for the problem of a free particle
in  a cavity, we choose this basis to be the free-particle states at the given
energy  $E=\hbar^2 k^2/2m$.  In polar coordinates $r,\theta$ such a basis is
$\rho_j(\bbox{r})  \propto J_j(k r) e^{i j\theta}$ ($j=0,\pm 1,\pm 2\ldots$),
where  $J_j$ are Bessel functions of the first kind. Using the addition
theorem  for the Bessel functions, we obtain
\begin{equation}\label{spatial-correlation}
\overline{\Psi^\ast(\bbox{r}) \Psi(\bbox{r'})} = {\cal A}^{-1} J_0(k|\bbox{r}
-  \bbox{r'}|) \;.
\end{equation}
Similar results are obtained if $\rho_j$ are chosen to be plane waves with fixed energy but random orientation of momentum $\bbox k$.  Equation (\ref{spatial-correlation}) was first derived by
Berry (1977), assuming that the  Wigner function of
an ergodic system is {\em microcanonical} on the energy
surface and averaging over a spatial region whose linear extension
is large compared  with the particle's wavelength.
It follows that in the point-contact model, the channel correlation matrix of a chaotic dot
is  given by
 $M_{cc^\prime} = \left(\sqrt{\alpha_c \alpha_{c'}} \Delta/ \pi \right)
         J_0(k| \bbox{r}_c - \bbox{r}_{c^\prime}|)$.

In $d$ dimensions the spatial correlations of eigenfunctions are
\begin{eqnarray}\label{spatial-correlation-d}
\overline{\Psi^\ast(\bbox{r}) \Psi(\bbox{r'})}/\overline{|\Psi(\bbox{r})|^2}
& = &  2^{d/2-1} \Gamma(d/2) {J_{d/2-1}(k |\bbox{r} - \bbox{r'}|) \over
(k|\bbox{r}  - \bbox{r'}|)^{d/2-1}} \nonumber \\ & \equiv &f_d(|\Delta \bbox{r}|) \;.
\end{eqnarray}
 The envelope of $f_d(|\Delta \bbox{r}|)$
decays  as a power law $(k|\Delta \bbox{r}|)^{-(d-1)/2}$.
  For weakly disordered systems, $f_d$ contains an additional
factor  of $e^{-|\Delta \bbox{r}|/2l}$, resulting in an exponential cutoff
of the spatial correlations beyond $l$
(Mucciolo, Prigodin, and Altshuler, 1995; Prigodin, 1995).

\subsubsection{Width distribution}
\label{width-distributions}

 We next determine the level-width distribution (note that for a symmetric dot $\Gamma^l  = \Gamma^r$, and the conductance peak height is proportional to the
width). Using   $\Gamma=\sum_c|\gamma_c |^2=
\bbox{\gamma}^\dagger\bbox{\gamma}$ and the Gaussian nature
(\ref{joint-dist}) of the partial width amplitudes, we can easily calculate
the  characteristic function of the width distribution $P(u) \equiv \int_0^\infty d\Gamma
\exp(iu\Gamma)  P(\Gamma)= [\det(I-2i M u/\beta)]^{-\beta/2}$.
 This width distribution $P(\Gamma)$ is then (Alhassid and Lewenkopf, 1995)
\begin{eqnarray}
\label{Gamma-dist}
 P(\Gamma) = \frac{1}{2 \pi} \int^{\infty}_{-\infty}
    du \,\frac{e^{-i u \Gamma}}{\left[\det (I -
       2 i u M/\beta)\right]^{\beta/2}} \; 
\end{eqnarray}
and depends only on the eigenvalues $w_c^2$ of the  positive-definite
correlation  matrix  $M$.

Equation (\ref{Gamma-dist}) can be evaluated by  contour integration.
 All the singularities of the integrand are along the negative imaginary axis
 $u=-i \tau$ at $\tau=1/w_c^2$. For the GOE case the singularities are of the
type
$(\tau -1/2w_c^2)^{-1/2}$, leading to
\begin{eqnarray}
\label{Gamma-dist-GOE}
 P_{GOE}(\Gamma) && = \frac{1}{\pi 2^{\Lambda/2}}
  \left(\prod_c  \frac{1}{w_c} \right)
 \sum_{m=1}^{\Lambda} (-)^{m+1} \int_{1/2 w^2_{2m-1}}^{1/2 w^2_{2m}}
 d \tau \nonumber \\ \times\, &&\frac{\mbox{e}^{-\Gamma \tau}}
 {\sqrt{\prod_{r=1}^{2m-1}   (\tau - \frac{1}{2 w^2_r})
    \prod_{s=2m}^\Lambda (\frac{1}{2 w^2_{s}} - \tau)}} \;,
\end{eqnarray}
 where the eigenvalues of $M$ are arranged in ascending order and we
have  defined
$1/2 w^2_{\Lambda+1} \rightarrow \infty $ for an odd number of channels.
For the GUE statistics, all the singularities are
poles.  If the eigenvalues of $M$ are nondegenerate, then the poles are
simple  and
\begin{eqnarray}
\label{Gamma-dist-GUE}
 P_{GUE}(\Gamma) =
   \left(\prod_c\frac{1}{w_c^2}\right)
   \sum_{c=1}^\Lambda && \left[\prod_{c^\prime \neq c}
   (\frac{1}{w_{c^\prime}^2} - \frac{1}{w_c^2})\right]^{-1} \nonumber \\&&\times\,
   \mbox{e}^{-\Gamma/w_c^2} \;.
\end{eqnarray}

 In the special case of uncorrelated and equivalent channels, all the
eigenvalues  of $M$ are degenerate, $w_c^2 = w^2$, and the width distribution
can  be found directly from Eq. (\ref{Gamma-dist}) to be the $\chi^2$ distribution
in  $\beta\Lambda$ degrees of freedom.

\subsubsection{Peak-height distributions}
\label{peak-distributions}

 In a closed dot we define $g$ to be the (dimensionless) conductance peak height $G^{\rm peak}$ measured in units of $(e^2/h)(\pi \Gamma/4k T)$ [see Eqs. (\ref{peak-height})]. Assuming that the widths $\Gamma^l$ and $\Gamma^r$ are uncorrelated,
 $P(g)$ can be computed using  Eq. (\ref{g-peak})
and the known width distributions of Sec. \ref{width-distributions}.
  In the simple case of one-channel symmetric leads ($\Lambda=1; \;
\bar{\Gamma}^l  = \bar{\Gamma}^r$), we find (Jalabert, Stone, and Alhassid, 1992; Prigodin, Efetov, and Iida, 1993)
\begin{mathletters}
\begin{eqnarray}\label{Pg-closed}
P_{GOE}(g) & = &  \sqrt{{2 / \pi g}}  e^{-2 g} \label{PgGOE} \;,\\
P_{GUE}(g) & = & 4 g  e^{-2g}  \left[K_0(2g) +K_1(2g) \right]
\;, \label{PgGUE}
\end{eqnarray}
\end{mathletters}
where $K_0$ and $K_1$ are the
modified  Bessel functions.

  The peak-height distributions were measured independently
by Chang~{\em et~al.} (1996) and by
Folk~{\em et~al.} (1996) for  dots with single-channel tunneling leads.
The results of Chang~{\em et~al.} [(1996), Figs.
\ref{fig:Exp-peaks}(a) and \ref{fig:Exp-peaks}(b)] are from dots  with
$\sim 100$ electrons.
 The histograms are the experimental
results at  $T=75 \;$mK; 72 peaks were collected for $B=0$ and
216 peaks for $B \neq 0$.  The solid lines are the predicted
theoretical distributions (\ref{PgGOE})  and (\ref{PgGUE}) for
$B=0$ and $B \neq 0$, respectively. The conversion  from the
measured conductance peak $G^{\rm peak}$ to the dimensionless
conductance $g$ requires an unknown parameter -- the average width  $\bar \Gamma$ of a resonance in the dot. Since $\bar \Gamma$
 is independent of the magnetic field, the
theoretical distributions in Figs. \ref{fig:Exp-peaks}(a) and \ref{fig:Exp-peaks}(b)
represent a one-parameter  fit $\bar \Gamma \approx 0.086\;kT$
to {\em both} curves.  The measured  distributions are
non-Gaussian and characterized by a large  number of small
peaks with more small peaks for the $B=0$ case, in agreement
with the theoretical predictions. The inset of Fig. \ref{fig:Exp-peaks}(a) shows a $B=0$ peak sequence versus gate voltage at $T=75\;$mK
(lower trace); three  of the peaks are too small to be observed, but can be seen  at a higher temperature
trace ($T=600\;$mK) of the same peaks (upper trace).

 Figures \ref{fig:Exp-peaks}(c) and \ref{fig:Exp-peaks}(d) show the experimental results of Folk~{\em et~al.} (1996) at $T = 70 \pm
20\;$mK.  Their dots are larger ($\sim 1000$ electrons) than
in  the previous experiment,  so that $kT/\Delta \sim 0.3-0.5 \;\Delta$ is higher. Using
shape-distorting  gates, larger statistics could be collected.  Each
distribution  includes $\sim 600$ peaks,
although only $\sim 90$ are statistically independent. The solid lines are
fits to the RMT predictions, (\ref{PgGOE}) and (\ref{PgGUE}). The insets show
the same distributions on a linear-log scale, where good agreement between
theory and experiment is seen over two to three orders of magnitude.
We remark, however, that the strong correlations observed between heights of neighboring peaks [see, for example, Fig. \ref{fig:closed-dot}(c)] is at variance with RMT, and we shall return to this point in Sec. \ref{G-finite-T:closed}.

It is also possible to calculate the peak-height distributions for a general
leads  configuration for both the orthogonal ({Alhassid and Lewenkopf, 1995, 1997) and
unitary ( Alhassid and Lewenkopf, 1995, 1997; Mucciolo, Prigodin and Altshuler, 1995) symmetries.
The RMT predictions for the peak-height distributions were tested in a
model  of a ballistic dot -- the conformal billiard (see Sec.
\ref{disordered})  -- in its chaotic regime. Good agreement is found for dots
with  single-channel (Stone and Bruus, 1993; Bruus and Stone, 1994) and multimode leads
(Alhassid and Lewenkopf, 1995, 1997).

  The Coulomb-blockade peaks exhibit a weak-localization effect.  The average conductance peak height for symmetric leads with $\Lambda$ channels in each lead is
\begin{eqnarray}\label{average-g-closed}
\bar{g}_\Lambda = \left\{ \begin{array}{cc}  {\Lambda^2
\over 2(\Lambda+1)} & ({\rm GOE})
\\   {\Lambda^2 \over 2\Lambda+1} & ({\rm GUE})
\end{array}  \right. \;.
\end{eqnarray}
Thus $\bar g_\Lambda$ for GOE is smaller by an amount
$\Lambda^2/[2(\Lambda+1)(2\Lambda+1)]$  than its GUE value -- a
weak-localization effect. The relative reduction of the
average conductance  $\delta g_\Lambda/\bar g_\Lambda^{\rm GUE}=
1/(2\Lambda+2)$ (where $\delta g_\Lambda = \bar g_\Lambda^{\rm GUE} - \bar g_\Lambda^{\rm GOE}$) is 
\begin{figure}[h!]
\epsfxsize= 6.3 cm
\centerline{\epsffile{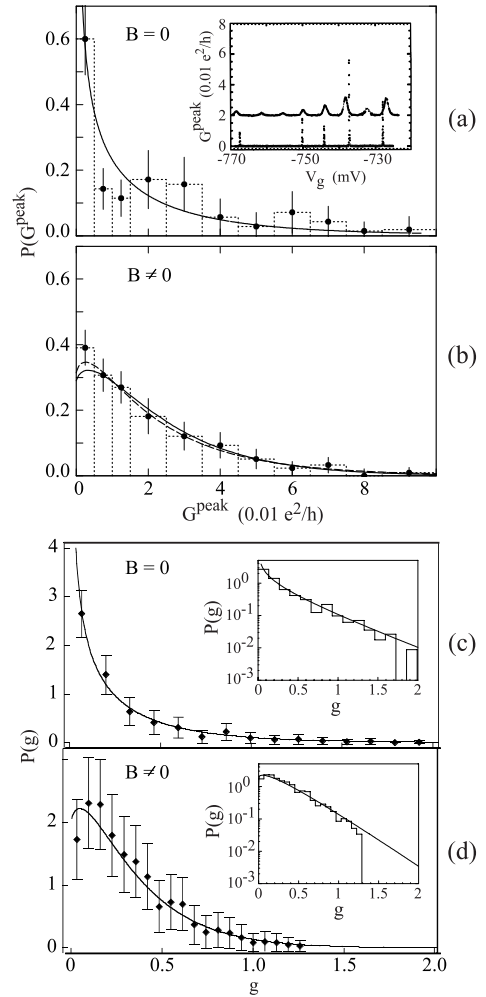}}
\vspace*{2 mm}
\caption{Conductance peak-height distributions in small and large
Coulomb-blockade quantum dots. Panels (a) and (b): measured distributions of Chang~{\em et~al.} (1996) [panel (a)] at $B=0$  and [panel (b)] at $B \neq 0$  in small dots of
effective size $0.25\;\mu$m$\times 0.25\;\mu$m at $T=75\;$K
(${\cal N} \sim 100$ electrons and $kT/\Delta \sim 0.15$); solid
lines, the theoretical  predictions (\protect\ref{PgGOE}) and
(\protect\ref{PgGUE}) of Jalabert, Stone, and Alhassid (1992)  for
conserved and broken time-reversal symmetry,  respectively.
 Inset in panel (a):
lower trace, a sequence  of $B=0$ Coulomb-blockade peaks as a
function of gate voltage at $T=75  \;$mK;  upper trace,
 the same peak series but at $T=660\;$mK.  Notice the large
fluctuations in the peak heights and the presence of very small
peaks (three of them are unobserved) at low temperature.
 From Chang~{\em et~al.} (1996).
Panels (c) and (d):   measured
peak-height  distributions of Folk~{\em et~al.} (1996) in Coulomb-blockade
dots  at $T =70 \pm 20 \;$mK, compared with the theoretical predictions
(solid  lines) for [panel (c)] $B=0$  and [panel (d)] $B \neq 0$.  The two dots used in the experiment, with areas
$0.32\;\mu$m$^2$  and $0.47\; \mu$m$^2$, are relatively large (${\cal N} \sim
1000$).   This allows for a larger number of peaks ($\sim 40$) to be observed
in  each sweep of the gate voltage,  but $kT/\Delta$ is only $\sim 0.3-0.5$.
The insets show the same distributions on a log-linear scale (histograms are
from  the experiment).  Adapted from Folk~{\em et~al.} (1996).
}
\label{fig:Exp-peaks}
\end{figure}
\noindent smaller by a factor of
$(2\Lambda+1)/(2\Lambda+2)$  than its corresponding value for
open dots [see Eq. (\ref{average-g})]. For example, in single-channel leads the
weak-localization  correction in closed dots is $\delta g/\bar
g^{\rm GUE}= 1/4$ compared  with $1/3$ in open
dots.

 In the limit of large $\Lambda$, the GUE variance of the conductance peak heights is smaller by a
factor of two  than the GOE variance (Alhassid, 1998). A similar
behavior was found in open  dots [see  Eq. (\ref{variance-g})].
 For single-channel symmetric leads, the variance of
the  conductance peak reduces from $1/8$ in the GOE to $4/45$ in the GUE.

\subsection{Parametric correlations of the conductance peaks}
\label{conductance-correlator}

  In closed dots one can follow a specific conductance peak as a function of
an  external parameter such as magnetic field or shape [see, for example, Fig. \ref{fig:peak-corr}(a)], and calculate the
correlation  between peak heights at different values of the external
parameter.  Alhassid and Attias (1996a) calculated this conductance peak
correlator using the framework of Gaussian
processes  discussed in Sec. \ref{Gaussian-processes} and found it to be universal upon an appropriate
scaling  of the external parameter.

The parametric width correlator $c_\Gamma(x-x') = \overline{\delta \Gamma(x) \delta \Gamma(x')}/
\left\{\overline{[\delta \Gamma(x)]^2}\;\; \overline{[\delta
\Gamma(x')]^2}\right\}^{1/2}$,
where $\delta \Gamma(x) \equiv \Gamma(x)- \bar \Gamma(x)$  can be calculated in the
framework of the Gaussian process and is universal upon the
scaling (\ref{scaling}) of the external parameter.  It can also
be shown to be independent of the channel correlation  matrix
$M$ (Alhassid and Attias, 1996b) and is therefore determined by the
symmetry  class alone.  The width correlator coincides with the
overlap correlator  (\ref{overlap-corr}) of Sec.
\ref{Gaussian-processes} in the limit $N \to \infty$ and is
thus well approximated by Eq. (\ref{approx-overlap-corr}),  a
Lorentzian in the Gaussian orthogonal process and a  squared Lorentzian in  the Gaussian unitary process.

The conductance peak-height correlator
\begin{equation}
c_g(x-x') = { \overline{\delta G(x) \delta G(x')} / \left\{\overline{[\delta
G(x)]^2}\;\;  \overline{[\delta G(x')]^2}\right\}^{1/2}}
\end{equation}
 depends on the number of channels in each lead $\Lambda^l$ and $\Lambda^r$
and  on the eigenvalues $(w_c^{l,r})^2$ of the correlation
matrices $M^l$ and $M^r$  in the left and right leads,
respectively.
 The case most relevant to experiments is that of single-channel leads. When the
leads are symmetric ($\bar\Gamma^l = \bar\Gamma^r$), the deviation of the corresponding
$c_g(\Delta x)$ from the width correlator is the largest. This
correlator is well fitted by the form of Eq. (\ref{approx-overlap-corr}), i.e., a
Lorentzian for the Gaussian orthogonal process and a squared Lorentzian for the Gaussian unitary process [see inset in Fig. \ref{fig:peak-corr}(b)], but with $\alpha_1=0.37
\pm  0.04$ and $\alpha_2 = 0.54 \pm 0.04$, respectively.  The universality of the conductance correlator
was  verified in Anderson-model simulations (Alhassid and Attias, 1996a) as well as in
billiard-model  calculations (Bruus, Lewenkopf, and Mucciolo, 1996). For multichannel symmetric leads
it  was found that the conductance correlator approaches the width correlator
as  the number of channels increases (Alhassid and Attias, 1996b).

\begin{figure}
\epsfxsize= 8 cm
\centerline{\epsffile{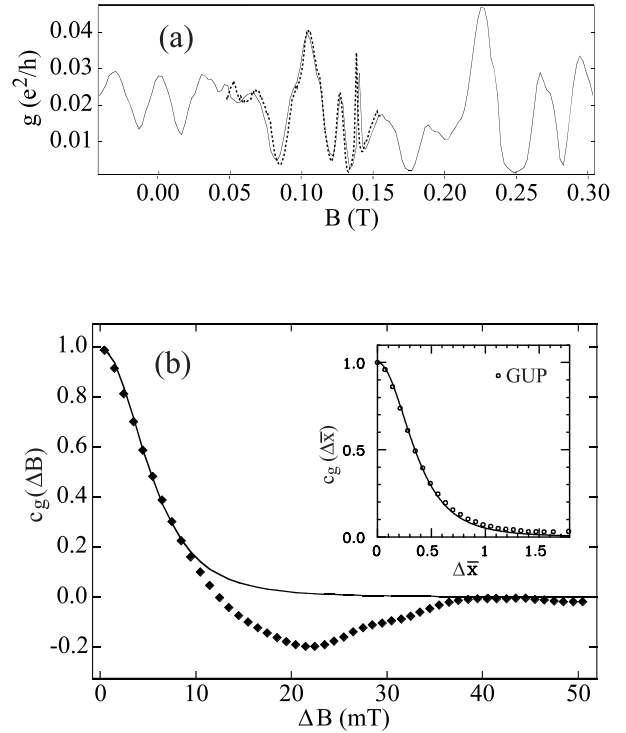}}
\vspace*{2 mm}
\caption{
Conductance peak-height fluctuations vs magnetic field in
closed dots: (a)
the measured
peak  height of a single Coulomb-blockade peak as a function of magnetic
field $B$ (solid line) and $-B$ (dashed line), showing the symmetry and reproducibility of the data; 
(b) the peak-height autocorrelation function in a magnetic field $c_g(\Delta B)$; diamonds, the experimentally determined
peak-height  autocorrelation function in a magnetic field $c_g(\Delta B)$ at
$T=70  \pm 20 \;$mK, averaged over many
traces  of the type shown in panel (a) (Folk~{\em et~al.}, 1996);  solid line, the fitted  squared-Lorentzian line shape
(\protect\ref{approx-overlap-corr})  predicted by
Alhassid and Attias (1996a);  inset,  the universal Gaussian unitary process correlator  calculated by random-matrix process simulations (circles) compared with its
squared-Lorentzian  fit (\protect\ref{approx-overlap-corr}) with $\alpha_2 =
0.54$  (solid line).
Adapted from Folk~{\em et~al.} (1996) and
Alhassid and Attias (1996a).
}
\label{fig:peak-corr}
\end{figure}

The correlator $c_g(\Delta x)$ can be calculated perturbatively to leading
order  in $\Delta \bar x$ (Alhassid and Attias, 1996a, 1996b),
\begin{eqnarray}\label{peak-corr-pert}
   c_g(\Delta x)\approx \left\{\begin{array}{ll}    1  -
b_1 \mid\Delta \bar{ x}\mid^2  & \\
\;\times\left( -{\pi^2  \over 6} \ln \mid\Delta \bar{x}\mid + {\rm const}
\right) & \;(\beta=1)\\
   1- b_2 {\pi^2 \over 3}
   \mid \Delta\bar{x}\mid^2 & \;(\beta=2)
\end{array} \right. \;,
  \end{eqnarray}
and for the Gaussian orthogonal process it is nonanalytic at $\Delta \bar x=0$.
The constant $b_\beta$ in Eq. (\ref{peak-corr-pert}) depends on the leads.
 For single-channel symmetric leads $b_1=7/4$ and $b_2=3$.

 The peak-height autocorrelation versus a magnetic field was measured by Folk~{\em et~al.} (1996). The
results  are shown in  Fig. \ref{fig:peak-corr}(b) (diamonds)
and  compared with the theoretical prediction (\ref{approx-overlap-corr}) of
Alhassid and Attias (1996a; solid line). Random-matrix theory  predicts a universal correlator
after  the parameter is scaled. However, the scaling
factor  itself -- for a particular choice of the parameter -- cannot be
computed  in RMT.  When $x$ is a magnetic field, we can rewrite
Eq. (\ref{approx-overlap-corr})  as
\begin{equation}\label{GUP-corr}
c_g(\Delta B) \approx \left[1+ (\Delta B/B_c)^2\right]^{-2} \;,
\end{equation}
where $B_c$ is the correlation field.
The curve in Fig. \ref{fig:peak-corr}(b) represents a one-parameter fit (i.e.,
$B_c$)  to the data. It is found that $B_c = 8.1 \pm 0.5 \;$mT or $\Phi_c
\approx  0.8 \Phi_0$.

 The correlation field in closed dots can be estimated semiclassically,
similarly  to the open dot case (see Sec. \ref{weak-localization:open}).
Since  the decay time of a resonance in a closed dot is much longer than the
Heisenberg  time, the latter becomes the relevant time scale for the diffusive
area  accumulation. The estimate for $B_c$ in a closed dot is then obtained
by replacing $\tau_{\rm escape}$ in
(\ref{crossover-Phi})  with $\tau_H$:
\begin{equation}\label{B-correlation-closed}
{B_c {\cal A} / \Phi_0} = \kappa \left(2\pi{\tau_c / \tau_H}\right)^{1/2}=
\kappa   g_T^{-1/2} = \kappa {4\pi^2 \cal N}^{-1/4}
\;,
\end{equation}
where $g_T$ is the ballistic Thouless conductance. A semiclassical
derivation can  be found in Bohigas~{\em et~al.} (1995). In the
conformal billiard with a flux line (Berry and Robnik, 1986) $\Phi_c \approx 0.1
\Phi_0$, but for a stadium in a uniform magnetic  field
(Bohigas~{\em et~al.}, 1995) flux  is accumulated less efficiently   and
$\Phi_c\approx 0.3 \Phi_0$  (Alhassid, Hormuzdiar, and Whelan, 1998). This is still
  below  the experimental value $\Phi_c \approx 0.8 \Phi_0$, indicating that  the single-particle
picture is inadequate for estimating the correlation  field.
We shall return to this problem in Sec. \ref{peaks:interactions}.

  \subsection{Crossover from conserved to broken time-reversal symmetry}
\label{crossover:closed}

 Following Alhassid, Hormuzdiar, and Whelan (1998), we derive in this section the
peak-height  statistics in the crossover between GOE and GUE.
We use the crossover random-matrix ensemble (\ref{transition-ens}), which is
characterized  by the transition parameter $\zeta$.  When time-reversal
symmetry  is broken by a magnetic field,
$\zeta = \Phi/\Phi_{cr}$, where $\Phi_{cr}$ is a characteristic crossover
flux of the same order as the correlation
field in Eq. (\ref{B-correlation-closed}).

\subsubsection{Conductance peak distributions}
\label{crossover-distributions:closed}

   Using the method of Sec. \ref{crossover-ensembles}, we decompose the
partial  amplitudes of an eigenfunction $\bbox{\psi}$ in the principal frame:
$\gamma_c  = \gamma_{cR} + i \gamma_{cI} =
\langle \bbox{\phi}_c | \bbox{\psi}_R \rangle +
 i  \langle \bbox{\phi}_c | \bbox{\psi}_I \rangle$. The
joint  partial-width amplitude distribution is then given by
\begin{eqnarray}\label{joint-dist-crossover}
&& P_\zeta(\bbox{\gamma}) =  \left\langle   P(\bbox{\gamma} |t )
 \right\rangle \;, \nonumber \\
 && P(\bbox{\gamma} |t )   =  \left ({ 1+t^2  \over 2\pi t}\right)^\Lambda
 (\det M)^{-1} \nonumber \\ 
&& \times\, \exp \left( - {1+ t^2 \over 2} \bbox{\gamma}^T_R M^{-1}
\bbox{\gamma}_R  - { 1+ t^2 \over 2 t^2} \gamma^T_I M^{-1}
 \bbox{\gamma}_I \right) \;,
\end{eqnarray}
where the brackets $\langle \ldots\rangle$ denote an average
over the distribution $P_\zeta(t)$ in Eq. (\ref{shape-dist}). Here 
$M$  is the correlation matrix [Eq. (\ref{M-matrix})] and is assumed to be
independent  of  $\zeta$. This is correct as long as the dot-leads geometry is held fixed as the magnetic
field is changed.

   The width for decay into a one-channel lead is given by
$\Gamma= |\gamma|^2 = \gamma_R^2 + \gamma_I^2$.
Using Eq. (\ref{joint-dist-crossover}) we find
 $P_\zeta(\hat{\Gamma}) = \langle a_+ e^{-a_+^2 \hat{\Gamma}}
I_0( a_+a_- \hat{\Gamma})\rangle$,
where $a_\pm  \equiv (t^{-1} \pm t)/2$,
and $I_0$ is the modified Bessel function of order zero.

For the general case of $\Lambda$ inequivalent and/or correlated channels, we
note  that  the joint conditional distribution $P(\bbox{\gamma} | t)$ in Eq.
(\ref{joint-dist-crossover})  is identical
to the joint partial-width amplitude distribution for a GOE problem of
$2\Lambda$  channels
with partial amplitudes $\gamma_{cR}, \gamma_{cI}$ and an extended
correlation matrix
${\cal M}$ composed of four $\Lambda \times \Lambda$ blocks:
\begin{eqnarray}
{\cal M} = \left ( \begin{array}{cc}
                 {1 \over 1+t^2} M  & 0\\
                    0  &  {t^2 \over 1+t^2} M
         \end{array} \right) \;. \nonumber
\end{eqnarray}
 We can therefore use the known GOE width and conductance peak distributions
from  Secs. \ref{width-distributions} and \ref{peak-distributions}.
 The $2\Lambda$ eigenvalues of ${\cal M} $ are given by
$ \{ \omega^2_j \} = \{ {1 \over 1+t^2} w_c^2, {t^2 \over 1+t^2} w_c^2 \}$,
where
 $w_c^2$ are the $\Lambda$ eigenvalues of $M$. Sorting the  inverse
 eigenvalues of ${\cal M}$ in
ascending order, $\omega_1^{-2}< \omega_2^{-2} < \ldots$, we find for the
crossover  width distribution
\begin{eqnarray}\label{Gamma_cr}
 P_\zeta(\Gamma)&& = \left\langle \frac{1}{\pi 2^{\Lambda}}
  \left(\prod_c  \frac{1}{\omega_c} \right)
 \sum_{m=1}^{\Lambda}  (-)^{m+1} \int_{1/2 \omega^2_{2m-1}}^{1/2
\omega^2_{2m}}
 d \tau \right.\nonumber \\ 
&& \left. \times \, \frac{\mbox{e}^{-\Gamma \tau}}
 {\sqrt{\prod_{r=1}^{2m-1}   (\tau - \frac{1}{2 \omega^2_r})
    \prod_{s=2m}^{2\Lambda} (\frac{1}{2 \omega^2_{s}} - \tau)}}
\right\rangle
 \;.
\end{eqnarray}

   In the crossover regime
$\Gamma^l$  and $\Gamma^r$ are no longer statistically independent, i.e.,
$P(\Gamma^l, \Gamma^r) = \langle P(\Gamma^l | t ) P(\Gamma^r | t) \rangle
 \neq \langle P(\Gamma^l |t) \rangle \langle P( \Gamma^r |t) \rangle = P(\Gamma^l) P(\Gamma^r)$. This
is  just the manifestation of  the long-distance correlations in
the transition  statistics discovered by Fal'ko and Efetov (1996).
On the other hand, at fixed $t$,  $\Gamma^l$  and $\Gamma^r$ are
independent,  and $P(g | t)$ is calculated in closed form by
following the same steps as for the GOE case
(Alhassid, Hormuzdiar, and Whelan, 1998). The peak-height distributions $P_\zeta(\ln g)$ in the crossover from GOE
to  GUE are shown in Fig. \ref{fig:Pg-trans}. Here $\ln g$ is chosen as the
variable  in order to show the behavior at small intensities over several
orders  of magnitude (Alhassid and Levine, 1986).
 The left inset confirms the RMT
predictions for the conformal billiard.

\begin{figure}
\epsfxsize= 8  cm
\centerline{\epsffile{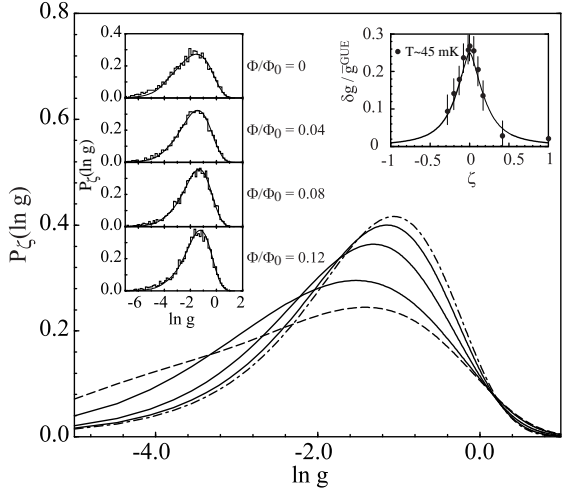}}
\vspace*{2 mm}
\caption{Coulomb-blockade conductance peak-height distributions
in the crossover from conserved  to broken time-reversal
symmetry. The distributions for single-mode leads  are shown
vs $\ln g$ for $\zeta=0$ (GOE, dashed line),
$\zeta=0.1,0.25,0.5$  (solid lines), and $\zeta \gg 1$ (GUE,
dot-dashed line). Left inset, distributions obtained from simulations of
the conformal billiard threaded by  different amounts of
magnetic flux (histograms) are compared with theoretical
distributions calculated for $\zeta \approx 4\Phi/\Phi_0$ (solid lines).
Right inset, the analytic weak-localization line shape $\delta
g(\zeta)/\bar g^{\rm GUE}$ for a dot with single-channel
symmetric leads (solid line), compared with a recent
experiment (solid circles) by
Folk~{\em et~al.} (2000). The quantity $\bar g^{\rm GUE}$ is measured away from $B=0$ and $\zeta=B/B_{cr}$ with $B_{cr} \approx 6$ mT. Adapted from
Alhassid, Hormuzdiar and Whelan (1998) and
Alhassid (1998). }
 \label{fig:Pg-trans}
\end{figure}

\subsubsection{Weak localization}

We already know from  Sec. \ref{peak-distributions} that a weak-localization
correction  is predicted for the average conductance peak height.  The
complete  dependence of $\bar g$ on $\zeta$ (for symmetric leads with $\Lambda$ channels in each lead) 
was calculated in closed form by Alhassid (1998).
 The calculation is simpler than the corresponding one in
 open dots (see Sec. \ref{weak-localization:open}) and can be done
  within the framework of RMT.   For a dot with single-channel symmetric leads ($\Lambda=1$),
\begin{eqnarray}\label{wlocal1}
\bar{g}(\zeta) = \frac{1}{4} +  \left\langle
 \left(\frac{t}{1-t^2}\right)^2 \left({2 t^2 \over 1- t^4} \ln t
+\frac{1}{2}  \right)
\right\rangle \;.
\end{eqnarray}

 As in the case of open dots, we define
$\delta g_\Lambda(\zeta) \equiv \bar{g}^{\rm GUE}_\Lambda -
\bar{g}_\Lambda(\zeta)$. The quantity $\delta g_\Lambda(\zeta)$ is largest at
$\zeta=0$ (GOE limit) where $\delta g_\Lambda(0)/\bar
g^{\rm GUE}_\Lambda = 1 - \bar{g}^{\rm GOE}_\Lambda/\bar{g}^{\rm
GUE}_\Lambda = 1/[2(\Lambda+1)]$, and approaches zero for $\zeta  \to
\pm \infty$. The right inset of Fig.
\ref{fig:Pg-trans} shows that the predicted  weak-localization
line shape $\delta g_\Lambda(\zeta)/\bar g_\Lambda^{\rm GUE}$ for a dot with one-channel symmetric leads
(solid line) agrees well with a recent experiment (Folk~{\em et~al.}, 2000)
after scaling of $B$.

    In closed dots, the full width at half maximum (FWHM) of the weak-localization line shape is almost
independent  of the number of channels $\Lambda$,  in contrast
with open dots, where the FWHM behaves as $\sim \sqrt{\Lambda}$ for large
$\Lambda$. This difference can be understood in terms of the different
time scales involved. In open dots, the crossover in the average conductance
occurs  when  $\tau_{\rm escape}/\tau_{\rm mix} = 4 \pi^2 \zeta^2/\Lambda^2 \sim 1 $,  leading to
$\zeta^{\rm  open}_{cr} \sim \sqrt{\Lambda} /2\pi$ [see the discussion following Eq. (\ref{RMT-WL})]. In closed dots, on the other hand, the
``escape'' time (by tunneling) is much longer than the Heisenberg time $\tau_H$, and it is the latter  that competes with the mixing time. Since $\tau_H$ is longer
by  a factor of $\Lambda$ than the escape time in an open dot with $\Lambda$
ideal  channels, we conclude that the crossover in the average conductance
 in closed dots occurs when $\zeta^{\rm closed}_{cr} \sim 1$, independent of the number
of  channels.

 The variance of the conductance peak height in the crossover from GOE to GUE
  can also be calculated in closed form (Alhassid, 1998). For single-channel symmetric leads, 
\begin{eqnarray}
\overline{g^2}(\zeta)  = && {3 \over 16} + {27 \over 2}
\left\langle \left( {t \over 1+t^2} \right)^2  \left( {t \over 1-t^2}
\right)^4 \left[ {1+t^2 \over 1-t^2} \ln t \right. \right. \nonumber \\ &&\left.\left. + 1 + \frac{1}{12}
 \left( { 1-t^2 \over t}\right)^2 -\frac{2}{27}
\left( { 1-t^2 \over t}\right)^4 \right]
 \right\rangle \;.
\end{eqnarray}

\subsection{Peak-spacing statistics}
\label{peak-spacing:closed}

 The spacings $\Delta_2$ between successive Coulomb-blockade peaks are
observed  to fluctuate around an average charging energy that
changes smoothly as  more electrons are added to the dot
(Sivan~{\em et~al.}, 1996). In the constant-interaction model, $\Delta_2 = e \alpha
(V_g^{{\cal N}+1} - V_g^{{\cal N}})=  E_{{\cal N}+1} - E_{\cal
N} +e^2/C$ [see Eq. (\ref{peak-spacing}]. Thus if the single-particle states are not
spin degenerate (but without treating  the spin-up and spin-down
manifolds as statistically independent), we  expect a (shifted)
Wigner-Dyson distribution of the peak spacings.   A
detailed discussion of the experimental results is postponed to
Sec.  \ref{peak-spacing:interactions}. Here we remark only that
 the observed distribution does not have a Wigner-Dyson shape but is
closer to a Gaussian (Sivan~{\em et~al.}, 1996; Simmel, Heinzel and Wharam, 1997; Patel, Cronenwett, {\em et~al.}, 1998).

Can Gaussian-like spacing distributions be explained in the constant-interaction--plus--RMT approach?  It was pointed out
by  Vallejos, Lewenkopf and Mucciolo (1998) that deviations from a Wigner-Dyson distribution
may  be due to shape deformation of the dot as the gate voltage changes.  At gate voltage $V_g^{\cal N}$, corresponding to the ${\cal N}$th conductance peak, the shape of the dot is 
 $x_{\cal N}$. However, at the degeneracy point $V_g^{{\cal N}+1}$ of
the  next peak, the shape of the dot has changed to $x_{{\cal N}+1}$. The
spacing  $\Delta_2$ is now given by
$\Delta_2 -  e^2/ C = E_{{\cal N}+1}(x_{{\cal N}+1}) - E_{\cal N}(x_{\cal
N})$,
where $E_\lambda(x)$ are the  single-particle energies of the dot with shape
$x$. The generic variation of the energy levels with $x$ can be studied in the framework of
the  Gaussian process (see Sec. \ref{Gaussian-processes}). Measuring all energies in units of the average level
spacing  $\Delta$, we have for $\tilde\Delta_2 \equiv (\Delta_2 - { e^2 / C})/
\Delta$
\begin{eqnarray}\label{tilde-Delta_2}
\tilde\Delta_2 & = & [\epsilon_{{\cal N}+1}(x_{{\cal N}+1}) - \epsilon_{\cal
N}(x_{{\cal  N}+1})] + [\epsilon_{\cal N}(x_{{\cal N}+1}) - \epsilon_{\cal
N}(x_{\cal  N})] \nonumber \\ & \equiv & \Delta \epsilon^{({\cal N}+1)} + \Delta \epsilon_{\cal
N}\;,
\end{eqnarray}
where $\Delta \epsilon^{({\cal N}+1)}$ denotes the  spacing between
successive levels in a dot with a fixed shape
$x_{{\cal  N}+1}$, while $\Delta \epsilon_{\cal N}$ denotes the parametric
variation  of the ${\cal N}$-th level as the shape of the dot changes between
peaks.  The parametric fluctuations of
the  levels are universal once $x$ is scaled according to Eq. (\ref{scaling}).
Defining $\delta x_{\cal N} \equiv x_{{\cal N}+1} - x_{\cal N}$, we assume
that  the scaled $\delta \bar x_{\cal N} \approx \delta \bar x$ is independent of ${\cal
N}$ ($\delta \bar x$ is the parametric distance $\delta x$ measured in units of the average distance between avoided level crossings).  
  The distributions $P(\tilde \Delta_2)$ are then universal and depend
only  on  $\delta \bar x$ and the symmetry class.
Figure \ref{fig:spacings-par}(c) shows the standard deviation of the
spacings $\sigma(\tilde\Delta_2)$  versus $\delta \bar
x$. It increases with
$\delta \bar x$, i.e., the peak-spacing fluctuations are larger when the single-particle spectrum changes faster
with  the addition of electrons into the dot. The inset in Fig. \ref{fig:spacings-par}(c)
shows the GOE $\sigma(\tilde\Delta_2)$  (solid line) in comparison with
$\left[\overline{(\Delta  \epsilon_{\cal N})^2}\right]^{1/2}$ (the level-diffusion correlator calculated in
Sec. \ref{Gaussian-processes}). The $\sigma(\tilde\Delta_2)$
curve interpolates  between the Wigner-Dyson value at $\delta
\bar x=0$ and $\left[\overline{(\Delta  \epsilon_{\cal N})^2}\right]^{1/2}$
at large $\delta \bar x$.

For small $\delta \bar x$, the distributions can be calculated in closed
form.  Using the Gaussian process (\ref{simple-GP}) near $x=0$,
 we find that $\Delta \epsilon_{\cal N} \approx \langle \psi_{\cal N}
|  H_2 | \psi_{\cal N} \rangle \delta x$, where $\psi_{\cal N}$ is an
eigenfunction  of $H_1$. At fixed $H_1$,  $\Delta \epsilon_{\cal N}$ is thus a Gaussian variable with zero mean and variance of $(\delta \bar x)^2$, i.e.,
$P(\Delta  \epsilon_{\cal N}) = (2 \pi)^{-1/2}(\delta \bar x)^{-1}
\exp[-(\Delta \epsilon_{\cal N})^2/2(\delta \bar x)^2]$. Following Eq. (\ref{tilde-Delta_2}), we convolute
this  Gaussian with the Wigner-Dyson distribution $P_{\rm WD}(s)$ [see Eq.
(\ref{Wigner})] to find
\begin{equation}\label{peak-spacing-pert}
P(\tilde\Delta_2)\approx (2 \pi)^{-1/2} \int_0^\infty ds P_{\rm WD}(s)
(\delta  \bar x)^{-1} e^{-{(\tilde\Delta_2 -s)^2 \over 2 (\delta \bar
x)^2}}\;.
\end{equation}
The distributions for finite $\delta \bar x$ are easily
calculated by simulations.    Figures 
\ref{fig:spacings-par}(a) and \ref{fig:spacings-par}(b) show the  peak-spacing
distributions (histograms) for $\delta \bar x=0.75$. Each of the distributions is
compared  with the Wigner-Dyson distribution of the same
symmetry class (solid lines).  The distributions are more
Gaussian-like and  have tails extending to negative spacings.

\begin{figure}
\epsfxsize= 8 cm
\centerline{\epsffile{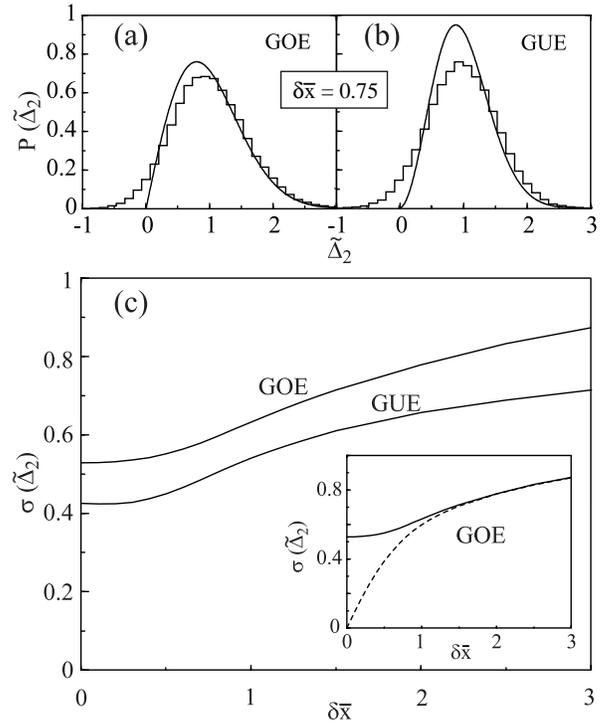}}
\vspace*{2 mm}
\caption{
Peak-spacing statistics and the parametric motion of energy levels:
panels (a) and (b), the peak-spacing distribution $P(\tilde\Delta_2)$ (histograms) at
$\delta  \bar x=0.75$ for the (a) orthogonal and (b) unitary symmetries
are  compared with the Wigner-Dyson nearest-neighbor spacing distribution for
the  respective symmetry (solid lines); (c) the standard deviation of the
peak  spacing $\sigma(\tilde\Delta_2)$ as a function of $\delta \bar x$ for
both  the GOE and GUE statistics.  Inset in panel (c), $\sigma(\tilde \Delta_2)$ (solid
line) compared with $\left[\overline{(\Delta \epsilon_{\cal N})^2}\right]^{1/2}$, the rms
of  the change of a given energy level (dashed line). The latter is just the level-diffusion correlator shown in Fig.
\protect\ref{fig:univ-corr}(c).
}
\label{fig:spacings-par}
\end{figure}

 We shall return to the subject of peak-spacing statistics in the context
of  finite temperature (Sec. \ref{peak-spacing-T}) and interaction effects
(Sec.  \ref{interactions}).

\subsection{Finite-temperature statistics}
\label{finite-T:closed}

   At finite temperatures that are not much smaller than $\Delta$, several
resonances  in the dot may contribute to the same conductance peak owing to
the  thermal smearing of the electron energy in the leads.  The charging energy $E_C$ plays an important role.
 When $E_C \gg \Delta$, only two manifolds of the
many-electron energy levels with ${\cal N}$ and ${\cal N}-1$
electrons  in the dot contribute
significantly since all other manifolds are pushed away by the
charging energy.  The rate-equations theory of Beenakker (1991)  discussed  in Sec.
\ref{Coulomb-blockade} takes into account this charging energy effect, and was used by Alhassid, G\"{o}k\c{c}eda\u{g}, and Stone (1998) to extend the statistical theory of closed dots to finite temperature.

\subsubsection{Conductance peaks}
\label{G-finite-T:closed}

   In Sec. \ref{Coulomb-blockade} we saw that the finite-temperature conductance is a weighted average of the single-level
conductances: $g=\sum_\lambda  w_\lambda(T,\tilde E_F)
g_\lambda$, with thermal weights   $w_\lambda$  given by
Eq. (\ref{thermal-weight}) for $T \ll e^2/C$. These  thermal
weights depend on the canonical free energy  $F_{\cal N}$ and
canonical occupation numbers $\langle n_\lambda \rangle_{_N}$.
The latter are calculated  exactly using particle-number
projection (Ormand~{\em et~al.}, 1994):
\begin{mathletters}
\begin{eqnarray}
 Z_{\cal N} & = &  e^{-F_{\cal N}/T} = { e^{-\beta E_0 } \over N_{sp}} \nonumber \\ && \times\,
\sum\limits_{m=1}^{N_{sp}}
\prod_{i=1}^{N_{sp}}\left(1 + e^{-\beta |E_i- \mu|} e^{i \sigma_i \phi_m}
\right) \;,  \label{partition} \\
\langle n_\lambda \rangle_{\cal N} & = & { e^{-\beta E_0 }
 \over N_{sp} Z_{\cal N}}\sum\limits_{m=1}^{N_{sp}} \left[ \prod_{i =1}^{N_{sp}}
\left(1 + e^{-\beta |E_i - \mu|} e^{i \sigma_i \phi_m} \right) \right]  \nonumber \\
&& \times\,{1 \over 1 + e^{\beta(E_\lambda - \mu)} e^{i\phi_m}} \label{occupation} \;,
\end{eqnarray}
\end{mathletters}
where the quadrature points are $\phi_m = 2\pi m/N_{sp}$ ($N_{sp}$ is the
number of single-particle states), and $E_0 =\sum_i E_i$.  The quantity $\mu$ is a chemical potential chosen  anywhere in the
interval  $E_{\cal N} \leq \mu < E_{{\cal N}+1}$; $\sigma_i=1$ for a hole
 ($ E_i \leq \mu$) and $-1$ for a
particle ($ E_i > \mu$).

 The effect of energy-level fluctuations on the
conductance  statistics is small, and one can use a
picket-fence spectrum  to demonstrate the results. An example is
shown in panels (a) and (b) of Fig. \ref{fig:finite-T} for
$T=0.5\;\Delta$. The canonical occupation numbers as  a function
of $E_\lambda$ follow a curve that is similar to a Fermi-Dirac
distribution  with a chemical potential of $(E_{\cal N} +
E_{{\cal N}+1})/2$ but with an effective  temperature smaller by
almost a factor of 2 in the vicinity of  the  chemical
potential [see inset of Fig. \ref{fig:finite-T}(b)]. These results are in agreement with
estimates  by Kamenev and Gefen (1997).  The thermal weights
[Eq. (\ref{thermal-weight}] are shown in
Fig. \ref{fig:finite-T}(a) as a function  of the effective Fermi
energy  for several levels  in the vicinity of the central level
 through which the tunneling occurs at low
temperatures (denoted in the following by $\lambda=0$). The ratio of the thermal weights $w_\lambda(T,
E_0)$ to the noninteracting weights [Eq. (\ref{LIB-weights})] is
shown  in Fig. \ref{fig:finite-T}(b).  For all $\lambda \neq 0$, we
have $w_\lambda/  w^{(0)}_\lambda \approx 1/2$ to within 20\% or
better. On the other  hand, the ratio for the central level
$w_0/w_0^{(0)} = \langle n_0 \rangle  >1/2$ is enhanced with
respect to that for other levels.  This enhancement  causes the
distribution of the conductance peaks (see below) to be  less
sensitive to temperature than what we would expect from a
non-interacting  theory. For temperatures above $\sim 2-3 \;
\Delta$, $\langle  n_0 \rangle \approx 1/2$ and the conductance peak
approaches  $G \approx G^{(0)}/2$
(Beenakker, 1991), where $G^{(0)}$ is the classical conductance in the absence of Coulomb blockade (see the discussion at the end of Sec. \ref{Coulomb-blockade}). 

\begin{figure}
\epsfxsize= 8 cm
\centerline{\epsffile{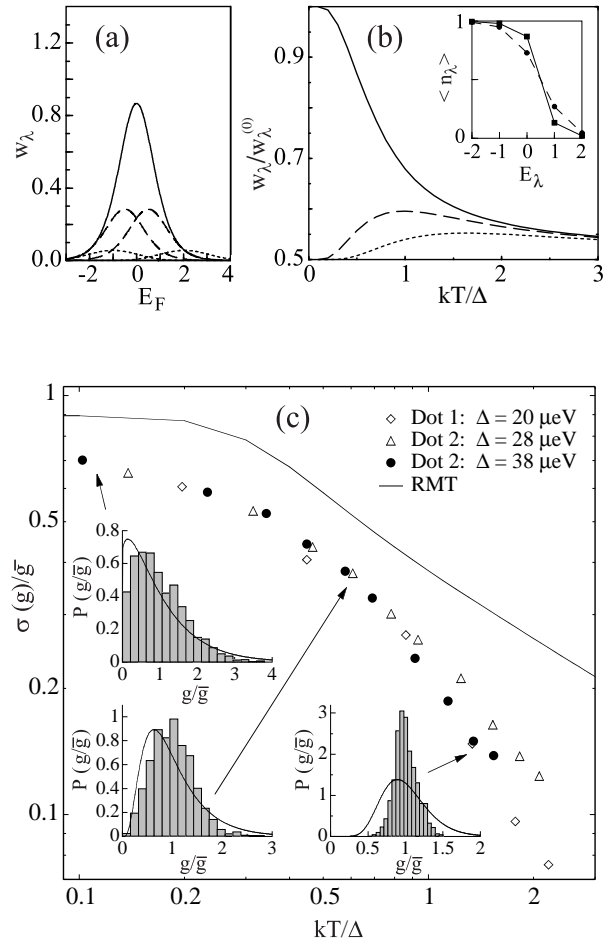}}
\vspace*{2 mm}
\caption{Thermal weights of level conductances and finite-temperature statistics of
the  conductance peaks: (a) thermal weights $w_\lambda(T,\tilde E_F)$ of several level conductances
vs  Fermi energy $\tilde E_F$ at $kT=0.5\;\Delta$; (b) ratio of
thermal  weights to their non-interacting values $w_\lambda /w_\lambda^{(0)}$  at $\tilde E_F =E_0$ vs $kT/\Delta$. Solid lines, 
$\lambda=0$;  dashed lines, $\lambda=\pm 1$; dotted lines, 
$\lambda=\pm  2$. The results are for a picket-fence spectrum. The inset in (b)
shows the canonical occupations (solid squares) compared with the Fermi-Dirac
distribution at $kT=0.5 \Delta$ (dashed line).
Adapted from Alhassid, G\"{o}k\c{c}eda\u{g} and Stone (1998);
(c) the measured ratio
$\sigma(g)/\bar  g$ between the standard deviation of the peak-height
fluctuations and the average peak height for three dots (symbols), compared with the RMT results
(solid  curve). The insets compare several of the measured distributions
(histograms)  with the finite-temperature RMT distributions (solid lines).
From  Patel, Stewart, {\em et~al.} (1998).
}
\label{fig:finite-T}
\end{figure}

\noindent{\em a. Distributions}

 Complete RMT simulations show that spectral fluctuations have a small
  effect on the finite-temperature peak-height distribution $P(g)$ (Alhassid, G\"{o}k\c{c}eda\u{g}, and Stone, 1998). Closed expressions for  $P(g)$ can be obtained if these fluctuations are
ignored.  In RMT, eigenfunctions that belong to different
levels  are uncorrelated and  $P(g_1, g_2, \ldots) =\prod_\lambda
P(g_\lambda)$, where $P(g_\lambda)$ is the distribution of a single-level conductance, derived in Sec.
\ref{conductance-peaks}.  It follows that for a fixed sequence of energy levels, the
characteristic  function of the conductance peak distribution $P(u) \equiv
\int_0^\infty dg  e^{i u g} P(g)$ factorizes. Using the known single-level conductance
distributions  Eqs. (\ref{PgGOE}) and (\ref{PgGUE}), we find
\begin{eqnarray}\label{finite-T-dist}
  P(u)  =
  \left\{\begin{array}{ll} \prod\limits_{\lambda}{\left( 1 - {i u w_\lambda \over 2 }
 \right)^{-1/2} } & \mbox{(GOE)}  \\
\prod\limits_\lambda  {1 \over 2 (1 - {iu w_\lambda \over 4})  }  & \\ \;\;\times 
\left[ 1 + { \arcsin({i u w_\lambda \over 4})^{1/2} \over
 ( {iu w_\lambda \over 4})^{1/2} ( 1 - {i u w_\lambda \over 4})^{1/2}}
\right]  & \mbox{(GUE)}
\end{array}  \right.
\;.
\end{eqnarray}

 Figure \ref{fig:finite-T}(c) compares recent experimental results by
Patel, Stewart, {\em et~al.} (1998) for the temperature dependence of $\sigma(g)/\bar g$
with  the RMT predictions (solid line). The observed fluctuations exhibit a
similar  temperature dependence, but are smaller than the RMT predictions.
Also  shown are some of the experimental distributions (histograms) in
comparison   with the RMT distributions (solid lines). The
deviations  are larger at higher temperature, suggesting that they might be due
 to decoherence effects. Finite-temperature phase-breaking
effects  on the conductance of closed dots have not yet been studied.\\

\noindent {\em b. Peak-to-peak correlations}

The measured distributions of the Coulomb-blockade peak heights at low
temperatures  have confirmed the predictions of the statistical theory (see
Sec.  \ref{peak-distributions}). However,  one of these experiments
(Folk~{\em et~al.}, 1996) also produced a puzzle:
neighboring peaks are observed to be correlated [see Fig. \ref{fig:closed-dot}(c)] although in RMT  different
eigenfunctions  are uncorrelated.  Since the temperature in this experiment is only
$\sim 0.3 - 0.5 \;\Delta$, some of these correlations might be due to the
finite  temperature, where several resonances contribute to the same peak.
We define the peak-to-peak  correlator
\begin{equation}
c(n) =  { \overline{\delta G _{\cal N} \delta G_{{\cal N}+n}} /
 \overline{(\delta G_{\cal N})^2} } \;,
\end{equation}
where $\delta G_{\cal N}= G_{\cal N}- \bar G_{\cal N}$ is the fluctuation of
the  ${\cal N}$th conductance peak around its average.
An approximate expression for $c(n)$ is obtained by assuming that the
location  of the ${\cal N}$th peak is fixed at its low-temperature value, 
$\tilde{E}_F  = E_{\cal N}$ [note that $\tilde E_F$ is measured relative to $({\cal N}-1/2) e^2/C$].
Since eigenvectors and eigenvalues are uncorrelated in RMT,
and using $\overline{g_\lambda g_\mu} = \overline{g_\lambda^2}
\delta_{\lambda \mu} + \overline{g}_\lambda^2 (1 - \delta_{\lambda \mu} )$,
 we find
\begin{equation}\label{peak-corr}
c(n) \approx  {\overline{ \sum_\lambda w_\lambda({\cal N}+n) w_\lambda({\cal
N})}  \over
\overline{\sum_\lambda w^2_\lambda({\cal N})} }\;,
\end{equation}
where $w_\lambda({\cal N}) \equiv w_\lambda(T, E_{\cal N})$, and
the remaining average is over the  spectrum. Equation (\ref{peak-corr}) can be simplified  for a picket-fence spectrum:
$c(n) \approx { \sum_\lambda w_{\lambda -n} w_\lambda   / \sum
w_\lambda^2}$,
where $w_\lambda$ are the weights for a fixed number of electrons in the dot. The number $n_c$ of correlated peaks, defined as the FWHM of the correlator
$c(n)$,  is shown in Fig. \ref{fig:corr-T-exp}(c) versus $T/\Delta$ and compared with a recent
experiment by Patel, Stewart, {\em et~al.} (1998) for three dots of different size.  [The measured correlators $c(n)$ are shown in Figs.  \ref{fig:corr-T-exp}(a) and (b) for two of the dots.]  While the qualitative increase  of
correlations with temperature is confirmed, we see
enhanced correlations in the low temperature data.
 This is not fully  understood, although several
possible explanations were suggested: 

\noindent (i) The
correlations are enhanced because of spin-paired levels. Such
levels can  be identified by their similar magnetoconductance
traces.

\begin{figure}
\epsfxsize= 8 cm
\centerline{\epsffile{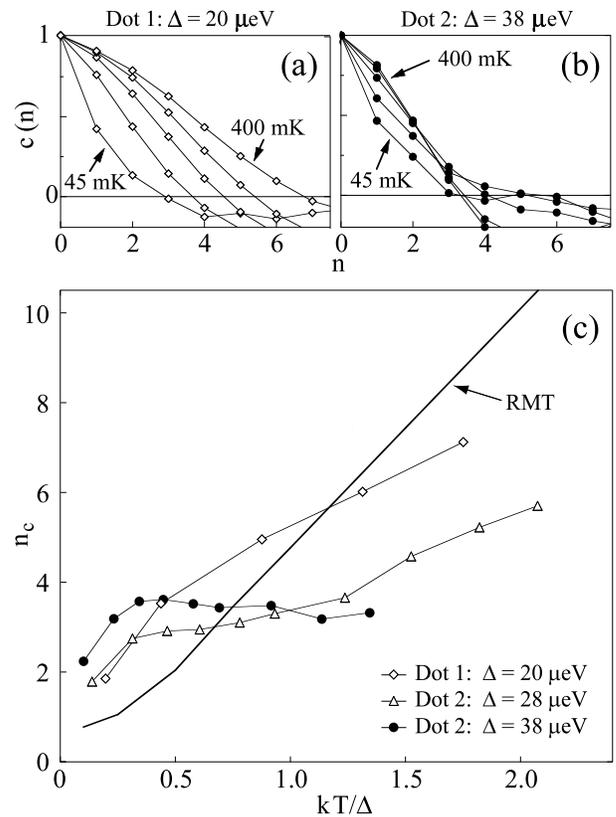}}
\vspace*{2 mm}
\caption{Finite-temperature peak-to-peak correlations and spectral scrambling (experiment).  The peak-to-peak correlator $c(n)$ is shown at different
temperatures  for (a) a larger dot with $\Delta =20 \;\mu$eV  and
(b) a smaller dot with $\Delta =38 \;\mu$eV (right). Notice that the correlator
saturates  sooner in the smaller dot. Compare with panels (a) and (b) of Fig.
\protect\ref{fig:corr-T}.  (c) The number of correlated peaks $n_c$ as a
function  of $T/\Delta$  in three dots of different sizes.
Also  shown (solid line) are the RMT results  for an unchanging spectrum. 
 From Patel, Stewart, {\em et~al.} (1998).
}
\label{fig:corr-T-exp}
\end{figure}

\noindent (ii)
A mechanism was suggested (Hackenbroich, Heiss, and Weidenm\"uller, 1997; Baltin~{\em et~al.}, 1999) whereby the change in 
deformation  of the confining potential upon the addition of an
electron into the  dot  results in a level crossing between
successive Coulomb-blockade peaks such that the  next added electron is
essentially filling the same state. This can lead to  a series
of strongly correlated peaks. The mechanism assumes certain
geometries  (e.g., harmonic potentials) that are more suitable
to nearly integrable  dots. It remains  to be
seen whether this model can also explain enhanced correlations
in  ``generic'' chaotic dots. 

\noindent (iii) A semiclassical theory of
the Coulomb-blockade peak heights was discussed by
Narimanov~{\em et~al.} (1999). The level width to decay into one of the leads is  expanded as a sum over periodic orbits that are well coupled to the lead. A periodic modulation of the peak-height envelope is expected with a period of $\sim \hbar/\tau \Delta$, where $\tau=L/v_F$ is the period of the shortest orbit. For the dot used in the experiment  of Patel, Stewart, {\em et~al.} (1998), the estimated period corresponds to $\sim 12$ peaks, close to the experimental value of $\sim 15$.  The peak heights have Porter-Thomas fluctuations only locally around the semiclassical envelope (Kaplan and Heller, 1998), but the resulting peak-height distribution is still found to be very close to the RMT distribution.  The modulation also leads to enhanced
correlations of adjacent peaks. However, this explanation requires
certain geometries with periodic orbits that are strongly
coupled to at least one of the leads.

An intriguing effect in the experimental results of Fig. \ref{fig:corr-T-exp}
is  the saturation of $n_c$ vs temperature at a value that depends on the dot's size.  This effect will be explained in Sec. \ref{scrambling}.

\subsubsection{Peak spacings}
\label{peak-spacing-T}

 The finite-temperature statistical theory can also be used to calculate the
temperature  dependence of the peak-spacing distribution.  Unlike the
conductance peaks, the peak spacings are sensitive to the
fluctuations of  both the spectrum and the wave functions.  While for $T \ll
\Delta$  the peak height is located at $E_{\cal N}$,  at temperatures of order
$\Delta$   several levels contribute to a given peak, and fluctuations of the
individual  level conductances $g_\lambda$ may shift the peak location away
from  $E_{\cal N}$.

 While we do not expect to reproduce the observed functional form of the
distribution  using a single-particle spectrum that is unchanged with the
addition  of electrons into the dot (see Sec. \ref{peak-spacing:closed}),
it  is still of interest to understand the dependence of its width
$\sigma(\tilde\Delta_2)$  on $T/\Delta$. Figure  \ref{fig:spacings-T}(a) shows a typical sequence of  peak spacings at two different temperatures, demonstrating the decrease in peak-spacing fluctuations with temperature.   Figure  \ref{fig:spacings-T}(b) compares the
 RMT result (solid line) for $\sigma_{\rm GUE}(\tilde\Delta_2)$  (Alhassid and Malhotra, 1999)  with the
experimental  results of Patel, Cronenwett, {\em et~al.} (1998) for $B \neq 0$.   Above
$T/\Delta  \sim 0.5$, we observe a sharp decrease of $\sigma(\tilde\Delta_2)$,
in  agreement with the experimental results. The inset is the calculated ratio
$\sigma_{\rm  GOE}(\tilde\Delta_2)/\sigma_{\rm GUE}(\tilde\Delta_2)$ as a
function  of $T/\Delta$. The experimental ratio ($\sim 1.2 - 1.3$) measured at $T \sim 100\;$mK is consistent with the
calculations.

\begin{figure}
\epsfxsize= 8  cm
\centerline{\epsffile{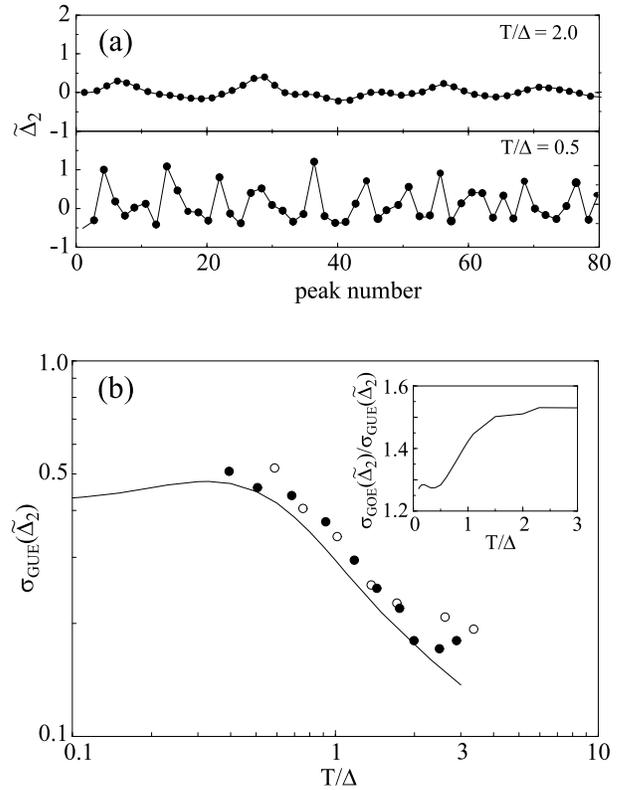}}
\vspace*{2 mm}
\caption{Temperature dependence of the peak-spacing statistics in closed dots:
(a) a sequence of peak spacings vs peak index at $T/\Delta=0.5$
(lower  trace) and $T/\Delta=2$ (upper trace) obtained from one random-matrix realization;
(b) the standard deviation $\sigma(\tilde\Delta_2)$ in the GUE statistics
is  shown versus $T/\Delta$ on a log-log scale. The theoretical RMT results (solid
line)  are compared with recent experimental data by
Patel, Cronenwett, {\em et~al.} (1998)  taken at $B \neq 0$ for two dot configurations: solid circles,
  $\Delta =21\;\mu$eV; open circles, $\Delta=14\;\mu$eV.  The charging energy is  $E_C=590 \;\mu$eV. The results are
expressed  in units of the mean level spacing $\Delta$.  Inset,  the ratio $\sigma_{\rm  GOE}(\tilde\Delta_2)/\sigma_{\rm GUE}(\tilde\Delta_2)$ as a function of $T/\Delta$.
From Alhassid and Malhotra (1999).
}
\label{fig:spacings-T}
\end{figure}

\subsection{Spectral scrambling}
\label{scrambling}

   In Sec. \ref{finite-T:closed}, the finite-temperature statistics were
discussed  assuming that the single-particle spectrum is unchanged as
electrons  are added to the dot. However, in Sec.
\ref{peak-spacing:closed}  we saw that a changing electronic spectrum has
important  effects on the $T \ll \Delta$ peak-spacing distribution.   The single-particle 
 spectrum is expected to change with the addition of electrons
not  only because of changes in the dot's shape, but, more importantly,
because  of electron-electron interactions that lead to charge rearrangement
on  the dot.   A detailed discussion of this point is postponed to Secs.
\ref{Koopmans-theorem} and \ref{parametric-mean-field},  and here we simply assume that the changing spectrum
 can be modeled by a parametric dependence of an effective single-particle
 potential: the dot's Hamiltonian $H(x)$ depends on a parameter $x$
that assumes a discrete set of values ${x_{\cal N}}$ as  electrons are
added  to the dot.  From the theory of Gaussian processes (Sec. \ref{Gaussian-processes}), we expect the peak-to-peak correlator to be
determined  universally (for a fixed $T/\Delta$)  by the value of the scaled
parametric change $\delta \bar{x}$ between two successive peaks.

\begin{figure}
\epsfxsize= 8.3 cm
\centerline{\epsffile{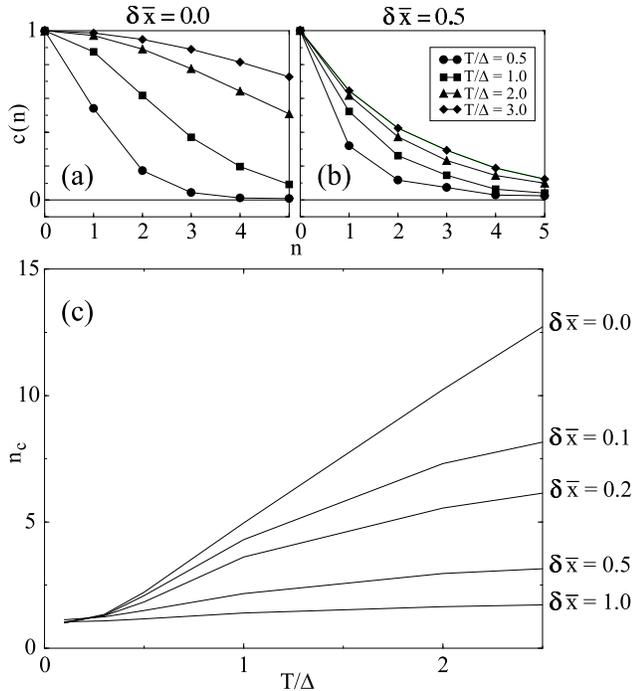}}
\vspace*{2 mm}
\caption{Finite-temperature peak-to-peak correlations and spectral scrambling (theory).  The peak-to-peak correlator $c(n)$ at
$T/\Delta=0.5,1,  2,$ and $3$ is shown for (a) a closed dot with a fixed spectrum ($\delta  \bar x=0$) and  (b) a dot whose spectrum changes with the addition of
electrons  ($\delta \bar x=0.5$). Here $\bar \delta x$ is a parameter characterizing the degree of statistical scrambling upon the addition of one electron to the dot. Notice the saturation of the
correlator  with temperature for the $\delta \bar x \neq 0$ case. (c) The
number of correlated peaks $n_c$ [FWHM of $c(n)$] as a function of $T/\Delta$ for
$\delta  \bar x= 0, 0.1, 0.2, 0.5$, and $1$. The correlation length
saturates  sooner in a dot whose spectrum scrambles faster  (i.e., in a dot with larger $\delta \bar x$).
From Alhassid and Malhotra (1999).
}
\label{fig:corr-T}
\end{figure}

  Peak-to-peak correlations should be sensitive to a changing spectrum. For a
fixed  spectrum, the number of correlated peaks $n_c$ increases
approximately linearly  with $T$, as the number of levels that
contribute to each peak is $\sim  T/\Delta$. However, if the
addition of each electron changes the spectrum,  then $n_c$ is expected to saturate at  a
certain value $\sim m$ that measures the number of added electrons
required to scramble the spectrum completely.  These
expectations are confirmed both experimentally  and
theoretically. Panels (a) and (b) of Fig. \ref{fig:corr-T} show  the
calculated peak-to-peak correlator $c(n)$ at several
temperatures for  $\delta \bar x=0$ and $\delta \bar x
=0.5$. We see that the  correlator's width saturates in
the $\delta \bar x \neq 0$ case. In Fig. \ref{fig:corr-T}(c) we show the number of correlated peaks $n_c$ as a function of temperature for several values of
the scrambling parameter $\delta  \bar x$. We observe that $n_c$ saturates at a value $m$ that decreases
rapidly with increasing $\delta \bar  x$.

 Similar qualitative results are found in experiment (Patel, Stewart, {\em et~al.}, 1998).
Figure \ref{fig:corr-T-exp}(c) shows the experimental $n_c$ vs temperature for dots of different sizes. Also shown in Figs. \ref{fig:corr-T-exp}(a) and \ref{fig:corr-T-exp}(b) is the  temperature dependence of the correlator $c(n)$ for two of the
dots.  Notice the similarity to the theoretical correlators of Figs. \ref{fig:corr-T}(a) and \ref{fig:corr-T}(b).  We see that in the smallest device,
saturation  occurs at $m\sim 3-4$ already for $T \agt 0.5 \; \Delta$, while for the
larger  dots, $n_c$ continues to increase with temperature.  This suggests
that  spectral scrambling is slower in the larger dots.
Indeed, it is argued in Sec. \ref{parametric-mean-field} that $\delta \bar x$  is smaller for a dot with a larger ballistic Thouless conductance $g_T$ [see Eqs. (\ref{Delta-x-g}) below]. Since
$g_T  \propto \sqrt{\cal N}$, we expect the spectrum of the larger dots to have a smaller $\delta \bar x$ and thus to be
less sensitive to the addition of electrons.

\subsection{Correlations between the addition and excitation spectra}
\label{addition-excitation}

 In the constant-interaction  model, the ground state of a dot with ${\cal
N}+n$  electrons is obtained by adding $n$ electrons to the first excited
single-particle  states of the ${\cal N}$-electron dot. On the other hand, we
have  seen in Sec. \ref{scrambling} that interactions scramble the single-particle
spectrum  when electrons are added to the dot, leading to the loss of correlations between the addition and excitation spectra.  The measured
finite-temperature  peak-to-peak correlations shown in Fig. \ref{fig:corr-T-exp} suggest that a  complete scrambling of the single-particle
spectrum  occurs only after several electrons  ($\sim m$) are added to the dot. We
thus expect that for a small $n$ ($\alt m$), the ground state of a dot with ${\cal N}+n$ electrons is still
correlated  with the $n$th excited state of the dot with ${\cal N}$ electrons. The low-lying excited states in the dot can be observed 
through  non-linear transport experiments in the single-charge tunneling
regime (see Sec . \ref{non-linear-transport}).

Stewart~{\em et~al.} (1997) observed large correlations between the addition and
excitation  spectra up to $m \sim 4$. The strongest evidence
for  such correlations was observed in the magnetoconductance traces of the
ground  and excited levels in the dot. The height and position (in gate
voltage  $V_g$)  of the differential conductance peak at finite source-drain
voltage  $V_{sd}$ could be followed as a function of magnetic field and
compared  with similar traces of the ground state in the linear Coloumb-blockade measurements. Stewart~{\em et~al.} found that the magnetoconductance trace of the $n$th
 excited states of an ${\cal N}$-electron dot  was
similar  to the trace of the ground state of an
 ${\cal N} +n$-electron dot for $n < 4$.

  Another important result of the above experiment is the absence of spin
degeneracy,  contrary to the results observed in a few-electron dot
(Tarucha~{\em et~al.}, 1996; Kouwenhoven, Oosterkamp, {\em et~al.}, 1997).  An excited level appears in the excitation
spectrum  for every electron that is removed from the dot.

\section{Interaction Effects}
\label{interactions}

 Electron-electron interactions -- beyond the average interaction energy
${\cal  N}^2e^2/2C$ of the constant-interaction  model --  are expected to play
an  important role in closed dots.   Theoretical studies of interaction
effects on the mesoscopic fluctuations in closed dots have been
largely  motivated by experiments showing deviations from the
 constant-interaction--plus--RMT model:

\noindent  (i) The peak-spacing distributions
(Sivan~{\em et~al.}, 1996; Simmel, Heinzel, and Wharam, 1997; Patel, Cronenwett, {\em et~al.}, 1998; Simmel~{\em et~al.}, 1999)  do not have
the  Wigner-Dyson form and their width is larger than expected from the constant-interaction  model (see Sec. \ref{peak-spacing:closed}). 

\noindent (ii) The measured correlation flux of a conductance peak height is larger than its single-particle estimate (see Sec. \ref{conductance-correlator}).

\noindent (iii) Correlations between the addition and excitation spectra diminish after the addition of a small number of electrons (see Sec. \ref{addition-excitation}). 

\noindent (iv) The saturation of the peak-to-peak correlator with increasing temperature indicates spectral scrambling due to interactions (see Sec. \ref{scrambling}).

  One way to include interaction effects while retaining a single-particle
picture is in the Hartree-Fock approximation. Assuming that the Hartree-Fock single-particle wave functions do not change upon the addition of an electron to the dot, Koopmans' theorem (Koopmans, 1934) states that the addition energy is given by the Hartree-Fock single-particle energy of the added electron.
It is then possible to relate the peak spacing to the change in a single-particle Hartree-Fock level. This change is dominated by a certain diagonal interaction matrix element, which fluctuates due to the fluctuations of the single-particle wave functions.
 Blanter, Mirlin, and Muzykantskii (1997) used the random-phase approximation (RPA) to construct
an  effective screened potential from the bare Coulomb interaction in systems with finite geometries.   They estimated the variance of a diagonal interaction matrix element, and find peak-spacing fluctuations that are larger but still of the  order of the mean level spacing.

The RPA breaks down at strong interactions. Sivan~{\em et~al.} (1996) used an Anderson model of a disordered dot with
electron-electron  interactions to calculate numerically  the the
 peak-spacing distribution.  These calculations can be done only for a very small number of
electrons  (much fewer than in the experiments), but they explain the Gaussian shape of the distributions and yield larger widths for these distributions at stronger interactions.
Berkovits and Sivan (1998) used the same model to study
interaction  effects on the peak-height statistics. Their results indicate that the peak-height distributions are only
weakly  sensitive to interactions but that the
correlation  field increases with
interaction  strength.

How does the electron's spin manifest itself in quantum dots? In the absence of interactions, the single-particle states come in spin-degenerate pairs and the peak-spacing distribution is expected to be bimodal. No bimodality was seen in the experiments, an effect explained by strong electron-electron interactions (Berkovits, 1998). However, a recent experiment in dots with higher electron densities (where the Coulomb interactions are effectively weaker) showed spin-pairing effects in both the peak spacings and the parametric dependence of the peaks (L\"{u}scher~{\em et~al.}, 2000). The spin of the ground state and how it is affected by disorder or one-body chaos was the subject of recent theoretical investigations, and experimental results are expected in the near future.

 The gas parameter $r_s$ measures the strength of the Coulomb
interaction at an average distance between the electrons relative to their kinetic energy.  It is universally determined
by  the density $n_s$ of the electron gas. In 2D,
$\pi (r_s a_B)^2  = {1 / n_s}$,
where $a_B=\hbar^2/m^\ast e^2$ is the Bohr radius. The gas parameter $r_s$ is
thus the radius, in atomic units,  of the circle that encloses one unit of
electron  charge. The Fermi
momentum  is given by  $k_F a_B = \sqrt{2}/r_s$, while the Fermi energy is $E_F
=  (\hbar^2/2m^\ast a_B^2)(2/r_s^2)$. The ratio between a typical Coulomb
interaction  energy $e^2/2 r_s a_B$ and the average kinetic energy $E_F/2$ is
thus given by $r_s$.  $r_s$ can also be expressed in terms of the Fermi
velocity  and the electron charge: $r_s=e^2/\hbar v_F$. The RPA is valid for $r_s < 1$,
but  in typical semiconductor quantum dots $r_s \sim 1 -2$.

 In Sec. \ref{peak-spacing:interactions} we discuss interaction effects on
the  peak-spacing statistics using mean-field approximations and exact simulations. Spin effects are reviewed in Sec. \ref{spin-effects} and  interaction effects on the peak-height statistics are discussed in Sec. \ref{peaks:interactions}. A random interaction matrix model is discussed in Sec. \ref{random-interaction}.

\subsection{Peak-spacing statistics and interactions}
\label{peak-spacing:interactions}

The first experiment to measure the peak-spacing distribution in
Coulomb-blockade  quantum dots was carried out by Sivan~{\em et~al.} (1996).
 The spacing between successive Coulomb-blockade peaks is related to a second
difference  of the ground-state energy as a function of the number of
electrons  ${\cal N}$. To see that, we denote by ${\cal E}^{({\cal N})}_j$ the
ground-state  energy of a dot with ${\cal N}$ electrons at a gate voltage $V_g(j)$
that  corresponds to the degeneracy point of the $j -1 \to j$ transition. Since
the  average one-body potential induced by the gate is linear in the gate
voltage  $V_g$, the total energy of ${\cal N}$ electrons at the $j$th peak is
${\cal  E}^{({\cal N})}_j - e {\cal N} \alpha V_g(j)$, where $\alpha$ is the ratio between the
capacitance  of the dot  with respect to its gate and the total capacitance (see Sec. \ref{Coulomb-blockade}).
The  degeneracy condition for the ${\cal N}+1$ peak is then $E_F + e \alpha
V_g({\cal  N}+1) = {\cal E}^{({\cal N}+1)}_{{\cal N}+1} - {\cal E}^{({\cal
N})}_{{\cal  N}+1}$. The spacing between two consecutive peaks $\Delta V_g=
V_g({\cal  N}+1) -V_g({\cal N})$ is  (Sivan~{\em et~al.}, 1996)
\begin{equation}\label{exact-spacing}
\Delta_2({\cal N}+1) \equiv e \alpha \Delta V_g =
{\cal E}^{({\cal N}+1)}_{{\cal N}+1} + {\cal E}^{({\cal N}-1)}_{{\cal N}} -
{\cal  E}^{({\cal N})}_{{\cal N}+1} - {\cal E}^{({\cal N})}_{{\cal N}} \;.
\end{equation}
 If the ground-state energy of the dot is independent of the gate voltage,
then  $\Delta_2 = {\cal E}^{({\cal N}+1)}+ {\cal E}^{({\cal N}-1)} - 2 {\cal
E}^{({\cal  N})}$. However, generally the change in the gate voltage is
accompanied  by a deformation of the dot so that ${\cal E}^{({\cal
N})}_{{\cal  N}+1} \neq {\cal E}^{({\cal N})}_{{\cal N}}$.

 In the constant-interaction  model, the ground-state energy is  ${\cal E}^{({\cal N})}_j = {\cal N}^2e^2/2C + \sum_{k=1}^{{\cal N}} E_k$ (where $E_k$
are  the single-particle energies), and $\Delta_2({\cal N}+1) = E_{{\cal N}+1}
-  E_{\cal N} +e^2/C$.  Thus if we ignore the spin degrees of freedom, we expect
$\tilde\Delta_2  \equiv (\Delta_2 -e^2/C)/\Delta$ to have a Wigner-Dyson
distribution  $P_{WD}(\tilde\Delta_2)$. The variance of $\Delta_2$ would then
be  $0.52 \,\Delta$ in the GOE and $0.42\,\Delta$ in the GUE. This is the standard constant-interaction--plus--RMT model that we have used in earlier sections.

In the case of spin-degenerate single-particle states, we expect in the constant-interaction  model $\Delta_2({\cal  N}+1)-e^2/C=0$ for {\em odd} ${\cal N}$ (since two electrons with spin up and down can occupy the same level $E_{({\cal
N}+1)/2}$), but  $\Delta_2({\cal N}+1)-e^2/C=E_{{\cal
N}/2+1}-E_{{\cal  N}/2}$ for {\em even} ${\cal N}$. The resulting distribution of $\tilde \Delta_2$ is bimodal:
\begin{equation}\label{bimodal-dist}
 P(\tilde \Delta_2) = (1/2) \left[\delta(\tilde\Delta_2) + {1 \over 2}
P_{WD}\left(\tilde\Delta_2  \over 2 \right)\right]\;,
\end{equation}
where the $\delta$ function and the Wigner-Dyson distribution originate in odd and even ${\cal N}$'s, respectively. This is the constant-interaction--plus--spin-degenerate-RMT (CI + SDRMT)
model,  where the spacing for even ${\cal N}$ is on average larger by
$2\Delta$  than the spacing for odd ${\cal N}$, leading to a larger variance
than in the constant-interaction--plus--RMT model: $\sigma(\tilde\Delta_2)=1.24\,\Delta$ for GOE and $1.16\,\Delta$  for GUE.

Finally, another simple model considered in the literature is the
constant-interaction--plus--spin-resolved-RMT  model (CI+SRRMT). Here the
assumption  is that, because of exchange interactions, the different spin
states are nondegenerate. However, the spin is a good quantum
\begin{figure}
\epsfxsize= 8 cm
\centerline{\epsffile{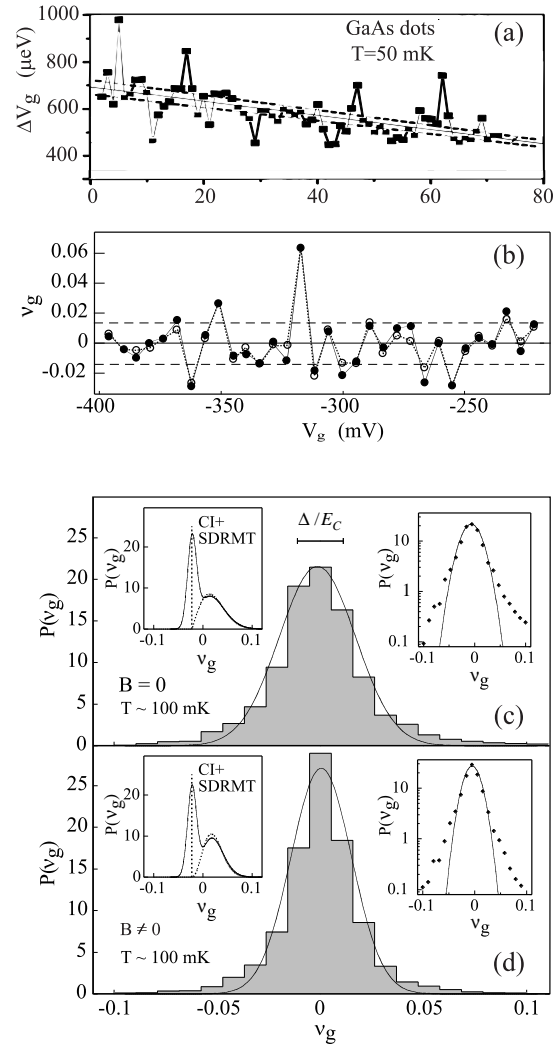}}
\vspace*{2 mm}
\caption{Measured peak-spacing statistics in Coulomb-blockade
 dots:  (a) peak-spacing series from
Sivan~{\em et~al.} (1996) at $B=0$ showing $\Delta V_g$ between
neighboring peaks vs the number of electrons
${\cal N}$
  in the dot; solid line, linear fit describing the
increase in capacitance with ${\cal N}$; (b) sequence of  peak-spacing fluctuations $\nu_g \equiv (\Delta V_g -
\overline{\Delta  V_g})/ \overline{\Delta V_g}$ vs $V_g$ from Patel, Cronenwett, {\em et~al.} (1998); solid symbols, 
$B=30  \;$mT; open symbols, $B=-30 \;$mT. The dashed lines in panels (a) and (b) show  the standard RMT deviation.
Panels (c) and (d), spacing distributions $P(\nu_g)$  for $B=0$ and $B \neq 0$. The shaded histograms are the measured
distributions  at $T\sim 100\;$mK. The data contain 4300 peaks for $B=0$ of
which  720 are statistically independent, and 10800 peaks for $B \neq 0$, of
which  1600 are statistically independent. The
data were collected at $T \sim
100\;$mK from three dots with $\Delta = 9 -11 \;\mu$eV so that $T/\Delta \sim 0.8$.  The solid lines
are  Gaussian fits. The right insets in (c) and (d) present the data and the fit on a
log-linear  scale to show deviations from Gaussians at the tails. The left
insets in (c) and (d) are the bimodal constant-interaction--plus--spin-degenerate-RMT distributions before (dotted lines) and after
(solid lines)  convolution with a Gaussian noise.
From Patel, Cronenwett, {\em et~al.} (1998).
}
\label{fig:spacings}
\end{figure}
 \noindent number, and therefore  the subspaces of spin up and
 spin down are described by two
independent  random-matrix ensembles. The statistics of a superposition of two
uncorrelated spectra were discussed by Dyson (1962b). The
corresponding level-spacing distribution (see Mehta, 1991, Appendix
A.2,  p. 402) is characterized by a nonzero value at zero spacing. The corresponding widths
$\sigma(\tilde\Delta_2)=0.70 \, \Delta$ for GOE and
$0.65 \, \Delta$ for GUE are in between the values predicted by the CI+RMT and
CI+SDRMT  models.

Sequences of peak spacings measured in gate voltage $\Delta V_g$ are shown in
  Figs. \ref{fig:spacings}(a) and \ref{fig:spacings}(b). The sequence in panel (a) is from
Sivan~{\em et~al.} (1996), and the sequence in panel (b) is from Patel, Cronenwett, {\em et~al.} (1998), where the  quantity drawn is $\nu_g \equiv(\Delta V_g -\overline{\Delta
V_g})/\overline{\Delta  V_g}$. The dashed lines describe the standard
deviation of the Wigner-Dyson distribution. The fluctuations in Fig. \ref{fig:spacings}(a) are significantly larger than in RMT, while those in \ref{fig:spacings}(b)
are  somewhat larger than in RMT. The difference between the experiments in the magnitude of the fluctuations is not
currently  understood.  The peak-spacing distribution is found to be more symmetric than in any of the above simple
models  and closer to a Gaussian. In particular, no bimodal structure is observed, suggesting the absence of spin degeneracy.  An example
of  the observed distributions (histograms) is shown in panels (c) and (d) of
Fig.  \ref{fig:spacings} (Patel, Cronenwett, {\em et~al.}, 1998), where the
solid lines are Gaussian fits. The scale of the mean level spacing is shown
in  units of the charging energy $E_C$.  The ratio between the $B=0$ and $B \neq 0$ widths is
$\sigma_{B=0}(\Delta_2)/\sigma_{B\neq  0}(\Delta_2)\approx 1.23$.

  The discussion here is limited to almost-closed dots.  The statistics of the peak spacings for a partially open dot were recently discussed by Kaminski and Glazman (2000), but only within the constant-interaction  model. For a partially open dot, the peak position is affected by the dot-lead couplings, and the randomness of the electronic wave functions at the point contacts contribute to the peak-spacing fluctuations. This contribution   increases with the strength of the dot-lead couplings and becomes comparable to the fluctuations of the single-particle spacing while Coulomb-blockade peaks can still be observed.

  In the following discussions of peak spacings and interactions,  we ignore spin. Spin effects will be discussed in Sec. \ref{spin-effects}.

\subsubsection{Hartree-Fock approximation and Koopmans' theorem}
\label{Koopmans-theorem}

A consideration of interaction effects while retaining a single-particle picture is best done in the Hartree-Fock 
approximation. We start from a Hamiltonian of interacting electrons:
$H= \sum_{ij} h_{ij} a^\dagger_i a_j +{1\over 4} \sum_{ijkl}
 v^A_{ijkl}a^\dagger_i a^\dagger_j a_l a_k$,
where $h = t + V $ is the one-body part ($V$ describes one-body disorder or a confining potential) and $v^A_{ij;kl} \equiv \langle i j | v| k l \rangle - \langle i j | v| l k
\rangle$  are the antisymmetrized matrix elements of the two-body interaction.
The single-particle Hartree-Fock energies and wave functions are determined self-consistently by solving the Hartree-Fock equations (see, e.g., Kittel, 1987). Denoting by $E_k^{({\cal N})}$ the energy of the $k$th single-particle state in a dot with ${\cal N}$ electrons, we have in the self-consistent basis 
\begin{equation}
E_k^{({\cal N})} = h_{kk} + \sum\limits_{i=1}^{\cal N} v^A_{ki;ki}  \;,
\end{equation}
where the sum is over the ${\cal N}$ lowest occupied single-particle states.  
The Hartree-Fock ground-state energy for ${\cal N}$ electrons is then given by ${\cal E}_{HF}^{({\cal N})} =
\sum_{k=1}^{{\cal  N}} h_{k k} + (1/2) \sum_{i,j=1}^{{\cal N}} v^A_{ij;ij}$.

Numerical solutions of the Hartree-Fock equations will be discussed in Sec. \ref{Anderson:interactions}.  Here we use the framework of Koopmans' theorem (Koopmans, 1934), enabling us to relate the peak
spacing to the single-particle Hartree-Fock energies. Koopmans' theorem states that
\begin{equation}\label{Koopmans}
{\cal E}_{HF}^{({\cal N}+1)} - {\cal E}_{HF}^{({\cal N})} \approx E_{{\cal
N}+1}^{({\cal  N})} \;.
\end{equation}
Its  basic assumption is
that  the single-particle wave functions do not change when an electron is
added.  This assumption should hold only for large systems (Kittel, 1987),
but  it might be a good starting point for a dot with several hundred
electrons.
Under this assumption we find ${\cal E}_{HF}^{({\cal N}+1)} - {\cal E}_{HF}^{({\cal N})} \approx h_{{\cal
N}+1,{\cal  N}+1} + \sum_{i=1}^{\cal N} v^A_{{\cal N}+1, i; {\cal
N}+1,  i}
= E_{{\cal N}+1}^{({\cal N})}$, which is just Eq. 
(\ref{Koopmans}). Under the conditions of Koopmans' theorem we also have
$E_{{\cal N}+1}^{({\cal  N})} = E_{{\cal
N}+1}^{({\cal  N}+1)}$, and the peak spacing can be written as
\begin{equation}\label{Delta-sum}
\Delta_2({\cal N}+1) \approx E_{{\cal N}+1}^{({\cal N}+1)} - E_{\cal N}^{({\cal
N})}=   \Delta E^{({\cal N}+1)} + \Delta E_{\cal N} \;.
\end{equation}
Here $\Delta E^{({\cal N}+1)} \equiv E_{{\cal N}+1}^{({\cal N}+1)} -
E_{{\cal  N}}^{({\cal N}+1)}$ measures the spacing between the ${\cal N}$th
and  ${\cal N}+1$st levels for a {\em fixed} number of electrons (${\cal
N}+1$)  and is of the order of the mean level spacing $\Delta$ (both levels are occupied and their spacing is expected to follow Wigner-Dyson statistics), while $\Delta
E_{\cal  N} \equiv E_{{\cal N}}^{({\cal N}+1)} - E_{\cal N}^{({\cal N})}$ is
the  change in energy of the same ${\cal N}$th level when the number of
electrons  in the dot is increased from ${\cal N}$ to ${\cal N}+1$:
\begin{eqnarray}\label{delta-E}
\Delta E_{\cal N}  & \approx & v^A_{{\cal N}+1,{\cal N};{\cal N}+1,{\cal N}} \nonumber \\ & =  &
 \int d\bbox{r} d \bbox{r'}
\left[|\psi_{\cal N}(\bbox{r})|^2  v(\bbox{r}, \bbox{r'}) |\psi_{{\cal
N}+1}(\bbox{r'})|^2  \right. \nonumber \\ & -  &   \left. \psi^\ast_{\cal
N}(\bbox{r})  \psi_{\cal N}(\bbox{r'})    v(\bbox{r}, \bbox{r'}) \psi_{{\cal
N}+1}(\bbox{r})\psi^\ast_{{\cal  N}+1}(\bbox{r'}) \right] \;.
\end{eqnarray}
 The quantity $\Delta E_{\cal N}$ is of the order of the charging energy, and is a constant   $e^2/C$ in the constant-interaction  model. However, the fluctuations of the single-particle wave functions in Eq. (\ref{delta-E}) lead to fluctuations of $\Delta E_{\cal N}$ and can modify the peak-spacing statistics.

\subsubsection{Random-phase approximation in disordered dots}
\label{RPA:interactions}

Blanter, Mirlin, and Muzykantskii (1997) calculated the fluctuations of $\Delta E_{\cal N}$ in Eq.
(\ref{delta-E})  using the RPA and assuming a disordered dot in its metallic
regime.  The derivation consists of (i) calculating an effective
interaction  in the RPA to replace the bare Coulomb interaction in Eq. 
(\ref{delta-E}),  and (ii) calculating the variance of the matrix element in Eq. 
(\ref{delta-E})  from the fluctuations of the  single-particle eigenfunctions.

 The Coulomb interaction $v_0(\bbox{r}_1 - \bbox{r}_2) = e^2/\epsilon
|\bbox{r}  - \bbox{r'}|$ (where $\epsilon$ is the dielectric constant) is
screened.   An effective screened potential
$v(\bbox{r},\bbox{r'})$  in the finite geometry of the dot can be found in the RPA (Fetter and Walecka, 1971):
\begin{eqnarray}\label{RPA}
 v(\bbox{r},\bbox{r'}) = && v_0(\bbox{r} -\bbox{r'}) - \int d\bbox{r}_1
d\bbox{r}_2  v_0(\bbox{r} - \bbox{r}_1) \nonumber \\ && \times\,\Pi_0(\bbox{r}_1, \bbox{r}_2)
v(\bbox{r}_2,\bbox{r'})\;,
\end{eqnarray}
where $\Pi_0$ is the static polarization function relating the electric
potential  fluctuation $\delta V$ to the charge density fluctuations $\delta
\rho$  through $\delta \rho(\bbox{r}) =-2e \int d \bbox{r'}
\Pi_0(\bbox{r},\bbox{r'})  \delta V(\bbox{r'})$. In the limit where the
screening  length is large compared to the Fermi wavelength,  the polarization
function  can be approximated by a localized function
$\Pi_0 (\bbox{r},\bbox{r'}) \approx \nu [\delta(\bbox{r}- \bbox{r'}) - {\cal
A}^{-1}]$,
 where $\nu$ is the density of states per unit area and ${\cal A}$ is the
area  of the dot.  In an infinite system,  Eq. (\ref{RPA})  is easily solved by a
Fourier  expansion. In a finite geometry, the equations are solved  by
expanding  in a complete set of eigenfunctions $\chi_\lambda(\bbox{r})$ of the
Laplacian  in the dot  with eigenvalues $q_\alpha^2$ that include the zero
mode  $\chi_0 = 1/{\cal A}$.  Three contributions to the
effective  potential are found:
\begin{equation}\label{effective-int}
v(\bbox{r},\bbox{r'}) = {e^2 / C} +
[V_\kappa(\bbox{r}) + V_\kappa(\bbox{r'})] +  v_\kappa(\bbox{r},\bbox{r'})
\;.
\end{equation}
The first term is the usual charging energy, and the third term is the 2D
screening  potential
$v_\kappa(\bbox{r},\bbox{r'})=
 (2\pi e^2/ \epsilon) \sum_{\alpha \neq 0} (q_\alpha + \kappa)^{-1}
\chi_\alpha(\bbox{r})  \chi^\ast_\alpha(\bbox{r'})$, where
$\kappa = 2 \pi e^2 \nu / \epsilon$ is the inverse screening
length. The  new contribution in a finite geometry is a one-body
potential, which for a  disk of radius $R$ is
$V_\kappa(\bbox{r}) = -(e^2 / 2 \kappa R) (R^2 - r^2)^{-1/2}$.
This potential is the result of excess charge that is
pushed to the boundaries  of the dot: the
added electron attracts a positive cloud around it, generating
excess negative charge at the  boundaries.

  Using the effective interaction [Eq. (\ref{effective-int})]  in Eq.
(\ref{delta-E}),  the variation $\Delta E_{\cal N}$ of the  ${\cal N}$th Hartree-Fock
level  when an electron is added to the dot is  composed of three
contributions, 
\begin{equation}\label{delta-E-012}
\Delta E_{\cal N} = {e^2 / C} + \Delta E_{\cal N}^{(1)} + \Delta E_{\cal
N}^{(2)}\;,
\end{equation}
corresponding to the three parts of the effective potential [Eq. 
(\ref{effective-int})].   Fluctuations of $\Delta E_{\cal N}$  originate from fluctuations of the single-particle
wave functions, and the variances $\sigma(\Delta E_{\cal N}^{(i)})$ (where $i=1,2$) can then be
expressed  in terms of wave-function correlations. For example,
\begin{equation}\label{sig-E-1}
\sigma^2(\Delta E_{\cal N}^{(1)}) = 2\int d\bbox{r} d\bbox{r'}
V_\kappa(\bbox{r})  V_\kappa(\bbox{r'}) \overline{|\psi_{\cal N}(\bbox{r})|^2
|\psi_{\cal  N}(\bbox{r'})|^2}\;,
\end{equation}
where we have ignored the exchange term in Eq. (\ref{delta-E}).

 The universal RMT result for the correlation of the intensities of two
eigenfunctions in chaotic systems is ${\cal A}^2  \overline{|\psi_\lambda
(\bbox{r})|^2  |\psi_\mu(\bbox{r'})|^2} -1 = \delta_{\lambda \mu}(2/\beta)
f_d^2(|\bbox{r}  - \bbox{r'}|)$, where $f_d$ is  defined in Eq.
(\ref{spatial-correlation-d}).   In diffusive systems ($l \ll L$) this
correlation  is short-range (with a range of $\sim l$) because of an
additional  exponential factor $e^{-|\Delta \bbox{r}|/2l}$ in
$f_d$.  Long-range (diffuson) correlations are weaker by $1/g_T$ but become important when
integrated over in Eq. (\ref{sig-E-1}). Wave-function correlations in disordered
systems  were studied to order $1/g_T$   by Blanter and Mirlin (1997) and reviewed
by  Mirlin (1997, 2000). For $\beta=1,2$  and eigenfunctions $\psi_\lambda,\;\psi_\mu$ ($\lambda \neq \mu$) whose energy separation is smaller than $E_c$, they find 
\begin{mathletters}\label{wavefunction-corr}
\begin{eqnarray}
&&{\cal A}^2  \overline{|\psi_\lambda (\bbox{r})|^2
|\psi_\lambda(\bbox{r'})|^2}  -1 \nonumber \\ && \; =  {2 \over\beta}\left[ f_d^2(|\bbox{r} -
\bbox{r'}|)  \left( 1 + {2 \over \beta} \Pi_D(\bbox{r}, \bbox{r'})\right) +
\Pi_D(\bbox{r},\bbox{r'})  \right] \label{auto-corr} \;,\\
&& {\cal A}^2  \overline{|\psi_\lambda (\bbox{r})|^2 |\psi_\mu(\bbox{r'})|^2} -1 \nonumber \\
& &\; =  {2 \over \beta}  f_d^2(|\bbox{r} - \bbox{r'}|)
\Pi_D(\bbox{r},\bbox{r'})   \label{cross-corr} \;,\\
&&{\cal A}^2  \overline{\psi^\ast_\lambda (\bbox{r}) \psi_\mu(\bbox{r}) \psi_\lambda (\bbox{r'}) \psi^\ast_\mu(\bbox{r'})} \nonumber \\ && \; =  f_d^2(|\bbox{r} - \bbox{r'}|)[ 1 + (2 -\beta) \Pi_D(\bbox{r},\bbox{r'})] +  \Pi_D(\bbox{r},\bbox{r'})  \;, \label{exchange-corr}
\end{eqnarray}
\end{mathletters}
where $\Pi_D(\bbox{r},\bbox{r'}) = (\pi \nu)^{-1} \sum_{\alpha \neq 0}
\chi_\alpha(\bbox{r})  \chi_\alpha(\bbox{r'})/D q^2_\alpha$ is the diffusion
propagator  expressed in terms of the eigenfunctions $\chi_\alpha(\bbox{r})$
and  eigenvalues $D q^2_\alpha$ of the diffusion operator.  Here $\Pi_D \sim 1/g_T$
is  negligible at short distances compared with $f_d^2$, but dominates at long distances.

 Using the correlator (\ref{auto-corr}) in Eq. (\ref{sig-E-1}), we find for the
variance  of $\Delta E_{\cal N}^{(1)}$
\begin{eqnarray}\label{sig-E-1a}
\sigma^2(\Delta E_{\cal N}^{(1)}) = &&
{4 \over \beta {\cal A}^2 } \int d\bbox{r} d\bbox{r'}
V_\kappa(\bbox{r})[f_d^2(\bbox{r}-\bbox{r'}) \nonumber \\ &&  +  \Pi_D(\bbox{r},\bbox{r'})]
V_\kappa(\bbox{r'})  \;.
\end{eqnarray}
We now estimate the contribution to Eq. (\ref{sig-E-1a}) of the short-range and long-range parts of
the  wave-function correlator. In 2D, $V_\kappa \sim
e^2/\kappa  {\cal A}$ over the whole dot while $\int d\bbox{r} d\bbox{r'} f^2_2(\bbox{r} -
\bbox{r'})  \sim {\cal A}l/k_F$ and $\Pi_D \sim 1/g_T$. The contribution of  $f_2^2$ to Eq. (\ref{sig-E-1a}) is then $\sim \beta^{-1} (e^2/\epsilon
\kappa  {\cal A})^2 {\cal A} l/k_F \sim \beta^{-1}(l/L)^2 \Delta^2/k_F
l$,  while the contribution of $\Pi_D$ is $\sim \beta^{-1} \Delta^2 /g_T$.
Since  $g_T \sim k_F l$, we conclude that the short-range contribution is
suppressed  by a factor of $(l/L)^2$ relative to the long-range
contribution, and
\begin{eqnarray}\label{sig-E-1b}
\sigma^2(\Delta E_{\cal N}^{(1)}) & = &
{4 \over \beta {\cal A}^2 } \int d\bbox{r} d\bbox{r'} V_\kappa(\bbox{r})
\Pi_D(\bbox{r},\bbox{r'})  V_\kappa(\bbox{r'}) 
\nonumber \\ & \sim & \frac{1}{\beta}{\Delta^2
\over  g_T} \;.
\end{eqnarray}
A similar calculation of $\Delta E_{\cal N}^{(2)}$ using the two-body screened
potential  $v_\kappa(\bbox{r},\bbox{r'})$ in Eq. (\ref{delta-E}) and the
correlator  (\ref{cross-corr}) gives
\begin{eqnarray}\label{sig-E-2}
\sigma^2(\Delta E_{\cal N}^{(2)}) 
& \approx & {4 \Delta^2 \over \beta^2 {\cal A}^2 } \int d\bbox{r} d\bbox{r}'
\Pi_D^2(\bbox{r},\bbox{r}')  \nonumber \\
  & \sim & \frac{1}{\beta^2} \left({\Delta \over g_T}\right)^2
 \;.
\end{eqnarray}

The charging energy $e^2/C$ also fluctuates  because of small deviations
($\sim 1/\kappa R$) of the capacitance from its purely geometrical
value  (Berkovits and Altshuler, 1997; Blanter and Mirlin, 1998), leading to
$\sigma\left({e^2  / C}\right) \sim (\ln g_T/\beta)^{-1/2} \Delta/g_T$.

 The largest contribution to $\sigma(\Delta E_{\cal N})$ comes from the surface charge:
$\sigma(\Delta E_{\cal N}) \sim \Delta /\sqrt{\beta g_T}$ [see Eq. (\ref{sig-E-1b})]. In its absence (e.g., using periodic boundary conditions), the dominating contribution is from the usual 2D screened interaction $\sigma(\Delta E_{\cal N}) \sim \Delta /{\beta g_T}$ [se Eq. (\ref{sig-E-2})].  It was assumed that
the  added-electron charge is spread over the complete dot. In practice, this
will  depend on the geometry of the dot and the gates. An opposite extreme is
when  the added-electron charge is confined to an area whose linear dimension
is  $\sim \lambda_F$, leading to $\sigma(\Delta E_{\cal N}) \sim \Delta
/\sqrt{\beta}$.

The standard deviation  $\sigma(\Delta_2)$ is  obtained by
combining the usual RMT fluctuations  ($\sim \Delta$)  with the
fluctuations of $\Delta E_{\cal N}$ [see Eq. (\ref{Delta-sum})].  This results
in  peak-spacing fluctuations that are enhanced with respect to
RMT, but still of the  order of $\Delta$.

 The dots used in the experiments are ballistic rather than diffusive, requiring a new estimate of the fluctuations using
wave-function correlations in chaotic systems. The leading universal 
contribution to the correlations in 2D is  $f_2^2$, whose envelope
decays slowly like $1/k_F|\bbox{r} - \bbox{r'}|$ over the whole  area
of the dot, and leads to $\sigma(\Delta E_{\cal
N}^{(1)}) \sim  \Delta/\sqrt{\beta k_F L}$. A similar estimate for $\sigma(\Delta E_{\cal
N}^{(1)})$ is obtained if we replace $g_T$ in the diffusive result [Eq. (\ref{sig-E-1b})] by its ballistic analog $k_F L \sim {\cal N}^{1/2}$ (since $l > L$). How important  are nonuniversal
contributions to the wave-function correlator in the  chaotic
case?  Blanter, Mirlin, and Muzykantskii (1998) derived the wave-function
statistics in a billiard  with diffusive surface scattering.
They found ${\cal A}^2  \overline{|\psi_\lambda (\bbox{r})|^2
|\psi_\lambda(\bbox{r'})|^2}  -1 = (2 /\beta)
\Pi_B(\bbox{r},\bbox{r'})$, where $\Pi_B$ is the ballistic
analog of the diffusion propagator.  The contribution  to
$\Pi_B(\bbox{r},\bbox{r'})$ from straight-line trajectories
connecting  $\bbox{r}$ to $\bbox{r'}$ gives  $\Pi_B =1/k_F
|\bbox{r} - \bbox{r'}|$,  which is just the smoothed version of the universal
correlations $f_2^2(\bbox{r}  - \bbox{r'})$. Evaluation of $\Pi_B$ beyond the universal part
requires knowledge of the classical  dynamics. Scars along periodic orbits can enhance $\Pi_B$ and  lead to
larger peak-spacing  fluctuations (Stopa, 1998) in
self-consistent density-functional calculations (Stopa, 1996).
Analogous results for the wave-function correlations were derived
for a purely ballistic chaotic dot using semiclassical methods
(Hortikar and Srednicki, 1998).

\subsubsection{Parametric variation of the mean field}
\label{parametric-mean-field}

As  electrons are added to the dot, the Hartree-Fock potential changes owing to charge
rearrangement  caused by the two-body interaction. We denote the Hartree-Fock 
Hamiltonian for ${\cal N}$ electrons by $H(x_{\cal N})$.  Rather then solving the microscopic Hartree-Fock equations, we can adopt a ``macroscopic'' approach  assuming
that,  for a chaotic dot, $H(x_{\cal N})$ describes a discrete Gaussian
process (Attias and Alhassid, 1995). In Koopmans' limit we can relate the peak spacing to the single-particle levels through Eq. (\ref{Delta-sum}).  In particular, the quantity $\Delta E_{\cal N}$ describes  a
discrete  parametric variation of the ${\cal N}$th eigenvalue when the
Hamiltonian  changes from $H(x_{\cal N})$ to $H(x_{{\cal N}+1})$. Since a
discrete  Gaussian process can be embedded in a continuous Gaussian process $H(x)$ (Sec.  \ref{Gaussian-processes}), we obtain the  formulation already
discussed  in Sec. \ref{peak-spacing:closed}. However, the conceptual
difference  is that here the parametric variation of the single-particle
spectrum  is due to interaction effects,
while  in Sec. \ref{peak-spacing:closed} it originated from a
deformation  of the dot's shape.  Experimental
results  (Patel, Stewart, {\em et~al.}, 1998) suggest that the primary cause of a changing
single-particle  spectrum is interactions.

The variation of the spectrum with the addition of one electron to the dot is described by the scrambling parameter $\delta \bar x$ (see Sec. \ref{scrambling}). This parameter can be determined from the dot's properties in the limit of Koopmans' theorem where the single-particle wave functions are unchanged.   In the parametric approach, this limit corresponds to first-order perturbation theory, where  $\sigma^2(\Delta
E_{\cal  N}) = \Delta^2 (\delta \bar x)^2$.  Since $\Delta E_{\cal N}$ in the parametric approach corresponds to $\Delta E_{\cal N}^{(1)} + \Delta E_{\cal N}^{(2)}$ in the microscopic approach of Sec. \ref{RPA:interactions} ($e^2/C$ has been subtracted in both cases), we can compare $\sigma^2(\Delta
E_{\cal  N})$  with Eq. (\ref{sig-E-1b}) [or Eq. (\ref{sig-E-2})] to find that (Alhassid and Malhotra, 1999; Alhassid and Gefen, 2000)
\begin{mathletters}\label{Delta-x-g}
\begin{eqnarray}
& \delta\bar x \sim (\beta g_T)^{-1/2} \sim \beta ^{-1/2} {\cal N}^{-1/4} \label{Delta-x-g1} \\
{\rm or}\;\;\; & \delta \bar x \sim (\beta g_T)^{-1} \sim \beta ^{-1} {\cal N}^{-1/2}  \;, \label{Delta-x-g2}
\end{eqnarray}
\end{mathletters}
where Eq. (\ref{Delta-x-g2}) holds in the absence of surface charge.  Equations (\ref{Delta-x-g}) are valid in the regime where $\sigma^2(\Delta
E_{\cal  N})$ is linear in
$(\delta  \bar x)^2$, i.e., for $\delta \bar x \alt 0.3$.  Complete scrambling (upon the addition of $m$ electrons) is expected when the overall parametric change is one avoided crossing, i.e., $m \delta \bar x \sim 1$. We conclude  that $m \sim (\beta g_T)^{1/2}$ or $m \sim \beta g_T$ for Eqs. (\ref{Delta-x-g1})  or (\ref{Delta-x-g2}), respectively.

\subsubsection{Anderson model with interactions}
\label{Anderson:interactions}

 The RPA estimate in interacting disordered systems gives a peak-spacing
standard deviation  of the order of $\Delta$. However, the RPA is valid in the limit of high densities ($r_s \ll 1$) and breaks down at the lower electron densities in
semiconductor  quantum dots (where $r_s \sim 1-2$). Yet $r_s$ is still substantially below the limit of Wigner crystallization, and additional insight into this regime of intermediate
interactions can be gained by numerical calculations. Exact numerical diagonalizations are possible only for dots with a very small number of electrons
($\sim  10$) on a lattice with $m \sim 20$ sites (compared with several hundred electrons in the experiment).
Berkovits and Sivan (see Sivan~{\em et~al.}, 1996) used a 2D Anderson model with
on-site  disorder and Coulomb interactions.  The Hamiltonian is $H=H_1 + H_2$, where $H_1$ is a one-body Anderson
Hamiltonian  [Eq. (\ref{AndModel})], and $H_2 ={1\over 2} U_c  \sum_{i \neq j} (|\bbox{r}_i
-\bbox{r}_j|/a)^{-1}  a^\dagger_i a_i a^\dagger_j a_j$
 describes the two-body
Coulomb  interaction between electrons on the lattice, where
$U_c=e^2/a$ is the interaction strength over one lattice spacing.
They used small lattices ($\sim 4\times 5$) with ${\cal N}
\leq  13$ to study the distribution and variance of $\Delta_2$  as a function of the interaction strength $U_c$.  As $r_s \sim (\pi m/{\cal N})^{1/2} (U_c/4t)$  increases
from  zero, the spacing distribution deviates from the Wigner-Dyson
distribution  and becomes approximately a Gaussian for $r_s \agt 1$.
The variance continues to increase with $r_s$, and for $r_s \gg 1$
they  find $\sigma(\Delta_2) \propto \langle \Delta_2 \rangle \approx e^2/C$ with a proportionality constant of about $0.1-0.2$.

  Larger dots can be solved under certain approximations.  The self-consistent Hartree-Fock approximation was used to calculate the ground state of dots  with up to $\sim 100$ spinless electrons  (Cohen, Richter, and Berkovits, 1999; Levit and Orgad, 1999; Walker, Montambaux, and Gefen, 1999).
 The single-particle orbits are computed self-consistently, allowing for configuration rearrangement as an electron is added to the dot.
The dominant contribution to the peak spacing comes from the direct matrix element $v_{{\cal N}+1, {\cal N};{\cal N}+1, {\cal N}}$ (which  in Koopmans' limit gives $\Delta E_{{\cal N}}$).
The self-consistent Hartree-Fock calculations confirm that, for increasing $r_s$, the occupied (and unoccupied) levels exhibit Wigner-Dyson statistics, while the peak-spacing distribution evolves rapidly into a Gaussian-like distribution (at $r_s \sim 1$) with a width that is
enhanced compared with the noninteracting picture. The gap in the Hartree-Fock spectrum between the highest filled level and lowest empty level is dominated by an interaction matrix element $v_{{\cal N}, {\cal N}+1;{\cal N}, {\cal N}+1}$ and is found to have a Gaussian-like distribution for large values of $r_s$.

 Screening may not be very effective in the small dots used in the exact numerical simulations. Furthermore, screening can also be generated by external charges on nearby metallic plates. Consequently, Hartree-Fock simulations were also carried out for short-range (e.g., nearest-neighbor) interactions (Walker, Montambaux, and Gefen, 1999).

The dependence of the width $\sigma(\Delta_2)$ on the size of the dot is not
fully understood. The RPA estimates predict $\sigma(\Delta_2)
\propto  \Delta $ for $r_s\ll 1$, while classical calculations by
Koulakov, Pikus, and Shklovskii (1997)  give  $\sigma(\Delta_2) \propto e^2/C \approx
\langle \Delta_2 \rangle$  for $r_s \gg 1$. The RPA scaling $\Delta \propto 1/L^2$
has  a different size dependence than the classical scaling $\langle \Delta_2 \rangle \sim
1/L$,  a difference due to the delocalized character of the
wave functions.   It is not clear which is the correct scaling in the
intermediate  regime relevant to the experiments ($r_s \sim 1-2$), although the self-consistent Hartree-Fock calculations (for Coulomb interactions) suggest typical fluctuations of $\sigma(\Delta_2) \sim 0.52 \,\Delta + a\langle \Delta_2\rangle/\sqrt{g_T} + {\cal O}(r_s^2)$, where $a$ is a constant (Bonci and Berkovits, 1999; Walker, Montambaux, and Gefen, 1999). 
For short-range interactions, a scaling of $\sim r_s \Delta$ is observed for $r_s \alt 1$.

  The measured width of the spacing
fluctuations shows substantial variation between different experiments. The  experiments  of Sivan~{\em et~al.} (1996), and Simmel, Heinzel, and Wharam (1997) gave $\sigma(\Delta_2) \approx
(0.10  -0.15) e^2/C \sim (2 -3) \Delta$, while Patel, Cronenwett, {\em et~al.} (1998) found smaller fluctuations $\sigma(\Delta_2)
\approx  0.05 e^2/C \sim \Delta$ for similar GaAs dots ($r_s\sim 1$).
 
For short-range interactions Koopmans' theorem breaks down at $r_s \agt 1$, and the self-consistent Hartree-Fock ground state develops charge-density modulations and increased short-range density correlations (Walker, Gefen, Montambaux, 1999). The addition spectrum shows nonuniversal features, and $\langle \Delta_2 \rangle$ exhibits sharp maxima at certain magic numbers of ${\cal N}$, in agreement with a classical model of interacting charges. This result is quite surprising since $r_s$ is still much smaller than the Wigner crystal limit. A model of classical charges in a parabolic confining potential (Koulakov and Shklovskii, 1998) has explained the capacitance experiments of Zhitenev~{\em et~al.} (1997), where bunching was observed in the addition spectrum.

Nonuniversal effects were found in the addition spectra of clean chaotic dots for strong Coulomb interactions (Ahn, Richter, and Lee, 1999). They were explained by charge rearrangement that forms geometry-dependent ordered states localized at the edge of the dot.

\subsection{Spin effects and interactions}
\label{spin-effects}

 In the presence of spin degrees of freedom, the single-particle levels come in degenerate pairs of spin up and spin
down  with identical spatial wave functions. Though the Coulomb interaction does not depend explicitly on spin, the Pauli principle leads to an exchange interaction that favors larger spin values. Indeed, in a spin-polarized state, the orbital part of the wave function (having permutation symmetry conjugate to that of the spin part) is less symmetric, thus lowering the Coulomb repulsion.   In the limit $g_T \gg 1$, it is possible to write a simple Hamiltonian for the dot (Kurland, Aleiner, and Altshuler, 2000). The fluctuations of the matrix elements in the disorder basis are ${\cal O}(1/g_T)$ [see, for example, Eq. (\ref{sig-E-2})], and in the limit $g_T \to \infty$ only the average diagonal interaction survives.  We obtain
\begin{equation}\label{CI-spin}
H_{\rm dot} = \sum_{\lambda s} E_\lambda  a_{\lambda s}^\dagger
a_{\lambda s}^{} + {e^2 \over 2C} \hat{\cal N}^2  -  {1 \over 2}\xi {\bbox S}^2 \;,
\end{equation}
where $\xi/2 \equiv \overline{v}_{\alpha \beta; \beta \alpha}$ is the average exchange matrix element and $\bbox S$ is the total spin operator of the dot. $a^\dagger_{\lambda s}$ and $a_{\lambda s}$ are creation and annihilation operators of an electron at orbital state $\lambda$ and spin $s=\pm 1/2$. Both interaction terms in Eq. (\ref{CI-spin}) are invariant under a change of the single-particle basis (an additional Cooper-channel interaction  is possible for the orthogonal symmetry, and is not included here). The single-particle part of Eq. (\ref{CI-spin}) satisfies RMT statistics. Compared with the constant-interaction  model, the Hamiltonian (\ref{CI-spin}) contains a new parameter $\xi$. Its statistical properties are expected to be universal for a given value of $\xi$. An RPA estimate gives $\xi/2 \sim 0.3 \,\Delta$ for GaAs dots with $r_s \sim 1-2$.

    A quantity that might be sensitive to spin is the peak spacing.  In the absence of
interactions,    Eq. (\ref{Delta-sum}) for the peak spacing still
holds,  but the explicit expressions for $\Delta E^{({\cal N}+1)}$ and $\Delta
E_{\cal  N}$ depend on the  parity of ${\cal N}$: $\Delta E^{({\cal N}+1)}=
 E_{{\cal N}/2+1} - E_{{\cal N}/2}$ for ${\cal N}$ even
and
$\Delta E^{({\cal N}+1)}= 0$ for ${\cal N}$ odd
as in the CI+SDRMT model.  $\Delta E_{\cal N}$ can still be approximated by an interaction matrix element [see Eq. (\ref{delta-E})]. However, for even ${\cal
N}$,  $\Delta E_{\cal N} =
v_{{\cal N}/2 +1,{\cal N}/2\;;\; {\cal N}/2 +1,{\cal N}/2}$,  while for odd ${\cal N}$, $\Delta E_{\cal N}  =  \int d\bbox{r} d
\bbox{r'} |\psi_{({\cal N}+1)/2}(\bbox{r})|^2  v(\bbox{r}, \bbox{r'})
|\psi_{({\cal N}+1)/2}(\bbox{r'})|^2$, since the ${\cal N}+1$ electron
(with spin down) is now added to the same spatial  orbital
occupied by a spin-up electron. Using the effective RPA interaction (\ref{effective-int}),  we obtain a
decomposition of $\Delta  E_{\cal N}$ into
three parts, as in Eq. (\ref{delta-E-012}). While the charging
energy term and $\Delta E_{\cal N}^{(1)}$ are similar in
magnitude for  both odd and even ${\cal N}$, the term $\Delta
E_{\cal N}^{(2)}$, which originates from the two-body screened
interaction, is on average larger for odd ${\cal  N}$ than it is
for even ${\cal N}$, since in the former case the spin-up and
spin-down electrons occupy the same spatial wave function. We find (Mirlin, 1997)
\begin{eqnarray}\label{xi}
 2 \tilde \xi & \equiv & \overline{\Delta E_{\cal N}^{(2)}}\left.\right|_{{\rm odd}\;{\cal N}} - \overline{\Delta  E_{\cal N}^{(2)}\left.\right|}_{{\rm even}\;{\cal N}} \nonumber \\ & \approx & {2
\over  \beta {\cal A}^2} \int d \bbox{r} d\bbox{r'} f^2_d(|\bbox{r} -
\bbox{r'}|)  v_\kappa(\bbox{r}- \bbox{r'})\;.
\end{eqnarray}
 This shift moves the average peak spacing for odd ${\cal N}$ towards the
average  spacing for even ${\cal N}$ and is expected to reduce the degree of
bimodality.  In the RPA regime, the estimate for this shift, $\sim
\Delta  (\kappa/k_F) \ln(k_F/\kappa)$, is small compared with $\Delta$, since  the screening length is much larger than $\lambda_F$
($\kappa/k_F  \sim r_s \ll 1$). Thus, in the weak-interaction limit, the
odd-even  structure is expected to persist. However, as $r_s$ increases, the shift increases towards $\Delta$ and the bimodality is expected to be lost. We remark that the considerations leading to Eq. (\ref{xi}) are based on $S=0$ and $S=1/2$ ground states of the dot with even and odd ${\cal N}$, respectively. For stronger exchange interactions, we expect a spin distribution in the dot (see end of this section) that will further affect the spacings distribution.

Can signatures of spin pairing be observed in the peak-spacing distribution?  In a recent experiment, L\"{u}scher~{\em et~al.} (2000) used a GaAs quantum dot with a relatively high density,  $n_s =5.9 \times 10^{15}$ m$^{-2}$, corresponding to $r_s=0.72$, a value smaller than in previous experiments. While the measured peak-spacing distribution does not have a bimodal structure, it is found to be {\em asymmetric}.  It is interpreted as a superposition of two components that evolve from the noninteracting formula (\ref{bimodal-dist}): (i) The $\delta$ function in Eq. (\ref{bimodal-dist}) is shifted by an amount $\tilde\xi/\Delta$ [see Eq. (\ref{xi})] and broadened to a Gaussian of width $\sigma$ (due to fluctuations of the interaction matrix elements); (ii) The Wigner-Dyson distribution in Eq. (\ref{bimodal-dist}) is shifted to have an average of $2-\tilde\xi/\Delta$ (since the spacing from the upper level of a spin pair to the lower level  of the next spin pair is reduced from $2\Delta$ to $2\Delta - \tilde \xi$), and is convoluted with the Gaussian of width $\sigma$. Good fits are obtained with two fit parameters $\tilde \xi$ and $\sigma$. Of the two, $\tilde \xi$ is found to be smaller in the presence of magnetic field, in qualitative agreement with theory [see, for example, Eq. (\ref{xi})].

To study spin effects at intermediate and strong values of $r_s$,    Berkovits (1998) extended his exact diagonalization
calculations  of the Anderson model with interactions to include the spin
degrees  of freedom.  Denoting by $a_{i s}^\dagger$ and $a_{i s}$ the creation
and  annihilation operators of an electron with spin  $s=\pm 1/2$ at lattice
site  $i$, he included in the interaction part $H_2$ of the Hamiltonian an on-site interaction
between  electrons with opposite spins (in
addition  to the long-range Coulomb interaction among electrons at different sites):
\begin{eqnarray}\label{H-2-spin}
H_2 = &&{1\over 2} U_c  \sum\limits_{i \neq j \atop s,s'} {1 \over |\bbox{r}_i -\bbox{r}_j|/a}  a^\dagger_{i s} a_{i s} a^\dagger_{j s'} a_{j s'} \nonumber \\ 
&& + {1\over 2} U' \sum_{i ,s} a^\dagger_{i s} a_{i s} a^\dagger_{i -s} a_{i -s}
\;.
\end{eqnarray}
The on-site interaction  $U'=10 U_c/3$ was chosen to agree with Hubbard's
estimate  based on hydrogenlike orbitals (Hubbard, 1963).
 Calculations were done for up to ${\cal N}=9$ electrons on a $3\times 4$
lattice.  The standard deviation of $\Delta_2$ is shown in  Fig. \ref{fig:spacings-int}(a) as a function of $U_c$ and for both an odd and an even number of electrons.
 The odd-even asymmetry of $\sigma(\Delta_2)$, expected in the noninteracting limit,
 disappears above $U_c \sim 0.6$ ($r_s \sim 0.3$).
The peak-spacing distributions are
shown  in Fig. \ref{fig:spacings-int}(b) for several 
\begin{figure}
\epsfxsize= 8 cm
\centerline{\epsffile{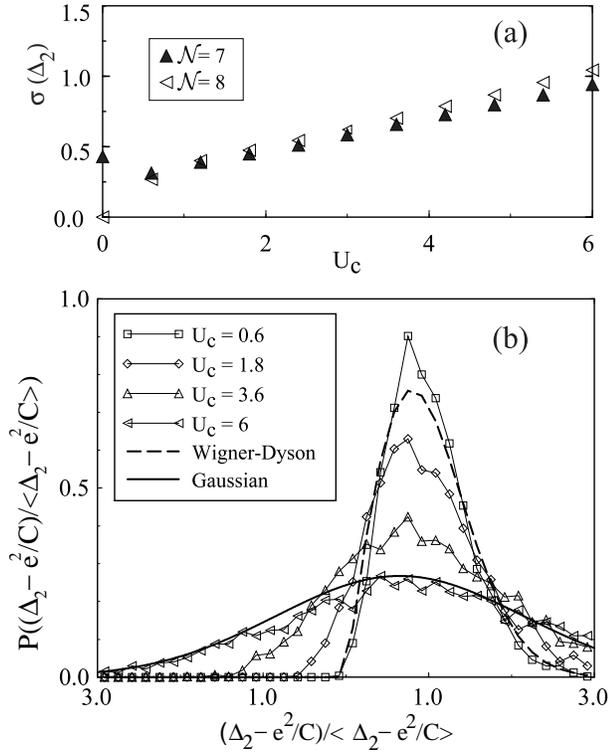}}
\vspace*{2 mm}
\caption{Peak-spacing statistics in the Anderson model with $w/t=3$ and Coulomb interactions [Eq. (\protect\ref{H-2-spin})]:
(a) the standard deviation $\sigma(\Delta_2)$ (in units of $\Delta$) as a function of the Coulomb interaction strength $U_c$ for ${\cal  N} =7$ and $8$ electrons in the dot;
(b) the distributions of $(\Delta_2-e^2/C)/\langle\Delta_2-e^2/C \rangle$
for  several values of $U_c$ (note that $\langle\Delta_2-e^2/C \rangle \approx 0.81 \Delta$). The dashed line is the Wigner-Dyson distribution
and  the solid line is a Gaussian fit to the $U_c=6$ distribution.  Notice the
absence  of a bimodal distribution already at weak interactions and the
crossover  to Gaussian shapes at stronger interactions.
From Berkovits (1998).
}
\label{fig:spacings-int}
\end{figure}
\noindent values of the  interaction parameter $U_c$. For $U_c=0$ (not shown) the distribution is
bimodal  [see Eq. (\ref{bimodal-dist})], as expected. However, already for
$U_c=0.6$  ($r_s \sim 0.3$)  this bimodal structure is lost.  The calculated
distributions  for weak interactions ($U_c \sim 0.5 -2$) are closer to the usual
spinless  constant-interaction--plus--RMT model (dashed line) rather than to the CI+SRRMT model (not
shown).  For $r_s \agt 1$, the distribution is approximately a Gaussian with a width  that continues to increase with $r_s$.

Another signature of spin and interactions is expected when a conductance peak position is followed as a function of an external parameter. Baranger, Ullmo, and Glazman (2000) suggested the appearance of kinks, i.e., abrupt changes in the parametric dependence of a Coulomb-blockade peak position.
This is illustrated in Fig.~\ref{fig:levels}.
Consider for simplicity an even number of electrons. In the absence of interactions, the single-particle levels are doubly degenerate and follow a Gaussian process as a function of the parameter $x$. Because of interactions, the top two electrons can either occupy the 
\begin{figure}
\epsfxsize= 8 cm
\centerline{\epsffile{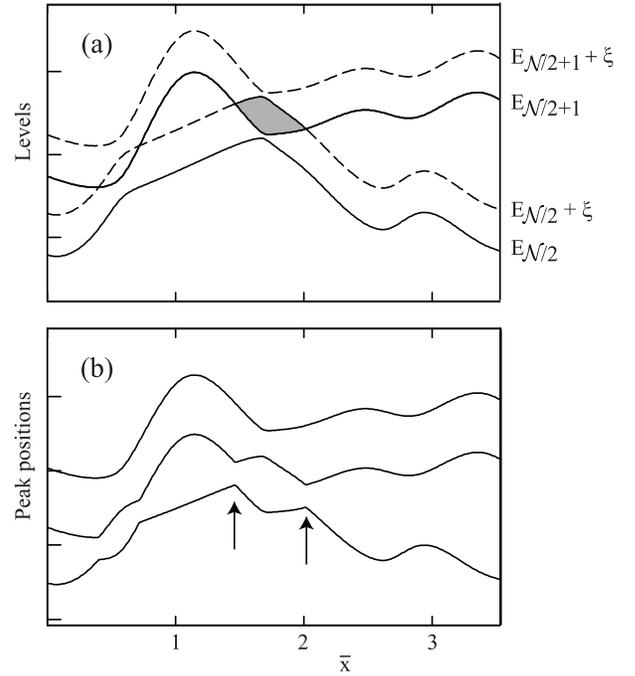}}
\vspace*{2 mm}
\caption{Spin pairing leading to kinks in the Coulomb-blockade peak positions: (a) the solid lines are two orbital levels  $E_{{\cal N}/2}$ and $E_{{\cal N}/2+1}$ from a  Gaussian process vs a scaled parameter $\bar x$ (assuming even ${\cal N}$).  When a second electron with opposite spin occupies such a level, its energy is displaced by an interaction energy $\bar\xi$. The displaced levels are shown by dashed lines. In the vicinity of the avoided crossing shown by the shaded area, the triplet state (top two electrons occupy levels  $E_{{\cal N}/2}$ and $E_{{\cal N}/2+1}$) has a lower energy than the singlet state (the electrons occupy levels $E_{{\cal N}/2}$ and $E_{{\cal N}/2}+ \bar\xi$). (b) Coulomb-blockade peak positions corresponding to the level diagram in panel (a) (traces are offset by the charging energy). Crossings of the singlet and triplet states lead to kinks in the peak position. A pair of such kinks in the vicinity of an avoided crossing is indicated by arrows.
From Baranger, Ullmo, and Glazman (2000).
}
\label{fig:levels}
\end{figure}
\noindent same spatial orbital, forming a singlet $S=0$ at the cost of an average interaction energy $\xi$, or fill two successive orbitals, forming a triplet $S=1$ at the cost of a kinetic energy $E_{{\cal N}/2+ 1}-E_{{\cal N}/2}$ [see Eq. (\ref{CI-spin})]. Suppose that $S=0$; we can imagine the top two electrons filling paired levels $E_{{\cal N}/2}$ and $E_{{\cal N}/2}+ \xi$. As we change the parameter $x$, the next single-particle level $E_{{\cal N}/2+1}$ can intersect $E_{{\cal N}/2}+ \xi$, in which case the triplet $S=1$ becomes the lowest state and the top two electrons will occupy the levels $E_{{\cal N}/2}$ and $E_{{\cal N}/2+1}$ [see, for example, the shaded region in Fig. \ref{fig:levels}(a)]. This configuration change causes a kink in the peak position [see Fig. \ref{fig:levels}(b)]. Kinks are more likely to occur near an avoided crossing of the levels and therefore appear in pairs versus the parameter $x$. We note that a kink occurs in the limit when only the ground-state level participates in the conductance. However, around a kink both the singlet and the triplet levels contribute to the conductance, so that the kinks in Fig. \ref{fig:levels}(b) are expected to become smoother. Recently L\"{u}scher~{\em et~al.} (2000) observed apparent signatures of kinks versus a magnetic field $B$. Spin-paired levels are identified by the correlation of their magnetoconductance traces. These correlations are interrupted at certain values of $B$ and an apparent kink in the peak position is observed, presumably due to a rearrangement of the ground-state spin. Baranger, Ullmo, and Glazman (2000) showed that the average density of the kinks and the distribution of their separation are sensitive to the breaking of time-reversal symmetry by a magnetic field.

  Another important question is the ground-state spin distribution in a disordered or chaotic dot.  The exchange interaction favors a spin-polarized  state, while the kinetic energy is minimized when the single-particle orbits are occupied pairwise, leading to an unpolarized $S=0$ state (for an even number of electrons). In a clean metal, the (unscreened) short-range part of the Coulomb interaction leads to a ferromagnetic (spin-polarized) instability  when $\xi/2 \agt \Delta$. This is known as the Stoner instability. Andreev and Kamenev (1998) studied the Stoner instability in a dot or metallic nanoparticle in the presence of disorder. They took into account only the disorder-averaged diagonal matrix elements of the interaction $\bar v_{\alpha \beta;\alpha \beta}$ and $\bar v_{\alpha \beta ;\beta \alpha}$ and found that the Stoner instability can develop even though the clean system is still paramagnetic. This can be traced to the effective enhancement of the interaction by the diffusive dynamics where the electrons spend more time together (Altshuler and Aronov, 1985).  Brouwer, Oreg, and Halperin (1999) considered a one-body RMT Hamiltonian with on-site Hubbard interaction and computed the spin distribution of the ground state in the mean-field approximation. They concluded that the probability of a nonzero spin state can be appreciable even for interaction strengths below the Stoner instability.
Jacquod and Stone (2000) pointed out that fluctuations of the off-diagonal interaction matrix elements (in the disorder basis) favor minimal spin for the ground state and therefore compete with exchange effects that favor large spin at strong interactions. The fluctuations of the interaction matrix elements determine the bandwidth of the many-body density of states and are largest for minimal spin. This effect is demonstrated using a random interaction matrix model (see Sec. \ref{random-interaction}). A similar model for nuclei also showed high probability of a zero-spin ground state in even-even nuclei (Johnson, Bertsch, and Dean, 1998)

Berkovits (1998) studied the ground-state spin in exact simulations  of the small Anderson model plus interactions [Eq. (\ref{H-2-spin})]. For $r_s < 1$ there is a finite probability for $S=1$ states for an even number of electrons but almost no $S=3/2$ states for an odd number of electrons. Indeed,  a spin flip costs a kinetic energy of $E_{{\cal N}/2+1} - E_{{\cal N}/2}$ for even ${\cal N}$ but $E_{({\cal N}+3)/2} - E_{({\cal N} -1)/2}$ for odd ${\cal N}$, and it is much less likely to find two consecutive small single-particle level spacings than one small spacing.  For $r_s >1$, higher spin values can occur.

Experimentally, the spin of a quantum dot is difficult to measure. A promising technique is conductance measurements in the presence of an in-plane magnetic field that leads to a Zeeman splitting of the ground state. For a theoretical study of the (in-plane) magnetic field dependence of the conductance peak position and height, see Kurland, Berkovits, and Altshuler (2000).  Nonzero spin has been observed in metallic nanoparticles (Ralph, Black, and Tinkham, 1997; Davidovi\'{c} and Tinkham, 1999) and in some carbon nanotube ropes (Cobden~{\em et~al.}, 1998).

\subsection{Peak-height statistics and interactions}
\label{peaks:interactions}

How do interactions affect the distributions of the conductance
peak heights and  their parametric correlations? Experimentally,  the predictions
of RMT were confirmed (Chang~{\em et~al.}, 1996; Folk~{\em et~al.}, 1996) for the
distributions and for the parametric correlations. However, a
semiclassical estimate of the correlation  flux gives $\Phi_c
\sim 0.3 \Phi_0$ (assuming a geometric factor of  order unity)
-- significantly lower than the experimental value of $\Phi \sim
0.8 \Phi_0$ (see Sec. \ref{conductance-correlator}).  Since
the geometrical factor is unknown, we cannot
completely rule out a single-particle theory, but the
discrepancy suggests that the interactions affect the
correlation flux. Here $\Phi_c$ is a nonuniversal  parameter that
depends on the details of the system's dynamics and  cannot be
calculated in a pure RMT.

Berkovits and Sivan (1998) used the spinless Anderson model with Coulomb interactions
 to study numerically the peak-height statistics and sensitivity
 to a magnetic flux. A $4\times 6$ lattice with disorder $w=3 t$ was
used for ${\cal N}=3$  and ${\cal N}=4$ electrons.

 At low temperatures, the conductance peak height is still given by Eq. (\ref{peak-height}),
but $\Gamma^{l(r)}$ are partial widths of the {\em many-particle} ground state of the dot.   $R$-matrix theory (Sec. \ref{R-matrix-formalism}) is also valid for a many-body system
(Lane and Thomas, 1958). The partial-width amplitude is expressed as in Eq. 
(\ref{partial-amplitude})  but with $\psi_\lambda$ replaced by the {\em
interacting}  many-body eigenfunction $\Psi_{\cal N}$ of ${\cal N}$ electrons
in  the dot, and the channel wave function $\Phi$  describing the transverse
wave function  of  one electron in the lead and the interacting wave function $\Psi_{{\cal N}-1}$ of
${\cal  N}-1$ electrons in the dot.  In second-quantized notation
(Berkovits and Sivan, 1998)
\begin{equation}\label{peak-int}
\Gamma^{l(r)} \propto |\sum\limits_{m \in \;l(r)} \langle \Psi_{{\cal N}+1} |
a^\dagger_{m}  | \Psi_{\cal N}\rangle |^2 \;,
\end{equation}
where $a^\dagger_m$ is the creation operator of the electron injected into
the  dot at the site $m$ belonging to the respective lead.

 The width average and variance are found to decrease with $r_s$. This suppression is explained by an interaction-induced
 short-range order.
The distributions of the dimensionless peak heights are shown in Figs.
\ref{fig:Pcorr-int}(a) and \ref{fig:Pcorr-int}(b) for several values of the interaction strength $\;\;U_c$.
\begin{figure}
\epsfxsize= 7.5 cm
\centerline{\epsffile{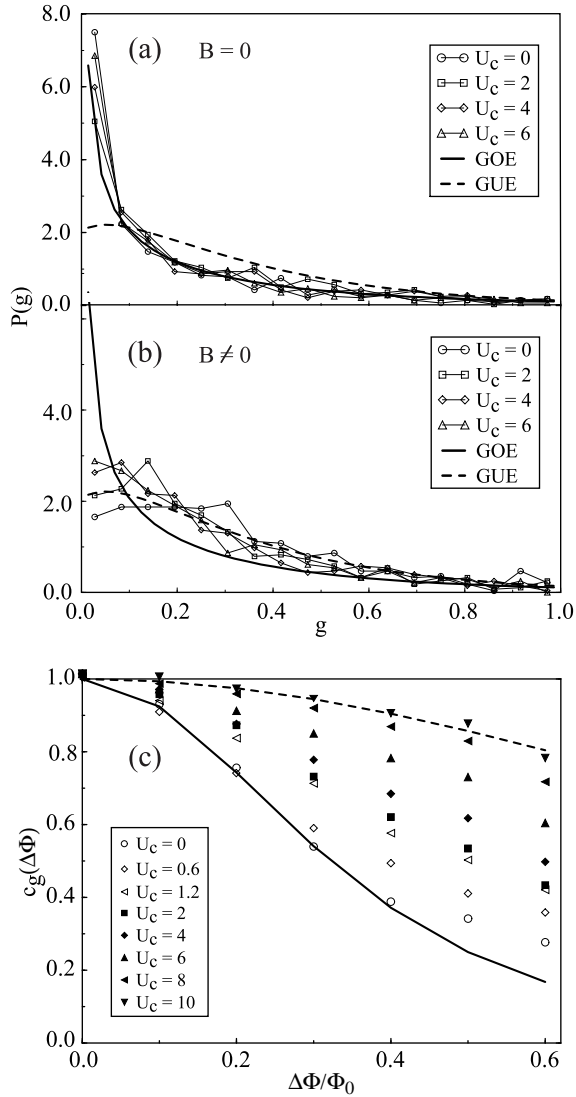}}
\vspace*{2 mm}
\caption{
Interaction effects on the conductance peak height statistics and parametric correlations: Panels (a) and (b), conductance peak-height distributions in the presence of
Coulomb  interactions.  The distributions $P(g)$ [where $g$ is defined in Eq. (\protect\ref{g-peak})]
are  calculated from simulations of a 2D Anderson model with
Coulomb  interactions. Results are shown for
several  values of the interaction parameter $U_c=0,2,4,6$ (a) without and  (b) with a magnetic field $B$;
solid lines, GOE prediction [Eq. (\protect\ref{PgGOE})]; dashed lines, GUE prediction 
[Eq. (\protect\ref{PgGUE})].  The distributions show only weak
sensitivity  to the strength of the interaction and are in agreement with the RMT  predictions.
(c) The autocorrelation function $c_g(\Delta
\Phi)$  of the conductance peak height vs magnetic flux $\Delta \Phi$ for several values of $U_c$.  The correlators at $U_c=0$ and
$U_c=10$  can be fitted to the RMT form
[Eq. (\protect\ref{B-correlation-closed})],  but with $\Phi_c=0.5 \;\Phi_0$ (solid
line)  and $\Phi_c=1.75\; \Phi_0$ (dashed line), respectively. Note the strong
sensitivity  of the correlation flux to the interaction strength.
 From Berkovits and Sivan (1998).
}
\label{fig:Pcorr-int}
\end{figure}
\noindent They  are rather insensitive to the two-body interaction, confirming the
predictions of RMT for both conserved ($\Phi=0$) and broken ($\Phi=0.4 \,
\Phi_0$)  time-reversal symmetry. At large values of $r_s$,  some enhancement
of  the small conductance probability is observed in the case of $\Phi \neq 0$, in agreement with the experimental results of Chang~{\em et~al.} (1996).  Similar conclusions are reached when spin is included in the model as in Sec. \ref{spin-effects} (Berkovits, 1999). 

 The parametric correlator  of the peak height $G$ as a function of
magnetic  flux, $c_g(\Delta \Phi)\equiv { \overline{\delta G(\Phi) \delta G(\Phi+\Delta \Phi)} / \left\{\overline{[\delta
G(\Phi)]^2}\;\;  \overline{[\delta G(\Phi+\Delta \Phi)]^2}\right\}^{1/2}}$, where $\delta G = G - \bar G$, is shown  in  Fig.
\ref{fig:Pcorr-int}(c) for different values of $U_c$.  At $r_s=0$ and $r_s \agt 1.4$, the RMT functional form, Eq. 
(\ref{approx-overlap-corr}),  of the correlator is reproduced. However, the
value  of the correlation flux is increased from its noninteracting value
$\Phi_c=0.5 \, \Phi_0$ to $\Phi_c=1.75\, \Phi_0$ at $r_s \sim 3.5$. Thus, in the
regime  relevant to semiconductor quantum dots, the correlation field is a factor of $2-3$  larger than its noninteracting value, in general agreement with the experiment.

\subsection{Random interaction matrix model}
\label{random-interaction}

 An important issue is whether RMT can be used to describe interacting systems.  RMT was originally developed to describe the
statistics of a strongly interacting system -- the compound nucleus -- at high excitations.
Calculations in interacting systems in nuclear, atomic, and condensed-matter  physics suggest that the Wigner-Dyson statistics are generic to complex
 many-body systems at sufficiently large excitations
(Montambaux~{\em et~al.}, 1993; Flambaum~{\em et~al.}, 1994; Zelevinsky~{\em et~al.}, 1996).
 In quantum dots, however, our interest is in the statistical behavior at
or  near the system's ground state as the number of electrons varies.  Standard RMT makes no explicit reference to interactions or number of particles.  Generic interaction effects on the statistics can be studied in a random-matrix model that contains interactions explicitly. A two-body random-interaction matrix model, introduced in nuclear physics by French and Wong (1970) and Bohigas and Flores (1971), was used together with a random single-particle spectrum to study thermalization (Flambaum, Gribakin, and Izrailev, 1996) and the transition from Poisson to Wigner-Dyson statistics (Jacquod and Shepelyansky, 1997) in many-body systems. However, the Poissonian single-particle statistics of this model are not suitable for studying dots whose single-particle dynamics is chaotic.  Alhassid, Jacquod, and Wobst (2000) introduced a random interaction matrix model (RIMM) to study generic fluctuations in chaotic dots with interactions.
The RIMM is an ensemble of interacting Hamiltonians, 
\begin{eqnarray}\label{RIMM}
H = \sum\limits_{ij} h_{ij} a^\dagger_i a_j +{1\over 4} \sum_{ijkl}
 u^A_{ijkl}a^\dagger_i a^\dagger_j a_l a_k
\;,
\end{eqnarray}
where the one-body matrix elements $h_{ij}$ are chosen from the appropriate
 Gaussian random-matrix ensemble, and the
 antisymmetrized two-body matrix elements $u^A_{ij;kl}  \equiv u_{ij;kl} - u_{ij;lk}$
form a GOE in the two-particle space \begin{eqnarray}\label{random-ensemble}
P(h) \propto e^{-{\beta\over 2a^2} {\rm Tr}\; h^2} \;;\;\;\; P(u^A)
\propto   e^{- {\rm Tr}\;(u^A)^2/2U^2}
\;.
\end{eqnarray}
 The variance of the diagonal (off-diagonal) interaction matrix elements is $U^2$ ($U^2/2$). The states
$|i\rangle  = a_i^\dagger |0\rangle$ describe a fixed basis of $m$
single-particle states. Here $h$ is an $m\times m$ GOE (GUE)
matrix when the single-particle dynamics conserve (break)
time-reversal symmetry, while the
two-body interaction preserves time-reversal
symmetry and forms a GOE, irrespective of the symmetry of the
 one-body Hamiltonian. In general, the two-body interaction can include a nonvanishing average part $\bar u$ that is invariant under orthogonal transformations of the single-particle basis. For spinless electrons, the only such invariant is the charging energy $e^2 {\cal N}^2/2C$, which is a constant and does not affect the statistical fluctuations of (\ref{RIMM}). In the presence of spin, an additional contribution to $\bar u$ is an exchange interaction $-\xi \bbox S^2/2$. We remark that in a physical model of a dot, the Coulomb interaction matrix elements are given in a fixed basis. Fluctuations of the interaction matrix elements were introduced in the RIMM to obtain a generic model that is independent of a particular interaction, in the original spirit of RMT (French and Wong, 1970; Bohigas and Flores, 1971).

 The model was used to study both the peak-spacing (Alhassid, Jacquod, and Wobst, 2000) and peak-height statistics (Alhassid and Wobst, 2000). The peak-spacing distribution describes a crossover from a Wigner-Dyson distribution to a Gaussian-like distribution as $U/\Delta$ increases [see Fig. \ref{fig:RIMM}(a)]. The partial level width is calculated from an expression analogous to Eq. (\ref{peak-int}), and for a GOE one-body Hamiltonian, its distribution $P(\hat\Gamma)$ is found to be a GOE Porter-Thomas distribution independent of $U/\Delta$. However, for a GUE one-body Hamiltonian, the width distribution makes a crossover from GUE to a GOE Porter-Thomas distribution as a function of $U/\Delta$ [see Fig. \ref{fig:RIMM}(b)]. The crossover distributions are well described by $P_\zeta(\Gamma)$ of Eq. (\ref{intensity-dist}), where $\zeta$ is a monotonically decreasing function of $U/\Delta$.  This is due to the competition between the asymptotic GUE symmetry of the one-body Hamiltonian $h$ and the GOE symmetry of the two-body interaction $u$.  In the range $U/\Delta \sim 0.7 - 1.5$, the peak-spacing distribution is already Gaussian-like, while the width statistics are still close to the GUE limit. In the RIMM, $U/\Delta$ is a free parameter, and reasonable values can be determined by comparing its results against physical models. Such a comparison was made to a small ($4\times 5$) Anderson model with Coulomb interactions and periodic boundary conditions. Apart from finite-size effects, similar behavior was observed where the range $U/\Delta \sim 0.7 -1.5$ in the RIMM corresponded to $U_c/t \sim 2 - 5$ in the Coulomb model.

\begin{figure}
\epsfxsize= 8 cm
\centerline{\epsffile{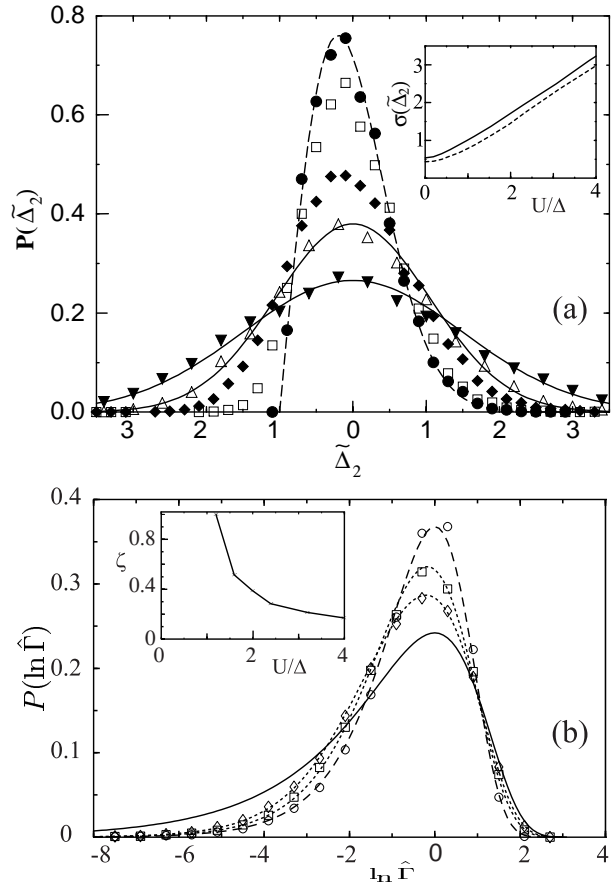}}
\vspace*{2 mm}
\caption{Peak-spacing and peak-height statistics in the random interaction matrix model [RIMM; Eq.(\protect\ref{RIMM})]. (a) Peak-spacing distributions $P(\tilde \Delta_2)$ in the RIMM with GOE one-body statistics for $m=12$, ${\cal N}=4$: solid circles, 
$U/\Delta=0$; open squares, $U/\Delta=0.35$; solid diamonds, $U/\Delta= 0.7$; open triangles, 
$U/\Delta=1.1$; solid triangles, $U/\Delta=1.8$. Notice the crossover from a Wigner-Dyson distribution at $U=0$ (dashed line) to Gaussian-like distributions (solid lines)  for $U/\Delta \protect\agt 1$. Inset to panel (a), standard deviation $\sigma(\tilde\Delta_2)$ of the peak spacings vs $U/\Delta$ for GOE (solid) and GUE (dashed) one-body statistics. (b) Width distributions in the RIMM with GUE one-body statistics.
  Distributions $P(\ln \hat \Gamma)$ vs $\ln \hat \Gamma$  are shown for
$m=12$, ${\cal N}=4$: open circles, $U/\Delta=0$;
 open squares, $U/\Delta=2.4$; open diamonds, $U/\Delta=4$; solid line, GOE Porter-Thomas distribution; dashed line, GUE Porter-Thomas distribution; short-dashed lines;  analytic width distributions  $P_\zeta(\ln \hat\Gamma)$ in the crossover between the GUE and GOE (see Section \protect\ref{crossover-distributions:closed}). Inset to panel (b), crossover parameter $\zeta$ vs $U/\Delta$.
Adapted from Alhassid, Jacquod and Wobst (2000) and Alhassid and Wobst (2000).}
\label{fig:RIMM}
\end{figure}

\section{Charging Energy Effects in Quantum Dots}
\label{charging-energy-effects}

  In this section we discuss other mesoscopic phenomena in quantum dots
where the charging energy plays an important role: fluctuations
of the off-resonance conductance in closed (or semiopen) dots and mesoscopic
Coulomb blockade in dots that are strongly coupled to a
single-channel lead.

  The dominant mechanism for the off-resonance conductance is cotunneling (see Sec.
\ref{cotunneling}).  In particular, for $k T < \sqrt{E_C
\Delta}$,  the dominant process is elastic cotunneling, describing
 the virtual tunneling of an electron or hole through a large number $\sim
E_C/\Delta$  of excited levels in the dot. The coherent superposition of a
large  number of weak amplitudes exhibits fluctuation properties that are
qualitatively  different from those of the conductance peaks. Since a virtual
transition  occurs across a gap of $\sim E_C/2$, we expect the charging energy
to  play an important role in determining the statistics of the minima. The
mesoscopic  fluctuations in elastic cotunneling were derived by
Aleiner and Glazman (1996) and observed by Cronenwett~{\em et~al.} (1997). They are
discussed  in Sec. \ref{mesoscopic:cotunneling}.

In the crossover from closed to open dots, the {\em classical}
Coulomb-blockade  oscillations (observed in the limit $\Delta \to 0$)
gradually  weaken.  Matveev (1995) showed that these oscillations
completely  disappear for a fully transmitting one-channel lead ($T_c=1$).
However, Aleiner and Glazman (1998) showed that for a dot with finite $\Delta$,
{\em  quantum} Coulomb blockade is not fully destroyed and the conductance
exhibits  mesoscopic fluctuations that are periodic in the gate voltage but
have  a random phase. The signatures of this mesoscopic Coulomb blockade can
be  seen in the correlation functions of various fluctuating observables. The
main  mechanism for these fluctuations is the backscattering of electrons from
the  boundaries of the dot into the strongly coupled channel.  An
experiment  by Cronenwett, {\em et~al.} (1998) confirms signatures of
mesoscopic  Coulomb blockade in dots with one fully open lead
$T^l  \sim 1$ and one weakly coupled lead $T^r \ll 1$. Among the striking
effects  seen is the strong suppression of Coulomb-blockade oscillations
at  finite magnetic fields, contrary to the behavior in closed dots.
Mesoscopic  Coulomb blockade is discussed in Sec.
\ref{mesoscopic-Coulomb-blockade}.

 In a partially open dot, the charge is not quantized and exhibits
mesoscopic  fluctuations, as does the differential capacitance  $d
Q/d  V_g$. The capacitance fluctuations in an open dot were derived by
Gopar, Mello, and B\"{u}ttiker (1996)  in the limit of noninteracting electrons using the
distribution  of the scattering time delays in the dot.  However, the charging energy should be taken into account in dots with a
partially open single channel.  The statistical properties of the capacitance fluctuations in the presence of charging energy were calculated by Kaminski, Aleiner, and Glazman (1998), and are discussed in
Sec.  \ref{capacitance-fluctuations}.

\subsection{Mesoscopic fluctuations in elastic cotunneling}
\label{mesoscopic:cotunneling}

 The mesoscopic fluctuations of the conductance minima were derived by Aleiner and Glazman (1996) in the diagrammatic approach. According  to Eq. (\ref{cotunneling-amplitude}), the off-resonance  conductance amplitude is determined by a large number of fluctuating terms, of which $\sim E_e/\Delta$ and  $\sim E_h/\Delta$ contribute significantly to the sum over particles and over holes, respectively. When $E_e$ and $E_h$ are below the Thouless energy, the contributing levels are in the universal regime where the partial widths $\gamma_{c\lambda}$ and energies $E_\lambda$ have RMT statistics.  In this regime, we can apply an RMT approach for both the on- and off-resonance  conductance.
In both cases the mesoscopic fluctuations are determined by the
same underlying statistics of the partial widths of the
resonance levels.  The difference between the statistics of the conductance
maxima and minima originates from their different transport mechanisms.

We first discuss the distribution of the conductance minima in the crossover between conserved and broken time-reversal symmetry. The
cotunneling  amplitude ${\cal T}$ in Eq. 
(\ref{cotunneling-amplitude}) is the sum  of  a large number of
terms, and we expect the central limit theorem to apply,
leading to Gaussian distributions for both the real and the 
imaginary parts  of ${\cal T}$.
  The distribution of  $G \propto  ({\rm Re} \; {\cal T})^2 +  ({\rm Im}\;
{\cal  T})^2$ (measured in units of $\bar{G}$) is then given by
\begin{eqnarray}\label{cotunneling-distribution}
P(\hat{G}) = (1-\chi)^{-1} e^{-{\hat{G}  \over  1 - \chi}}
 I_0\left(\frac{\sqrt{\chi}}{1-\chi} \hat{G} \right)
\;,
\end{eqnarray}
where $I_0$ is a Bessel function 
and  $\chi \equiv \left\{ {\overline{({\rm Re} \;  {\cal T} )^2}
/\overline{|{\cal  T}|^2}} -
{ \overline{({\rm Im} \;{\cal T})^2} / \overline{|{\cal T}|^2}} \right\}^2$.
 An expression of the form  (\ref{cotunneling-distribution}) was obtained by
Aleiner and Glazman (1996)  when the time-reversal symmetry was broken by a magnetic
field  $B$. They found $\chi = \Lambda(B/B^{\rm valley}_c)$, where $B^{\rm
valley}_c$  is a correlation (or crossover) field for the conductance ``valleys'' [see Eq.
(\ref{crossover-cotunneling})  below] and
\begin{eqnarray}\label{crossover-function}
\Lambda(x) \equiv && {1 \over \pi^2 x^4} \left[ \ln x^2 \ln(1+x^4) + \pi \arctan
x^2  \right. \nonumber \\ 
&& + \left. {\rm Li}_2(-x^4)/2 \right]^2 \;.
\end{eqnarray}
The function ${\rm Li}_2$ in Eq. (\ref{crossover-function}) is the second
polylogarithm  function.

 The average cotunneling conductance is found to be
\begin{eqnarray}\label{average-G:cotunneling}
 \bar{G} = \frac{e^2}{h}  \left( {\overline{\Gamma^l} \over \Delta} \; {
\overline{\Gamma^r}   \over \Delta} \right)  \Delta \left( \frac{1}{E_h} +
\frac{1}{E_e}  \right)\;,
\end{eqnarray}
where $E_e$ ($E_h$) is the distance between the Fermi energy and the closest state available for electron (hole) tunneling (see Sec. \ref{cotunneling}). 
 Thus, in striking contrast to the average conductance peak, the  average
off-resonance  conductance is {\em independent} of  magnetic field and does
not  exhibit a weak-localization effect.  This has been confirmed
experimentally  (Cronenwett~{\em et~al.}, 1997).

 The correlation field $B^{\rm valley}_c$ for the conductance valleys (which is of the same order as the crossover field)  can be
estimated  semiclassically as for the conductance peaks, except
that the Heisenberg time $\tau_H= h/\Delta$ is now replaced by the virtual
tunneling  time  $h/E$, where $E={\rm min}(E_e,E_h)$.  
 In analogy with Eq. (\ref{B-correlation-closed}) we  find
\begin{eqnarray}\label{crossover-cotunneling}
{ B_c^{\rm valley} {\cal A}/  \Phi_0} = \kappa \sqrt{E / E_T}  \;,
\end{eqnarray}
where $E_T \sim \hbar v_F/\sqrt{\cal A}$ is the ballistic Thouless energy, and
$\kappa$  is a geometric factor of the dot. The valley correlation field
 is seen to be larger than the peak correlation field
\begin{eqnarray}\label{Phi-ratio}
{B_c^{\rm valley} / B_c^{\rm peak}} = \sqrt{E / \Delta} \;.
\end{eqnarray}

We next turn to the parametric correlations of the cotunneling
conductance versus magnetic field (Aleiner and Glazman, 1996). Assuming a
point-contact  model, the cotunneling  amplitude [Eq. 
(\ref{cotunneling-amplitude})] can be rewritten as
\begin{eqnarray}\label{Green-cotunneling}
{\cal T} = &&  {\sqrt{\bar \Gamma^l \bar\Gamma^r }\over \Delta}{ 1\over \nu}
\int  {d\omega \over 2\pi i} \left[ G^A(\bbox r_l,\bbox r_r, \omega)\right. \nonumber \\ && - \left. G^R(\bbox
r_l,\bbox  r_r, \omega)\right] G^{\rm ret}(\omega) \;.
\end{eqnarray}
The quantity $\nu$ is the average density of states in the dot per unit area,  $G^R$ and $G^A$ are retarded and advanced Green's  functions of the non-interacting dot and $G^{\rm ret}(\omega) = - (|\omega| + E_e)^{-1} + (|\omega| + E_h)^{-1}$ is the retarded cotunneling Green's function of a dot with interactions (Baltin and Gefen, 2000).
 The calculation of the parametric
correlator  requires the ensemble averages of the corresponding products of
Green's  functions.  In the metallic regime and for
$E \gg \Delta$,  the latter can be calculated in the diagrammatic approach and expressed in terms of the diffuson ${\cal D}$ and
the  cooperon ${\cal C}$:
$\overline{G^R_B(\bbox r, \bbox r', \omega) G^A_{B'}(\bbox r', \bbox r,
\omega')}  = 2\pi \nu {\cal D}_{B,B'}(\bbox r,\bbox r',\omega-\omega')$ and
$\overline{G^R_B(\bbox r, \bbox r', \omega) G^A_{B'}(\bbox r, \bbox r',
\omega')}  = 2\pi \nu {\cal C}_{B,B'}(\bbox r,\bbox r',\omega-\omega')$.
If  $E < E_T$, a typical $\Delta \omega =\omega-\omega'$ that contributes to the correlator is
below  $E_T$ and one can use the zero-mode approximation:
\begin{eqnarray}\label{universal-correlators}
{\cal D}_{B,B'} & = &{1 \over {\cal A}} \left[ -i\Delta\omega + \left({\cal A} {B -B'\over
\Phi_0}\right)^2  E_T \right]^{-1}\;; \nonumber \\
{\cal C}_{B,B'} & = & {1 \over {\cal A}} \left[-i\Delta\omega + \left({\cal A} {B +B'\over
\Phi_0}\right)^2  E_T \right]^{-1} \;.
\end{eqnarray}
Equations (\ref{universal-correlators}) lead to a universal parametric correlator for the off-resonance conductance
\begin{eqnarray}\label{magnetic-correlator}
c_G(\Delta B) & = & {\overline{\delta G(B) \delta G(B')}/\bar G^2} \nonumber \\
& = & \Lambda\left({\Delta B \over B^{\rm valley}_c}\right) + \Lambda\left({2B +
\Delta  B \over B^{\rm valley}_c}\right)
\;,
\end{eqnarray}
where $\Lambda(x)$ is the scaling function (\ref{crossover-function})  and the correlation field $B^{\rm valley}_c$ is given by Eq. 
(\ref{crossover-cotunneling}), with $\kappa=1$. The GUE correlator is
obtained  from Eq. (\ref{magnetic-correlator}) in the limit $B \gg B^{\rm
valley}_c$,  where $c_G(\Delta B) = \Lambda(\Delta B/B^{\rm valley}_c)$.

\begin{figure}
\epsfxsize= 8 cm
\centerline{\epsffile{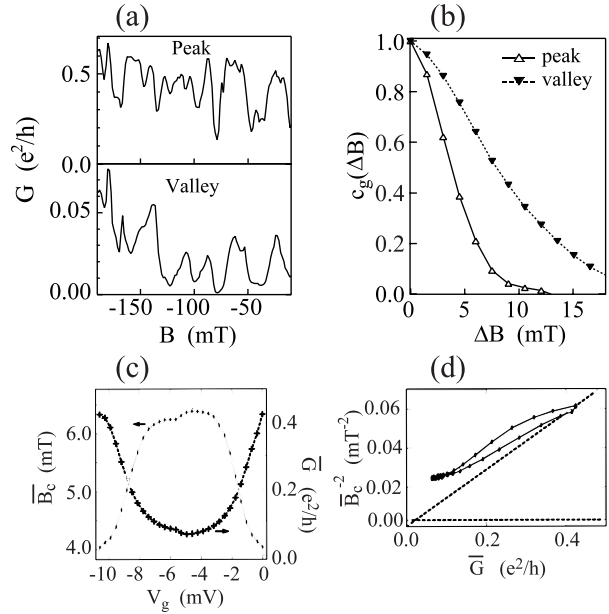}}
\vspace*{2 mm}
\caption{Elastic cotunneling in Coulomb-blockade dots: panels (a) and (b), conductance
fluctuations  of a peak vs a valley. Panel (a), a conductance
 peak and an adjacent conductance minimum (valley) as a
function  of a magnetic field $B$ (different scales are used for
$G$  at a peak or a valley). The valley fluctuates on a longer scale than the
peak. Panel (b), the measured conductance correlator vs $\Delta B$  for a peak (dashed line) and for a valley (solid line). Panels (c) and (d) describe an experimental test of
the  scaling relation (\protect\ref{scaling-relation}).  Panel (c), the  measured average correlation field $\bar B_c$ (solid line) and the
average  conductance $\bar G$ (dashed line) as the gate voltage varies between
two  neighboring peaks. The ensemble-averaged $\bar B_c$ is obtained from
$\sim  30$ statistically independent traces of the conductance.
Panel (d), $\bar B_c^{-2}$ as a function of $\bar G$ (solid line) using the same data as in (c).
The  diagonal dashed line is Eq. (\protect\ref{scaling-relation}).
From Cronenwett~{\em et~al.} (1997) and Marcus~{\em et~al.} (1997).
}
\label{fig:valleys-exp}
\end{figure}

The correlation field $B_c^{\rm valley}$ for the conductance valleys is
enhanced  by $\sqrt{E/\Delta}$ compared with the correlation field of the peaks
[see  Eq. (\ref{Phi-ratio})]. This enhancement was observed in the experimental results of
Cronenwett~{\em et~al.} (1997)  shown in Fig. \ref{fig:valleys-exp}.   In order to
measure  the weak conductance in the valleys, the dots used were  semiopen with
$\bar  \Gamma \sim 0.7 \Delta$.  Typical  fluctuations versus magnetic field
of  a conductance peak in comparison with a conductance valley are shown in Fig. \ref{fig:valleys-exp}(a). The valley
 fluctuates on a scale that is larger than the scale over which a
peak  fluctuates.  The conductance autocorrelation functions for the peaks and
the  valleys are shown in Fig. \ref{fig:valleys-exp}(b). The observed ratio $B_c^{\rm
valley}/B_c^{\rm  peak}\sim 1.6$ is smaller than the expected value of $\sim
4$.  The discrepancy is not fully understood, but we remark that (i) the peak
correlation field is larger than the single-particle estimate owing to
interaction effects (see Sec. \ref{peaks:interactions}), and (ii) the
theoretical estimate for the peak correlation field is for $T \ll \Delta$,
while  in the experiment $T \sim \Delta$. Both effects lead to enhancement of
the  peak correlation field and might explain the observed reduction in the
valley-to-peak  ratio of the correlation fields.

The enhancement of the correlation field for the off-resonance conductance
is  a charging energy effect.
According to (\ref{crossover-cotunneling}),  the largest correlation field is
obtained  at midvalley, where $E$ obtains its maximal value $E_C/2$. At this
gate  voltage the average conductance is minimal according to Eq. 
(\ref{average-G:cotunneling}).  Using the approximation $\bar{G} \propto
E^{-1}$  [see Eq. (\ref{average-G:cotunneling})], we obtain
\begin{equation}\label{scaling-relation}
B_c^2 \bar{G} \approx {\rm const.}
\end{equation}
Thus the correlation field is maximal when the average conductance is
smallest,  and vice versa.   Figure 
\ref{fig:valleys-exp}(c)  shows the measured correlation field $B_c$ (solid line),
averaged  over independent peak-valley-peak data sets, and the average
conductance  $\bar G$ (dashed line) plotted as a function of  gate voltage
 in the region between two Coulomb-blockade peaks. To test the
approximate  scaling relation (\ref{scaling-relation}), the same data are used
to  plot $\bar B^{-2}_c$ versus $\bar G$ in Fig. \ref{fig:valleys-exp}(d) (solid
lines).  The diagonal dashed line is the scaling relation
(\ref{scaling-relation}).

Baltin and Gefen (2000) have used Eq. (\ref{Green-cotunneling}) to calculate the normalized cotunneling conductance correlator between valleys ${\cal N}$ and ${\cal N} + n$. Parametrizing the fractional distance between two neighboring peaks by $y$ [so that $E_h= y E_C$ and $E_e=(1-y)E_C$], they find
\begin{eqnarray}\label{valley-correlator}
c_G(n,y) = && \left[{ y(1-y) \over \left( 1+ n \frac{\Delta}{E_C}\right) n\frac{\Delta}{E_C}}\right]^2 \left[\ln \left(1 + {n \over y}{\Delta \over E_C}\right) \right. \nonumber \\ && + \left. \ln
\left( 1 + {n \over 1- y}{\Delta \over E_C}\right) \right]^2\;.
\end{eqnarray}
The cotunneling amplitudes in two neighboring valleys contain a sum over similar contributions from intermediate states, except that one particle state becomes a hole state. The correlator (\ref{valley-correlator}) is then expected to decay slowly with $n$ on a scale set by $\sim E_C/\Delta$. The measured correlator (Cronenwett~{\em et~al.}, 1997) is found to decay faster, presumably due to scrambling of the single-particle spectrum when an electron is added to the dot.

  Another interesting issue is the phase change $\Delta \alpha$  of the transmission amplitude between two consecutive valleys. Two experiments (Yacoby~{\em et~al.}, 1995; Schuster~{\em et~al.}, 1997) have measured the phase of the transmission amplitude through a dot, employing an interferometer with two arms, one of which contains a quantum dot. A surprising result of these experiments was that $\Delta \alpha =0$ (mod $2\pi$) across all measured peaks. Since the phase changes by $\pi$ across a resonance,  an additional ``phase lapse'' of $\pi$ must occur.  This is contrary to what one expects in a dot with noninteracting electrons where the values of $\Delta \alpha$ across a series of peaks form some sequence of $0$ and $\pi$.  Various possible explanations have assumed specific  geometries that lead to preferred levels in the dot (see, for example, Baltin~{\em et~al.}, 1999). A more generic mechanism for a disordered or chaotic dot, suggested by Baltin and Gefen (1999), employs a formula that interpolates between the valley transmission amplitude [Eq. (\ref{cotunneling-amplitude})] and the Breit-Wigner peak amplitude (see also Oreg and Gefen, 1997). An approximate sign sum rule states that the number of $\pi$ changes of the phase between two neighboring valleys (each due to a resonance, a near-resonance phase lapse, or a valley phase lapse) is even. The probability to deviate from this sign sum rule is small ($\sim \Delta/E_C$). The sign sum rule relies on the strong correlations of the transmission phase in neighboring valleys, and spectral scrambling would lead to its breakdown. However, in the experiments only $\Delta \alpha=0$ was observed.

 At very low temperatures and in the strong-tunneling regime, the valley conductance that corresponds to an odd number of electrons can be enhanced due to the Kondo effect. Higher-order virtual tunneling processes that effectively flip the unpaired spin on the dot can lead to a coherent many-body resonance at the Fermi energy, known as the {\em Kondo resonance} (Glazman and Raikh, 1988; Ng and Lee, 1988; Meir, Wingreen, and Lee, 1993). It is formed between the spin of the dot and the delocalized electrons in the leads -- in analogy to the Kondo effect (Kondo, 1964),  which occurs when a magnetic impurity is placed in a metal and the unpaired electron in the impurity forms a singlet state with the metal's electrons. The Kondo effect in a quantum dot was only recently observed by Goldhaber-Gordon~{\em et~al.} (1998) and by Cronenwett, Oosterkamp, and Kouwenhoven (1998). The energy scale for observing the Kondo resonance is the Kondo temperature $T_K$, which is essentially the binding energy of the resonance. At midvalley $T_K \sim \sqrt{E_C \Gamma}e^{-\pi E_C /8\Gamma}$, and to bring $T_K$ within the range of experimentally accessible temperatures it was necessary to fabricate much smaller  dots  ($L \sim 100\;$nm). In a smaller dot, $\Delta$ is larger and $\Gamma$ can be increased substantially while the dot remains semiopen  (i.e., $\Gamma < \Delta$). The appearance of a Kondo resonance in the density of states enhances the conductance in the valley that corresponds to an odd number of electrons as the temperature decreases below $T_K$.

\subsection{Mesoscopic Coulomb blockade}
\label{mesoscopic-Coulomb-blockade}

 In Secs. \ref{open-dots} and \ref{closed-dots} we discussed 
mesoscopic  fluctuations of the conductance in two opposite limits -- open
dots and almost-closed dots. In the crossover from closed to
open  dots, the charge quantization and the Coulomb-blockade oscillations
gradually  disappear. An interesting case is a lead with one fully
transmitting  channel. Rather than discussing a dot with symmetric leads, it
is  easier to study the asymmetric limit where, e.g., the left lead has a
transmission  coefficient of $T^l \approx 1$, while  the right lead has a large
barrier  ($T^r \ll 1$).    The Hamiltonian of this system is different from
the  Hamiltonian (\ref{Hamiltonian}) used to describe cotunneling in that the
subsystem  of the dot plus the left lead is now described by a single
Hamiltonian  (because of the strong coupling): 
\begin{eqnarray}\label{dot-channel}
H = && \int d\bbox{r}  \; \psi^\dagger(\bbox{r})\left[ -\nabla^2/2m^\ast +
U(\bbox{r})  -\mu\right] \psi(\bbox{r})  +  {\hat Q^2 \over 2C}
\nonumber \\ && - \alpha V_g
\hat  Q + \sum\limits_{k, s \in r} E_{k s}
c^\dagger_{k s} c_{k s} \nonumber \\ &  & + \sum_{k,s \in r} [V_{k s}
c^\dagger_{k s} \psi(\bbox{r_c}) + H.c. ]
\;.
\end{eqnarray}
The first term on the right describes the dot-channel Hamiltonian with a
confining  potential $U(\bbox{r})$ and chemical potential $\mu$ 
[$\psi^\dagger(\bbox  r)$ is the creation operator of an electron at $\bbox
r$].  The term $\hat Q^2/2C$, with $\hat Q = e \hat {\cal N}  \equiv e
\int_{\rm  dot} d\bbox{r}\; \psi^\dagger(r) \psi(\bbox{r}) $ being the charge
operator  of the dot, describes the Coulomb interaction in the constant-interaction model. The last two terms describe the Hamiltonian of the right lead and the
tunneling Hamiltonian between the dot and the right lead (where $c^\dagger_{k s}$ is the creation operator of an electron in the leads with wave number $k$ and spin $s$, and $\bbox{r}_c$ is the right point contact).

Furusaki and Matveev (1995a, 1995b) calculated the conductance for the Hamiltonian (\ref{dot-channel}) in the  classical Coulomb-blockade limit $k T \gg
\Delta$. The calculation is nonperturbative  in the charging
energy. In the limit of perfect transmission ($T^l=1$),  the
dot-channel Hamiltonian can be effectively described by a
one-dimensional  fermionic Hamiltonian that is solved using
bosonization methods.  At perfect transmission,  Coulomb
blockade disappears, but even a small reflection (due to a weak backscatterer at the dot-channel interface) causes Coulomb-blockade oscillations with an amplitude that depends quadratically on temperature away from the charge degeneracy points, in analogy with
inelastic cotunneling.

Aleiner and Glazman (1998) generalized the model to the quantum regime where the
temperature  is comparable to $\Delta$ and found mesoscopic Coulomb-blockade
effects  that resemble elastic cotunneling.  The coherent backscattering of an
electron into the lead mimics the  backscatterer effects in the
classical  case and leads to mesoscopic Coulomb-blockade oscillations even at
perfect  transmission  ($T^l=1$).  However, their phase is random, and the
signature  of mesoscopic Coulomb blockade is thus best quantified in terms of conductance correlations at different values of the gate voltage.

 Since the dot is in 2D, it is not possible to describe the
backscattering  from the dot's boundaries by a one-dimensional Hamiltonian.
Instead,  an effective action that is nonlocal in time but expressed in terms
of  the one-dimensional channel variables is derived. The charging energy
is  then treated exactly using bosonization methods.  For spinless
electrons  and in the limit $T^l=1$, the
correlation function of the conductance fluctuations versus gate voltage is given by
\begin{eqnarray}
&&\overline{\delta G(V_{g 1}) \delta G(V_{g 2})}/{G^r}^2 \nonumber \\ && \;\;\;\; =  0.78
\beta^{-1}\left({\Delta  \over E_C}\right)^2  \cos\left( 2\pi {\alpha C \over e}  \Delta V_g\right) \;,
\end{eqnarray}
where $G^r$ is the conductance of the right point contact. 
The average conductance and its standard deviation are proportional to
$\Delta/E_C$, similarly  to elastic cotunneling. The periodicity
of the correlation function versus  gate voltage corresponds to
a period of one electron charge and is just  the manifestation
of mesoscopic Coulomb blockade.  The fluctuations are larger
in the absence of magnetic
field $(\beta=1)$, since  the constructive interference of
time-reversed trajectories enhances coherent backscattering.

  When the spin of the electrons is taken into account, the results are quite different:
\begin{eqnarray}\label{spin}
&&\overline{\delta G(V_{g 1}) \delta G(V_{g 2})}/{G^r}^2 \nonumber \\ && \; =  0.83
\beta^{-2}{\Delta  \over T} \left({\Delta \over E_C}\right)^2  \ln^3\left({E_C
\over  T}\right) \cos\left( 2\pi {\alpha C \over e} \Delta V_g\right)
\label{correlations-spin}  \;.
\end{eqnarray}
In particular, the mesoscopic fluctuations become
sensitive to temperature, and the suppression  of the fluctuations
with magnetic field is stronger.

 Some of the predicted characteristics of mesoscopic Coulomb blockade were
confirmed  qualitatively in an experiment by
Cronenwett~{\em et~al.} (1998).  Experimentally, it is more convenient to study the
power  spectrum $P_G(f)$ of $G(V_g)$, i.e., the Fourier transform of the
correlation  function (\ref{correlations-spin}). It is found to be centered in a narrow band around the Coulomb-blockade frequency, and the integrated  power is given by
\begin{equation}\label{power}
P(T) = 0.207 {G^r}^2 \beta^{-2} {\Delta \over T} \left({\Delta \over
E_C}\right)^2  \ln^3 \left({E_C \over kT} \right) \;.
\end{equation}
The crossover field from conserved to broken time-reversal symmetry is
similar  to the one found for the conductance minima.  For magnetic fields
that  are large compared with this correlation field ($\beta=2$), the power of
the  mesoscopic fluctuations is expected to be four times smaller than without
a  magnetic field ($\beta=1$).   Figure  \ref{fig:mesoscopic-CB} compares typical
Coulomb-blockade fluctuations as a function of gate voltage for the
one-channel regime (left) and the weak-tunneling regime (right),  both with [panels (c) and (d)] and without [panels (a) and (b)]  a magnetic field. A striking effect is the
strong  suppression of the fluctuations in the one-channel regime when a
magnetic  field is applied. This strong sensitivity to magnetic field is also
seen  in the integrated power spectrum  versus magnetic field [Figs. \ref{fig:mesoscopic-CB}(e) and \ref{fig:mesoscopic-CB}(f)]. The power spectrum for the one-channel
regime  is strongly peaked at $B=0$, while it does not show any particular
field  dependence in the weak-tunneling regime. The enhancement of the $B=0$
power  relative to its large-$B$ value is found to be $\sim 5.3 \pm 0.5$,
somewhat  larger than the predicted value of $4$. The experiment also
confirms
the sensitivity of the mesoscopic fluctuations to temperature.

\begin{figure}
\epsfxsize= 8 cm
\centerline{\epsffile{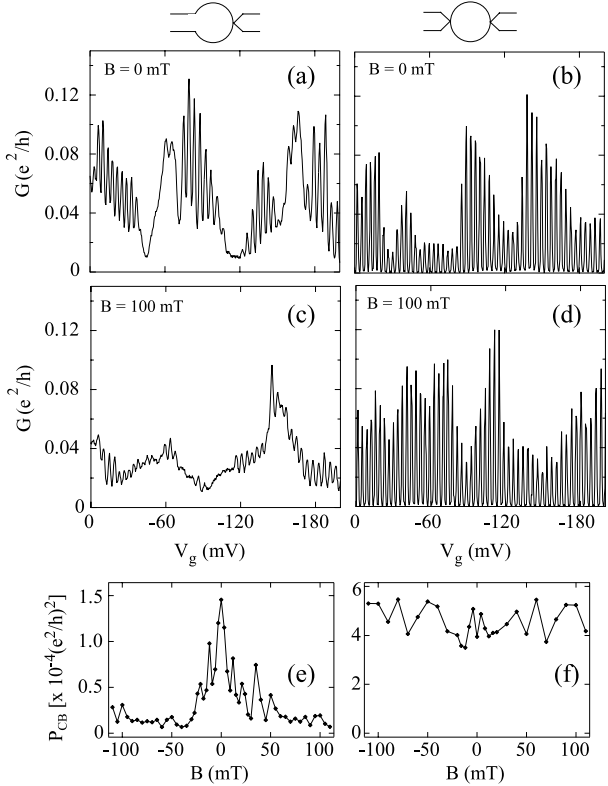}}
\vspace*{2 mm}
\caption{Conductance fluctuations in a one-channel dot (left)
and in a closed dot (right).
 Traces of conductance vs gate voltage are shown for [(a) and (b)] $B=0$  and [(c) and (d)] $B=100$ mT. Notice that the mesoscopic Coulomb-blockade oscillations  in the
one-channel dot are suppressed by a magnetic field, in contrast to
the closed dot. Panels (e) and (f) show the integrated power of
the  Coulomb-blockade oscillations as a function of magnetic
field. For a one-channel  dot  [panel (e)] the power decreases with
magnetic field, while for a closed  dot [panel (f)] there is only weak dependence
of the power on magnetic field. From
Cronenwett~{\em et~al.} (1998). }
\label{fig:mesoscopic-CB}
\end{figure}

  Brouwer and Aleiner (1999) showed that mesoscopic charge quantization also affects the conductance in {\em open} dots with ideal leads. A charging energy term in the Hamiltonian of an open dot plus leads does not affect the average conductance in the unitary case, but does enhance the weak localization correction in the orthogonal case $\Lambda/(4\Lambda+2)$ [see Eq. (\ref{average-g}) for $\beta=1$] by  
$(c_\Lambda/\Lambda)\Lambda\Delta/(8\pi^2 T)$, where $\Lambda$ is the number of open channels in each lead ($c_1\approx 3.18$ and $c_\infty=\pi^2/6$). Furthermore, the variance of $G$ in the unitary case, $\Lambda \Delta/(96 T)$ [see Eq. (\ref{high-T-corr})],  acquires an additional term $(c_\Lambda/\Lambda)\Lambda^2 \Delta^2 /(32 \pi^2 T^2)$ in the limit $T \gg \Lambda \Delta$ and with $c_\Lambda\approx 6.49$ for $\Lambda \gg 1$.

\subsection{Mesoscopic fluctuations of the differential capacitance}
\label{capacitance-fluctuations}

 In the weak-coupling regime (where the transmission coefficient $T_c \ll 1$), the charge on the dot is quantized
and  increases in a steplike manner as the gate voltage changes. As the
conductance  of the point contact increases (but is still small compared with
$e^2/h$),  the deviation of the average charge from its quantized value is
linear  in $T_c$ (Glazman and Matveev, 1990b; Matveev, 1991), except near the degeneracy
points  where the average charge increases sharply by about one unit $e$. A
related  quantity is the differential capacitance of the dot
$C_{\rm diff}(V_g) \equiv {\partial Q / \partial V_{\rm ext}}=
\alpha^{-1}{\partial Q / \partial V_g}$, which, 
 in the weak-coupling limit, exhibits sharp peaks at the degeneracy points of
$V_g$.    Matveev (1995) and Flensberg (1993)
showed that when a single-channel lead is connected to a dot and in the classical limit $\Delta  \to 0$,
Coulomb blockade  vanishes at
perfect transmission (i.e., $T_c=1$). In this limit, the average charge increases
linearly as a function of gate voltage and the differential
capacitance  is a constant.  However, for nearly perfect
transmission (i.e., $T_c$ slightly smaller than 1), the
average charge and differential capacitance exhibit weak Coulomb-blockade oscillations versus gate voltage.

  Aleiner and Glazman (1998) studied the mesoscopic fluctuations of the
differential  capacitance (of a dot with single-channel lead) for finite $\Delta$ assuming the dynamics in the dot
is  chaotic. In the case of perfect transmission,  $\overline{\delta C_{\rm
diff}(V_g)}=0$,  and the Coulomb-blockade periodicity is seen in the
correlation  function of the differential capacitance versus gate voltage.  The capacitance correlations are found to be
\begin{mathletters}
\begin{eqnarray}
&&\overline{\delta C_{\rm diff}(V_{g 1}) \delta C_{\rm diff}(V_{g 2})}/C^2 \nonumber \\ & & \;\; = 
5.59\beta^{-1}\left({\Delta  \over E_C}\right)  \cos\left( 2\pi {\alpha C \over
e} \Delta V_g\right) \\ &&
\overline{\delta C_{\rm diff}(V_{g 1}) \delta C_{\rm diff}(V_{g 2})}/C^2 \nonumber \\ & & \;\; =  0.54  \beta^{-1} \left({\Delta \over E_C}\right)  \ln^3\left({E_C \over
T}\right)\left[  \left({\Delta \over E_C}\right)\ln\left({E_C \over T}\right)
  \right] \nonumber \\ &&\;\;\;\; \; \times\,\cos\left( 2\pi {\alpha C \over e} \Delta V_g\right) \;,
\end{eqnarray}
\end{mathletters}
for the cases without and with spin, respectively.

  Capacitance fluctuations in the weak-coupling regime were studied by
Kaminski, Aleiner and Glazman (1998).  Away from the peaks, the capacitance
could be calculated by treating the tunneling Hamiltonian as a perturbation. The capacitance  fluctuations could then be related
to the fluctuations of the single-electron  Green's function of the dot at
the point contact. The latter are universal when $E = {\rm min}(E_e,E_h)$ (see Sec. \ref{cotunneling}) is below
the Thouless energy and could be expressed in terms of  the
diffuson and cooperon [Eqs. (\ref{universal-correlators})]. For example, the
standard deviation of the capacitance fluctuations in the absence of magnetic field was found to be
$\sigma(\delta C_{\rm diff}) = (C g_0 /\sqrt{6}\pi^2) (\Delta/
E)^{1/2} (E_C / E)$,
where $g_0 = G_0/(2e^2/h)$ ($\ll 1$) is the dimensionless point-contact conductance.
The correlation field of the capacitance fluctuations in a magnetic field had
the  same scale as the correlation field [Eq. (\ref{crossover-cotunneling})] for the
cotunneling  conductance fluctuations. The capacitance fluctuations were also studied in the strong-tunneling regime, and it was concluded that
 the maximal fluctuations are reached
for  a {\em partially} open channel ($G_0 < 2e^2/h$).

 Recently, high-sensitivity single-electron transistors were used to measure
a  dot's charge and capacitance, and fluctuations of the differential
capacitance  were observed (Berman~{\em et~al.}, 1999). However, the accuracy of these experiments is insufficient to quantify these fluctuations.   Kaminski and Glazman (1999) pointed out that certain mesoscopic fluctuations in a partially open dot can be more easily measured in a double-dot geometry, where each dot is weakly coupled to a lead. The interdot coupling can be adjusted, resulting in peak doublets as a function of the gate voltages on each of the dots. The spacing  between the doublets fluctuates because of the fluctuations  of the interdot tunneling amplitude. The rms-to-average ratio of the doublet spacing was found to be $\sim (2\Delta /\beta E_C)^{1/2}$.  This provides information on the mesoscopic fluctuations of the ground-state energy of a partially open dot, which are otherwise difficult to measure directly.

\section{Conclusion and Future Directions}

Since they were first produced about a decade ago, quantum dots have become
a  powerful tool for investigating the physics of small, coherent quantum
systems.  The ability to control their shape, size, number of electrons, and
coupling  strength has made them particularly attractive for experimental
studies.  This review has focused on the statistical regime of quantum dots, a
regime  characterized by quantum interference effects, chaotic dynamics of the
quasiparticles,  and electron-electron interaction effects.

Table I summarizes the main mesoscopic effects in quantum dots
versus the main theoretical techniques used to calculate them.
For each case, a reference is made to the equation (in
parentheses) and/or the figure 
\begin{figure}[]
\begin{center}
\epsfxsize=  9 cm
\centerline{\epsffile{table.eps}}
\end{center}
\end{figure}
\noindent  (in square brackets) that is
relevant to the effect. The table is restricted to effects that have been experimentally observed.

  Quantum dots have several energy scales. Among them are the mean level spacing $\Delta$, arising from the confinement of the electrons and inversely proportional
to  the dot's area, and the average level width $\bar\Gamma$, representing the strength
of  the dot-lead couplings. In an open dot, where $\bar\Gamma \gg \Delta$, the
electrons  can be treated as noninteracting quasiparticles, and
electron-electron interactions are considered indirectly through their effect
on the decoherence rate $\Gamma_\phi$. The limit
$\Gamma_\phi  \ll \bar\Gamma$ (i.e., $\tau_{\rm escape} \ll \tau_\phi$) is the limit of full phase coherence, where
quantum  interference effects dominate the mesoscopic fluctuations of the conductance. The universality of these fluctuations is determined by another energy scale $E_T$ -- the ballistic Thouless energy (or the
Thouless  energy in a disordered dot). For $\bar \Gamma \ll E_T$ (i.e., $\tau_{\rm escape} \gg \tau_c$, where $\tau_c$ is the ergodic time) the fluctuations are universal.  Finite temperature can reduce the fluctuations through thermal smearing and shorter dephasing times. In the absence of dephasing, the fluctuations are largest and temperature independent when $T \ll \bar\Gamma$.

Phase breaking becomes important at temperatures where $\Gamma_\phi$ is comparable to $\bar \Gamma$  and leads to deviations from universality.  Indeed, even at the lowest temperatures attained in the experiments,
dephasing  must be included to obtain good agreement between theory and
experiment.  

 With the introduction of tunnel barriers between the dot and the
leads,  the charge on the dot is quantized and an additional energy scale becomes relevant -- the charging energy $e^2/C$. In the almost-isolated
 dots typically used in experiments, all three scales, $\Gamma$,
$\Delta$,  and $e^2/C$, separate: $\bar\Gamma \ll \Delta \ll e^2/C$, and the
temperature  determines which energy scales are resolved. The regime of
interest  in this review is the quantum regime, where the temperature is around or smaller than $\Delta$ and mesoscopic fluctuations are observed.
In  typical experiments in closed dots, $\bar\Gamma \ll T $, and the level width is not directly resolved.

 The statistical theory of disordered or ballistic chaotic dots is  well
understood in the limit of noninteracting quasiparticles.  In closed dots this ``single-particle''
approach  includes a constant-interaction term ${\cal N}^2e^2/2C$ (the constant-interaction  model).
While the constant-interaction  model is the simplest way to
include  charging energy effects in closed dots, several  experiments indicate
that  electron-electron interactions play an important role in the statistical
properties  of such dots. This is not surprising considering that
the  Coulomb interaction is rather strong in semiconductor dots where
the  gas parameter $r_s$ is $\sim 1-2$. Understanding the effects of interactions
on  the statistical fluctuations is one of the major current  directions in
mesoscopic  physics in general, and in the statistical theory of quantum dots
in  particular. Recent progress includes Hartree-Fock and RPA estimates, but the experimental
values  of $r_s$ are in the range where it is necessary to go beyond the RPA.
Currently,  most of the results for  $r_s \agt 1$ are based on numerical
simulations  of small disordered systems that include interactions. The number
of  electrons used in the simulations is smaller than in the experiments.
The effects seen are believed to be independent of the number of
electrons,  but this is not proven. Except for $r_s\ll 1$, it is not known what is the parametric dependence of the fluctuations on the gas parameter $r_s$  and on properties of the dot such as its Thouless conductance $g_T$.

Another problem of current interest is the role played by  electron spin in closed dots. Indirect evidence of spin pairing was recently found in the statistical properties of the conductance peaks (L\"{u}scher~{\em et~al.}, 2000). In a chaotic or disordered dot we expect a spin distribution in the ground state, but its dependence on interactions and on the Thouless conductance $g_T$ is not yet understood (the limit $g_T \to \infty$ was recently studied by Kurland, Berkovits, and Altshuler, 2000).  Hund's rules are not expected to hold in chaotic dots that do not possess any particular symmetries, and we would like to understand how the total spin of these dots changes with the addition of electrons.  Several experimental groups are now working on measuring spin in dots. A promising technique is an in-plane magnetic field that couples to the spin but does not directly affect the orbital motion.

  While the limits of almost-closed dots ($\bar\Gamma \ll \Delta$) and open dots ($\bar\Gamma \gg \Delta$) have been studied extensively, less is known about the intermediate regime where the coupling to the leads is strong but some charge quantization remains. In this regime, transport becomes more complicated even within the constant-interaction  model, leading to cotunneling and mesoscopic Coulomb blockade. A recent experiment (Maurer~{\em et~al.}, 1999) investigated Coulomb-blockade fluctuations in dots with symmetric leads as a function of the dot-lead couplings.   Few theoretical results are available for this intermediate regime.  Some of the experimental results were not expected within the theory of cotunneling, e.g., a Kondo-like anomalous temperature dependence of the conductance valleys. In general, the coupling to the leads makes interaction effects within the dot more difficult to handle than in completely closed systems.  For example, at low temperatures Kondo-type resonance can be formed between an unpaired electron in the dot and the delocalized electrons in the leads, and a perturbative approach is not possible. A consistent formulation of transport in the presence of interactions can be done in the Keldysh formalism (Meir and Wingreen, 1992) but is difficult to implement.

Another important topic for future research is the statistical properties of
excited  states in quantum dots. Thus far, statistical studies have focused on
the  linear regime of a small source-drain voltage $V_{sd}$, where the
observed  Coulomb-blockade peaks probe the ground state of a dot with
different  numbers of electrons. Information about excited states in a
dot  with a fixed number of electrons can be obtained through nonlinear
measurements  (see Sec. \ref{addition-excitation}).  The experiment of
Stewart~{\em et~al.} (1997)  suggests certain similarities between the low excitation spectrum and the addition spectrum of a quantum dot. However, the statistical
properties  of these low-lying states and their manifestation in the
fluctuations of the nonlinear differential conductance have not been
investigated.

 A theory  of the quasiparticle lifetime (due to
electron-electron  interactions) in a quantum dot has suggested that, above a
critical  value of the excitation energy, the quasiparticle width becomes
finite (Altshuler~{\em et~al.}, 1997). This
behavior  is deduced by relating the problem to Anderson localization in real space and is the result of a delocalization phase transition in Fock space
where  the interacting wave function becomes fragmented over a large number of
noninteracting  states.  The crossover from Poisson to Wigner statistics at
high  excitations as a function of the interaction strength has been linked to
this  Fock-space delocalization (Berkovits and Avishai, 1998).  Numerical investigations  in a more realistic random-matrix model with one-body disorder and interactions (Mejia-Monasterio~{\em et~al.}, 1998) found that at the low excitations relevant to the experiments there was a smooth crossover from almost-localized to delocalized states but no Anderson-like localization. This model could explain the main features of an experiment by
Sivan~{\em et~al.} (1994) that found only a few ($\sim 7$) resolved excited levels
(of the order of $g_T$) before their width grew beyond the mean level spacing.
Recent numerical studies suggest a possible localization transition for very large values of $g_T$, where the transition occurs at very high excitations (Leyronas, Tworzydlo, and Beenakker, 1999).

 Several models have been proposed to explain the effect of dephasing on the
conductance  statistics in open dots. Dephasing in almost-closed dots and how it might affect the conductance is
not well understood. Theoretical estimates for the temperature dependence of the dephasing time in closed dots due to $e$-$e$ interactions were derived by Sivan, Imry, and Aronov (1994) and by Blanter (1996). Full phase coherence is expected in a closed dot below a temperature that is parametrically larger than $\Delta$  (Altshuler~{\em et~al.}, 1997). However, reduced values of the weak-localization correction are found experimentally at lower temperatures (Folk~{\em et~al.}, 2000). Also, deviations of the  experimental peak-height
distributions  from the theoretical phase-coherent predictions are found to increase
with  temperature (Sec. \ref{G-finite-T:closed}). These results suggest apparent phase breaking and are not understood. It is possible that $e$-$e$ interaction is not the main dephasing mechanism in the experimentally studied devices. Other suggested mechanisms are external radiation (Gershenson, 1999) and nuclear spins (Dyugaev, Vagner, and Wyder, 2000).

  Most investigations of transport in quantum dots have concentrated on
chaotic  dots. Much less is known about fluctuations in nonchaotic dots,
especially  in the Coulomb-blockade regime. In such systems the fluctuations
are  not expected to be universal and the semiclassical approach is the most
suitable  one. However, this approach encounters difficulties at long time
scales  of the order of the Heisenberg time.
 In a recent study of conductance fluctuations in an integrable cavity, the
conductance  fluctuations were found to increase with incoming energy
(Pichaureau and Jalabert, 1999),  unlike the universal fluctuations in chaotic cavities.  Ketzmerick (1996) argued that the conductance through a cavity with mixed phase space displays a fractal behavior as a function of an external parameter, e.g., a magnetic field. This behavior originates from trajectories that are trapped near the boundary between regular and chaotic regions, leading to an {\em algebraic} decay of the enclosed-area distribution. The change of the conductance $\Delta G$ with a magnetic field is expected to have a variance of $\overline{(\Delta G)^2} \propto (\Delta B)^\gamma$ where $1< \gamma < 2$ [in contrast with a chaotic dot where $\overline{(\Delta G)^2} \propto (\Delta B)^2$ for small $\Delta B$]. An experiment by Sachrajda~{\em et~al.} (1998) claims to have observed this fractal behavior in dots coupled to unusually wide ($\sim 0.7\;\mu$m) leads.

 In the work discussed in this review, transport through the dot is driven by an applied bias. Another way to produce a dc current through an open dot (at zero bias)  is by a cyclic change of its deformation or any other parameter that affects the interference pattern of the electron's wave function.  For low-frequency changes, the electrons maintain equilibrium and the device is known as an adiabatic quantum pump. In an open dot,  the electronic wave function  extends into the leads and an adiabatic cyclic change of at least two parameters can cause a net charge transport $Q$ per cycle.  The theory of parametric pumping was worked out by Zhou, Spivak, and Altshuler (1999) and by Brouwer (1998), and an experiment was carried out by Switkes~{\em et~al.} (1999). While confirming some of the theoretical predictions at weak pumping (e.g., $\langle Q^2 \rangle \propto S_A^2$ where $S_A$ is the area enclosed by the contour in the two-parameter space), there were unexplained quantitative differences at strong pumping (e.g., the dependence of $\langle Q^2 \rangle$ on $S_A$ was {\em slower} than linear). Recently the theory was generalized to the strong-pumping regime (Shutenko, Aleiner, and Altshuler, 2000), and it was found that $\langle Q^2 \rangle \propto l_A$ ($l_A$ being the length of the contour) -- slower than the naive expectation of $\langle Q^2 \rangle \propto S_A$. Issues of dissipation and dephasing, important for the temperature dependence of pumping, still need to be understood. Aleiner and Andreev (1998) showed that for an almost-open dot the charge transmitted in one cycle is quantized in the limit $T \to 0$.

Recent years have seen the fabrication of new conducting nanostructures
smaller  than quantum dots. These devices have similarities to quantum dots,
and  Coulomb-blockade peaks are observed in both linear and nonlinear $I$-$V$
measurements  versus gate voltage. A particularly interesting example is the
nanometer-scale  Al particle, part of a tunneling device that includes a gate
electrode  (Ralph, Black, and Tinkham, 1997).   The spectrum of the Al particle as a function of
the  number of electrons it contains can be determined from nonlinear
measurements  similar to those made on semiconductor quantum dots. Odd-even
effects  in the number of electrons can be understood in terms of the pairing
interaction.   Davidovi\'{c} and M. Tinkham (1999) measured the spectra of Au nanoparticles. At higher excitations the resonances of the particles overlapped to form broad resonances that eventually merged into the continuum around the Thouless energy, in overall agreement with the theory of  Altshuler~{\em et~al.} (1997). Another type of conducting nanostructure is the carbon nanotube,
which  also exhibits charging-energy effects (Bockrath~{\em et~al.}, 1997; McEuen, 1998).  Figure 
\ref{fig:nanotube}  shows Coulomb-blockade peaks in the measured
conductance  of a nanotube, where large fluctuations of the peak heights can
be  seen (Cobden~{\em et~al.}, 1998).  The statistical theory of quantum dots
should  find interesting applications in some of these novel nanoscale
devices.

\begin{figure}
\epsfxsize= 8 cm
\centerline{\epsffile{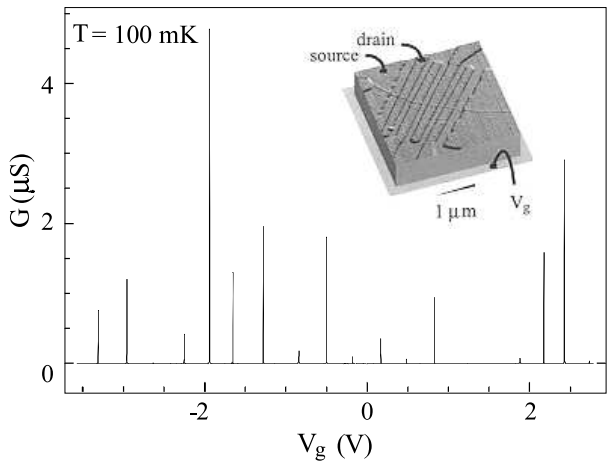}}
\vspace*{2 mm}
\caption{The conductance vs gate voltage in a carbon nanotube rope
 ($\sim 200\;$nm segment). Sharp
  Coulomb-blockade peaks are observed, each describing the tunneling through a
single  resonance level. Note the strong fluctuations in the peak heights.
 The inset is an image of the device with schematic wires drawn. From
Cobden~{\em et~al.} (1998).
}
\label{fig:nanotube}
\end{figure}

\section{Acknowledgments}

I  would like to thank H. Attias, Y. Gefen, M. G\"{o}k\c{c}eda\u{g}, J. N. Hormuzdiar, Ph. Jacquod, R. Jalabert,  S. Malhotra, S.  Patel, N. Whelan and A. Wobst for fruitful collaborations on various topics in this review.   Special thanks to A. D. Stone and C. M. Marcus for productive collaborations and useful discussions.  I am grateful to U. Alhassid, K. Aschheim, G. F. Bertsch, C. M. Marcus, A. Polkovnikov, A. D. Stone, T. Wettig, and in particular H. A. Weidenm\"uller} for their comments and suggestions on the manuscript.
 This work was supported in part by the Department of Energy grant No.\ DE-FG-0291-ER-40608.


\begin{references}

\harvarditem{Abrikosov, Gor'kov and Dzhyaloshinskii}{1963}{Abrikosov63}
Abrikosov, A.A., L.P. Gor'kov, and I.E. Dzhyaloshinskii, 1963, {\em Methods
of  Quantum Field Theory in Statistical Physics} (Prentice-Hall, New York).

\harvarditem{Agam, Altshuler, and Andreev}{1995}{Agam95}
Agam, O., B.L. Altshuler and A.V. Andreev, 1995, Phys. Rev. Lett. {\bf 75},
4389.

\harvarditem{Ahn, Richter and Lee}{1999}{Ahn99}
Ahn, K.-H., K. Richter, and  I.-H. Lee, 1999, Phys. Rev. Lett. {\bf 83}, 4144.

\harvarditem{Akkermans~{\em et~al.}}{1995}{Akkermans95}
Akkermans, E., G. Montambaux, J.-L. Pichard, and J. Zinn-Justin, eds., 1995,
{\em  Mesoscopic Quantum Physics} 
 (North-Holland,  Amsterdam).

\harvarditem{Aleiner, Altshuler and Gershenson}{1999}{Aleiner99}
Aleiner, I.L., B.L. Altshuler, and M.E. Gershenson, 1999, Waves Random Media {\bf 9}, 201.

\harvarditem{Aleiner and Andreev}{1998}{Aleiner98'}
Aleiner, I.L., and A.V. Andreev, 1998, Phys. Rev. Lett. {\bf 81}, 1286.

\harvarditem{Aleiner and Glazman}{1996}{Aleiner96}
Aleiner, I.L., and L.I. Glazman, 1996, Phys. Rev. Lett. {\bf 77}, 2057.

\harvarditem{Aleiner and Glazman}{1998}{Aleiner98}
Aleiner, I.L., and L.I. Glazman, 1998, Phys. Rev. B {\bf 57}, 9608.

\harvarditem{Alhassid}{1998}{Alhassid98''}
 Alhassid, Y., 1998, Phys. Rev. B {\bf 58}, R 13383.

\harvarditem{Alhassid and Attias}{1996a}{Alhassid96a}
 Alhassid, Y.,  and H. Attias, 1996a,  Phys. Rev. Lett. {\bf 76}, 1711.

\harvarditem{Alhassid and Attias}{1996b}{Alhassid96b}
 Alhassid, Y., and H. Attias, 1996b, Phys. Rev. B {\bf 54}, 2696.

\harvarditem{Alhassid and Attias}{1995}{Alhassid95}
 Alhassid, Y., and H. Attias, 1995, Phys. Rev. Lett. {\bf 74}, 4635.

\harvarditem{Alhassid and Gefen}{2000}{Alhassid00c}
 Alhassid, Y., and Y.Gefen, 2000, ``Spectral scrambling in Coulomb-blockade quantum dots,'' reprint.

\harvarditem{Alhassid, G\"{o}k\c{c}eda\u{g} and Stone}{1998}{Alhassid98'}
 Alhassid, Y., M. G\"{o}k\c{c}eda\u{g}, and A.D. Stone, 1998, Phys. Rev. B {\bf 58}, R7524.

\harvarditem{Alhassid, Hormuzdiar and Whelan}{1998}{Alhassid98}
Alhassid, Y., J.N. Hormuzdiar and N. Whelan, 1998, Phys. Rev. B {\bf 58}, 4866.

\harvarditem{Alhassid, Jacquod and Wobst}{2000}{Alhassid00a}
 Alhassid, Y., Ph. Jaquod, and A. Wobst, 2000, Phys. Rev. B {\bf 61}, R13357.

\harvarditem{Alhassid and Levine}{1986}{Alhassid86}
Alhassid, Y., and R.D. Levine, 1986, Phys. Rev. Lett. {\bf 57}, 2879.

\harvarditem{Alhassid and Lewenkopf}{1995}{Alhassid95'}
 Alhassid, Y., and C.H. Lewenkopf, 1995, Phys. Rev. Lett. {\bf 75}, 3922.

\harvarditem{Alhassid and Lewenkopf}{1997}{Alhassid97}
 Alhassid, Y., and C.H. Lewenkopf, 1997, Phys. Rev. B {\bf 55}, 7749.

\harvarditem{Alhassid and Malhotra}{1999}{Alhassid99}
 Alhassid, Y., and S. Malhotra, 1999, Phys. Rev. B {\bf 60}, R16315.

\harvarditem{Alhassid and Wobst}{2000}{Alhassid00b}
 Alhassid, Y., and A. Wobst, 2000, ``Interactioneffects on the conductance peak height statistics in quantum dots,'' e-print  cond-mat/0003255.

\harvarditem{Altland and Gefen}{1993}{Altland93}
 Altland, A., and Y. Gefen, 1993, Phys. Rev. Lett. {\bf 71}, 3339.

\harvarditem{Altshuler}{1985}{Altshuler85} Altshuler, B.L., 1985,
JETP Lett. {\bf 41}, 648.

\harvarditem{Altshuler and Aronov}{1985}{Altshuler85'}
Altshuler, B.L., and A. G. Aronov, 1985, in {\em Electron-Electron
Interactions in Disordered Systems}, edited by A.L. Efros and M. Pollak
(North-Holland, Amsterdam), p. 1.

\harvarditem{Altshuler {\em et al.}}{1997}{Altshuler97}
Altshuler, B.L., Y. Gefen, A. Kamenev, and L. S. Levitov, 1997, Phys. Rev.
Lett.  {\bf 78}, 2803.

\harvarditem{Altshuler~{\em et~al.}}{1980}{Altshuler80}
Altshuler, B.L., D.E. Khemlnitskii, A.I. Larkin, and P.A. Lee, 1980, Phys. Rev. B {\bf 20}, 5142.

\harvarditem{Altshuler, Lee and Webb}{1991}{Altshuler91}
 Altshuler, B.L., P.A. Lee, and R.A. Webb, editors, 1991, {\em
Mesoscopic Phenomena in Solids} (North-Holland, Amsterdam).

\harvarditem{Altshuler and Shklovskii}{1986}{Altshuler86} Altshuler, B.L.,
and  B.I. Shklovskii, 1986,
Sov. Phys. JETP {\bf 64}, 127.

\harvarditem{Altshuler and Simons}{1995}{Altshuler95}
 Altshuler, B.L., and B.D. Simons, 1995, in {\em Mesoscopic Quantum
Physics}, edited by E. Akkermans, G. Montambaux, J.-L. Pichard, and J.
Zinn-Justin (North-Holland, Amsterdam), p. 1.

\harvarditem{Anderson}{1958}{Anderson58}
 Anderson, P.W., 1958, Phys. Rev. {\bf 109}, 1492.

\harvarditem{Andreev {\em et al.}}{1996a}{Andreev96}
 Andreev, A.V., O. Agam, B.D. Simons, and B.L. Altshuler, 1996,
Phys. Rev. Lett. {\bf 76}, 3947.

\harvarditem{Andreev and Altshuler}{1995}{Andreev95}
Andreev, A.V., and B.L. Altshuler, 1995, Phys. Rev. Lett. {\bf 75}, 902.

\harvarditem{Andreev and Kamenev}{1998}{Andreev98}
Andreev, A.V., and A. Kamenev, 1998, Phys. Rev. Lett. {\bf 81}, 3199.

\harvarditem{Argaman}{1996}{Argaman96}
Argaman, N., 1996, Phys. Rev. B {\bf 53}, 7035.

\harvarditem{Argaman, Imry and Smilansky}{1993}{Argaman93} Argaman, N., J. Imry,  and U. Smilansky, 1993, Phys. Rev. B {\bf 47}, 4440.

\harvarditem{Ashoori}{1996}{Ashoori96}
Ashoori, R.C., 1996, Nature (London) {\bf 379}, 413.

\harvarditem{Attias and Alhassid}{1995}{Attias95}
 Attias, H.,  and Y. Alhassid, 1995,  Phys. Rev. E  {\bf 52}, 4476.

\harvarditem{Austin and Wilkinson}{1992}{Austin92}
Austin, E.A.,  and M. Wilkinson, 1992, Nonlinearity {\bf 5}, 1137.

\harvarditem{Averin and Korotkov}{1990}{Averin90'}
Averin, D.V., and A.N. Korotkov,  1990, Sov. Phys. JETP {\bf 70}, 937.

\harvarditem{Averin, Korotkov, and Likharev}{1991}{Averin91}
Averin, D.V.,  A.N. Korotkov, and K.K. Likharev, 1991, Phys. Rev. B {\bf 44}, 6199.

\harvarditem{Averin and Likarev}{1986}{Averin86}
Averin, D.V., and K.K. Likharev, 1986, J. Low Temp. Phys. {\bf 62}, 345.

\harvarditem{Averin and  Likharev}{1991}{Averin91'}
 Averin, D.V., and K.K. Likharev, 1991, in {\em Mesoscopic
Phenomena in Solids}, edited by B. L. Altshuler, P. A. Lee, and R. A. Webb
(North-Holland, Amsterdam), p. 173.

\harvarditem{Averin and Nazarov}{1990}{Averin90}
Averin, D.V., and Yu. N. Nazarov, 1990, Phys. Rev. Lett. {\bf 65}, 2446.

\harvarditem{Balian}{1968}{Balian68}
Balian, R., 1968, Nuovo Cimento B {\bf 57}, 183.

\harvarditem{Balian and Bloch}{1972}{Balian72}
Balian, R., and C. Bloch, 1972,  Ann. Phys. (N.Y.) {\bf 69}, 76.

\harvarditem{Baltin and Gefen}{1999}{Baltin99}
Baltin, R., and  Y. Gefen, 1999,
Phys. Rev. Lett. {\bf 83}, 5094.

\harvarditem{Baltin and Gefen}{2000}{Baltin00}
Baltin, R., and  Y. Gefen, 2000,
Phys. Rev. B {\bf 61}, 10247.

\harvarditem{Baltin~{\em et~al.}}{1999}{Baltin99}
Baltin, R., Y. Gefen, G. Hackenbroich, and H.A. Weidenm\"uller, 1999,
Eur. Phys. J. B {\bf 10}, 119.

\harvarditem{Baranger}{1998}{Baranger98}
 Baranger, H. U., 1998, in {\em Nanotechnology}, edited
by G. Timp (American Institute of Physics, New York), p.~537.

\harvarditem{Baranger, Jalabert and Stone}{1993a}{Baranger93a}
Baranger, H. U., R. A. Jalabert, and A. D. Stone, 1993a, Phys.
Rev. Lett. {\bf 70}, 3876.

\harvarditem{Baranger, Jalabert and Stone}{1993b}{Baranger93b}
 Baranger, H.U., R.A. Jalabert and A.D. Stone, 1993b,
 Chaos {\bf 3}, 665.

\harvarditem{Baranger and Mello}{1994}{Baranger94}
 Baranger, H.U., and P.A. Mello, 1994, Phys. Rev. Lett. {\bf 73}, 142.

\harvarditem{Baranger and Mello}{1995}{Baranger95}
 Baranger, H.U., and P.A. Mello, 1995, Phys. Rev. B {\bf 51}, 4703.

\harvarditem{Baranger and Mello}{1996}{Baranger96}
 Baranger, H.U., and P.A. Mello, 1996,
 Europhys. Lett. {\bf 33}, 465.

\harvarditem{Baranger and Stone}{1989}{Baranger89}
Baranger, H.U., and A.D. Stone, 1989, Phys. Rev. B {\bf 40}, 8169.

\harvarditem{Baranger, Ullmo and Glazman}{2000}{Baranger00}
Baranger, H.U., D. Ullmo, and L.I. Glazman, 2000, Phys.
Rev. B {\bf 61}, R2425.

\harvarditem{Beenakker}{1991}{Beenakker91}
  Beenakker, C.W.J., 1991,  Phys. Rev. B {\bf 44}, 1646.

\harvarditem{Beenakker}{1993}{Beenakker93}
Beenakker, C.W.J., 1993, Phys. Rev. Lett. {\bf 70}, 4126.

\harvarditem{Beenakker}{1997}{Beenakker97}
Beenakker, C.W.J., 1997, Rev. Mod. Phys.{\bf~69}, 731.

\harvarditem{Beenakker and van Houten}{1989}{Beenakker89}
  Beenakker, C.W.J., and H. van Houten, 1989, Phys. Rev. Lett. {\bf 63}, 1857.

\harvarditem{Beenakker and van Houten}{1991}{Beenakker91'}
 Beenakker, C.W.J., and H. van Houten, 1991, Solid State Phys.
 {\bf 44}, 1.

\harvarditem{Ben-Jacob and Gefen}{1985}{Ben-Jacob85}
 Ben-Jacob, E., and Y. Gefen, 1985,  Phys. Lett. {\bf 108A}, 289.

\harvarditem{Bergmann}{1984}{Bergmann84}
 Bergmann, G., 1984, Phys. Rep. {\bf 107}, 1.

\harvarditem{Berkovits}{1998}{Berkovits98}
Berkovits, R., 1998, Phys. Rev. Lett. {\bf 81}, 2128.

\harvarditem{Berkovits}{1999}{Berkovits99}
Berkovits, R., 1999, Mod. Phys. Lett. B {\bf 13}, 3841.

\harvarditem{Berkovits and Altshuler}{1997}{Berkovits97}
Berkovits, R., and B. L. Altshuler, 1997, Phys. Rev. B
{\bf 55}, 5297.

\harvarditem{Berkovits and Avishai}{1998}{Berkovits98''}
Berkovits, R., and Y. Avishai, 1998, Phys. Rev. Lett. {\bf 80}, 568.

\harvarditem{Berkovits and Sivan}{1998}{Berkovits98'}
Berkovits, R., and U. Sivan, 1998,
Europhys. Lett. {\bf 41}, 653.

\harvarditem{Berman~{\em et~al.}}{1999}{Berman99}
Berman, D., N.B.~Zhitenev, R.C.~Ashoori, and M.~Shayegan, 1999,
Phys.~Rev.~Lett. {\bf 82}, 161.

\harvarditem{Berry~{\em et~al.}}{1994}{Berry94}
Berry, M.J., J.H. Baskey, R.M. Westervelt, and A.C. Gossard, 1994a, Phys. Rev.
B  {\bf 50}, 8857.

\harvarditem{Berry~{\em et~al.}}{1994}{Berry94'}
 Berry, M.J., J.H. Katine, R.M. Westervelt, and A.C.
Gossard,  1994b, Phys. Rev. B  {\bf 50}, 17721.

\harvarditem{Berry}{1977}{Berry77}
Berry, M.V., 1977,  J. Phys. A {\bf 10}, 2083.

\harvarditem{Berry}{1985}{Berry85}
 Berry, M.V., 1985, Proc. R. Soc. London, Ser. A {\bf 400}, 229.

\harvarditem{Berry and Robnik}{1986}{Berry86}
 Berry, M.V., and M. Robnik, 1986, J. Phys. A {\bf 19}, 649.

\harvarditem{Bertsch}{1991}{Bertsch91}
Bertsch, G.F., 1991, J. Phys.: Condens. Matter {\bf 3}, 373.

\harvarditem{Blanter}{1996}{Blanter96}
 Blanter, Ya.M.,  1996, Phys. Rev. B {\bf 54}, 12807.

\harvarditem{Blanter and Mirlin}{1997}{Blanter97'}
 Blanter, Ya.M., and A.D. Mirlin, 1997, Phys. Rev. E {\bf 55}, 6514.

\harvarditem{Blanter and Mirlin}{1998}{Blanter98}
  Blanter, Ya.M., and A.D. Mirlin, 1998, Phys. Rev. B {\bf 57}, 4566.

\harvarditem{Blanter, Mirlin and Muzykantskii}{1997}{Blanter97}
Blanter, Ya.M., A.D. Mirlin, and B.A. Muzykantskii, 1997,
Phys. Rev. Lett. {\bf 78}, 2449.

\harvarditem{Blanter, Mirlin and Muzykantskii}{1998}{Blanter98'}
Blanter, Ya.M., A.D. Mirlin, and B.A. Muzykantskii, 1998,
Phys. Rev. Lett. {\bf 80}, 4161.

\harvarditem{Blatt and Weisskopf}{1952}{Blatt52}
  Blatt, J.M., and V.F. Weisskopf, 1952, {\em Theoretical Nuclear Physics}
(Wiley,  New York).

\harvarditem{Bl\"{u}mel and Smilansky}{1988}{Blumel88}
 Bl\"{u}mel, R., and U. Smilansky, 1988, Phys. Rev. Lett. {\bf
60}, 477.
   
\harvarditem{Bl\"{u}mel and Smilansky}{1989}{Blumel89}
Bl\"{u}mel, R., and U. Smilansky, 1989, Physica D {\bf 36}, 111.

\harvarditem{Bl\"{u}mel and Smilansky}{1990}{Blumel90}
Bl\"{u}mel, R., and U. Smilansky, 1990, Phys. Rev. Lett. {\bf 64}, 241.

\harvarditem{Bl\"{u}mel and Smilansky}{1992}{Blumel92}
Bl\"{u}mel, R., and U. Smilansky, 1992, Phys. Rev. Lett. {\bf
69}, 217.

\harvarditem{Bockrath~{\em et~al.}}{1997}{Bockrath97}
Bockrath, M., D.H. Cobden, P.L. McEuen, N.G. Chopra, A. Zettl, A. Thess, and
R.E.  Smalley, 1997, Science {\bf 275}, 1922.

\harvarditem{Bohigas}{1991}{Bohigas91}
Bohigas, O., 1991, in {\it Chaos and Quantum Physics},
edited by M.-J. Giannoni, A. Voros and J. Zinn-Justin (North-Holland,
   Amsterdam).

\harvarditem{Bohigas and Flores}{1971}{Bohigas71}
Bohigas, O., and J. Flores, 1971, Phys. Lett. B {\bf 34}, 261.

\harvarditem{Bohigas and Giannoni}{1984}{Bohigas84'}
 Bohigas, O., and M.-J. Giannoni, 1984, in {\em Mathematical and
Computational  Methods in Nuclear Physics}, edited by J.S. Dehesa, J.M.G. Gomez
and  A. Polls, Lectures Notes in Physics No. {\bf 209}
  (Springer, Berlin), p. 1.

\harvarditem{Bohigas~{\em et~al.}}{1995}{Bohigas95}
Bohigas, O., M-J. Giannoni, A.M. Ozorio de Almeida, and
C. Schmit, 1995, Nonlinearity {\bf 8}, 203.

\harvarditem{Bohigas, Giannoni, and  Schmit}{1984}{Bohigas84}
 Bohigas, O., M.-J. Giannoni, and C. Schmit, 1984, Phys. Rev. Lett. {\bf 52},
1.

\harvarditem{Bohigas, Haq and Pandey}{1983}{Bohigas83}
Bohigas, O., R.U. Haq, and A. Pandey, 1983, in {\em Nuclear Data for Science
and  Technology}, edited by K.H. B\"ochhoff (Reidel, Dordrecht), p. 809.

\harvarditem{Bohigas, Haq and Pandey}{1985}{Bohigas85}
Bohigas, O., R.U. Haq,  and A. Pandey, 1985, Phys. Rev. Lett. {\bf 54}, 1645.

\harvarditem{Bohr and Mottelson}{1969}{Bohr69}
Bohr, A., and B.R. Mottelson, 1969, {\em Nuclear Structure}, Vol. I (Benjamin,  Reading, MA); 1998 edition by World Scientific, Singapore.

\harvarditem{Bonci and Berkovits}{1999}{Bonci99}
Bonci, L., and R. Berkovits, 1999, Europhys. Lett. {\bf 47}, 708.

\harvarditem{Braun and Montambaux}{1995}{Braun95}
Braun, D., and G. Montambaux, 1995, Phys. Rev. B {\bf 52}, 13903.

\harvarditem{Brezin and Zee}{1993}{Brezin93}
Br\'{e}zin E., and A. Zee, 1993, Nucl. Phys. B {\bf 402}, 613.

\harvarditem{Brody~{\em et~al.}}{1981}{Brody81}
 Brody, T.A., J. Flores,
   J.B. French, P.A. Mello, A. Pandey, and S.S.M. Wong, 1981, Rev. Mod. Phys.
{\bf  53}, 385.

\harvarditem{Brookhaven}{1964}{Brookhaven64}
Brookhaven National Laboratory, 1964, {\em Neutron Cross Sections}, BNL 325, Suppl. 2 (Brookhaven National Laboratory, Upton, NY).

\harvarditem{Brouwer}{1995}{Brouwer95'}
Brouwer, P.W., 1995, Phys. Rev. B {\bf 51}, 16878.

\harvarditem{Brouwer}{1998}{Brouwer98}
Brouwer, P.W., 1998, Phys. Rev. B {\bf 58}, R10135.

\harvarditem{Brouwer and Aleiner}{1999}{Brouwer99}
 Brouwer, P.W., and I.L. Aleiner, 1999, Phys. Rev. Lett. {\bf
82}, 390.

\harvarditem{Brouwer and Beenakker}{1994}{Brouwer94}
Brouwer, P.W., and C.W.J. Beenakker, 1994, Phys. Rev. B {\bf
50}, 11263.

\harvarditem{Brouwer and Beenakker}{1995}{Brouwer95}
Brouwer, P.W., and C.W.J. Beenakker, 1995, Phys. Rev. B {\bf 51}, 7739.

\harvarditem{Brouwer and Beenakker}{1996}{Brouwer96}
Brouwer, P.W., and C.W.J. Beenakker, 1996, J. Math. Phys. {\bf
37}, 4904.

\harvarditem{Brouwer and Beenakker}{1997}{Brouwer97}
 Brouwer, P.W., and C.W.J. Beenakker, 1997, Phys. Rev. B {\bf
55}, 4695.

\harvarditem{Brouwer, Oreg and Halperin}{1999}{Brouwer99'}
Brouwer, P.W., Y. Oreg, and B.I. Halperin, 1999, Phys. Rev. B {\bf 60}, R13977.

\harvarditem{Bruus, Lewenkopf and Mucciolo}{1996}{Bruus96}
Bruus, H., C.H. Lewenkopf, and E.R. Mucciolo, 1996, Phys. Rev. B {\bf 53},
9968.

\harvarditem{Bruus and Stone}{1994}{Bruus94}
 Bruus, H.,  and A.D. Stone, 1994, Phys. Rev. B {\bf 50},
18275.

\harvarditem{B\"{u}ttiker}{1986a}{Buttiker86}
B\"{u}ttiker, M., 1986a,  Phys. Rev. Lett. {\bf 57}, 1761.

\harvarditem{B\"{u}ttiker}{1986b}{Buttiker86'}
B\"{u}ttiker, M., 1986b, Phys. Rev. B {\bf 33}, 3020.

\harvarditem{B\"{u}ttiker}{1988b}{Buttiker88}
B\"{u}ttiker, M., 1988a, IBM J. Res. Dev. {\bf 32}, 63.

\harvarditem{B\"{u}ttiker}{1988a}{Buttiker88'}
B\"{u}ttiker, M., 1988b, IBM J. Res. Dev. {\bf 32}, 317.

\harvarditem{Chakravarty and Schmid}{1986}{Chakravarty86}
 Chakravarty, S., and A. Schmid, 1986, Phys.\ Rep.\ {\bf 140}, 193.

\harvarditem{Chan~{et~al.}}{1995}{Chan95}
 Chan, I.H., R.M. Clarke, C.M. Marcus, K. Campman, and
 A.C. Gossard, 1995, Phys. Rev. Lett. {\bf 74}, 3876.

\harvarditem{Chang~{\em et~al.}}{1994}{Chang94}
 Chang, A.M., H.U. Baranger, L.N. Pfeiffer, and K.W. West, 1994,
Phys. Rev. Lett.~{\bf 73}, 2111.

\harvarditem{Chang~{\em et~al.}}{1996}{Chang96}
Chang, A.M., H.U. Baranger, L.N. Pfeiffer,
K.W. West, and T.Y. Chang, 1996, Phys. Rev. Lett. {\bf 76}, 1695.

\harvarditem{Choi, Tsui and Alavi}{1987}{Choi87}
Choi, K.K., D.C. Tsui, and K. Alavi, 1987, Phys. Rev. B {\bf 36}, 7551.

\harvarditem{Clarke~{\em et~al.}}{1995}{Clarke95}
Clarke, R.M., I.H. Chan, C.M. Marcus, C.I. Duru\"{o}z, J.S. Harris, Jr., K. Campman,
and  A.C. Gossard, 1995, Phys. Rev. B {\bf 52}, 2656.

\harvarditem{Cobden~{\em et~al.}}{1998}{Cobden98}
Cobden, D.H., M. Bockrath, P.L. McEuen, A.G. Rinzler, and R.E. Smalley, 1998,
Phys.  Rev. Lett. {\bf 81}, 681.

\harvarditem{Cohen, Richter and Berkovits}{1999}{Cohen99}
Cohen, A., K. Richter, and R. Berkovits, 1999, Phys. Rev. B {\bf 60}, 2536.

\harvarditem{Cronenwett~{\em et~al.}}{1998}{Cronenwett98}
 Cronenwett, S.M., S.M. Maurer, S.R. Patel, C.M. Marcus, C.I. Duru\"{o}z, and
J.S.  Harris, Jr., 1998, Phys. Rev. Lett. {\bf 81}, 5904.

\harvarditem{Cronenwett, Oosterkamp and Kouwenhoven}{1998}{Cronenwett98'}
 Cronenwett, S.M., T.H. Oosterkamp, and L.P. Kouwenhoven, 1998, Science {\bf 281}, 540.

\harvarditem{Cronenwett~{\em et~al.}}{1997}{Cronenwett97}
 Cronenwett, S.M., S.R. Patel, C.M. Marcus, K. Campman, and A.C. Gossard,
1997,  Phys. Rev. Lett. {\bf 79}, 2312.

\harvarditem{Datta}{1995}{Datta95}
 Datta, S., 1995, {\em Electronic Transport in Mesoscopic Systems}
(Cambridge University, Cambridge).

\harvarditem{Davidovi\'{c} and Tinkham}{1999}{Davidovic99}
Davidovi\'{c}, D., and M. Tinkham, 1999, Phys. Rev. Lett. {\bf 83}, 1644.

\harvarditem{Dittrich}{1996}{Dittrich96}
Dittrich, T., 1996, Phys. Rep. {\bf 271}, 268.

\harvarditem{Doron, Smilansky and Frenkel}{1990}{Doron90}
Doron, E., U. Smilansky, and A. Frenkel, 1990, Phys. Rev. Lett. {\bf 65}, 3072.
 
\harvarditem{Doron, Smilansky and Frenkel}{1991}{Doron91}
Doron, E., U. Smilansky, and A. Frenkel, 1991, Physica D {\bf 50}, 367.

\harvarditem{Dupuis and Montambaux}{1991}{Dupuis91}
 Dupuis, N., and G. Montambaux, 1991, Phys. Rev. B {\bf 43}, 14390.

\harvarditem{Dyugaev, Vagner, and Wyde}{2000}{Dyugaev00}
Dyugaev, A.M., I.D. Vagner,and P. Wyder, 2000, e-print cond-mat/0005005.

\harvarditem{Dyson}{1962a}{Dyson62a}
 Dyson, F.J., 1962a, J. Math. Phys. {\bf 3}, 140.

\harvarditem{Dyson}{1962b}{Dyson62b}
 Dyson, F.J., 1962b, J. Math. Phys. {\bf 3},  157, 166. 

\harvarditem{Dyson}{1962c}{Dyson62c}
 Dyson, F.J., 1962c, J. Math. Phys. {\bf 3}, 1191.

\harvarditem{Dyson}{1962d}{Dyson62d}
 Dyson, F.J., 1962d, J. Math. Phys. {\bf 3}, 1199.

\harvarditem{Dyson}{1970}{Dyson70}
Dyson, F.J., 1970, Commun. Math. Phys. {\bf 19}, 235.

\harvarditem{Dyson and Mehta}{1963}{Dyson63}
Dyson, F.J.,and M.L. Mehta, 1963, J. Math. Phys. {\bf 4}, 701.

\harvarditem{Echternach, Gersheson and Bozler}{1993}{Echternach93}
Echternach, P.M., M.E.  Gersheson, and H.M. Bozler, 1993, Phys. Rev. B {\bf
48},  11516.

\harvarditem{Economou and Soukoulis}{1981}{Economou81}
Economou, E.N., and C.M. Soukoulis, 1981, Phys. Rev. Lett. {\bf 46}, 618.

\harvarditem{Edwards and Thouless}{1972}{Edwards72}
Edwards, J.T., and D.J. Thouless, 1972, J. Phys. C {\bf 5}, 807.

\harvarditem{Efetov}{1983}{Efetov83}
 Efetov, K.B., 1983, Adv. Phys. {\bf 32}, 53.

\harvarditem{Efetov}{1995}{Efetov95}
  Efetov, K.B., 1995, Phys. Rev. Lett. {\bf 74}, 2299.

\harvarditem{Efetov}{1997}{Efetov97}
 Efetov, K.B., 1997, {\em Supersymmetry in Disorder and Chaos} 
(Cambridge University, Cambridge).

\harvarditem{Ericson}{1960}{Ericson60}
 Ericson, T., 1960, Phys. Rev. Lett. {\bf 5}, 430.

\harvarditem{Ericson}{1963}{Ericson63}
 Ericson, T., 1963, Ann. Phys. (N.Y.) {\bf 23}, 390.

\harvarditem{Ericson and Mayer--Kuckuk}{1966}{Ericson66}
Ericson, T., and T. Mayer--Kuckuk, 1966, Annu. Rev. Nucl.
Sci. {\bf 16}, 183.

\harvarditem{Fal'ko and Efetov}{1994}{Falko94}
 Fal'ko, V.I., and K.B. Efetov, 1994, Phys. Rev. B {\bf 50},
11267.

\harvarditem{Fal'ko and Efetov}{1996}{Falko96}
 Fal'ko, V.I., and K.B. Efetov, 1996, Phys. Rev. Lett. {\bf 77}, 912.

\harvarditem{Ferry and Goodnick}{1997}{Ferry97}
Ferry, D.K., and S.M. Goodnick, 1997, {\em Transport in Nanostructures}
(Cambridge University, Cambridge).

\harvarditem{Feshbach}{1992}{Feshbach92}
Feshbach, H., 1992, {\em Theoretical Nuclear Physics: Nuclear Reactions}
(Wiley,  New York).

\harvarditem{Fetter and Walecka}{1971}{Fetter71}
Fetter, A.L., and J.D. Walecka, 1971, {\em Quantum Theory of Many-Particle
Systems}  (McGraw-Hill, New York).

\harvarditem{Fisher and Lee}{1981}{Fisher81}
Fisher, D.S., and P.A. Lee, 1981, Phys. Rev. B {\bf 23}, 6851.

\harvarditem{Flambaum~{\em et~al.}}{1994}{Flambaum94}
Flambaum, V.V., A.A. Gribakina, G.F. Gribakin and M.G. Kozlov, 1994, Phys.
Rev.  A {\bf 50}, 267.

\harvarditem{Flambaum, Gribakin and Izrailev}{1996}{Flambaum96}
 Flambaum, V.V., G.F. Gribakin, and F.M. Izrailev, 1996, Phys.
Rev. E {\bf 53}, 5729.

\harvarditem{Flensberg}{1993,1994}{Flensberg93}
Flensberg, K., 1993, Phys. Rev. B {\bf 48}, 11156;
1994, Physica B {\bf 203}, 432.

\harvarditem{Folk~{\em et~al.}}{1996}{Folk96}
 Folk, J.A.,  S.R. Patel, S.F. Godijn, A.G. Huibers,
S.M. Cronenwett, C.M. Marcus, K. Campman, and A.C. Gossard, 1996,
Phys. Rev. Lett. {\bf 76}, 1699.

\harvarditem{Folk~{\em et~al.}}{2000}{Folk00}
 Folk, J.A., S.R. Patel, C.M. Marcus, C.I. Duru\"{o}z, and J.S. Harris, Jr., 2000, ``Decoherence in nearly-isolated quantum dots,'' e-print cond-mat/0008052.

\harvarditem{Frahm}{1995}{Frahm95}
 Frahm, K.M., 1995, Europhys.\ Lett.\ {\bf 30}, 457.

\harvarditem{French and Kota}{1982}{French82}
  French, J.B., and V.K.B. Kota, 1982, Ann. Rev. Nucl. Part. Sci. {\bf 32},
35.

\harvarditem{French~{\em et~al.}}{1985}{French85}
French, J.B., V.K.B. Kota, A Pandey and
 S. Tomsovic, 1985, Phys. Rev. Lett. {\bf 54}, 2313.

\harvarditem{French {\em et al.}}{1988}{French88}
French, J.B., V.K.B. Kota, A Pandey and
 S. Tomsovic, 1988, Ann. Phys. (N.Y.) {\bf 181}, 198.

\harvarditem{French and Wong}{1970}{French70}
French, J.B.,  and S.S.M. Wong, 1970, Phys. Lett. B {\bf 33}, 449.

\harvarditem{Friedman and Mello}{1985}{Friedman85}
Friedman, W.A., and P.A. Mello, 1985,  Ann. Phys. (N.Y.) {\bf 161}, 276.

\harvarditem{Fulton and  Dolan}{1987}{Fulton87}
Fulton, T.A., and G.J. Dolan, 1987, Phys. Rev. Lett. {\bf 59}, 109.

\harvarditem{Furusaki and Matveev}{1995a}{Furusaki95}
Furusaki, A., and K.A. Matveev, 1995a, Phys. Rev. Lett. {\bf 75}, 709.

\harvarditem{Furusaki and Matveev}{1995b}{Furusaki95'}
Furusaki, A., and K.A. Matveev, 1995b, Phys.
Rev.  B {\bf 52}, 16676.

\harvarditem{Fyodorov and Sommers}{1996a}{Fyodorov96a}
 Fyodorov, Y.V., and H.-J. Sommers, 1996a, Phys. Rev. Lett. {\bf 76}, 4709.

\harvarditem{Fyodorov and Sommers}{1996b}{Fyodorov96b}
 Fyodorov, Y.V., and H.-J. Sommers, 1996b, 
 JETP Lett. {\bf 63}, 1026.

\harvarditem{Fyodorov and Sommers}{1997}{Fyodorov97}
Fyodorov, Y.V., and H.-J. Sommers, 1997, J. Math. Phys. {\bf 38}, 1918.

\harvarditem{Garg~{\em et~al.}}{1964}{Garg64}
Garg, J.B.,  J. Rainwater, J.S. Petersen and W.W. Havens,
Jr.,  1964, Phys. Rev. {\bf 134}, B985.

\harvarditem{Gaudin}{1961}{Gaudin61}
Gaudin, M., 1961, Nucl. Phys. {\bf 25}, 447.

\harvarditem{Gershenson}{1999}{Gershenson99}
Gershenson, M.E., 1999, Ann. Phys. {\bf 8}, 559.

\harvarditem{Giaever and Zeller}{1968}{Giaever68}
Giaever, I., and H.R. Zeller, 1968, Phys. Rev. Lett. {\bf 20}, 1504.

\harvarditem{Giannoni, Voros and Zinn-Justin}{1991}{Giannoni91}
 Giannoni, M.-J., A. Voros, and J. Zinn-Justin, 1991, eds., {\em Chaos and Quantum Physics} (North-Holland, Amsterdam).

\harvarditem{Glazman and Matveev}{1988}{Glazman88}
Glazman, L.I., and K.A. Matveev, 1988,
 JETP Lett. {\bf 48}, 445.

\harvarditem{Glazman and Matveev}{1990a}{Glazman90a}
Glazman, L.I., and K.A. Matveev, 1990a,
 JETP Lett. {\bf 51}, 484.

\harvarditem{Glazman and Matveev}{1990b}{Glazman90b}
Glazman, L.I., and K.A. Matveev, 1990b, Sov. Phys. JETP {\bf 71}, 1031.

\harvarditem{Glazman and Raikh}{1988}{Glazman88}
Glazman, L.I., and M.E. Raikh , 1988,  JETP Lett. {\bf 47}, 452.

\harvarditem{Glazman and Shekther}{1989}{Glazman89}
Glazman, L.I., and  R.I. Shekther, 1989, J. Phys.: Condens. Matter {\bf 1}, 5811.

\harvarditem{Goldberg~{\em et~al.}}{1993}{Goldberg91}
 Goldberg, J., U. Smilansky, M.V. Berry, W. Schweizer,
G. Wunner, and G. Zeller, 1991, Nonlinearity {\bf 4}, 1.

\harvarditem{Goldhaber-Gordon~{\em et~al.}}{1998}{Goldhaber98}
Goldhaber-Gordon, D., H. Shtrikman, D. Mahalu, D. Abusch-Magder, U. Meirav, and M.A. Kastner, 1998, Nature (London) {\bf 391}, 156.

\harvarditem{Gopar, Mello and B\"{u}ttiker}{1996}{Gopar96}
Gopar, V.A., P.A. Mello, and M. B\"{u}ttiker, 1996, Phys. Rev. Lett. {\bf 77},
3005.

\harvarditem{Gor'kov and  Eliashberg}{1965}{Gorkov65}
 Gor'kov, L.P., and G.M. Eliashberg, 1965,
Sov.  Phys. JETP {\bf 21}, 940.

\harvarditem{Gor'kov, Larkin and Khmelnitskii}{1979}{Gorkov79}
 Gor'kov, L.P., A.I. Larkin,  and D.E. Khmelnitskii, 1979,
 JETP Lett. {\bf 30}, 228.

\harvarditem{Gossiaux, Pluha\u{r} and Weidenm\"{u}ller}{1998}{Gossiaux98}
 Gossiaux, P.B., Z. Pluha\u{r}, and H. A. Weidenm\"{u}ller, 1998, Ann. Phys. (N.Y.) {\bf 268}, 273.

\harvarditem{Greenwood}{1958}{Greenwood58}
Greenwood, D.A., 1958, Proc. Phys. Soc. London {\bf 71}, 585.

\harvarditem{Guhr, M\"{u}ller-Groeling  and Weidenm\"{u}ller}{1998}{Guhr98}
Guhr, T., A. M\"{u}ller-Groeling,  and H.A. Weidenm\"{u}ller, 1998, Phys. Rep. 
{\bf  299}, 190.

\harvarditem{Gutzwiller}{1967}{Gutzwiller67}
 Gutzwiller, M.C., 1967, J. Math. Phys. {\bf 8}, 1979.

\harvarditem{Gutzwiller}{1969}{Gutzwiller69}
 Gutzwiller, M.C.,
1969, J. Math. Phys. {\bf 10}, 1004.

\harvarditem{Gutzwiller}{1970}{Gutzwiller70}
 Gutzwiller, M.C., 
1970, J. Math. Phys. {\bf 11}, 1791.

\harvarditem{Gutzwiller}{1971}{Gutzwiller71}
 Gutzwiller, M.C., 1971, J. Math. Phys. {\bf 12}, 343.

\harvarditem{Gutzwiller}{1990}{Gutzwiller90}
 Gutzwiller, M.C., 1990,  {\em Chaos in Classical and Quantum Mechanics}
(Springer,  New York).

\harvarditem{Hackenbroich, Heiss and Weidenm\"uller}{1997}{Hackenbroich97}
  Hackenbroich, G., W.D. Heiss and H.A. Weidenm\"uller, 1997,
Phys. Rev. Lett. {\bf 79}, 127.

\harvarditem{Hackenbroich and Weidenm\"uller}{1995}{Hackenbroich95}
  Hackenbroich, G., and H.A. Weidenm\"uller, 1995,
Phys. Rev. Lett. {\bf 74}, 4118.

\harvarditem{Hannay and Ozorio de Almeida}{1984}{Hannay84}
Hannay, J.H., and A.M. Ozorio de Almeida, 1984, J. Phys. A {\bf 17}, 3429.

\harvarditem{Haq, Pandey and Bohigas}{1982}{Haq82}
Haq, R.U., A. Pandey, and O. Bohigas, 1982, Phys. Rev. Lett. {\bf 48}, 1086.

\harvarditem{Hauser and Feshbach}{1952}{Hauser52}
Hauser, W.,  and H. Feshbach, 1952, Phys. Rev. {\bf 87}, 366.

\harvarditem{Hortikar and Srednicki}{1998}{Hortikar98}
Hortikar, S., and M. Srednicki, 1998, Phys. Rev. Lett. {\bf 80}, 1646.

\harvarditem{Hua}{1963}{Hua63}
 Hua, L.K., 1963, {\em Harmonic Analysis of Functions of Several
Complex Variables in the Classical Domains} (American Mathematical Society,
Providence,  RI).

\harvarditem{Hubbard}{1963}{Hubbard63}
Hubbard, J., 1963, Proc. R. Soc. London, Ser. A {\bf 276},
238.

\harvarditem{Huibers~{et~al.}}{1998b}{Huibers98}
Huibers, A.G., M. Switkes, C.M. Marcus, P.W. Brouwer, C.I. Duru\"{o}z, and J.S.
Harris, Jr., 1998, Phys. Rev. Lett. {\bf 81}, 1917.

\harvarditem{Huibers~{et~al.}}{1998a}{Huibers98'}
Huibers, A.G., M. Switkes, C.M. Marcus, K. Campman, and A.C. Gossard, 1998,
Phys.  Rev. Lett. {\bf 81}, 200.

\harvarditem{Iida, Weidenm\"{u}ller and Zuk}{1990}{Iida90}
Iida, S., H.A. Weidenm\"{u}ller, and J.A. Zuk, 1990a, Phys.  Rev. Lett. {\bf 64}, 583.

\harvarditem{Iida, Weidenm\"{u}ller and Zuk}{1990}{Iida90'}
Iida, S., H.A. Weidenm\"{u}ller, and J. A. Zuk, 1990b,  Ann. Phys. (N.Y.) {\bf 200}, 219.

\harvarditem{Imry}{1986a}{Imry86a}
Imry, Y., 1986a, Europhys. Lett. {\bf 1}, 249.

\harvarditem{Imry}{1986b}{Imry86b}
Imry, Y., 1986b, in {\em Directions in Condensed Matter Physics}, vol. I, edited by G.
Grinstein  and G. Mazenko (World Scientific, Singapore).

\harvarditem{Imry}{1996}{Imry96}
 Imry, Y., 1996, {\em Introduction to Mesoscopic Physics} (Oxford
University, Oxford).

\harvarditem{Imry and Landauer}{1999}{Imry99}
Imry, Y., and R. Landauer, 1999, Rev. Mod. Phys. {\bf 71}, S306.

\harvarditem{Jacquod and Shepelyansky}{1997}{Jacquod97}
Jacquod, Ph.,  and D.L. Shepelyansky, 1997, Phys. Rev. Lett. {\bf 79}, 1837.

\harvarditem{Jacquod and Stone}{2000}{Jacquod00}
Jacquod, Ph., and A.D. Stone, 2000, Phys. Rev. Lett. {\bf 84}, 3938.

\harvarditem{Jalabert, Baranger and Stone}{1990}{Jalabert90}
Jalabert, R.A.,  H.U. Baranger, and A.D. Stone, 1990, Phys. Rev.
   Lett. {\bf 65}, 2442.

\harvarditem{Jalabert, Pichard, and Beenakker}{1994}{Jalabert94}
 Jalabert, R.
A.,  J.-L. Pichard, and C. W. J. Beenakker, 1994,
Europhys. Lett. {\bf 27}, 255.

\harvarditem{Jalabert, Stone and Alhassid}{1992}{Jalabert92}
 Jalabert, R.A., A.D. Stone, and Y. Alhassid, 1992, Phys. Rev. Lett. {\bf 68},
3468.

\harvarditem{Jensen}{1991}{Jensen91}
 Jensen, R.V., 1991, Chaos {\bf 1}, 101.

\harvarditem{Johnson~{\em et~al.}}{1992}{Johnson92}
Johnson, A.T., L.P. Kouwenhoven, W. de Jong, N.C. van der Vaart, C.J.P.M.
Harmans,  and C.T. Foxon, 1992, Phys. Rev. Lett. {\bf 69}, 1592.

\harvarditem{Johnson, Bertsch and Dean}{1998}{Johnson98}
 Johnson, C.W., G.F. Bertsch,  and D.J. Dean, 1998, Phys. Rev. Lett. {\bf 80}, 2749.

\harvarditem{Kamenev and Gefen}{1997}{Kamenev97}
 Kamenev, A.,  and Y. Gefen, 1997, Chaos, Solitons, and Fractals  {\bf 8},
1229.

\harvarditem{Kaminski, Aleiner and Glazman}{1998}{Kaminski98}
Kaminski, A., I.L. Aleiner, and L.I. Glazman, 1998, Phys. Rev.Lett. {\bf 81},
685.

\harvarditem{Kaminski and Glazman}{1999}{Kaminski99}
Kaminski, A., and L.I. Glazman, 1999, Phys. Rev. B {\bf 59}, 9798.

\harvarditem{Kaminski and Glazman}{2000}{Kaminski00}
Kaminski, A., and L.I. Glazman, 2000, Phys. Rev. B {\bf 61}, 15927.

\harvarditem{Kanzieper and Freilikher}{1996}{Kanzieper96}
Kanzepier, E., and V. Freilikher, 1996, Phys. Rev. B {\bf 54}, 8737.

\harvarditem{Kaplan and Heller}{1998}{Kaplan98}
Kaplan, L., and E.J. Heller, 1998, Phys. Rev. Lett. {\bf 80}, 2582.

\harvarditem{Kastner}{1992}{Kastner92}
 Kastner, M.A., 1992, Rev. Mod. Phys. {\bf 64}, 849.

\harvarditem{Kastner}{1993}{Kastner93}
Kastner, M.A., 1993, Phys. Today {\bf 46} (1), 24.

\harvarditem{Keller {\em et al.}}{1994}{Keller94}
 Keller, M.W., O. Millo, A. Mittal, D.E. Prober, and R.N. Sacks,
1994, Surf. Sci. {\bf 305}, 501.

\harvarditem{Keller~{\em et~al.}}{1996}{Keller96}
 Keller, M.W., A. Mittal, J.W. Sleight, R.G. Wheeler, D.E.
Prober, R.N. Sacks, and H. Shtrikmann, 1996, Phys. Rev. B {\bf 53}, 1693.

\harvarditem{Ketzmerick}{1996}{Ketzmerick96}
Ketzmerick, R., 1996, Phys. Rev. B {\bf 54}, 10841.

\harvarditem{Khmelnitskii}{1984}{Khmelnitskii84}
 Khmelnitskii, D.E., 1984, Physica B {\bf 126}, 235.

\harvarditem{Kittel}{1987}{Kittel87}
Kittel, C., 1987, {\em Quantum Theory of Solids}  (Wiley, New York).

\harvarditem{Kogan and Kaveh}{1995}{Kogan95}
Kogan, E., and M. Kaveh, 1995, Phys. Rev. B {\bf 51}, 16400.

\harvarditem{Kondo}{1964}{Kondo64}
Kondo, J., 1964, Prog. Theor. Phys. {\bf 32}, 37.

\harvarditem{Koopmans}{1934}{Koopmans34}
Koopmans, T., 1934, Physica (Amsterdam) {\bf 1}, 104.

\harvarditem{Koulakov, Pikus, and Shklovskii}{1997}{Koulakov97}
Koulakov, A.A.,  F.G. Pikus, and B.I. Shklovskii, 1997,
 Phys. Rev. B {\bf~55}, 9223.

\harvarditem{Koulakov and Shklovskii}{1998}{Koulakov98}
Koulakov, A.A., and B.I. Shklovskii, 1998,
 Phys. Rev. B {\bf~57}, 2352.

\harvarditem{Kouwenhoven~{\em et~al.}}{1997}{Kouwenhoven97b}
 Kouwenhoven, L.P., C.M. Marcus, P.L. Mceuen, S. Tarucha, R.M. Wetervelt, and
N.S.  Wingreen, 1997, in  {\em Mesoscopic
electron transport}, NATO Advanced Study Institute, Series E, No. 345, edited by  L. L. Sohn, L. P. Kouwenhoven, and G. Schoen
(Kluwer, Dordrecht).

\harvarditem{Kouwenhoven and McEuen}{1998}{Kouwenhoven98}
 Kouwenhoven, L.P., and P.L. McEuen, 1998, in  {\em Nanotechnology},
 edited by G. Timp (AIP, New York), p.~471.

\harvarditem{Kouwenhoven~{\em et~al.}}{1997}{Kouwenhoven97}
Kouwenhoven, L.P., T.H. Oosterkamp, M.W.S. Danoesastro, M. Eto, D.G. Austing,
T.  Honda, and S. Tarucha, 1997, Science {\bf 278}, 1788.

\harvarditem{Krieger and Porter}{1963}{Krieger63}
 Krieger, T.J.,  and C.E. Porter, 1963, J. Math. Phys. {\bf 4}, 1272.

\harvarditem{Kubo}{1957}{Kubo57}
Kubo, R., 1957, J. Phys. Soc. Jpn. {\bf 12}, 570.

\harvarditem{Kulik and Shekhter}{1975}{Kulik75}
Kulik, I.O., and R.I. Shekhter, 1975,
Sov.  Phys. JETP {\bf 41}, 308.

\harvarditem{Kurdak~{\em et~al.}}{1992}{Kurdak92}
Kurdak, C., A.M. Chang, A. Chin, and T.Y. Chang, 1992, Phys. Rev. B {\bf 46},
6846.

\harvarditem{Kurland, Aleiner and Altshuler}{2000}{Kurland00a}
Kurland, I.L., I.L. Aleiner, and B.L. Altshuler, 2000, ``Mesoscopic Stoner instability,'' e-print cond-mat/0004205.

\harvarditem{Kurland, Berkovits and Altshuler}{2000}{Kurland00b}
Kurland, I.L., R. Berkovits, and B.L. Altshuler, 2000, ``Conductance peak motion due to a magnetic field in weakly coupled chaotic quantum dots,'' e-print cond-mat/0005424.

\harvarditem{Landauer}{1957}{Landauer57}
Landauer, R.,
1957, IBM J. Res. Dev. {\bf 1}, 223.

\harvarditem{Landauer}{1970}{Landauer70}
Landauer, R.,  1970,
Philos. Mag. {\bf 21}, 863.

\harvarditem{Lane and Thomas}{1958}{Lane58}
Lane, A.M.,  and R.G. Thomas, 1958, Rev. Mod. Phys. {\bf 30}, 257.

\harvarditem{Langer and Neal}{1966}{Langer66}
Langer, J.S.,  and T. Neal, 1966, Phys. Rev. Lett. {\bf 16}, 1984.

\harvarditem{Lee and Ramakrishnan}{1985}{Lee85'}
Lee, P.A., and T.V. Ramakrishnan, 1985, Rev. Mod. Phys. {\bf 57},
287.

\harvarditem{Lee and Stone}{1985}{Lee85}
Lee, P.A., and A.D. Stone, 1985, Phys. Rev. Lett. {\bf 55}, 1622.

\harvarditem{Lee, Stone and Fukuyuma}{1987}{Lee87}
Lee, P.A., A.D. Stone, and H. Fukuyama, 1987, Phys.\ Rev.\ B
{\bf 35}, 1039.

\harvarditem{Levit and Orgad}{1999}{Levit99}
Levit, S., and D. Orgad, 1999, Phys. Rev. B {\bf 60}, 5549.

\harvarditem{Leyronas, Tworzydlo and Beenakker}{1999}{Leyronas99}
Leyronas, X., J. Tworzydlo, and C.W.J. Beenakker, 1999, Phys. Rev. Lett. {\bf 82}, 4894.

\harvarditem{Likharev and Zorin}{1985}{Likharev85}
Likharev, K.K., and A.B. Zorin, 1985, J. Low Temp. Phys. {\bf 59}, 347.

\harvarditem{Lin and Giordano}{1986}{Lin86}
Lin, J.J., and N. Giordano, 1986, Phys. Rev. B {\bf 33}, 1159.

\harvarditem{L\"{u}scher~{\em et~al.}}{2000}{Luscher00}
L\"{u}scher, S., T. Heinzel, K. Ensslin, W. Wegscheider, M. Bichler, 2000, ``Signatures of spin pairing in a quantum dot in the Coulomb-blockade regime,'' cond-mat/0002226.

\harvarditem{Mahaux and Weidenm\"{u}ller}{1969}{Mahaux69}
 Mahaux, C., and H.A. Weidenm\"{u}ller, 1969, {\em Shell-Model
Approach to Nuclear Reactions} (North-Holland, Amsterdam).

\harvarditem{Marcus {\em et al.}}{1997}{Marcus97}
 Marcus, C.M., S. R. Patel, A. G. Huibers, S.M. Cronenwett, M. Switkes, I.H.  Chan, R.M. Clarke, J.A. Folk, S.F. Godijn, K. Campman, and A.C.
Gossard,  1997, Chaos, Solitons, and Fractals {\bf 8}, 1261.

\harvarditem{Marcus~{\em et~al.}}{1992}{Marcus92}
 Marcus, C.M., A.J. Rimberg, R.M. Westervelt, P.F. Hopkins, and
A.C. Gossard, 1992, Phys. Rev. Lett. {\bf 69}, 506.

 \harvarditem{Marcus~{\em et~al.}}{1993}{Marcus93}
  Marcus, C.M., R.M. Westervelt, P.F. Hopkins, and A.C.
  Gossard, 1993a, Phys. Rev B {\bf 48}, 2460.

\harvarditem{Marcus {\em et al.}}{1993}{Marcus93'}
 Marcus, C.M., R.M. Westervelt, P.F. Hopkins, and
A. C. Gossard, 1993b, Chaos {\bf 3}, 643.

 \harvarditem{Marcus~{\em et~al.}}{1994}{Marcus94}
  Marcus, C.M., R.M. Westervelt, P.F. Hopkins and A.C.
  Gossard,  1994,
Surf. Sci. {\bf 305}, 480.

\harvarditem{Matveev}{1991}{Matveev91}
 Matveev, K.A., 1991, Sov. Phys. JETP {\bf 72}, 892.

\harvarditem{Matveev}{1995}{Matveev95}
 Matveev, K.A., 1995, Phys. Rev. B {\bf 51}, 1743.

\harvarditem{Maurer {\em et al.}}{1999}{Maurer99}
Maurer, S.M., S.R. Patel, C.M. Marcus, C.I. Duru\"{o}z, and
J.S.  Harris, Jr., 1999, Phys. Rev. Lett. {\bf 83}, 1403.

\harvarditem{McEuen}{1998}{McEuen98}
McEuen, P.L., 1998, Nature (London) {\bf 393}, 6680.

\harvarditem{McEuen {\em al.}}{1993}{McEuen93}
 McEuen, P.L., N.S. Wingreen, E.B. Foxman, J. Kinaret, U. Meirav, M.A.
Kastner,  and Y. Meir, 1993, Physica B {\bf 189}, 70.

\harvarditem{Mehta}{1960}{Mehta60}
Mehta, M.L., 1960, Nucl. Phys. {\bf 18}, 395.

\harvarditem{Mehta}{1971}{Mehta71}
Mehta, M.L., 1971, Commun. Math. Phys. {\bf 20}, 245.

\harvarditem{Mehta}{1991}{Mehta91}
Mehta, M.L., 1991, {\em Random Matrices},
   second edition (Academic, New York).

\harvarditem{Mehta and Gaudin}{1960}{Mehta60'}
M.L. Mehta, and M. Gaudin, 1960,
Nucl.  Phys. {\bf 18}, 420.

\harvarditem{Mehta and Pandey}{1983}{Mehta83}
 Mehta, M.L., and A. Pandey, 1983, 
J. Phys. A {\bf 16}  2655; L601.

\harvarditem{Meir and Wingreen}{1992}{Meir92}
 Meir, Y., and  N. S. Wingreen, 1992, Phys. Rev. Lett.
{\bf 68}, 2512.

\harvarditem{Meir, Wingreen, and Lee}{1991}{Meir91}
 Meir, Y., N.S. Wingreen, and P.A. Lee, 1991, Phys. Rev. Lett.
{\bf 66}, 3048.

\harvarditem{Meir, Wingreen, and Lee}{1993}{Meir93}
 Meir, Y., N.S. Wingreen, and P.A. Lee, 1993, Phys. Rev. Lett.
{\bf 70}, 2601.

\harvarditem{Meirav and Foxman}{1995}{Meirav95}
 Meirav, U., and E.B. Foxman, 1995, Semicond. Sci. Technol.
 {\bf 10}, 255.

\harvarditem{Meirav, Kastner and Wind}{1990}{Meirav90}
Meirav, U., M.A. Kastner and S.J. Wind, 1990, Phys. Rev. Lett. {\bf 65}, 771.

\harvarditem{Mejia-Monasterio~{\em et~al.}}{1998}{Mejia98}
Mejia-Monasterio, C.,  J. Richert, T. Rupp,
 and H.A. Weidenm\"{u}ller, 1998, Phys. Rev. Lett.
{\bf 81}, 5189.

\harvarditem{Mello}{1995}{Mello95}
 Mello, P.A., 1995, in {\em Mesoscopic Quantum Physics}, edited by
E. Akkermans, G. Montambaux, J.-L. Pichard, and J. Zinn-Justin
(North-Holland,  Amsterdam), p. 435.

\harvarditem{Mello and Baranger}{1995}{Mello95'}
Mello, P.A., and H.U. Baranger, 1995, Physica A {\bf 220}, 15.

\harvarditem{Mello and Baranger}{1999}{Mello99}
Mello, P.A., and H.U. Baranger, 1999, Waves Random Media {\bf 9}, 105.

\harvarditem{Mello, Pereyra, and  Seligman}{1985}{Mello85}
Mello, P.A., P. Pereyra, and T.H. Seligman, 1985, Ann. Phys.
(N.Y.) {\bf 161}, 254.

\harvarditem{Mirlin}{1997}{Mirlin97}
Mirlin, A.D., 1997, in {\em Supersymmetry and Trace Formulae}, Proceedings of the NATO Advanced Study Institute, Series B, No. 370, edited by I.V. Lerner, V.P. Keating, and D.E. Khmelnitskii (Kluwer Academic, New York).

\harvarditem{Mirlin}{2000}{Mirlin00}
Mirlin, A.D., 2000, Phys. Rep. {\bf 326}, 260.

\harvarditem{Mitchell, Alhassid and Kusnezov}{1996}{Mitchell96}
Mitchell, D., Y. Alhassid, and D. Kusnezov, 1996, Phys. Lett. A {\bf 215}, 21.

\harvarditem{P. Mohanty, Jariwala and Webb}{1995}{Mohanty97}
 Mohanty, P., E.M.Q. Jariwala, and R.A. Webb, 1997,
Phys. Rev. Lett. {\bf 78}, 3366.

\harvarditem{Montambaux}{1997}{Montambaux97}
Montambaux, G., 1997, in  {\em Quantum Fluctuations}, Proceedings of the Les
Houches  Summer  School, Session LXIII, edited by S. Reynaud, E. Giacobino,  and J.
Zinn-Justin (North-Holland,  Amsterdam), p. 387.

\harvarditem{Montambaux~{\em et~al.}}{1993}{Montambaux93}
Montambaux, G., D. Poilblanc, J. Bellissard, and C. Sire, 1993, Phys. Rev.
Lett.  {\bf 70}, 497.

\harvarditem{Mucciolo, Prigodin and Altshuler}{1995}{Mucciolo95}
 Mucciolo, E.R., V.N. Prigodin, and B.L. Altshuler, 1995, 
Phys. Rev. B {\bf 51}, 1714.

\harvarditem{Muzykantskii and Khmelnitskii}{1995}{Muzykantskii95}
Muzykantskii,  B.A., and D.E. Khmelnitskii, 1995,
JETP Lett. {\bf 62}, 76.

\harvarditem{Narayan and Shastry}{1993}{Narayan93}
Narayan, O., and B.S. Shastry, 1993, Phys. Rev. Lett. {\bf 71}, 2106.

\harvarditem{Narimanov~{\em et~al.}}{1999}{Narimanov99}
Narimanov, E.E., N.R. Cerruti, H.U. Baranger, and S. Tomsovic, 1999,
 Phys. Rev. Lett. {\bf 83}, 2640.

\harvarditem{Ng and Lee}{1988}{Ng88}
Ng, T.K., and P.A. Lee, 1988, Phys. Rev. Lett. {\bf 61}, 1768.

\harvarditem{Oosterkamp~{\em et~al.}}{1997}{Oosterkamp97}
Oosterkamp, T.H.,  L.P. Kouwenhoven, A.E.A. Koolen, N.C. van der Vaart, and
C.J.P.M.  Harmans, 1997, Phys. Rev. Lett. {\bf 78}, 1536.

\harvarditem{Oreg and Gefen}{1997}{Oreg97}
Oreg, Y, and Y. Gefen, 1997, Phys. Rev. B {\bf 55}, 13726.

\harvarditem{Ormand~{\em et~al.}}{1994}{Ormand94}
 Ormand, W.E., D.J. Dean, C.W. Johnson, G.H. Lang, 
 and S.E. Koonin, 1994, Phys. Rev. C {\bf  49}, 1422.

\harvarditem{Pandey}{1979}{Pandey79}
Pandey, A., 1979, Ann. Phys. (N.Y.) {\bf 119}, 170.

\harvarditem{Pandey and Mehta}{1983}{Pandey83}
 Pandey, A., and M.L. Mehta, 1983, Commun. Math.
Phys.  {\bf 87}, 49.

\harvarditem{Patel~{et~al.}}{1998a}{Patel98}
Patel, S.R., S.M. Cronenwett, D.R. Stewart, C. M. Marcus,
C. I. Duru\"{o}z, J. S. Harris, Jr., K. Campman, and A.C. Gossard, 1998,
Phys. Rev. Lett. {\bf 80}, 4522.

\harvarditem{Patel~{et~al.}}{1998b}{Patel98'}
Patel, S.R., D.R. Stewart, C. M. Marcus,
M. G\"{o}k\c{c}eda\u{g}, Y. Alhassid, A. D. Stone,
C. I. Duru\"{o}z, and J. S. Harris, Jr., 1998,
Phys. Rev. Lett. {\bf 81}, 5900.

\harvarditem{Pereyra and Mello}{1983}{Pereyra83}
Pereyra, P., and P.A. Mello, 1983, J. Phys. A {\bf 16}, 237.

\harvarditem{Pichaureau and Jalabert}{1999}{Pichaureau99}
Pichaureau, P., and R.A. Jalabert, 1999, Eur. Phys. J. B {\bf 9}, 299.

\harvarditem{Pines and Nozi\`{e}res}{1966}{Pines66}
Pines, D.,  and P. Nozi\`{e}res, 1966, {\em The Theory of Quantum Liquids}, Vol.
1 (Addison-Wesley, New York).

\harvarditem{Pluha\u{r} and Weidenm\"{u}ller}{1999}{Pluhar99}
 Pluha\u{r}, Z., and H.A. Weidenm\"{u}ller, 1999, Ann. Phys. (N.Y.) {\bf 272}, 295.

\harvarditem{Pluha\u{r}~{\em et~al.}}{1994}{Pluhar94}
 Pluha\u{r}, Z., H.A. Weidenm\"{u}ller, J. A. Zuk, and C. H.
Lewenkopf, 1994, Phys. Rev. Lett. {\bf 73}, 2115.

\harvarditem{Pluha\u{r} {\em et al.}}{1995}{Pluhar95}
 Pluha\u{r}, Z., H.A. Weidenm\"{u}ller, J.A. Zuk, C.H. Lewenkopf,
and F.J. Wegner, 1995, Ann. Phys. (N.Y.) {\bf 243}, 1.

\harvarditem{Porath and Millo}{1996}{Porath96}
 Porath, D., and O. Millo, 1996, J. Appl. Phys. {\bf 81}, 2241.

\harvarditem{Porter}{1965}{Porter65}
Porter, C.E., 1965, {\em Statistical Theory of Spectra:
  Fluctuations} (Academic, New York).

\harvarditem{Porter and Rosenzweig}{1960}{Porter60}
Porter, C.E.,  and N. Rosenzweig, 1960,  Phys. Rev. {\bf 120}, 1698.

\harvarditem{Prigodin}{1995}{Prigodin95}
 Prigodin, V.N., 1995, Phys. Rev. Lett. {\bf 74}, 1566.

\harvarditem{Prigodin, Efetov, and Iida}{1993}{Prigodin93}
Prigodin, V.N., K.B. Efetov, and S. Iida, 1993,
Phys. Rev. Lett. {\bf 71}, 1230.

\harvarditem{Ralph, Black and Tinkham}{1997}{Ralph97}
Ralph, D.C., C.T. Black, and M. Tinkham, 1997, Phys. Rev. Lett.
{\bf 78}, 4087.

\harvarditem{Reed~{\em et~al.}}{1988}{Reed88}
Reed, M.A., J.N. Randall, R.J. Aggarwall, R.J. Matyi, T.M. Moore, and A.E.
Wetsel, 1988, Phys. Rev. Lett. {\bf 60}, 535.

\harvarditem{Reichel}{1992}{Reichel92}
Reichel, L.E., 1992, {\em The Transition to Chaos in Conservative Classical Systems: Quantum Manifestations} (Springer, New York).

\harvarditem{Robnik}{1983}{Robnik83}
Robnik, M.,1983,  J. Phys. A {\bf 17}, 1049.

\harvarditem{Sachrajda~{\em et~al.}}{1998}{Sachrajda98}
Sachrajda, A.S., R. Ketzmerick, C. Gould, Y. Feng, P.J. Kelly, A. Delage and Z. Wasilewski, 1998, Phys. Rev. Lett. {\bf 80}, 1948.

\harvarditem{Schuster~{et~al.}}{1997}{Schuster97}
Schuster, R., E. Buks, M. Heiblum, D. Mahalu, V. Umansky, and H. Shtrikman,
1997,  Nature (London) {\bf 385}, 417.

\harvarditem{Scott-Thomas~{\em et~al.}}{1989}{Scott-Thomas89}
Scott-Thomas, J.H.F., S.B. Field, M.A. Kastner, H.I. Smith, and D.A.
Antonadis,  1989, Phys. Rev. Lett. {\bf 62}, 583.

\harvarditem{Shannon}{1948}{Shannon48}
Shannon, C.E., 1948, Bell Syst. Tech. J. {\bf 27}, 397; 623.

\harvarditem{Shekhter}{1972}{Shekhter72}
Shekhter, R.I., 1972,
 Sov. Phys. JETP {\bf  36}, 747.

\harvarditem{Shutenko, Aleiner and Altshuler}{2000}{Shutenko00}
Shutenko, T.A., I.L. Aleiner, and B.L. Altshuler, 2000, Phys. Rev. B {\bf 61}, 10366.

\harvarditem{Simmel~{\em et~al.}}{1999}{Simmel99}
 Simmel, F., D. Abusch-Magder, D.A. Wharam, M.A. Kastner,and J.P. Kotthaus, 1999, Phys. Rev. B {\bf 59}, R10441.

\harvarditem{Simmel, Heinzel and Wharam}{1997}{Simmel97}
 Simmel, F., T. Heinzel, and D.A. Wharam, 1997, Europhys. Lett.
{\bf 38}, 123.

\harvarditem{Simons and Altshuler}{1993a}{Simons93}
 Simons, B.D., and B.L. Altshuler, 1993a, Phys. Rev. Lett.
   {\bf 70}, 4063.

\harvarditem{Simons and Altshuler}{1993b}{Simons93'}
 Simons, B.D., and B.L. Altshuler, 1993b, Phys. Rev. B {\bf 48}, 5422.

\harvarditem{Sivan {\em et al.}}{1996}{Sivan96}
 Sivan, U., R. Berkovits, Y. Aloni, O. Prus, A. Auerbach, and  G. Ben-Yoseph,
1996,  Phys. Rev. Lett. {\bf 77}, 1123.

\harvarditem{Sivan~{\em et~al.}}{1994}{Sivan94}
 Sivan, U., F.P. Milliken, K. Milkove, S. Rishton, Y. Lee, J.M.
Hong, V. Boegli, D. Kern, and M. DeFranza, 1994,
Europhys. Lett. {\bf 25}, 605.

\harvarditem{Smilansky}{1990}{Smilansky90}
Smilansky, U., 1990, in {\em Chaos and Quantum Physics}, edited by
M.-J. Giannoni, A. Voros, and J. Zinn-Justin (North-Holland, Amsterdam), p.
371.

\harvarditem{Sommers and Iida}{1994}{Sommers94}
Sommers, H.-J., and S. Iida, 1994, Phys. Rev. E {\bf 49}, R 2513.

\harvarditem{Srednicki}{1996}{Srednicki96}
 Srednicki, M., 1996, Phys. Rev. E {\bf 54}, 954.

\harvarditem{Stewart~{\em et~al.}}{1997}{Stewart97}
Stewart, D.R., D. Sprinzak, C.M. Marcus, C.I. Duru\"{o}z, and J.S. Harris, Jr.,
1997,  Science {\bf 278}, 1784.

\harvarditem{Stone}{1985}{Stone85}
Stone, A.D., 1985, Phys. Rev. Lett. {\bf 54}, 2692.

\harvarditem{Stone}{1995}{Stone95}
 Stone, A.D., 1995, in {\em Mesoscopic Quantum
Physics}, edited by E. Akkermans, G. Montambaux, J.-L. Pichard, and J.
Zinn-Justin (North-Holland, Amsterdam).

\harvarditem{Stone and Bruus}{1993}{Stone93}
 Stone, A.D., and H. Bruus, 1993, Physica B {\bf 189}, 43.

\harvarditem{Stone and Bruus}{1994}{Stone94}
 Stone, A.D., and H. Bruus, 1994, Surf. Sci. {\bf 305}, 490.

\harvarditem{Stone~{\em et~al.}}{1991}{Stone91}
 Stone, A.D., P.A. Mello, K.A. Muttalib, and J.-L. Pichard, 1991,
in {\em Mesoscopic Phenomena in Solids}, edited by B.L. Altshuler, P.A.
Lee,  and R.A. Webb (North-Holland, Amsterdam), p. 369.

\harvarditem{Stone and Szafer}{1988}{Stone88}
Stone, A.D., and A. Szafer, 1988, IBM J. Res. Dev. {\bf 32}, 384.

\harvarditem{Stopa}{1996}{Stopa96}
Stopa, M., 1996, Phys. Rev. B {\bf 54}, 13767.

\harvarditem{Stopa}{1998}{Stopa98}
Stopa, M., 1998, Physica B {\bf 251}, 228.

\harvarditem{Switkes~{\em et~al.}}{1999}{Switkes99}
Switkes, M., C.M. Marcus, K. Campman, and A.C. Gossard, 1999,
 Science {\bf 283}, 1905.

\harvarditem{Szafer and Altshuler}{1993}{Szafer93}
 Szafer, A., and B.L. Altshuler, 1993, Phys. Rev. Lett. {\bf 70}, 587.

\harvarditem{Taniguchi, Andreev and Altshuler}{1995}{Taniguchi95}
Taniguchi, N.,  A.V. Andreev, and B.L. Altshuler, 1995,
   Europhys. Lett. {\bf 29}, 515.

\harvarditem{Tans~{\em et~al.}}{1997}{Tans97}
Tans, S.J., M.H. Devoret, H. Dai, A. Thess, R.E. Smalley, L.J. Geerligs, and
C.  Dekker, 1997, Nature (London) {\bf 386}, 474.

\harvarditem{Tarucha~{\em et~al.}}{1996}{Tarucha96}
Tarucha, S., D.G. Austing, T. Honda, R.J. van der hage, and L.P. Kouwenhoven,
1996,  Phys. Rev. Lett. {\bf 77}, 3613.

\harvarditem{Thomas and Porter}{1956}{Thomas56}
Thomas, R.G., and C.E. Porter, 1956, Phys. Rev. {\bf 104}, 483.

\harvarditem{Thouless}{1974}{Thouless74}
Thouless, D.J., 1974, Phys. Rep. {\bf 13}, 93.

\harvarditem{Thouless}{1977}{Thouless77}
Thouless, D.J., 1977, Phys. Rev. Lett. {\bf 39}, 1167.

\harvarditem{Ullah}{1963}{Ullah63}
Ullah, N., 1963, J. Math. Phys. {\bf 4}, 1279.

\harvarditem{Ullah}{1967}{Ullah67}
Ullah, N., 1967, J. Math. Phys. {\bf 8}, 1095.

\harvarditem{Vallejos, Lewenkopf and Mucciolo}{1998}{Vallejos98}
Vallejos, R.O.,  C.H. Lewenkopf and E.R. Mucciolo, 1998,
 Phys. Rev. Lett. {\bf 81}, 677.

\harvarditem{van Houten, Beenakker and  Staring}{1992}{vanHouten92}
 van Houten, H, C.W.J. Beenakker, and A.A.M. Staring, 1992,
in {\em Single-Charge Tunneling}, edited by H. Grabert and M.H. Devoret
(Plenum, New York), p. 167.

\harvarditem{van Kampen}{1981}{vanKampen81}
van Kampen, N.G., 1981, {\em Stochastic Processes in Physics and Chemistry}
(North-Holland,  Amsterdam).

\harvarditem{van Langen, Brouwer, and Beenakker}{1997}{vanLangen97}
van Langen, S.A., P.W. Brouwer, and C.W.J.
Beenakker, 1997, Phys. Rev. E {\bf 55}, R1.

\harvarditem{Verbaarschot, Weidenm\"{u}ller, and Zirnbauer}{1985}{Verbaarschot85}
 Verbaarschot, J.J.M., H.A. Weidenm\"{u}ller, and M.R. Zirnbauer,
1985, Phys. Rep. {\bf 129}, 367.

\harvarditem{von Brentano~{\em et~al.}}{1964}{vonBrentano64}
von Brentano, P., J. Ernst, O. Hausser, T. Mayer-Kuckuk, A. Richter, and W.
von  Witsch, 1964, Phys. Lett. {\bf 9}, 48.

\harvarditem{Walker, Gefen, Montambaux}{1999}{Walker99}
Walker, P.N., Y. Gefen, and G. Montambaux, 1999, Phys. Rev. Lett. {\bf 82}, 5329.

\harvarditem{Walker, Montambaux and Gefen}{1999}{Walker99a}
Walker, P.N., G. Montambaux, and Y. Gefen, 1999, Phys. Rev. B {\bf 60}, 2541.

\harvarditem{Washburn and Webb}{1993}{Washburn93}
Washburn, S.,  and R.A. Webb, 1993, Rep. Prog. Phys. {\bf 55}, 1311.

\harvarditem{Weisskopf}{1937}{Weisskopf37}
Weisskopf, V., 1937, Phys. Rev. {\bf 52}, 295.

\harvarditem{Westervelt}{1998}{Westervelt98}
 Westervelt, R.M., 1998, in  {\em Nanotechnology},
 edited by G. Timp (AIP, New York), p.~589.

\harvarditem{Wigner}{1951}{Wigner51}
Wigner, E.P., 1951, Ann. Math. {\bf
53},  36.

\harvarditem{Wigner}{1955}{Wigner55}
Wigner, E.P., 1955, Ann. Math. {\bf 62}, 548.

\harvarditem{Wigner}{1957}{Wigner57}
Wigner, E.P., 1957, Ann. Math. {\bf 65}, 203.

\harvarditem{Wigner}{1958}{Wigner58}
Wigner, E.P., 1958, Ann. Math. {\bf 67}, 325.

\harvarditem{Wigner}{1959}{Wigner59}
Wigner, E.P., 1959, {\em Group Theory and its Applications to the Quantum
Mechanics  of Atomic Spectra} (Academic, New York).

\harvarditem{Wigner and Eisenbud}{1947}{Wigner47}
 Wigner, E.P., and L. Eisenbud, 1947, Phys. Rev. {\bf 72}, 29.

\harvarditem{Wilkinson}{1989}{Wilkinson89}
Wilkinson, M., 1989, J. Phys. A {\bf 22}, 2795.

\harvarditem{Wilkinson and Walker}{1995}{Wilkinson95}
Wilkinson, M., and P.N. Walker, 1995, J. Phys. A {\bf 28}, 6143.

\harvarditem{Yacoby~{\em et~al.}}{1995}{Yacoby95}
Yacoby, A., M. Heiblum, D. Mahalu, and H. Shtrikman, 1995, Phys. Rev. Lett. {\bf 74}, 4047.

\harvarditem{Zelevinsky~{\em et~al.}}{1996}{Zelevinsky96}
Zelevinsky, V.,  B.A. Brown, N. Frazier, and M. Horoi, 1996, Phys. Rep. {\bf
276},  85.

\harvarditem{Zhitenev~{\em et~al.}}{1997}{Zhitenev97}
Zhitenev, N.B., R.C. Ashoori, L.N. Pfeiffer, and K.W. West, 1997, Phys. Rev. Lett. {\bf 79}, 2308.

\harvarditem{Zhou, Spivak and Altshuler}{1999}{Zhou99}
Zhou, F., B. Spivak, and B.L. Altshuler, 1999, Phys. Rev. Lett. {\bf 82}, 608.

\harvarditem{Zyczkowski and  Lenz}{1991}{Zyckowski91}
Zyczkowski, K., and G. Lenz, 1991, Z. Phys. B  {\bf 82}, 299.

\end{references}
\end{document}